\PassOptionsToPackage{colorlinks=true, linkcolor=blue, anchorcolor=blue, citecolor=blue}{hyperref}
\documentclass[3p,12pt]{elsarticle}

\usepackage{amssymb}
\usepackage{amsmath}

\usepackage[utf8]{inputenc}
\usepackage{subfigure}
\usepackage{graphicx}
\usepackage{bm}
\usepackage{threeparttable}
\usepackage{CJK}
\usepackage{color}
\usepackage{booktabs}
\usepackage{bookmark}
\usepackage{array}
\usepackage{multirow}
\usepackage{natbib}
\usepackage{setspace}

\usepackage{multirow}

\usepackage{makecell}

\usepackage{xcolor}

\usepackage{caption}
\usepackage{subcaption}

\journal{Aerospace Science and Technology}

\begin{document}
\begin{CJK*}{UTF8}{gbsn}

\begin{frontmatter}

\title{A simplified unified wave-particle method for diatomic gases with rotational and vibrational non-equilibrium}

\author[address1]{Sirui Yang}
\ead{ysr1997@mail.nwpu.edu.cn}

\author[address1,address2,address3]{Chengwen Zhong}
\ead{zhongcw@nwpu.edu.cn}

\author[address1]{Ningchao Ding}
\ead{zhuocs@nwpu.edu.cn}

\author[address1]{Junzhe Cao}
\ead{caojunzhe@mail.nwpu.edu.cn}

\author[address4]{He Zhang}
\ead{zhanghecalt@163.com}

\author[address1,address2,address3]{Congshan Zhuo}
\ead{zhuocs@nwpu.edu.cn}

\author[address1,address2,address3]{Sha Liu\corref{cor1}}
\ead{shaliu@nwpu.edu.cn}

\affiliation[address1]{organization={School of Aeronautics},
            addressline={Northwestern Polytechnical University}, 
            city={Xi'an},
            state={Shaanxi},
            postcode={710072},
            country={P.R. China}}
\affiliation[address2]{organization={National Key Laboratory of Aircraft Configuration Design},
            addressline={Northwestern Polytechnical University}, 
            city={Xi'an},
            state={Shaanxi},
            postcode={710072},
            country={P.R. China}}
\affiliation[address3]{organization={Institute of Extreme Mechanics},
            addressline={Northwestern Polytechnical University}, 
            city={Xi'an},
            state={Shaanxi},
            postcode={710072},
            country={P.R. China}}
            
\affiliation[address4]{organization={Science and Technology on Space Physics Laboratory},
            city={Beijing},
            postcode={100076},
            country={P.R. China}}
            
\cortext[cor1]{Corresponding author}

\begin{abstract}
The hypersonic flow around near-space vehicles constitutes a multi-scale flow problem.  Due to insufficient molecular collisions to achieve equilibrium, rarefied gas effects are present in the flow field. Thus, numerical methods capable of accurately resolving multi-scale flows are required. Furthermore, high-temperature gas effects in hypersonic flows mean vibrational excitation of polyatomic molecules. Consequently, numerical methods accounting for non-equilibrium in rotational and vibrational internal energy modes are required. This study derives a quantified model-competition (QMC) mechanism for diatomic gases with rotational and vibrational non-equilibrium, starting from integral solutions of kinetic model equations with rotational and vibrational energy. The QMC mechanism categorize collisional and free-transport particles in cell, applying computational weighting based on their local scale regimes. We developed a simplified unified wave-particle (SUWP) method for diatomic gases based on QMC mechanism. For the macroscopic of the method, a three-temperature model accounting for rotational and vibrational energy is incorporated into both the kinetic inviscid flux scheme and {Navier-Stokes} solvers. For the microscopic of the method, a collisionless DSMC solver is employed to resolve non-equilibrium flow physics. This work validates the proposed SUWP method with rotational and vibrational non-equilibrium through benchmark cases, including shock tube, shock structures, flow past a cylinder, Apollo 6 command module and space station Mir. Compared to the DSMC and deterministic methods, the SUWP method exhibits favorable computational efficiency while maintaining accuracy.
\end{abstract}




\end{frontmatter}
\end{CJK*}

\begin{CJK*}{UTF8}{gbsn}
\section{\label{sec:introduction}Introduction}
Multi-scale hypersonic flows are a critical issue in the research and design of hypersonic vehicles in near-space and atmospheric reentry spacecraft~\cite{schmisseur2015hypersonics, schouler2020survey}. In multi-scale flows, different domains of the flow field correspond to different levels of rarefied gas effects~\cite{gijare2019effect}. The insufficient molecular collisions in rarefied flow regimes result in a non-equilibrium state. Consequently, the gas properties differ significantly from those in continuum regimes, leading to issues such as velocity slip and temperature jump on the surface of the vehicle~\cite{li2009gas}. Additionally, it also causes a non-equilibrium state between the translational and rotational energies of polyatomic gases~\cite{Rykov1978}. For hypersonic flows, the energy of high-speed gases is transferred into internal molecular energy through molecular collisions and interactions with the vehicle surface~\cite{boyd2017nonequilibrium}. This process leads to the excitation of vibrational internal energy of gas molecules, affecting surface quantities such as heat flux.  At even higher energy levels, complex phenomena such as dissociation and ionization may occur within the flow field~\cite{anderson1989hypersonic, peng2025numerical}. To simulate and analyze hypersonic vehicles operating in near-space, it is essential to develop numerical methods capable of solving multi-scale flows while accounting for the non-equilibrium behavior of rotational and vibrational energy modes.

Traditional numerical methods for flow simulation are primarily based on the Navier{-}Stokes (N{-}S) equations ~\cite{blazek2015computational}. With the aid of techniques such as slip boundary conditions, these methods are capable of simulating rarefied flows in the slip flow regime~\cite{maxwell1879vii, KIF2}. However, for more rarefied flows, such as transitional and free molecular regimes, simulations based on the N{-}S equations exhibit inaccuracies. A series of macroscopic extended methods have been developed based on the N{-}S methods, including Grad's moment method~\cite{grad1949kinetic}, the nonlinear coupled constitutive relation method~\cite{myong2004generalized, nccr1}, and others. While these methods can simulate weakly non-equilibrium rarefied flows, they remain inadequate for accurately modeling flows exhibiting stronger non-equilibrium effects. The direct simulation Monte Carlo (DSMC) is among the most crucial numerical methods for simulating rarefied flows~\cite{Bird_Molecular}. The DSMC method employs a limited number of simulated molecules to represent real molecules during collisions and transport processes, effectively capturing the evolution of rarefied flows. However, the DSMC method's decoupled transport and collision require that the cell size and time step be smaller than the mean molecular free path and collision time, respectively~\cite{bird2013dsmc}. Consequently, the computational effort of DSMC grows exponentially for near-continuum flow simulations. In summary, both the N{-}S method and the DSMC method face challenges when applied to the simulation of multi-scale flows.

Simulating multi{-}scale flows requires multi{-}scale numerical methods, and developing numerical methods necessitates corresponding multi{-}scale model equations~\cite{Liu2022_AMS, yang2021direct}. The Boltzmann equation is a fundamental equation in the gas kinetic theory, which can describe gas molecular collisions and transport processes from continuum flow to free molecular flow at the kinetic scale~\cite{chapman1990mathematical}. However, the collision term in the Boltzmann equation is a high dimensional nonlinear term that is difficult to solve~\cite{cercignani1988boltzmann}. Similar to the DSMC method, both the cell size and time step are subject to stringent restrictions~\cite{Bird_Molecular}. Therefore, direct simulation of flows based on the Boltzmann equation remains a formidable challenge. The Bhatnagar{-}Gross{-}Krook (BGK) model~\cite{bhatnagar1954model} provides an elegant simplification of the collision term in the Boltzmann equation. With the BGK model, we can efficiently simulate the rarefied flows on larger time steps and grid sizes. However, the Prandtl (Pr) number of the BGK model is an unchangeable unit value (Pr = 1), while the exact value for a monoatomic gas is {2/3}. The Shakhov-BGK model~\cite{1968Generalization} and the ES-BGK model~\cite{Holway1966New} are Prandtl-number-correct monoatomic gas multi{-}scale models developed based on the BGK model. Rykov~\cite{Rykov1978} and Holway~\cite{Holway1966New}, based on the Shakhov-BGK and ES-BGK models respectively, developed multi{-}scale models for polyatomic gases that considering rotational energy non-equilibrium. Based on the Rykov and ES-BGK models, the models considering molecular vibrational energy non-equilibrium have been developed and applied to the simulation of hypersonic flows~\cite{wang2017unified, titarev2018application, todorova2020modeling}. Zhang et al. proposed a kinetic model equation for diatomic gases including vibrational degrees of freedom based on the Rykov model~\cite{ZHANG2023107079}. The collision operator of the vibrational kinetic model consists of an elastic collision term describing translational energy relaxation and an inelastic collision term describing rotational and vibrational energy relaxation. It demonstrates good capability in simulating both rotational and vibrational energy non-equilibrium. 

Based on BGK-type models and the Fokker-Planck (FP) type models (which also possess multi{-}scale characteristics)~\cite{fokker1914mittlere}, various multi{-}scale numerical simulation methods have been developed. These methods can be categorized into stochastic particle methods, deterministic methods, and hybrid methods~\cite{Liu2022_AMS}. The stochastic particle methods, similar to DSMC, represent the evolution of gas molecules in cell based on simulated particles~\cite{Macrossan2001_nu}. Stochastic particle methods require less memory and have high computational efficiency~\cite{pfeiffer2018particle}. However, similar to the DSMC method, they also suffer from problems such as statistical noise. When calculating low-speed flows or unsteady flows, this will lead to expensive computational costs~\cite{zhang2019particle}. The main multi{-}scale stochastic particle methods include: stochastic particle-BGK method (SP-BGK)~\cite{Macrossan2001_nu}, unified stochastic particle-BGK (USP-BGK) method~\cite{Fei2020A}, stochastic particle-FP method (SP-FP)~\cite{jenny2010solution, Gorji2011_fokker}, the multi-scale stochastic particle (MSP) method, the multi{-}scale temporal discretization FP method (MTD-FPM)~\cite{fei2017particle}, etc. In addition, the asymptotic-preserving Monte Carlo (APMC) method~\cite{ren2014asymptotic}, moment-guided Monte Carlo method~\cite{degond2011moment} and the expansion of the Boltzmann equation (time relaxed Monte Carlo Methods~\cite{Pareschi2001_TRMC}) also fulfilled the asymptotic-preserving properties. Currently, multi{-}scale stochastic particle methods have been extended to multi{-}species and polyatomic gases~\cite{pirner2024consistent, kim2025stochastic}. Deterministic methods employ a discrete velocity space to represent the molecular velocity distribution function in cells. Therefore, these methods are free from statistical noise and can utilize acceleration techniques such as implicit algorithms to enhance computational efficiency~\cite{yuan2020conservative}. The main multi{-}scale deterministic methods include: the unified gas kinetic scheme (UGKS)~\cite{xu2010unified, liu2014unified, LiuA2016, Li2018A, Chen2012A}, the discrete unified gas kinetic scheme (DUGKS)~\cite{guo2013discrete, Chen2019Conserved, zhao2023numerical, ZHANG2023107079}, the gas kinetic unified algorithm (GKUA)~\cite{li2009gas, peng2016implicit, Wu2020On}, the improved discrete velocity method (IDVM)~\cite{yang2018improved}, the multi{-}scale discrete velocity method\cite{yuan_novel_2021, zhang_multiscale_2023}, and the general synthetic iterative scheme (GSIS) method~\cite{su2020fast}, etc. At the present stage, they can deal with multi{-}scale flows effectively, like micro{-}nano flows~\cite{wang2022investigation}, jet flows~\cite{chen2020compressible, zhao2024interaction} and radiative transport~\cite{sun2015asymptotic}. 

The hybrid methods mainly include the N-S/DSMC hybrid method and the wave-particle method. A representative method of the N-S/DSMC hybrid method is the modular particle-continuum (MPC) method~\cite{wang2003assessment}. These methods perform the decomposition of the flow field into N-S domains, DSMC domains, and overlapping domains, and utilize different solvers for simulation in different domains~\cite{zhang2019particle}. In the ideal case, the N-S/DSMC hybrid method can maintain reasonable simulation accuracy and computational cost in both continuum and rarefied flow regimes~\cite{hash1995hybrid}. However, the robustness of overlapping domains and the selection of continuum breakdown parameters significantly impact the stability and computational efficiency of such methods~\cite{boyd1995predicting}. The wave-particle methods combine the advantages of the deterministic and the stochastic methods~\cite{wei2024unified}. Such methods use stochastic particles rather than velocity space to characterize the non-equilibrium of flows~\cite{suwp2}. The representative of wave-particle methods is the UGKWP method proposed by Xu et al. based on the framework of the UGKS~\cite{LiuUnified2020}. The wave-particle categorization in UGKWP makes the scheme adaptively become a particle method in highly rarefied flow regime and a hydrodynamic flow solver in the continuum flow regime. In the continuum flow limit at a small cell's Knudsen number, the UGKWP gets back to the gas kinetic scheme (GKS) for the N{-}S solution~\cite{xu2001gas}. Thus, the UGKWP method could achieve high efficiency both in the continuum and rarefied regimes~\cite{wei2024unified}. Currently, the UGKWP method has been implemented in radiation transport~\cite{liu2023implicit}, plasma simulation~\cite{PU2025113918}, and phonon transport~\cite{liu2025unified}. In the UGKWP method, molecules are categorized into three types: \uppercase\expandafter{\romannumeral1}. free transport, \uppercase\expandafter{\romannumeral2}. colliding before leaving its origin cell, \uppercase\expandafter{\romannumeral3}. colliding after leaving its origin cell~\cite{ChenY2020A}. A theoretical analysis of this time integral solution was conducted by Liu et al. and a quantified model-competition (QMC) mechanism was found~\cite{Liu2020Simplified}. This mechanism provides a rational and physically meaningful hybrid between continuum and rarefied flow models within the context of multi{-}scale flow. Based on the QMC mechanism, the simplified unified wave-particle (SUWP) method is proposed~\cite{Liu2020Simplified}. The SUWP method classifies the flow molecules into two categories: colliding molecules and free transport molecules, simulated by the N-S solver and the particle solver respectively. Therefore, the form of SUWP method is much simpler than UGKWP.

In the previous works, the SUWP method considering rotational non-equilibrium of diatomic gas has been developed based on the Rykov model~\cite{suwp2}. However, the SUWP-Rykov method cannot effectively simulate the excitation of vibrational degrees of freedom in high-speed gases. In this study, we propose the SUWP method considering rotational and vibrational non-equilibrium of diatomic gases. The vibrational kinetic model developed by Zhang et al. is used to describe the relaxation process from the non-equilibrium state to the equilibrium state. The three equilibrium state distribution functions in the model take into account elastic and inelastic collisions, as well as energy exchanges between translational, rotational, and vibrational degrees of freedom. The QMC mechanism for diatomic gases is derived based on the vibrational kinetic model. It is proved that colliding particles can be described by N-S equations. For the N-S solver, the translational-rotational-vibrational three-temperature N-S equations based on the 2nd order Chapman-Enskog (C-E) expansion of the vibrational kinetic model are obtained, and the kinetic inviscid flux (KIF) scheme~\cite{KIF1, KIF2} for three-temperature N-S equations are constructed. For the particle solver, sample particles from the distribution function of the vibrational kinetic model equation~\cite{Liu1996Metropolized}, and a suitable particle transport tracking technique is also implemented within the finite volume method framework. Based on the above work, the SUWP method for diatomic gases is established.

The rest of this paper is organized as follows. The kinetic model equation for diatomic gases involving rotational and vibrational degrees of freedom is introduced in Section \ref{sec:model}. In Section \ref{sec:method}, the basic algorithm of SUWP and QMC mechanism for diatomic molecules is described in detail. A series of numerical test cases are performed and discussed to validate the proposed method in Section \ref{sec:Test}. Finally, the concluding remarks are given in Section \ref{sec:conclusion}. The C-E expansion of the vibrational kinetic model is illustrated in \ref{sec:CEex}.

\section{\label{sec:model}Kinetic model for diatomic gas}
\subsection{Distribution function and moments}
In this paper, we consider the kinetic description of molecular gas flows, wherein each molecule has translational, rotational, and vibrational degrees of freedom. The state of the gas is described by the molecular number density distribution function $f\left(\boldsymbol{x}, \boldsymbol{u}, \eta_{\rm{rot}}, \eta_{\rm{vib}}, t \right)$,  where $\boldsymbol{x}$ and $\boldsymbol{u}$ are physical space position and particle velocity, respectively. The continuous variables $\eta_{\rm{rot}}$ and $\eta_{\rm{vib}}$ are molecular rotational and vibrational energies, respectively, and $t$ is the time.
The macroscopic flow variable $\boldsymbol{W}=\left(\rho, \rho \boldsymbol{U}, \rho E, \rho E_{\rm{rot}}, \rho E_{\rm{vib}}\right)^{\mathrm{T}}$ is defined as the moment of $f$ in the phase space ${\rm{d}} \boldsymbol{\Xi}= {\rm{d}}\boldsymbol{u} {\rm{d}}\eta_{\rm{rot}} {\rm{d}}\eta_{\rm{vib}}$,
\begin{equation}  \label{eq:realMacroVar}
\boldsymbol{W}=\int \boldsymbol{\psi} f\left(\boldsymbol{x}, \boldsymbol{u}, \eta_{\rm{r o t}}, \eta_{\rm{v i b}}, t\right) {\rm{d}} \boldsymbol{\Xi}, 
\end{equation} 
where $\boldsymbol{\psi}=\left(m, m \boldsymbol{u} , \frac{1}{2} m\left(|\boldsymbol{u}|^2 \right)+\eta_{\rm{r o t}}+\eta_{\rm{v i b}}, \eta_{\rm{r o t}}, \eta_{\rm{v i b}}\right)^{\mathrm{T}}$ is the moment vector. $m$  is the molecular mass, $\rho$ is the density, and $\boldsymbol{u}=(U, V, W)$ is the velocity of fluid. At the macroscopic level, the total energy $\rho E$ comprises the kinetic energy of the fluid and the molecular translational, rotational, and vibrational energies,
\begin{equation}  \label{eq:totalEnergy}
\rho E=\frac{1}{2} \rho|\boldsymbol{U}|^2+\rho E_{\rm{t r}}+\rho E_{\rm{r o t}}+\rho E_{\rm{v i b}}. 
\end{equation} 
The translational, rotational and vibrational temperatures $T_{\rm{tr}}$, $T_{\rm{rot}}$ and $T_{\rm{vib}}$ are defined assuming that their energies have the same distributions associated to their degrees of freedom $K_{\rm{tr}}$, $K_{\rm{rot}}$ and $K_{\rm{vib}}$, respectively,
\begin{equation}  \label{eq:freeDegree_Energy}
E_{\rm{t r}}=\frac{K_{\rm{t r}}}{2} R T_{\rm{t r}}, \quad E_{\rm{r o t}}=\frac{K_{\rm{r o t}}}{2} R T_{\rm{r o t}}, \quad E_{\rm{v i b}}=\frac{K_{\rm{v i b}}}{2} R T_{\rm{v i b}}.
\end{equation} 
Here $R$ is the specific gas constant. There are three translational degrees of freedom and two rotational degrees of freedom for the diatomic molecule. According to the harmonic oscillator model, the vibrational degrees of freedom at temperature $T$ can be determined by the following formula~\cite{Bird_Molecular},
\begin{equation}  \label{eq:freeDegree_Vib}
K_{\rm{vib}}(T)=\frac{2 \Theta_{\rm{v i b}} / T}{e^{\Theta_{\rm{v i b}} / T}-1}
\end{equation} 
where $\Theta_{\rm{v i b}}$ is the vibrational characteristic temperature. The temperature $T_{\rm{eq}}$ which correspond to equilibrium between the
translational, rotational and vibrational energy exchanges is defined as:
\begin{equation}  \label{eq:equilibrium_T}
T_{\rm{eq}}=\frac{K_{\rm{t r}} T_{\rm{t r}}+K_{\rm{r o t}} T_{\rm{r o t}}+K_{\rm{v i b}}(T_{\rm{v i b}}) T_{\rm{v i b}}}{K_{\rm{t r}}+K_{\rm{r o t}}+K_{\rm{vib}}(T_{\rm{eq}})}.
\end{equation} 
Similarly, the joint translational$-$rotational equilibrium temperature $T_{\rm{tr,rot}}$ is defined as:
\begin{equation}  \label{eq:trAndRot_T}
T_{\rm{tr,rot}}=\frac{K_{\rm{t r}} T_{\rm{t r}}+K_{\rm{r o t}} T_{\rm{r o t}}}{K_{\rm{t r}}+K_{\rm{r o t}}}.
\end{equation} 

The corresponding equilibrium pressures is $p=\rho R T_{\rm{eq}}$ and the pressure of translational motion is $p_{\rm{tr}}=\rho R T_{\rm{tr}}$. The stress tension $\mathbf{P}$ and
the heat flux $\boldsymbol{q}$ are also calculated by the moment of $f$,
\begin{equation}  \label{eq:matrixP}
 \mathbf{P}=\int \boldsymbol{c} \otimes \boldsymbol{c} \, m f\left(\boldsymbol{x}, \boldsymbol{u},  \eta_{\rm{r o t}}, \eta_{\rm{v i b}}, t\right) {\rm{d}} \boldsymbol{\Xi},
\end{equation} 
\begin{equation}  \label{eq:vectorQ}
\boldsymbol{q}=\int \boldsymbol{c}\left(m \frac{|\boldsymbol{c}|^2}{2}+\eta_{\rm{r o t}}+\eta_{\rm{v i b}}\right) f\left(\boldsymbol{x}, \boldsymbol{u},  \eta_{\rm{r o t}}, \eta_{\rm{v i b}}, t\right) \textbf{d} \boldsymbol{\Xi},
\end{equation} 
where $\boldsymbol{c}=\boldsymbol{u}-\boldsymbol{U}$ is the peculiar velocity. In particular, the translational, rotational, and vibrational heat fluxes $\boldsymbol{q}_{\rm{t r}}$, $\boldsymbol{q}_{\rm{r o t}}$, $\boldsymbol{q}_{\rm{v i b}}$ are calculated respectively as:
\begin{equation}  \label{eq:vectorQtr}
\boldsymbol{q}_{\rm{t r}}=\int \boldsymbol{c} \, \frac{|\boldsymbol{c}|^2}{2} m f\left(\boldsymbol{x}, \boldsymbol{u}, \eta_{\rm{r o t}}, \eta_{\rm{v i b}}, t\right) {\rm{d}} \boldsymbol{\Xi},
\end{equation} 
\begin{equation}  \label{eq:vectorQrot}
\boldsymbol{q}_{\rm{r o t}}=\int \boldsymbol{c} \, \eta_{\rm{r o t}} f\left(\boldsymbol{x}, \boldsymbol{u}, \eta_{\rm{r o t}}, \eta_{\rm{v i b}}, t\right) {\rm{d}} \boldsymbol{\Xi},
\end{equation} 
\begin{equation}  \label{eq:vectorQvib}
\boldsymbol{q}_{\rm{v i b}}=\int \boldsymbol{c} \, \eta_{\rm{v i b}} f\left(\boldsymbol{x}, \boldsymbol{u}, \eta_{\rm{r o t}}, \eta_{\rm{v i b}}, t\right) {\rm{d}} \boldsymbol{\Xi}.
\end{equation} 

\subsection{Kinetic model with rotational and vibrational non-equilibrium}
The dynamics of the number density distribution function $f$ is described by the phenomenological Boltzmann model equation~\cite{ZHANG2023107079}, which takes into account different time scales in the translational, rotational and vibrational relaxations,
\begin{equation}  \label{eq:dfuncVib}
\frac{\partial f}{\partial t}+\boldsymbol{u} \cdot \frac{\partial f}{\partial \boldsymbol{x}}=\frac{f^{\rm{t r}}-f}{\tau}+\frac{f^{\rm{r o t}}-f^{\rm{t r}}}{Z_{\rm{r o t}} \tau}+\frac{f^{\rm{v i b}}-f^{\rm{t r}}}{Z_{\rm{v i b}} \tau} = \frac{g-f}{\tau}.
\end{equation} 
where $g$ is total equilibrium states distribution function. The elastic collision process of molecular translation and the inelastic collision process of internal energy relaxation are described by the right-hand side of Eq.(\ref{eq:dfuncVib}) with three equilibrium states $f^{\rm{tr}}$, $f^{\rm{rot}}$ and $f^{\rm{vib}}$ volving different temperatures,
\begin{equation}  \label{eq:thefuncTr}
\begin{aligned} 
f^{\rm{t r}} & =n\left(\frac{1}{2 \pi R T\rm{_{t r}}}\right)^{\frac{3}{2}} \exp \left(-\frac{|\boldsymbol{c}|^2 }{2 R T\rm{_{t r}}}\right) \frac{1}{m R T\rm{_{r o t}}} \exp \left(-\frac{\eta\rm{_{r o t}}}{m R T\rm{_{r o t}}}\right) \Re\left(T\rm{_{v i b}}\right) \\ 
& \times\left\{1+\frac{\boldsymbol{c} \cdot \boldsymbol{q}{\rm{_{t r}}}}{15 R T{\rm{_{t r}}} p\rm{_{t r}}}\left[\frac{\left(|\boldsymbol{c}|^2 \right)}{R T\rm{_{t r}}}-5\right]+(1-\delta) \frac{\boldsymbol{c} \cdot \boldsymbol{q}\rm{_{r o t}}}{R T{\rm{_{t r}}} p\rm{_{r o t}}}\left(\frac{\eta\rm{_{r o t}}}{m R T\rm{_{r o t}}}-1\right)\right\},  
\end{aligned}
\end{equation} 
\begin{equation}  \label{eq:thefuncRot}
\begin{aligned} 
f^{\rm{r o t}} & =n\left(\frac{1}{2 \pi R T_{\rm{tr,rot}}}\right)^{\frac{3}{2}} \exp \left(-\frac{|\boldsymbol{c}|^2 }{2 R T_{\rm{tr,rot}}}\right) \frac{1}{m R T_{\rm{tr,rot}}} \exp \left(-\frac{\eta\rm{_{r o t}}}{m R T_{\rm{tr,rot}}}\right) \Re\left(T\rm{_{v i b}}\right) \\ 
& \times\left\{1+\omega_0 \frac{\boldsymbol{c} \cdot \boldsymbol{q}\rm{_{t r}}}{15 R T_{\rm{tr,rot}} p_{\rm{tr,rot}}}\left[\frac{\left(|\boldsymbol{c}|^2 \right)}{R T_{\rm{tr,rot}}}-5\right]+\omega_1(1-\delta) \frac{\boldsymbol{c} \cdot \boldsymbol{q}\rm{_{r o t}}}{R T_{\rm{tr,rot}} p_{\rm{tr,rot}}}\left(\frac{\eta\rm{_{r o t}}}{m R T_{\rm{tr,rot}}}-1\right)\right\},
\end{aligned}
\end{equation} 
\begin{equation}  \label{eq:thefuncVib}
\begin{aligned} 
f^{\rm{v i b}} & =n\left(\frac{1}{2 \pi R T_{\rm{eq}}}\right)^{\frac{3}{2}} \exp \left(-\frac{|\boldsymbol{c}|^2 }{2 R T_{\rm{eq}}}\right) \frac{1}{m R T_{\rm{eq}}} \exp \left(-\frac{\eta\rm{_{r o t}}}{m R T_{\rm{eq}}}\right) \Re(T_{\rm{eq}}) \\ & \times\left\{1+\omega_2 \frac{\boldsymbol{c} \cdot \boldsymbol{q}\rm{_{t r}}}{15 R T_{\rm{eq}} p}\left[\frac{\left(|\boldsymbol{c}|^2 \right)}{R T_{\rm{eq}}}-5\right]+\omega_3(1-\delta) \frac{\boldsymbol{c} \cdot \boldsymbol{q}\rm{_{r o t}}}{R T_{\rm{eq}} p}\left(\frac{\eta\rm{_{r o t}}}{m R T_{\rm{eq}}}-1\right)\right\},
\end{aligned}
\end{equation} 
with
\begin{equation}  \label{eq:theReOfTheFunc}
\Re(T)=\frac{\eta{\rm{_{v i b}}} ^{K{{\rm{_{v i b}}}(T)} / 2-1}(m R T)^{-K{{\rm{_{v i b}}}(T)} / 2}}{\Gamma\left(K{{\rm{_{v i b}}}(T)} / 2\right)} \exp \left\{-\frac{\eta{\rm{_{v i b}}}}{m R T}\right\},
\end{equation} 
where $\Gamma$ is the gamma function, and $n$ is the molecular number density. In Eq.(\ref{eq:dfuncVib}), $\tau$ is the characteristic relaxation time determined by the dynamic viscosity $\mu$ and translational pressure $p_{\rm{tr}}$ with $\tau=\mu/p_{\rm{tr}}$ . The dynamic viscosity µ is related to the inter-molecular interactions. For variable hard-sphere (VHS)
molecules, the dynamic viscosity is
\begin{equation}  \label{eq:muEqn}
\mu=\mu_{\rm{ref}}\left(\frac{T_{\rm{tr}}}{T_{\rm{ref}}} \right)^{\omega},
\end{equation}
where $\omega$ is the viscosity index. $\mu_{\rm{ref}}$ is the reference viscosity at the reference temperature $T_{\rm{ref}}$. The relationship between the mean free path $\lambda$ and the dynamic viscosity $\mu$ for VHS molecules is
\begin{equation}   \label{eq:knNumber}
\lambda=\frac{2\mu(7-2\omega)(5-2\omega)}{15\rho(2\pi R T_{\rm{tr}})^{1/2}}.
\end{equation} 
And the define of Knudsen number is 
\begin{equation}   \label{eq:shock_lambda1}
{\rm{Kn}}=\frac{\lambda}{L_{\rm{ref}}}.
\end{equation} 
where $L_{\rm{ref}}$ is characteristic length of flows.
The $Z_{\rm{rot}}$ and $Z_{\rm{vib}}$ are rotational and vibrational collision numbers, respectively. The rotational collision number $Z_{\rm{rot}}$ can be introduced from the DSMC as follows~\cite{wu2021derivation}
\begin{equation}   \label{eq:Z_ROT_EQN}
Z_{\rm{rot}}=\frac{\rm{Pr}}{30}(7-2\omega)(5-2\omega)\frac{K_{\rm{tr}}}{K_{\rm{tr}}+K_{\rm{rot}}}Z_{R},
\end{equation} 
where $\rm{Pr}$ is Prandtl number, and the value of $Z_{R}$ can be determined by approximating the theoretical formulas and comparing with the experimental data~\cite{parker1959rotational, boyd1990rotational}. The Parker formula~\cite{parker1959rotational} is adopted as follows
\begin{equation}   \label{eq:Z_R_wu_EQN}
Z_R=\frac{Z_R^{\infty}}{1+\left(\pi^{3 / 2} / 2\right) \sqrt{T^* / T_{\rm{t r}}}+\left(\pi^2 / 4+\pi\right)\left(T^* / T_{\rm{t r}}\right)},
\end{equation} 
In the present work, the $Z_R^{\infty}$=15.7 and $T^*$=80.0K~\cite{ZHANG2023107079}.

In the spatial homogeneous case, the relaxation rate of the heat fluxes are~\cite{ZHANG2023107079}
\begin{equation}   \label{eq:relaxRate_Qtr}
\frac{\partial \boldsymbol{q}_{\rm{t r}}}{\partial t}=-\left[\frac{2}{3}+\frac{1}{3 Z_{\rm{r o t}}}\left(1-\omega_0\right)+\frac{1}{3 Z_{\rm{v i b}}}\left(1-\omega_2\right)\right] \frac{1}{\tau} \boldsymbol{q}_{\rm{t r}}, 
\end{equation} 
\begin{equation}   \label{eq:relaxRate_Qrot}
\frac{\partial \boldsymbol{q}_{\rm{r o t}}}{\partial t}=-\left[\delta+\frac{1}{Z_{\rm{r o t}}}\left(1-\omega_1\right)(1-\delta)+\frac{1}{Z_{\rm{v i b}}}\left(1-\omega_3\right)(1-\delta)\right] \frac{1}{\tau} \boldsymbol{q}_{\rm{r o t}},
\end{equation} 
\begin{equation}   \label{eq:relaxRate_Qvib}
\frac{\partial \boldsymbol{q}_{\rm{v i b}}}{\partial t}=-\frac{1}{\tau} \boldsymbol{q}_{\rm{v i b}} .
\end{equation} 
The values of the parameters $\delta$, $\omega_0$, $\omega_1$, $\omega_2$ and $\omega_3$ are chosen to achieve proper relaxation of heat fluxes. For nitrogen gas, $\delta$=1/1.55, $\omega_0$=0.2354, $\omega_1$=0.2354, $\omega_2$=0.3049, $\omega_3$=0.2354~\cite{ZHANG2023107079}.

According to the moments of distribution function $f$ and the definition of the equilibrium distribution functions, it is easy to verify that the collision operator satisfy the following formulas:
\begin{equation}   \label{eq:macroOfZhang}
\int \boldsymbol{\psi} \frac{ g - f }{\tau} {\rm{d}} \boldsymbol{\Xi} = \left(
\begin{array}{ccccc}
0 \\
\boldsymbol{0} \\
0 \\
\frac{1}{Z_{\rm{rot}}\tau}\left( \rho R T_{\rm{tr,rot}} - \rho R E_{\rm{rot}}\right) + \frac{1}{Z_{\rm{vib}}\tau}\left( \rho R T_{\rm{eq}} - \rho R E_{\rm{rot}}\right)  \\
\frac{1}{Z_{\rm{vib}}\tau}  \left( \frac{K_{\rm{vib}}(T_{\rm{eq}})}{2}\rho R T_{\rm{eq}}- \rho R E_{\rm{vib}} \right)  
\end{array}
\right).
\end{equation} 

\section{\label{sec:method}Simplified unified wave-particle method}

\subsection{Quantified model-competition mechanism and hybrid framework}
The SUWP method classifies gas molecules into two categories: free transport molecules (Type-F) and colliding molecules (Type-C). The Type-F molecules are freely transported without colliding with other molecules, and are simulated by the particle solver. The Type-C molecules are colliding with other molecules, and simulated by the N-S solver. The QMC mechanism is a effective mechanism for determining the ratio of both types of molecules~\cite{Liu2020Simplified}. The derivation of QMC mechanism is as follows.
The time integral solution for BGK-type equations is
\begin{equation}  \label{eq:timeIntegralSolution}
\begin{aligned}
f(\boldsymbol{x}, \boldsymbol{u}, \eta_{\rm{rot}}, \eta_{\rm{vib}}, t)&={{e}^{-\frac{\Delta t}{\tau }}}f(\boldsymbol{x}_{0}, \boldsymbol{u}, \eta_{\rm{rot}}, \eta_{\rm{vib}}, 0)\\
&+\frac{1}{\tau }\int_{0}^{\Delta t}{g\left( \boldsymbol{x}', \boldsymbol{u}, \eta_{\rm{rot}}, \eta_{\rm{vib}}, t' \right){{e}^{\frac{t'}{\tau }}}{\rm{d}}t'},\text{ }
\end{aligned}
\end{equation} 
where the $\boldsymbol{x}_{0}$ is the coordinate of particle at time $0$. $\Delta t$ is the time step. The initial velocity of a group of molecules with velocity $\boldsymbol{u}$ is represented by the velocity distribution function $f({{\boldsymbol{x}}_{0}},\boldsymbol{u}, \eta_{\rm{rot}}, \eta_{\rm{vib}}, 0)$, where the subscript $0$ denotes the initial time. $\boldsymbol{x}{\,}' = {{\boldsymbol{x}}_{0}} + \boldsymbol{u}t'$ is the trace of molecules, with $t'$ ranging from $0$ to $\Delta t$. 

In the first term on the right hand side of Eq.(\ref{eq:timeIntegralSolution}), ${{e}^{-\frac{\Delta t}{\tau }}}$ represents the ratio of Type-F molecules, but maintain their velocity distribution from the initial time $f({{\boldsymbol{x}}_{0}},\boldsymbol{u}, \eta_{\rm{rot}}, \eta_{\rm{vib}}, 0)$, and move along the path defined by ${{\boldsymbol{x}}_{0}}+\boldsymbol{u}\Delta t$. Correspondingly, $1-{{e}^{-\frac{\Delta t}{\tau }}}$ represents the ratio of Type-C molecules. Collision changes their velocities and internal energies, and then follows the distribution $g\left( \boldsymbol{x}',\boldsymbol{u}, \eta_{\rm{rot}}, \eta_{\rm{vib}}, t' \right)$. By using the Taylor expansion to expand $g\left( \boldsymbol{x}',\boldsymbol{u}, \eta_{\rm{rot}}, \eta_{\rm{vib}},t' \right)$ to the second order in both space and time, then calculating the integral, the second term of the right hand side is denoted by $h(\boldsymbol{x},\boldsymbol{u}, \eta_{\rm{rot}}, \eta_{\rm{vib}}, \Delta t)$,
\begin{equation}  \label{eq:timeIntegralTerm}
h(\boldsymbol{x},\boldsymbol{u}, \eta_{\rm{rot}}, \eta_{\rm{vib}},\Delta t)=\left\{ 
\begin{aligned}
\left( 1-{{e}^{-\frac{\Delta t}{\tau }}} \right)\left[ g-\tau \left( \boldsymbol{u}\cdot \frac{\partial g}{\partial \boldsymbol{x}}+\frac{\partial g}{\partial t} \right) \right] \\
+{{e}^{-\frac{\Delta t}{\tau }}}\Delta t\left( \boldsymbol{u}\cdot \frac{\partial g}{\partial \boldsymbol{x}}+\frac{\partial g}{\partial t} \right) \\
+\left( \Delta t-{{e}^{-\frac{\Delta t}{\tau }}}\Delta t \right)\frac{\partial g}{\partial t} 
\end{aligned}
\right\}
_{(\boldsymbol{x},\boldsymbol{u}, \eta_{\rm{rot}}, \eta_{\rm{vib}} ,0)}.\text{ }
\end{equation}

Dropping the high-order term, Eq.(\ref{eq:timeIntegralTerm}) can be expressed as:
\begin{equation}  \label{eq:hydroEqn01}
h(\boldsymbol{x},\boldsymbol{u}, \eta_{\rm{rot}}, \eta_{\rm{vib}},\Delta t)=\left( 
1-{{e}^{-\frac{\Delta t}{\tau }}} \right){{\left\{
\begin{aligned}
 \left[ g -\tau \left( \boldsymbol{u}\cdot \frac{\partial g}{\partial \boldsymbol{x}}+\frac{\partial g}{\partial t} \right) \right] \\
 + \frac{\Delta t{{e}^{-\frac{\Delta t}{\tau }}}}{1-{{e}^{-\frac{\Delta t}{\tau }}}}\left( \boldsymbol{u}\cdot \frac{\partial g}{\partial \boldsymbol{x}}+\frac{\partial g}{\partial t} \right)
\end{aligned}
 \right\}}_{(\boldsymbol{x},\boldsymbol{u}, \eta_{\rm{rot}}, \eta_{\rm{vib}},0)}}.
\end{equation}
The the first term on the right hand side of Eq.(\ref{eq:hydroEqn01}) corresponds to the second-order C-E expansion of the total equilibrium states distribution function, and the second term is a scale-dependent anti-dissipative term. The first term in square brackets corresponds to the convective term of the N-S equation(see the \ref{sec:CEex}), and the second term correspondes to the dissipation term. In the N-S solver, the convective term and dissipation term are corresponded to the inviscid and viscous flux, respectively. The anti-dissipative term has the same form as the dissipative term, and it can be solved by the viscous flux with an appropriate viscous weight. Therefore, we express Eq.(\ref{eq:hydroEqn01}) in the following form
\begin{equation}  \label{eq:newTypeHeqn}
h(\boldsymbol{x},\boldsymbol{u}, \eta_{\rm{rot}}, \eta_{\rm{vib}},\Delta t)=\left( 
1-{{e}^{-\frac{\Delta t}{\tau }}} \right){{\left\{
\begin{aligned}
g -\tau \left( 1 - \left( \frac{\Delta t}{\tau}\right)\frac{{{e}^{-\frac{\Delta t}{\tau }}}}{1-{{e}^{-\frac{\Delta t}{\tau }}}} \right)\left( \boldsymbol{u}\cdot \frac{\partial g}{\partial \boldsymbol{x}}+\frac{\partial g}{\partial t} \right)
\end{aligned}
 \right\}}_{(\boldsymbol{x},\boldsymbol{u}, \eta_{\rm{rot}}, \eta_{\rm{vib}},0)}},
\end{equation}
where the scale-dependent viscous coefficient $c_{\rm{vis}}$ is defined as
\begin{equation}  \label{eq:coeffVis}
{{c}_{\rm{vis}}}=1-\left( \frac{\Delta t}{\tau } \right)\frac{{{e}^{-\frac{\Delta t}{\tau }}}}{1-{{e}^{-\frac{\Delta t}{\tau }}}}.
\end{equation} 

The Eq.(\ref{eq:timeIntegralSolution}) can be expressed as:
\begin{equation}  \label{eq:fOfTheTimestep}
\begin{aligned}
f(\boldsymbol{x},\boldsymbol{u},\eta_{\rm{rot}}, \eta_{\rm{vib}},\Delta t)&={{e}^{-\frac{\Delta t}{\tau }}}f(\boldsymbol{x},\boldsymbol{u},\eta_{\rm{rot}}, \eta_{\rm{vib}},0) \\
&+\left( 1-{{e}^{-\frac{\Delta t}{\tau }}} \right) {{\left[g -\tau  {c}_{\rm{vis}}   \left( \boldsymbol{u}\cdot \frac{\partial g}{\partial \boldsymbol{x}}+\frac{\partial g}{\partial t} \right) \right]}_{(\boldsymbol{x},\boldsymbol{u}, \eta_{\rm{rot}}, \eta_{\rm{vib}},0)}},
\end{aligned}
\end{equation} 
where the terms in the square bracket is consistent with the N-S equation, so its flux is calculated by a three-temperature N-S solver.
The macroscopic flux can be calculated from $\int \left(\boldsymbol{u} \cdot \boldsymbol{n}\right) \boldsymbol{\psi} f d \boldsymbol{\Xi}$, where $\boldsymbol{n}$ is the normal direction of a cell interface, $f$ is defined at the central point of the interface. Therefore, set $\boldsymbol{x}$ the central point of the interface whose normal direction is $\boldsymbol{n}$, the macroscopic flux $\mathbf{F}_{\rm{hydro}}$ caused by the hydrodynamic molecules is expressed as follows:
\begin{equation}  \label{eq:nsFlux}
\begin{aligned} \mathbf{F}_{\text {hydro }}= & \int \left(\boldsymbol{u} \cdot \boldsymbol{n}\right) \boldsymbol{\psi} h {\rm{d}} \boldsymbol{\Xi} \\ 
= & \left(1-e^{-\frac{\Delta t}{\tau}}\right)
{\left\{
{\int \left(\boldsymbol{u} \cdot \boldsymbol{n}\right) \boldsymbol{\psi} h {\rm{d}} \boldsymbol{\Xi} + c_{\text {vis }} \int \left(\boldsymbol{u} \cdot \boldsymbol{n}\right) \boldsymbol{\psi} \tau  \left(\boldsymbol{\xi} \cdot \frac{\partial g}{\partial \mathbf{x}}+\frac{\partial g}{\partial t}\right)  {\rm{d}} \boldsymbol{\Xi}  }
\right\}}.
\end{aligned}
\end{equation} 
Since the two integrals in Eq.(\ref{eq:nsFlux}) are the inviscid flux and viscous flux of the N-S equation. $\mathbf{F}_{\text {hydro }}$ can be finally written as
\begin{equation}  \label{eq:nsFlux2}
\begin{aligned} \mathbf{F}_{\text {hydro }}= \left(1-e^{-\frac{\Delta t}{\tau}}\right)
{\left\{
\mathbf{F}_{\text {NS, inv }} + c_{\text {vis }}  \mathbf{F}_{\text {NS, vis }}
\right\}}.
\end{aligned}
\end{equation} 
Here, “inv” stands for inviscid. Since $\left(1-e^{-\frac{\Delta t}{\tau}}\right)$ is the portion of the hydrodynamic molecules, Eq. (\ref{eq:nsFlux2}) means that the flux caused by Type-C molecules is in the form of N-S flux except a scale dependent coefficient $c_{\rm{vis}}$ is multiplied to the viscous flux. The weight of the rarefied model and the weight of the continuum model can be defined as follows, which are actually the proportions of free-transport and hydrodynamic molecules, respectively:
\begin{equation}  \label{eq:weightOfCAndF}
{{w}_{\rm{free}}}={{e}^{-\frac{\Delta t}{\tau }}},\quad {{w}_{\rm{hydro}}}=1-{{e}^{-\frac{\Delta t}{\tau }}}.
\end{equation} 

In order to describe the hybrid framework of SUWP method for diatomic gases with rotational and vibrational non-equilibrium (SUWP-vib), the mass, momentum, energy, rotational energy and vibrational energy in cell are defined as $\boldsymbol{Q}$:
\begin{equation}  \label{eq:consVar}
\boldsymbol{Q}=\boldsymbol{W} \Omega,
\end{equation} 
where $\Omega $ is the cell volume. And the total variables in the cell are denoted as ${{\boldsymbol{Q}}_{\rm{total}}}$, and its evolution equation can be written as:
\begin{equation}  \label{eq:macroVar}
\boldsymbol{Q}{{_{\rm{total}}^{n+1}}}=\boldsymbol{Q}{_{\rm{total}\text{ }}^{n}}-\Delta t \sum^{k_{\rm{max}}}_{k}{{{{\mathbf{F}}}{\rm_{hydro\text{ }}}}A_{{k}} +\boldsymbol{Q}{_{\rm{micro}}^{n}} +\boldsymbol{S},}
\end{equation} 
where $n$ denotes the present time step. The $A$ is the area of the cell interface. The $k$ is the index of cell interface, $k_{\rm{max}}$ is the number of interface in this cell. Without loss of generality, the normal directions of all interfaces are pointing outside. The composition of $\mathbf{F}_{\rm{hydro}}$ has been described previously. The flux of the macroscopic solver will be described in Sec. \ref{subsec:macroSolver}. 

The contribution of the Type-F molecules transport through the cell interfaces is denoted by $\boldsymbol{Q}_{\rm{micro}}^{n}$: 
\begin{equation}  \label{eq:Qmicro}
\boldsymbol{Q}_{\rm{micro}}^{n} = \boldsymbol{Q}{_{\rm{free}}^{n+\delta}} - \boldsymbol{Q}{_{\rm{free}}^{n}},
\end{equation} 
where $\boldsymbol{Q}{_{\rm{free}}^{n}}$ is the macroscopic variables of free-transport molecules at the present time step after the categorization. $\boldsymbol{Q}{_{\rm{free}}^{n+\delta}}$ free is the molecules belong to this cell at the end of the present time step. The particle transport and reconstruction processes will be described in Sec. \ref{subsec:particleSolver}. 

In the Eq.(\ref{eq:macroVar}), $\boldsymbol{S}$ is source term caused by internal energy relaxation. During the collision process, inelastic collisions will happen, which lead to energy exchange between the degrees of freedom of molecular translation, rotation and vibration. As a result, source terms appear in the macroscopic governing equations,
\begin{equation}  \label{eq:sourceTermInt}
\boldsymbol{S}= \int^{t^{n+1}}_{t^{n}}\frac{g-f}{\tau}\boldsymbol{\psi}{\rm{d}}\boldsymbol{\Xi}{\rm{d}}t=\int^{t^{n+1}}_{t^{n}}\boldsymbol{s}{\rm{d}}t,
\end{equation} 
where $\boldsymbol{s}$ is
\begin{equation}  \label{eq:sourceTerm}
\boldsymbol{s}= \left(0, \boldsymbol{0} ,0, {s}_{\rm{rot}}, {s}_{\rm{vib}} \right)^{\mathrm{T}} = \left(0, \boldsymbol{0} ,0, 
\frac{ \rho E_{\rm{rot}}^{\rm{aim,rot}} -\rho E_{\rm{rot}}}{Z_{\rm{rot}} \tau}+\frac{\rho E_{\rm{rot}}^{\rm{aim,vib}} -\rho E_{\rm{rot}}}{Z_{\rm{vib}} \tau}, 
\frac{\rho E_{\rm{vib}}^{\rm{aim}}-\rho E_{\rm{vib}}}{Z_{\rm{vib}} \tau}
\right)^{\mathrm{T}},
\end{equation} 
where $\rho E_{\rm{rot}}^{\rm{aim,rot}}= \frac{K_{\rm{rot}}}{2} \rho R T_{\rm{tr, rot}}$ is the target rotational energy of the molecules which relax to equilibrium states $f^{\rm{rot}}$, and $\rho E_{\rm{rot}}^{\rm{aim,vib}} = \frac{K_{\rm{rot}}}{2} \rho R T_{\rm{eq}}$ is the target rotational energy of the molecules which relax to equilibrium states $f^{\rm{vib}}$. The $\rho E_{\rm{vib}}^{\rm{aim}} = \frac{K_{\rm{vib}}(T_{\rm{eq}})}{2} \rho R T_{\rm{eq}}$ is the target vibrational energy of the molecules which relax to equilibrium states $f^{\rm{vib}}$.
With consideration of numerical stability, the source term is usually treated in an implicit way, such as the trapezoidal rule for rotational and vibrational energies
\begin{equation}  \label{eq:sourceTermAll}
\begin{aligned} 
S_{\rm{rot}} & =\frac{\Delta t}{2}
\left[  
\frac{(\rho E_{\rm{rot}}^{\rm{aim,rot}})^n-(\rho E_{\rm{rot}})^n}{Z_{\rm{rot}} \tau}+\frac{(\rho E_{\rm{rot}}^{\rm{aim,vib}})^n-(\rho E_{\rm{rot}})^n}{Z_{\rm{vib}} \tau}
\right] \\ 
& +\frac{\Delta t}{2}\left[
\frac{(\rho E_{\rm{rot}}^{\rm{aim,rot}})^{n+1}-(\rho E_{\rm{rot}})^{n+1}}{Z_{\rm{rot}} \tau}+\frac{(\rho E_{\rm{rot}}^{\rm{aim,vib}})^{n+1}-(\rho E_{\rm{rot}})^{n+1}}{Z_{\rm{vib}} \tau}
\right], \\ 
S_{\rm{vib}} & =\frac{\Delta t}{2}\left[
\frac{(\rho E_{\rm{vib}}^{\rm{aim}})^{n}-(\rho E_{\rm{vib}})^{n}}{Z_{\rm{vib}} \tau}+\frac{(\rho E_{\rm{vib}}^{\rm{aim}})^{n+1}-(\rho E_{\rm{vib}})^{n+1}}{Z_{\rm{vib}} \tau}
\right].
\end{aligned}
\end{equation} 
Based on the Eq.(\ref{eq:macroVar}), the conservative flow variables $\rho^{n+1} $, $(\rho \boldsymbol{U})^{n+1}$ and $ (\rho E)^{n+1}$ can be updated directly.  Then $(\rho E_{\rm{rot}}^{\rm{aim,vib}})^{n+1}$ and $(\rho E_{\rm{vib}}^{\rm{aim}})^{n+1}$ can be obtained from the updated conservative flow variables, and the vibrational energy $(\rho E_{\rm{vib}})^{n+1}$ with implicit source term can be solved in an explicit way without iterations
\begin{equation}  \label{eq:rhoEvPlus1}
(\rho E_{\rm{vib}})^{n+1}=\left( 1 + \frac{\Delta t}{2Z_{\rm{vib}}\tau} \right)^{-1} \left[ (\rho E_{\rm{vib}})^{+} + \frac{\Delta t}{2}\left( s_{\rm{vib}}^{n} + \frac{(\rho E_{\rm{vib}}^{\rm{aim}})^{n+1}}{Z_{\rm{vib}} \tau} \right)  \right].
\end{equation} 
The $(\rho E_{\rm{rot}}^{\rm{aim,rot}})^{n+1}$ can be obtained by the updated $(\rho E_{\rm{vib}})^{n+1}$, then the rotational energy 
\begin{equation}  \label{eq:rhoErPlus1}
(\rho E_{\rm{rot}})^{n+1}=\left( 1 + \frac{\Delta t}{2Z_{\rm{rot}}\tau} \right)^{-1} \left[ (\rho E_{\rm{rot}})^{+} + \frac{\Delta t}{2}\left( s_{\rm{rot}}^{n} + \frac{(\rho E_{\rm{rot}}^{\rm{aim,rot}})^{n+1}}{Z_{\rm{rot}} \tau} + \frac{(\rho E_{\rm{vib}}^{\rm{aim,vib}})^{n+1} - (\rho E_{\rm{rot}}^{\rm{aim,rot}})^{n+1}}{Z_{\rm{vib}} \tau} \right)  \right].
\end{equation} 
Here $(\rho E_{\rm{vib}})^{+}$ and $\rho E_{\rm{rot}})^{+}$ are the updated intermediate vibrational and rotational energies with inclusion of the fluxes only. The $K_{\rm{vib}}(T_{\rm{eq}}^{\,n+1})$ is solved by the iteration method~\cite{ZHANG2023107079},
\begin{equation}  \label{eq:iterationT}
\begin{aligned} 
T_{\rm{eq}}^{\,n+1,0} & =\frac{\left(2(\rho E)^{n+1}-\rho^{n+1}\left|\boldsymbol{U}^{n+1}\right|^2\right)}{\rho^{n+1} R\left(5+K_{\rm{v i b}}\left(T_{\rm{eq}}^n\right)\right)}, \\ 
T_{\rm{eq}}^{\,n+1, i} & =\frac{\left(2(\rho E)^{n+1}-\rho^{n+1}\left|\boldsymbol{U}^{n+1}\right|^2\right)}{\rho^{n+1} R\left(5+K_{\rm{vib}}(T_{\rm{eq}}^{n+1,i+1}) \right)},
\end{aligned}
\end{equation}
where $i$ is the iteration step.

\begin{figure*}[h!t]
    \centering
    \includegraphics[width=0.8\textwidth]{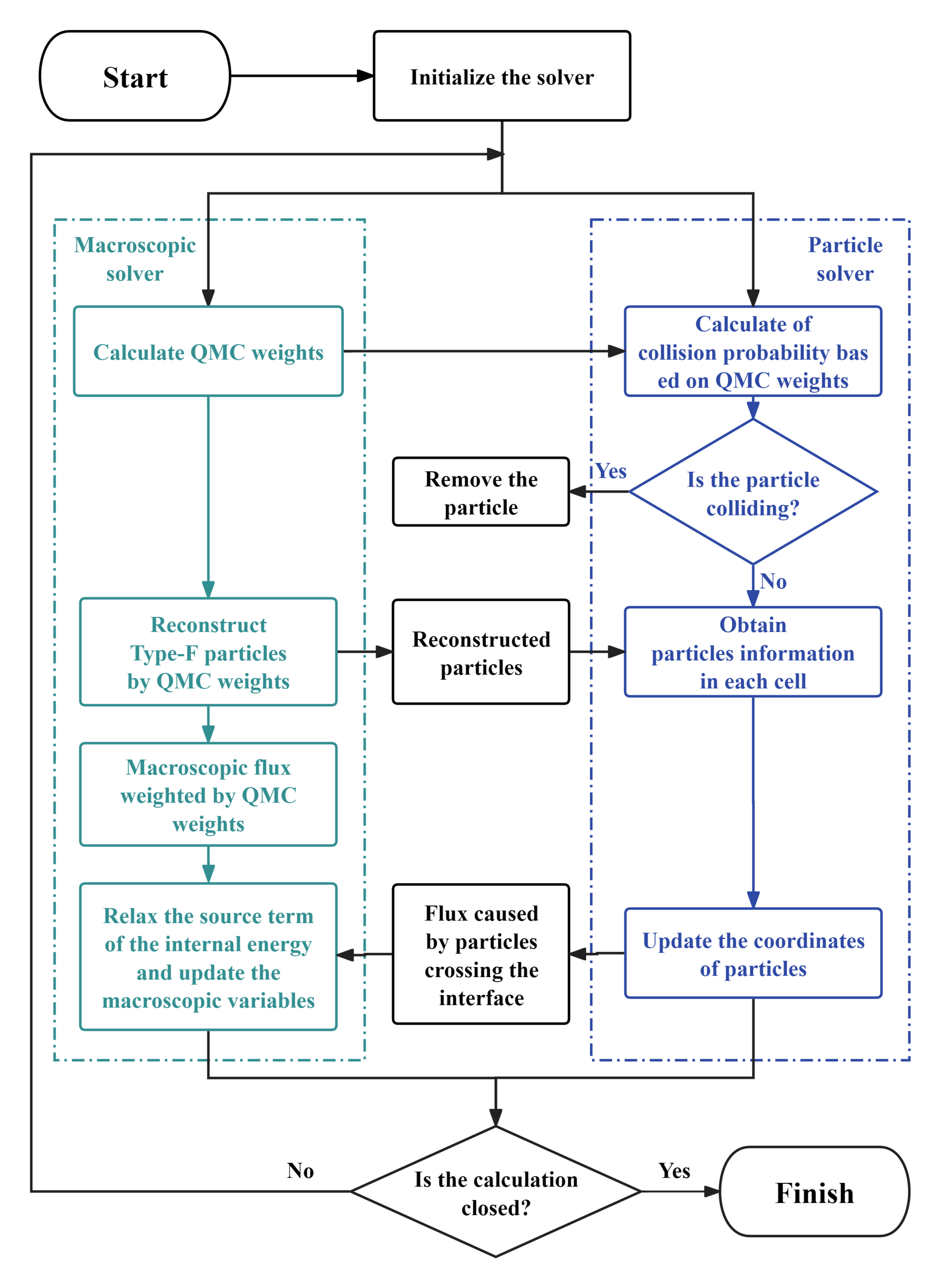}
    \caption{\label{Fig:solverProcess} The route of SUWP-vib method. (The blue region is the macroscopic solver, and the green region is the DSMC solver.)}
\end{figure*}

The computational process of the SUWP-vib method is illustrated in Fig.\ref{Fig:solverProcess}.
\begin{itemize}
\item[$\bullet$]Initially at each time step, each cell holds both particle and macroscopic information. Then, calculate the weights for Type-C molecules and Type-F molecules at the present step by the QMC mechanism.
\item[$\bullet$]New Type-C particles are removed based on QMC weights (absorb them back into the macroscopic information). New Type-F particles denote reconstructed particles also based on QMC weights by sampling from the equilibrium distribution $g$.
\item[$\bullet$]The particle solver moves the existing particles to their new cells, updates their coordinates and records the flux caused by particles crossing cell interfaces.
\item[$\bullet$]The macroscopic solver calculates the macroscopic flux at cell interfaces by multiplying the N-S flux by the hydrodynamic molecules weights.
\item[$\bullet$]The cell variables are updated by their changes obtained from the statistics in the particle solver and the flux calculated by the macroscopic solver (Eq.(\ref{eq:macroVar})).
\item[$\bullet$]The updating of particle information has already been completed during the particle transport step and requires no additional processing.
\end{itemize}

\subsection{Particle solver} \label{subsec:particleSolver}
In the SUWP-vib method, the particle solver is responsible for particle removal, reconstruction, transport, and counting the flux of particles across cell interfaces.

The particle removal and reconstruction processes are performed synchronously. Based on $w_{\rm{hydro}}$, new Type-C molecules are removed from the Type-F molecules. A random number $R_{\rm{n}}$ that is uniformly distributed in the interval $[0,1]$ is used to test all the particles in the cell. If the random number of the particle is less than the weight of the Type-C molecules, i.e. $R_{\rm{n}}<{w_{\rm{hydro}}}$, then this particle is removed. 

\begin{figure*}[h!t]
    \centering
    \includegraphics[trim=12 215 25 60, clip, width=0.9\textwidth]{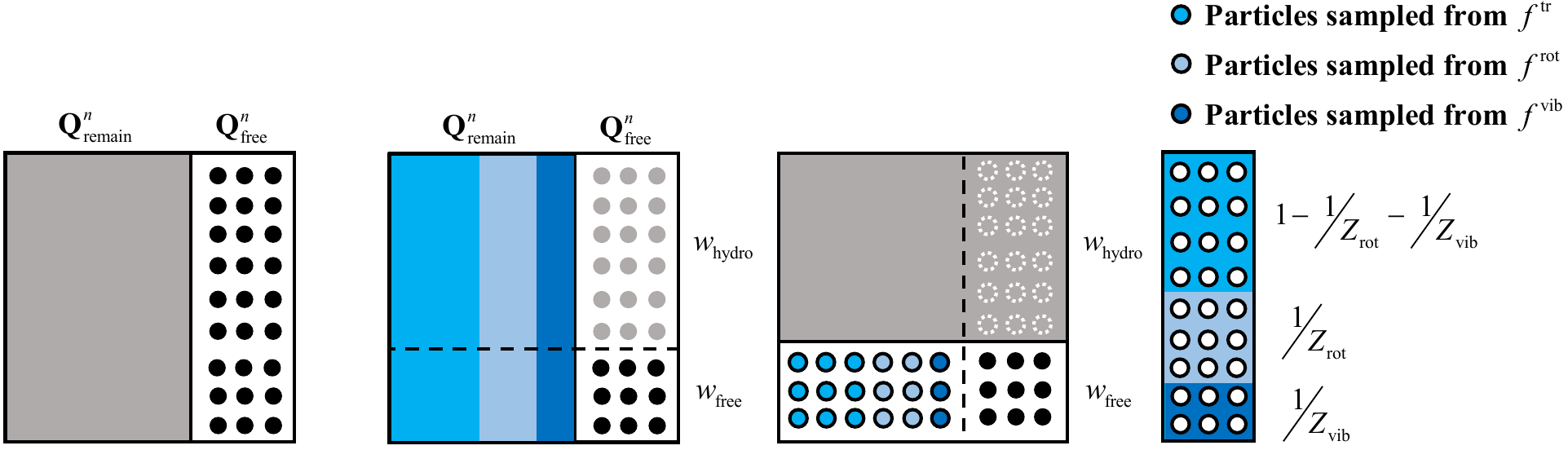}
    \caption{\label{Fig:vibWpEvalution} The particle reconstruction of SUWP-vib method.}
\end{figure*}

For reconstructed particles, as shown in Fig.~\ref{Fig:vibWpEvalution},using the remaining non-particle variables $\boldsymbol{Q}_{\rm{remain}}^{n}=\boldsymbol{Q}_{\rm{total}}^{n} - \boldsymbol{Q}_{\rm{free}}^{n}$ and the weight of the Type-F molecules are multiplied to give the macroscopic variables ${{w}_{\rm{free}}}\cdot \boldsymbol{Q}_{\rm{remain}}^{n}$ for reconstructed particles. A random number $R_n$ is used to determine whether the reconstructed particle follows the distribution $f^{\rm{vib}}$ ($R_n<1/Z_{vib}$), $f^{\rm{rot}}$ ($1/Z_{vib} \leq R_n < 1/Z_{\rm{vib}} + 1/Z_{\rm{rot}}$) or $f^{\rm{tr}}$ ($R_n \geq 1/Z_{\rm{vib}} + 1/Z_{\rm{rot}}$) of Eq.(\ref{eq:thefuncVib}), Eq.(\ref{eq:thefuncRot}) and Eq.(\ref{eq:thefuncTr}). After the reconstruction process, all particles in the cell are Type-F particles at present step.

For each particle in cell, the transport process are as follows:
\begin{equation}  \label{eq:transportPoint}
{{\boldsymbol{x}}^{n+1}}={\boldsymbol{x}}^{n}+\boldsymbol{u}\cdot \Delta t.
\end{equation} 
This process is simple and easy to understand, and will not be repeated here. It has been fully elaborated in DSMC and other particle methods. The particle transport in unstructured mesh is consistent with previous study~\cite{Liu2020Simplified}.The particle transport process can be expressed as the following three steps:

\textbf{{Step 1:}} Assign the time step $\Delta t$ as the remaining transport time for a particle.

\textbf{{Step 2:}} Check the transport path for intersections with any cell interface. If there is no intersection, update the particle coordinates directly to the endpoint coordinates. If there is an intersection, move the particle to the corresponding adjacent cell or update the particle's information according to the boundary conditions, and then allow the particle to continue its transport.

\textbf{{Step 3:}} Repeat the above steps until all particles in the field have completed their transport.

\subsection{Macroscopic solver} \label{subsec:macroSolver}
In the SUWP-vib method, the macroscopic solver is responsible for calculating the macroscopic fluxes $\mathbf{F}_{\rm{hydro}}$ at the cell interfaces and updating the macroscopic variables in the cells. The present SUWP-vib method uses the KIF~\cite{KIF1, KIF2} as the inviscid flux scheme in the macroscopic solver. This scheme combines the kinetic flux-vector splitting (KFVS) method and the totally thermalized transport (TTT) method through a weighted coupling~\cite{Xu1998GaskineticSF}. The SUWP-vib method adds rotational energy equation and vibrational energy equation to the basic KIF method. It is similar to the way of adding the rotational energy equation to the KIF~\cite{suwp2}. The rotational and vibrational energy equation can be obtained by integrating the rotational and vibrational energy of three-temperature equations. It should be noted that the pressure in the momentum equation of the three-temperature KIF is the $p_{tr}$.

The KFVS flux in this work can be written as:
\begin{equation}  \label{eq:momOfKFVS}
\boldsymbol{K}=\left\langle(\boldsymbol{u} \cdot \boldsymbol{n}) {\boldsymbol{\psi}} g_L\right\rangle_L+\left\langle(\boldsymbol{u} \cdot \boldsymbol{n}) {\boldsymbol{\psi}} g_R\right\rangle_R,
\end{equation} 
where the left half moment $\left\langle\cdot\right\rangle_L$ and right half moment $\left\langle\cdot\right\rangle_R$ are defined as
\begin{equation}  \label{eq:momOfKFVS_LR}
\begin{aligned}\langle\cdot\rangle_L & =\int_{-\infty}^0 {\rm{d}} u_1 \int_{-\infty}^{+\infty} {\rm{d}} u_2 \int_{-\infty}^{+\infty} {\rm{d}} u_3 \int_{0}^{+\infty} {\rm{d}} {\eta_{\rm{rot}}} \int_{0}^{+\infty}(\cdot) {\rm{d}} {\eta_{\rm{vib}}}, \, \\
\langle\cdot\rangle_R & =\int_0^{+\infty} {\rm{d}} u_1 \int_{-\infty}^{+\infty} {\rm{d}} u_2 \int_{-  \infty}^{+\infty} {\rm{d}} u_3 \int_{0}^{+\infty} {\rm{d}} {\eta_{\rm{rot}}} \int_{0}^{+\infty}(\cdot) {\rm{d}} {\eta_{\rm{vib}}}.\end{aligned}
\end{equation} 

According to the Eq.(\ref{eq:momOfKFVS}), the KFVS flux $\boldsymbol{K}$ can be represented in an explicit form:
\begin{equation}  \label{eq:kfvsFlux}
\begin{aligned}
{{K}_{\rm{mass}}} & =\frac{1}{2}\left( {{\rho }_{L}}{{U}_{L}}+{{\rho }_{R}}{{U}_{R}} \right)+\frac{1}{2}\left( {{\rho }_{L}}{{U}_{L}}{{\delta }_{L}}-{{\rho }_{R}}{{U}_{R}}{{\delta }_{R}} \right)+\frac{1}{2}\left( {{\rho }_{L}}{{\theta }_{L}}-{{\rho }_{R}}{{\theta }_{R}} \right), \\ 
{{K}_{\rm{xmon}}} & =\frac{1}{2}\left[ \left( {{\rho }_{L}}U_{L}^{2}+{{p}_{{\rm{tr}},L}} \right)+\left( {{\rho }_{R}}U_{R}^{2}+{{p}_{{\rm{tr}},R}} \right) \right]+\frac{1}{2}\left[ \left( {{\rho }_{L}}U_{L}^{2}+{{p}_{{\rm{tr}},L}} \right){{\delta }_{L}} \right.\left. -\left( {{\rho }_{R}}U_{R}^{2}+{{p}_{{\rm{tr}},R}} \right){{\delta }_{R}} \right] \\
 & +\frac{1}{2}\left( {{\rho }_{L}}{{U}_{L}}{{\theta }_{L}}-{{\rho }_{R}}{{U}_{R}}{{\theta }_{R}} \right), \\ 
{{K}_{\rm{ymon}}} & =\frac{1}{2}\left( {{\rho }_{L}}{{U}_{L}}{{V}_{L}}+{{\rho }_{R}}{{U}_{R}}{{V}_{R}} \right)+\frac{1}{2}\left( {{\rho }_{L}}{{U}_{L}}{{V}_{L}}{{\delta }_{L}}-{{\rho }_{R}}{{U}_{R}}{{V}_{R}}{{\delta }_{R}} \right)+\frac{1}{2}\left( {{\rho }_{L}}{{V}_{L}}{{\theta }_{L}}-{{\rho }_{R}}{{V}_{R}}{{\theta }_{R}} \right),\\ 
{{K}_{\rm{zmon}}} & =\frac{1}{2}\left( {{\rho }_{L}}{{U}_{L}}{{W}_{L}}+{{\rho }_{R}}{{U}_{R}}{{W}_{R}} \right)+\frac{1}{2}\left( {{\rho }_{L}}{{U}_{L}}{{W}_{L}}{{\delta }_{L}}-{{\rho }_{R}}{{U}_{R}}{{W}_{R}}{{\delta }_{R}} \right)+\frac{1}{2}\left( {{\rho }_{L}}{{W}_{L}}{{\theta }_{L}}-{{\rho }_{R}}{{W}_{R}}{{\theta }_{R}} \right), \\ 
{{K}_{\rm{energy}}} & ={{K}_{\rm{tr}}} + {{K}_{\rm{rot}}} + {{K}_{\rm{vib}}}, \\ 
{{K}_{\rm{tr}}} & =\frac{1}{2}\left( {{\rho }_{L}}{{U}_{L}}{{h}_{L}}+{{\rho }_{R}}{{U}_{R}}{{h}_{R}} \right)+\frac{1}{2}\left( {{\rho }_{L}}{{U}_{L}}{{h}_{L}}{{\delta }_{L}}-{{\rho }_{R}}{{U}_{R}}{{h}_{R}}{{\delta }_{R}} \right)  \\
& +\frac{1}{2}\left[ \left( {{\rho }_{L}}{{h}_{L}}-\frac{{{p}_{{\rm{tr}},L}}}{2} \right){{\theta }_{L}}-\left( {{\rho }_{R}}{{h}_{R}}-\frac{{{p}_{{\rm{tr}},R}}}{2} \right){{\theta }_{R}} \right], \\ 
{{K}_{\rm{rot}}} & =\frac{1}{2}\left( {{\rho }_{L}}{{E}_{{\rm{rot}},L}}{{U}_{L}}+{{\rho }_{R}}{{E}_{{\rm{rot}},R}}{{U}_{R}} \right)+\frac{1}{2}\left( {{\rho }_{L}}{{E}_{{\rm{rot}},L}}{{U}_{L}}{{\delta }_{L}}-{{\rho }_{R}}{{E}_{{\rm{rot}},R}}{{U}_{R}}{{\delta }_{R}} \right)  \\
& +\frac{1}{2}\left( {{\rho }_{L}}{{E}_{{\rm{rot}},L}}{{\theta }_{L}}-{{\rho }_{R}}{{E}_{{\rm{rot}},R}}{{\theta }_{R}} \right), \\ 
{{K}_{\rm{vib}}} & =\frac{1}{2}\left( {{\rho }_{L}}{{E}_{{\rm{vib}},L}}{{U}_{L}}+{{\rho }_{R}}{{E}_{{\rm{vib}},R}}{{U}_{R}} \right)+\frac{1}{2}\left( {{\rho }_{L}}{{E}_{{\rm{vib}},L}}{{U}_{L}}{{\delta }_{L}}-{{\rho }_{R}}{{E}_{{\rm{vib}},R}}{{U}_{R}}{{\delta }_{R}} \right)  \\
& +\frac{1}{2}\left( {{\rho }_{L}}{{E}_{{\rm{vib}},L}}{{\theta }_{L}}-{{\rho }_{R}}{{E}_{{\rm{vib}},R}}{{\theta }_{R}} \right), \\ 
\end{aligned}
\end{equation} 
where $U$ is the velocity component in the normal direction of the interface, $V$ and $W$ are the velocity component in the tangential direction of the interface. And the expression of $\delta$ and $\theta$ is
\begin{equation}  \label{eq:deltaAndTheta}
\begin{aligned}
{{\delta }_{\alpha }}&=\operatorname{erf}\left( \frac{{{U}_{\alpha }}}{\sqrt{2R{{T}_{\rm{tr},\alpha }}}} \right), \\
{{\theta }_{\alpha }}&=\sqrt{\frac{2R{{T}_{\rm{tr},\alpha }}}{\pi }}\exp \left( -\frac{U_{\alpha }^{2}}{2R{{T}_{\rm{tr},\alpha }}} \right), 
\end{aligned}
\end{equation} 
where the subscript $\alpha$ can be $L$ or $R$.

The TTT flux $\boldsymbol{G}$ in the present KIF is a simple Euler flux using the kinetic averaged values
\begin{equation}  \label{eq:tttFlux}
\begin{aligned}
 {{G}_{\rm{mass}}}&=\overline{\rho }\overline{U}, \\ 
 {{G}_{\rm{xmon}}}&=\overline{\rho }{{{\overline{U}}}^{2}}+\overline{p_{\rm{tr}}}, \\ 
 {{G}_{\rm{ymon}}}&=\overline{\rho }\overline{U}\overline{V}, \\ 
 {{G}_{\rm{zmon}}}&=\overline{\rho }\overline{U}\overline{W}, \\
 {{G}_{\rm{energy}}}&={{G}_{\rm{tr}}} + {{G}_{\rm{rot}}} + {{G}_{\rm{vib}}}, \\ 
 {{G}_{\rm{tr}}}&=\overline{\rho }{\overline{U}}\overline{h_{\rm{tr}}}, \\ 
 {{G}_{\rm{rot}}}&=\overline{\rho }{\overline{U}}\overline{{{E}_{\rm{rot}}}}, \\
 {{G}_{\rm{vib}}}&=\overline{\rho }{\overline{U}}\overline{{{E}_{\rm{vib}}}},
\end{aligned}
\end{equation} 
where the kinetic averaged values can be obtained from the following TTT process:
\begin{equation}  \label{eq:tttProcess}
\begin{aligned}
\overline{\rho }&=\frac{1}{2}\left( {{\rho }_{L}}+{{\rho }_{R}} \right)+\frac{1}{2}\left( {{\rho }_{L}}{{\delta }_{L}}-{{\rho }_{R}}{{\delta }_{R}} \right), \\ 
\overline{\rho u}&=\frac{1}{2}\left( {{\rho }_{L}}{{U}_{L}}+{{\rho }_{R}}{{U}_{R}} \right)+\frac{1}{2}\left( {{\rho }_{L}}{{\theta }_{L}}-{{\rho }_{R}}{{\theta }_{R}} \right) \\
 &+\frac{1}{2}\left( {{\rho }_{L}}{{U}_{L}}{{\delta }_{L}}-{{\rho }_{R}}{{U}_{R}}{{\delta }_{R}} \right), \\ 
\overline{\rho V}&=\frac{1}{2}\left( {{\rho }_{L}}{{V}_{L}}+{{\rho }_{R}}{{V}_{R}} \right)+\frac{1}{2}\left( {{\rho }_{L}}{{V}_{L}}{{\delta }_{L}}-{{\rho }_{R}}{{V}_{R}}{{\delta }_{R}} \right), \\ 
\overline{\rho w}&=\frac{1}{2}\left( {{\rho }_{L}}{{W}_{L}}+{{\rho }_{R}}{{W}_{R}} \right)+\frac{1}{2}\left( {{\rho }_{L}}{{W}_{L}}{{\delta }_{L}}-{{\rho }_{R}}{{W}_{R}}{{\delta }_{R}} \right), \\ 
\overline{\rho E_{\rm{tr}}}&=\frac{1}{2}\left( {{\rho }_{L}}{{E}_{{\rm{tr}},L}}+{{\rho }_{R}}{{E}_{{\rm{tr}},R}} \right)+\frac{1}{4}\left( {{\rho }_{L}}{{U}_{L}}{{\theta }_{L}}-{{\rho }_{R}}{{U}_{R}}{{\theta }_{R}} \right) \\
 &+\frac{1}{2}\left( {{\rho }_{L}}{{E}_{L}}{{\delta }_{L}}-{{\rho }_{R}}{{E}_{R}}{{\delta }_{R}} \right), \\ 
\overline{\rho {{E}_{\rm{rot}}}}&=\frac{1}{2}\left( {{\rho }_{L}}{{E}_{{\rm{rot}},L}}+{{\rho }_{R}}{{E}_{{\rm{rot}},R}} \right) +\frac{1}{2}\left( {{\rho }_{L}}{{E}_{{\rm{rot}},L}}{{\delta }_{L}}-{{\rho }_{R}}{{E}_{{\rm{rot}},R}}{{\delta }_{R}} \right), \\
\overline{\rho {{E}_{{\rm{vib}}}}}&=\frac{1}{2}\left( {{\rho }_{L}}{{E}_{{\rm{vib}},L}}+{{\rho }_{R}}{{E}_{{\rm{vib}},R}} \right) +\frac{1}{2}\left( {{\rho }_{L}}{{E}_{{\rm{vib}},L}}{{\delta }_{L}}-{{\rho }_{R}}{{E}_{{\rm{vib}},R}}{{\delta }_{R}} \right), \\
\end{aligned}
\end{equation}
where $\overline{U}=\overline{\rho U}/\overline{\rho }$, $\overline{V}=\overline{\rho V}/\overline{\rho }$, $\overline{W}=\overline{\rho W}/\overline{\rho }$, 
$\overline{e_{\rm{tr}}}=\overline{\rho E_{\rm{tr}}}/\overline{\rho} - 1/2(\overline{U}^2+\overline{V}^2+\overline{W}^2)$, 
$\overline{p_{\rm{tr}}}={(\gamma-1)\overline{\rho}\overline{e_{\rm{tr}}}}$ and $\overline{h_{\rm{tr}}}={\overline{\rho E_{\rm{tr}}} /\overline{\rho} +\overline{p} /\overline{\rho}}$.

In KIF, the KFVS fluxes and TTT fluxes are coupled as
\begin{equation}  \label{eq:coupleKFVSandTTT}
{\mathbf{F}_{\text{NS,inv}}}=\beta \boldsymbol{K}+\left( 1-\beta  \right)\boldsymbol{G},
\end{equation} 
where $\beta$ is a kind of weight (the weight of the free transport mechanism at a cell interface):
\begin{equation}  \label{eq:kifbeta}
\beta=F_{\rm{smooth}}\left(\Delta P,0,1\right)F_{\rm{smooth}}(\rm{Ma},0.5\rm{Ma}_{back},\rm{Ma}_{back}),
\end{equation} 
where $F_{\rm{smooth}}$ is a smooth function consisting of sine function. The expressions of $F_{\rm{smooth}}$, $\Delta P$ and $\rm{Ma}_{back}$ are in Ref.~\cite{KIF2}.
The coupling approach and proper chosen of $\beta$ ensure that most continuum regions in the flow field adopt the TTT scheme. Meanwhile, the KFVS scheme possesses the capability to capture discontinuities. By using the expression of $\beta$, the KFVS scheme dominates near shock waves, while the TTT scheme dominates near the boundary layer.

In this paper, the gradients of macroscopic variables are computed using the least squares method with the application of the Venkatakrishnan gradient limiter~\cite{venkatakrishnan1993implicit}.
The viscous flux ${{\boldsymbol{F}}_{\rm{vis}}}$ is calculated by the central scheme. The physical variables and their gradients at the cell interfaces are computed by taking the weighted average of the center values of two adjacent cells. These values are then utilized in calculating the viscous flux. The computational framework for the flux is the same as that of traditional N-S equations, except the viscous flux is slightly modified by Eq.(\ref{eq:nsFlux2}) to match the multi-scale physics. By treating the translational energy and rotational energy separately, the flux of the N-S solver is in good consistency with the evolution of colliding particles.

\subsection{Boundary condition}
The boundary condition of the SUWP-vib method is also hybrid. For the inflow boundary, the particle solver and the macroscopic solver compute the detailed information of inflow particles and macroscopic net flux, respectively, according to weights ${{w}_{\rm{free}}}$ and ${{w}_{\rm{hydro}}}$. For the wall boundary, the particle solver updates the particle velocity and energy of the particles hitting the wall according to wall speed and temperature, while the macroscopic solver calculates the flux by directly multiplying the N-S one by weight ${{w}_{\rm{hydro}}}$.
The macroscopic boundary condition is constructed similarly to the classical N-S solver. Therefore, the following discussion focuses on the boundary condition of the particle solver.

~\\

\textbf{Inflow and outflow boundary conditions}:

The inflow boundary of the particle solver is consistent with the DSMC method, except the inflow mass flux is set according to the scale weights as follows:
\begin{equation}  \label{eq:boundaryInflow}
\begin{aligned}
&{{{\boldsymbol{F}}}_{\rm{DSMC}}}=\left\{ \rho \exp \left( -\frac{U_{\rm{inflow}}^{2}}{2RT_{\rm{inflow}}} \right)\sqrt{\frac{RT_{\rm{inflow}}}{2\pi }} \right.\left. +\frac{{{U}}_{\rm{inflow}}}{2}\left[ 1+{\rm{erf}}\left( \frac{{{U}}_{\rm{inflow}}}{\sqrt{2RT_{\rm{inflow}}}} \right) \right] \right\}, \\ 
&{{{\boldsymbol{F}}}_{\rm{inflow}}}={{w}_{\rm{free}}}\cdot {{{\boldsymbol{F}}}_{\rm{DSMC}}},
\end{aligned}
\end{equation} 
where $U_{\rm{inflow}}$ is the normal component of inflow velocity relative to the boundary interface, and the $T_{\rm{inflow}}$ is the inflow temperature.
The final mass of particles entering from the inflow boundary is ${{\boldsymbol{F}}_{\rm{inflow}}}\cdot \Delta t \cdot A$, where $A$ is the area of the boundary interface. The remaining transport time for particles entering from the interface is $R_n\cdot \Delta t$, and their coordinates are randomly distributed on the inflow boundary. 

\textbf{Wall boundary condition}:

Particles should update their velocities and rotational energies after hitting the solid wall according to the Maxwell distribution at the wall. Therefore, the bounce back velocity in the face coordinate system of the wall can be sampled:
\begin{equation}  \label{eq:boundaryWall}
\boldsymbol{u}={{\boldsymbol{u}}_{\rm{wall}}}+{} \sqrt{2R{{T}_{\rm{wall}\text{ }}}}  \left( \begin{aligned}
  & \sqrt{-\log \left( R_{n1} \right)} \\ 
 & \sqrt{-\log \left( R_{n2} \right)}\cos \left( 2\pi \cdot R_{n3} \right) \\ 
 & \sqrt{-\log \left( R_{n2} \right)}\sin \left( 2\pi \cdot R_{n3} \right) \\ 
\end{aligned} \right),
\end{equation} 
where ${{\boldsymbol{u}}_{\rm{wall}}}$ is the wall movement speed. $T_{\rm{wall}}$ is the wall temperature. The expression for the rotational energy and vibrational energy of particles that hit the wall is
\begin{equation}  \label{eq:boundaryRotE}
\eta_{\rm{rot}} =-\log \left( R_n \right)R \, {{T}_{\rm{wall}}},\, \,
\eta_{\rm{vib}} =-\log \left( R_n \right)R \, {\Theta_{\rm{vib}}}/ \left({\exp \left({\frac{\Theta_{\rm{vib}}}{{{T}_{\rm{wall}}}}}\right) - 1} \right).
\end{equation} 

\section{\label{sec:Test}Numerical result}
For all cases, the non-dimensional of initial conditions is completed through the freestream density $\rho_{\rm{\infty}}$, freestream temperature 
$T_{\rm{\infty}}$ and reference velocity $U_{\rm{ref}}=\sqrt{2RT_{\rm{\infty}}}$. When the Knudsen number is prescribed, the freestream density is calculated by
\begin{equation}   \label{eq:case_noDim}
\rho_{\rm{\infty}}=\frac{2}{15}(5-2\omega)(7-2\omega)\sqrt{\frac{1}{2\pi R T_{\rm{\infty}}}}\frac{\mu}{L_{\rm{ref}}\rm{Kn}},
\end{equation} 
where $L_{\rm{ref}}$is the reference length of the case. The dynamic viscosity $\mu$ is calculated from the translational temperature by Eq.(\ref{eq:muEqn}). Then the following dimensionless quantities can be obtained:
\begin{equation}   \label{eq:case_noDim02}
\hat{\rho} = \frac{\rho}{ \rho_{\rm{\infty} }},\, \hat{\boldsymbol{U}} = \frac{\boldsymbol{U}}{ {U}_{\rm{ref} }},\, \hat{T} = \frac{T}{ T_{\rm{\infty} }},\, \hat{p} = \frac{p}{ \rho_{\rm{\infty} }{U}^{2}_{\rm{ref} }},\, \hat{\boldsymbol{q}} = \frac{\boldsymbol{q}}{\rho_{\rm{\infty}} {U}^{3}_{\rm{ref} }}.
\end{equation} 
The working gas for all test cases is nitrogen, employing the Variable Hard Sphere (VHS) collision model, with its parameters specified in Table \ref{table:1Dcase_Sod_Nitrogen_Table}.
\begin{table}[ht]
    \centering
    \caption{The parameters of Nitrogen gas.}  \label{table:1Dcase_Sod_Nitrogen_Table}%
    \begin{tabular}{ c c c  c  c  c  c  }
        \hline
        Gas & $R(\rm{J/(kg \cdot K)})$ & $T_{\rm{ref}} (\rm{K})$ & $\Theta_{\rm{vib}} (\rm{K})$ & $\mu_{\rm{ref}} (\rm{Nsm^{-2}})$ & $\alpha$ & $\omega$  \\ 
        \hline
        Nitrogen & 296.91 & 273.15 & 3371 & $1.656 \times 10^{-5}$ & 1.0 & 0.74 \\
        \hline
    \end{tabular}
\end{table}

\subsection{Sod's shock tube}
The Sod's shock tube problem under different conditions serves as classical flow benchmarks. This test case is employed to validate the present method for one-dimensional unsteady applications. The computational domain spans $[-0.5, 0.5]$ and is uniformly discretized into 200 cells. Since these are unsteady test cases, the particle number of cell $N_{\rm{p}} =2 \times 10^{5}$. By increasing the number of particles to reduce random errors, the averaging processing of the flow field is avoided. Other involved parameters are specified in Table \ref{table:1Dcase_Sod_Table}. For the three test cases at $\rm{Kn}=10,0.1,0.001$, the non-dimensional initial conditions are given below.
\begin{equation}  \label{eq:case1Dsod_initial}
(\rho, U, T_{tr}, T_{\rm{rot}}, T_{\rm{vib}})=
\left\{ 
    \begin{tabular}{ l l }
        (1, 0, 2, 2, 2), &  x $\leq$ 0, \\
        (0.125, 0, 1.6, 1.6, 1.6), &  x $>$ 0. \\
    \end{tabular}
\right.
\end{equation} 

\begin{table}[ht] 
    \centering
    \caption{The parameters of Sod's shock tube cases.} \label{table:1Dcase_Sod_Table}
    \begin{tabular}{ c  c  c  c }
        \hline
        $L_{\rm{ref}}$ & $T_{\rm{\infty}}(\rm{K})$ & $Z_{\rm{rot}}$ & $Z_{\rm{vib}}$ \\ 
        \hline
        1 & 273.15 & 3.5 & 10 \\
        \hline
    \end{tabular}
\end{table}

\begin{figure*}[h!t]
\centering
\subfigure[\label{Fig:case1D_sod_kn10_rho}]{
\includegraphics[trim=30 25 62 55, clip, width=0.3\textwidth]{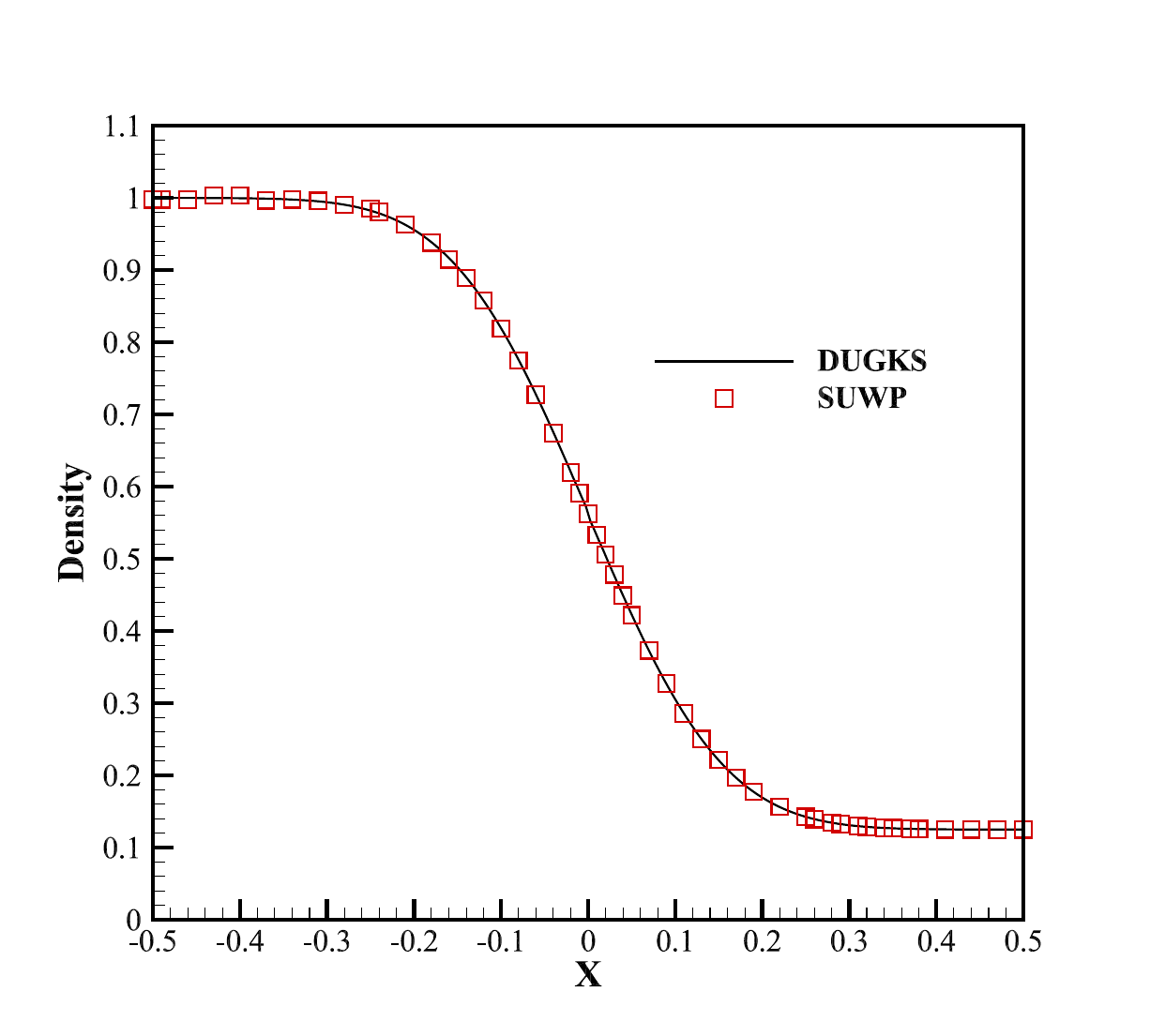}
}\hspace{0.01\textwidth}%
\subfigure[\label{Fig:case1D_sod_kn10_U}]{
\includegraphics[trim=30 25 62 55, clip, width=0.3\textwidth]{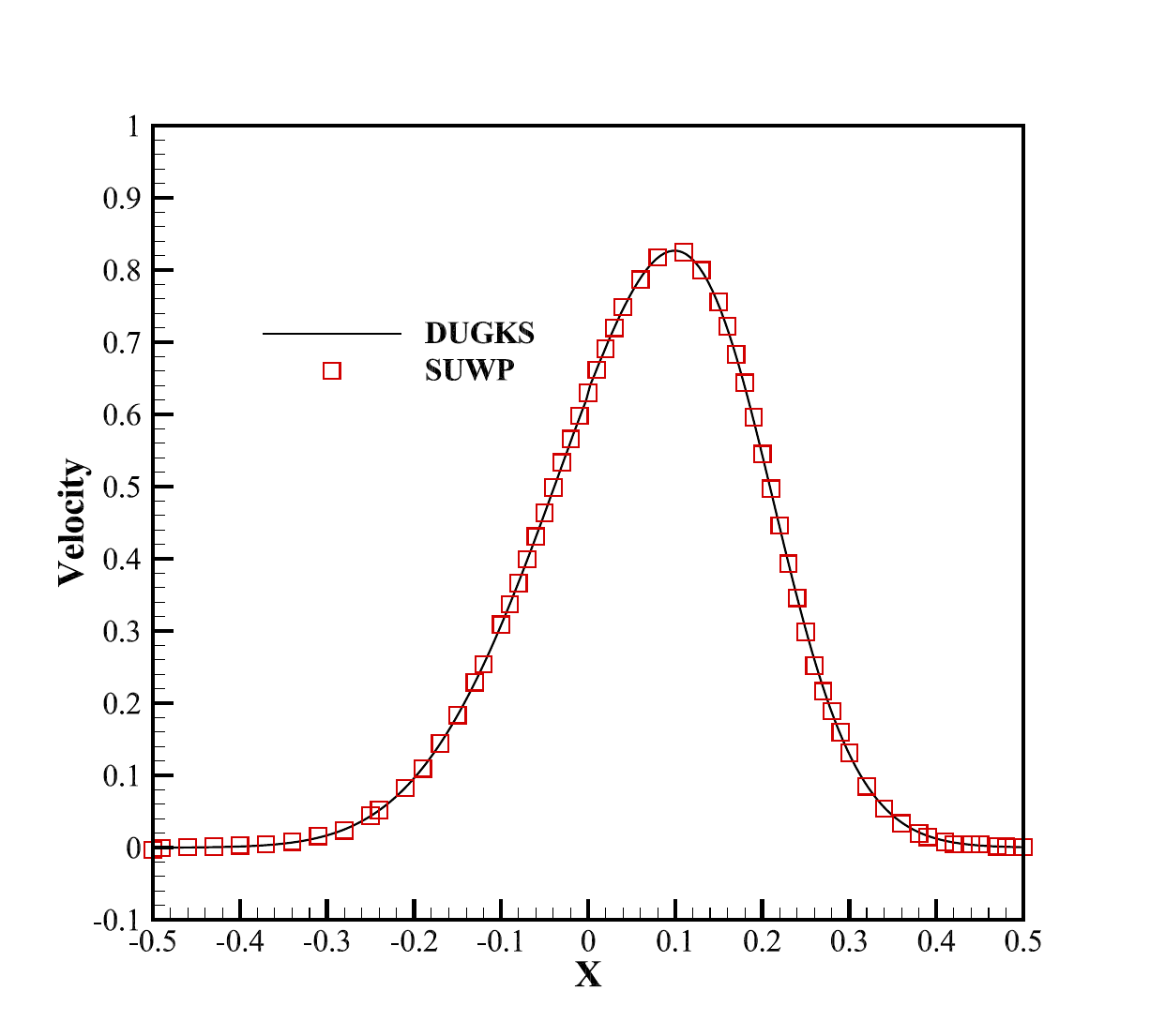}
}\hspace{0.01\textwidth}%
\subfigure[\label{Fig:case1D_sod_kn10_T}]{
\includegraphics[trim=30 25 62 55, clip, width=0.3\textwidth]{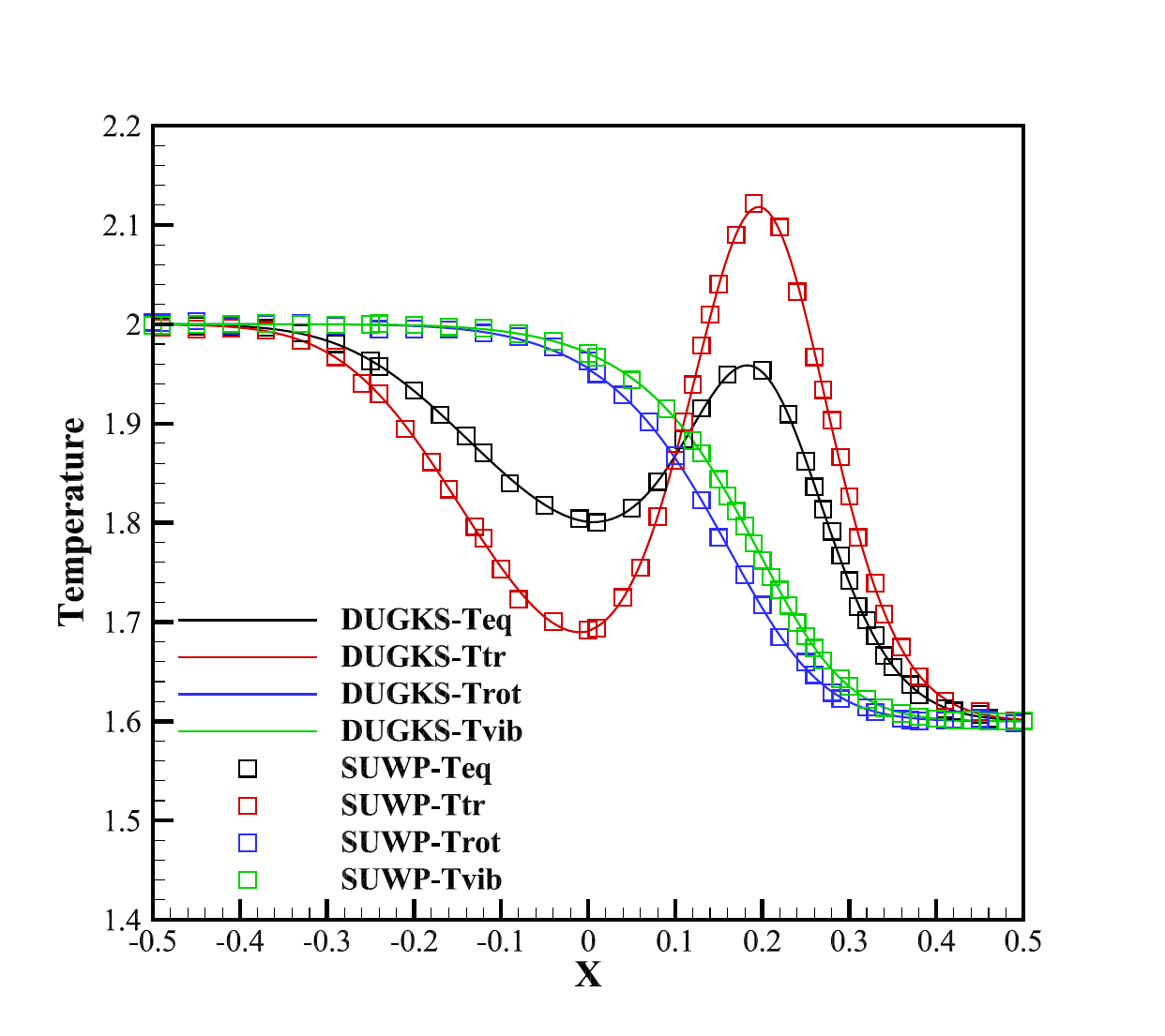}
}\\
\caption{\label{Fig:case1D_sod_kn10}The (a) desity, (b) velocity and (c) temperature profiles of the Sod's shock tube at Kn=10.}
\end{figure*}

\begin{figure*}[h!t]
\centering
\subfigure[\label{Fig:case1D_sod_kn0.1_rho}]{
\includegraphics[trim=30 25 62 55, clip, width=0.3\textwidth]{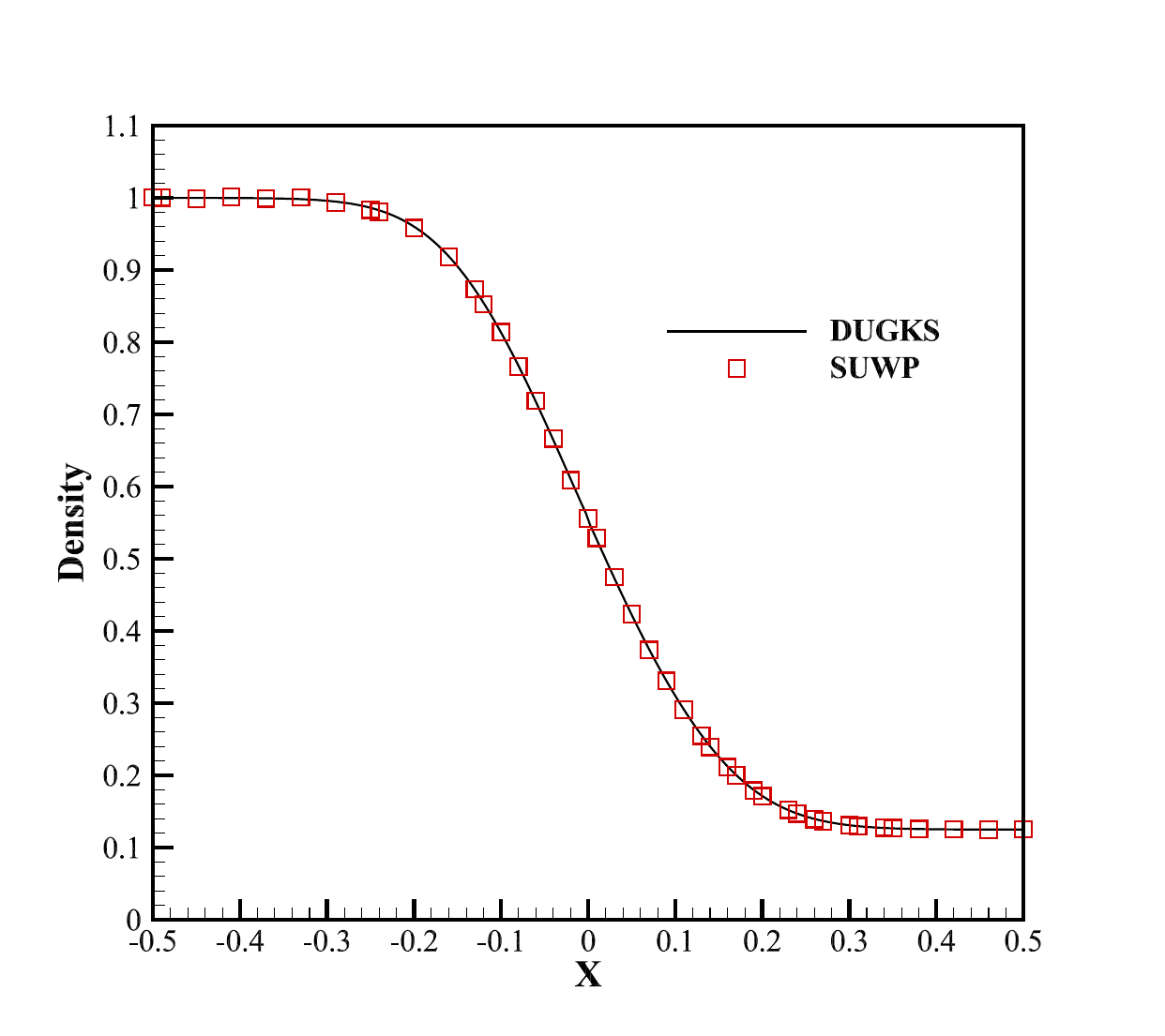}
}\hspace{0.01\textwidth}%
\subfigure[\label{Fig:case1D_sod_kn0.1_U}]{
\includegraphics[trim=30 25 62 55, clip, width=0.3\textwidth]{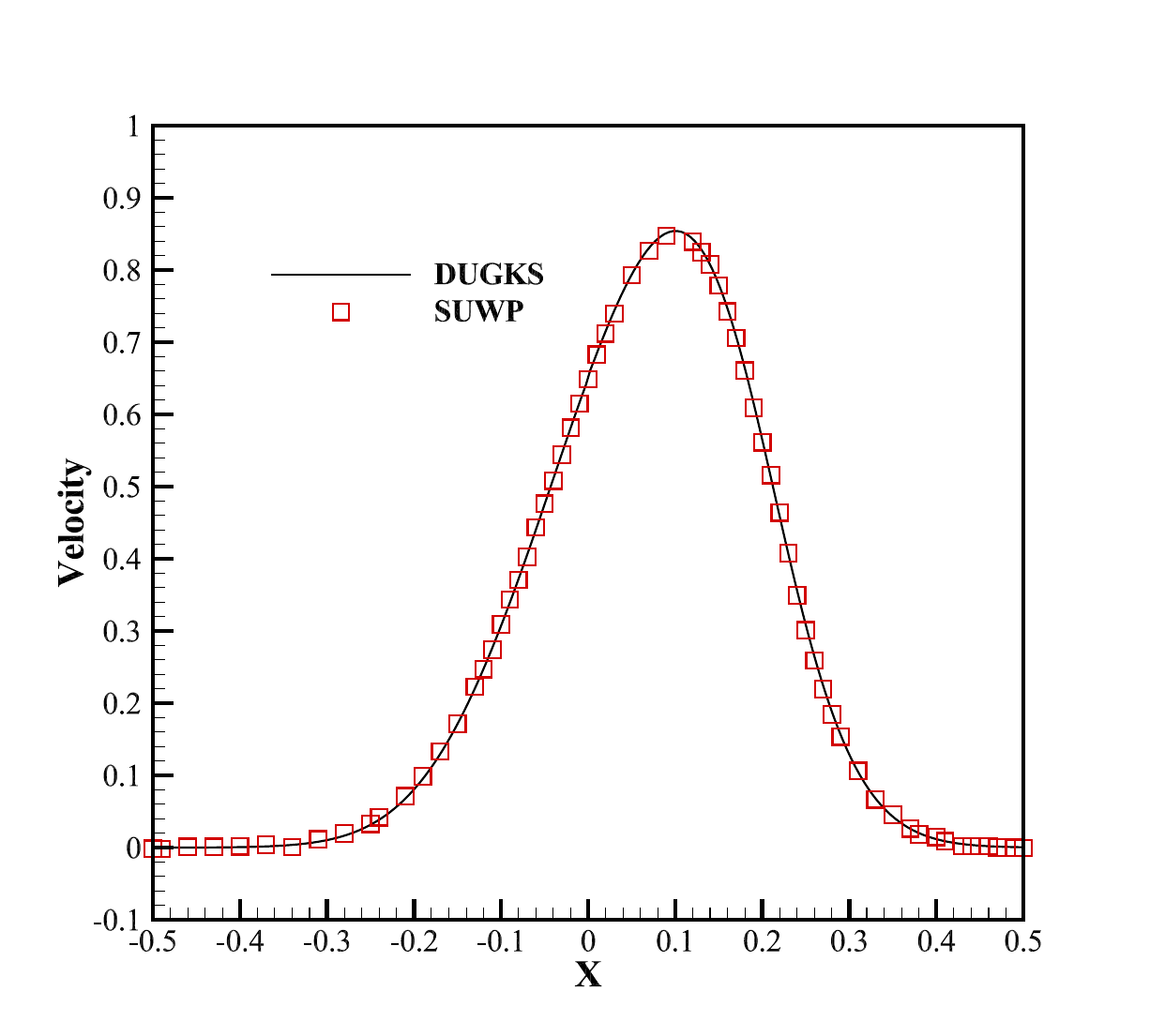}
}\hspace{0.01\textwidth}%
\subfigure[\label{Fig:case1D_sod_kn0.1_T}]{
\includegraphics[trim=30 25 62 55, clip, width=0.3\textwidth]{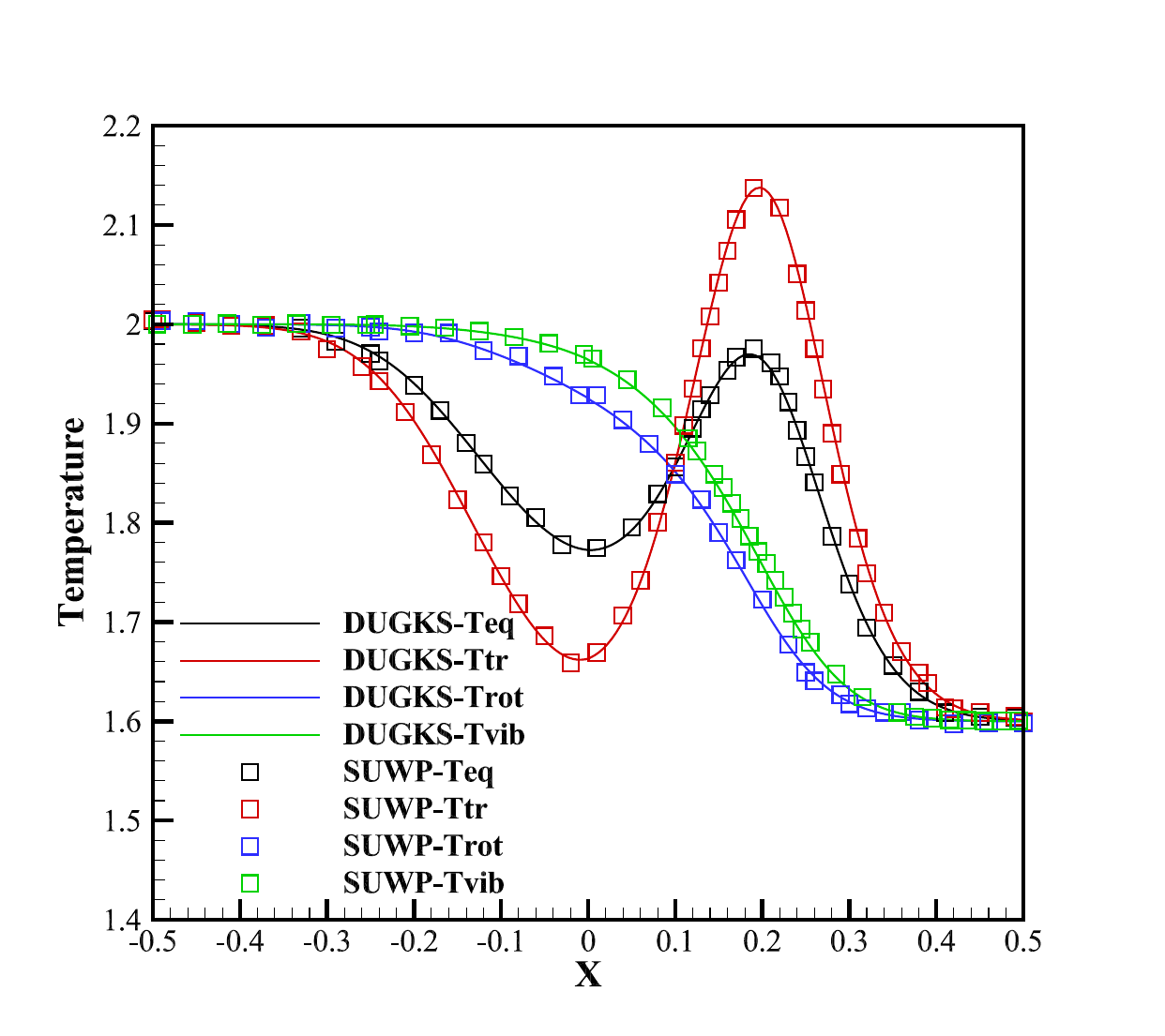}
}\\
\caption{\label{Fig:case1D_sod_kn0.1}The (a) desity, (b) velocity and (c) temperature profiles of the Sod's shock tube at Kn=0.1.}
\end{figure*}

\begin{figure*}[h!t]
\centering
\subfigure[\label{Fig:case1D_sod_kn0.001_rho}]{
\includegraphics[trim=30 25 62 55, clip, width=0.3\textwidth]{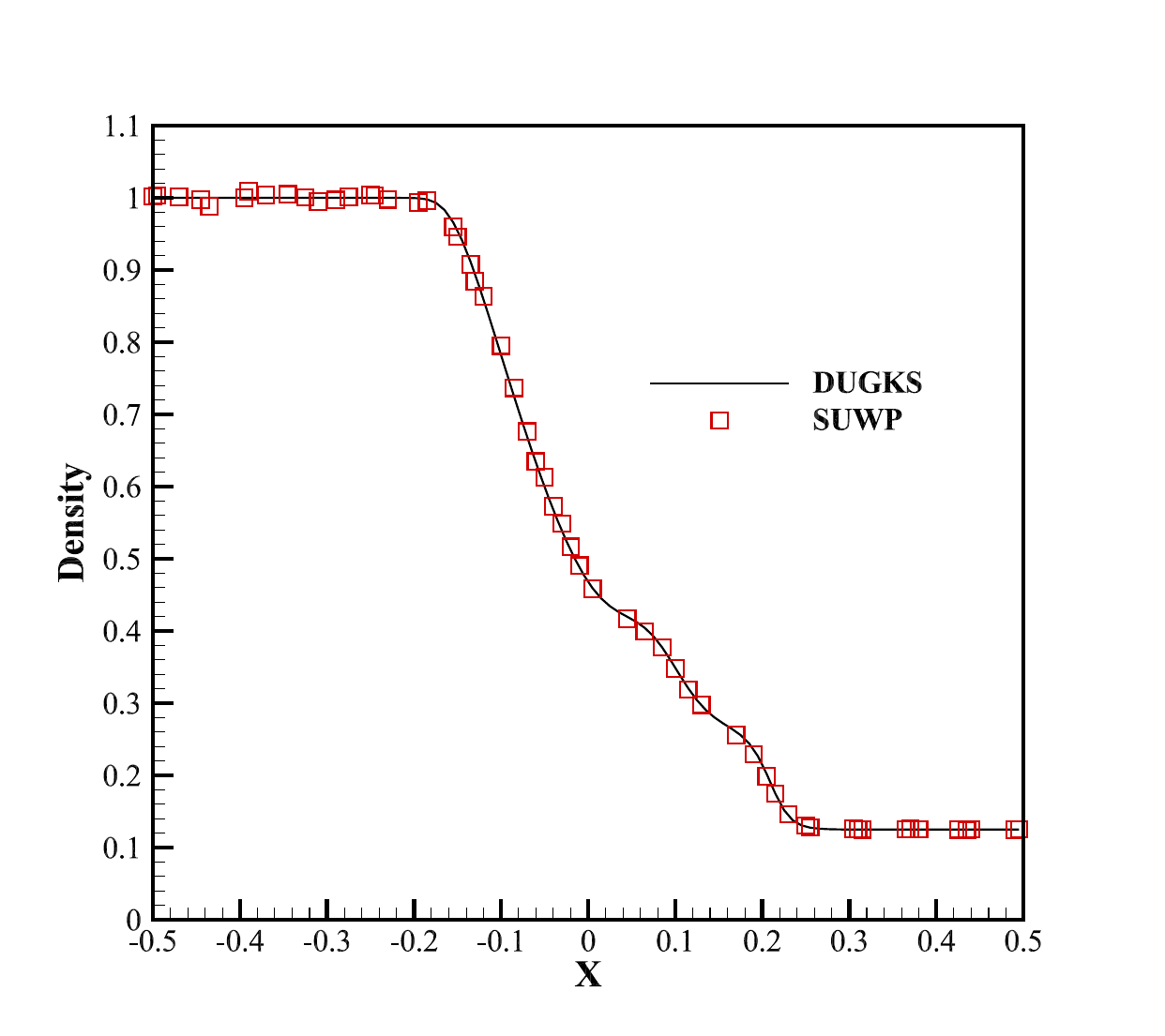}
}\hspace{0.01\textwidth}%
\subfigure[\label{Fig:case1D_sod_kn0.001_U}]{
\includegraphics[trim=30 25 62 55, clip, width=0.3\textwidth]{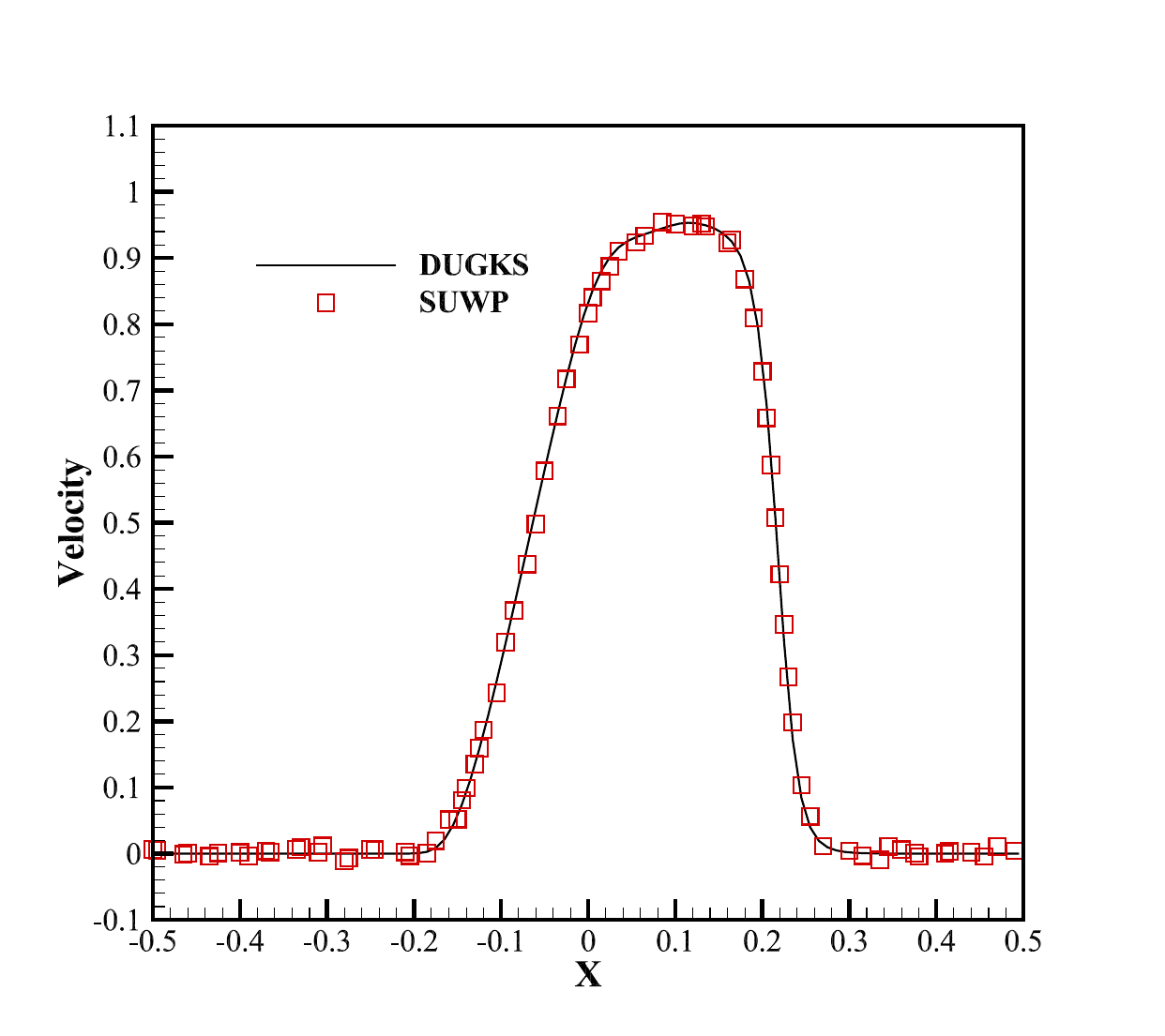}
}\hspace{0.01\textwidth}%
\subfigure[\label{Fig:case1D_sod_kn0.001_T}]{
\includegraphics[trim=30 25 62 55, clip, width=0.3\textwidth]{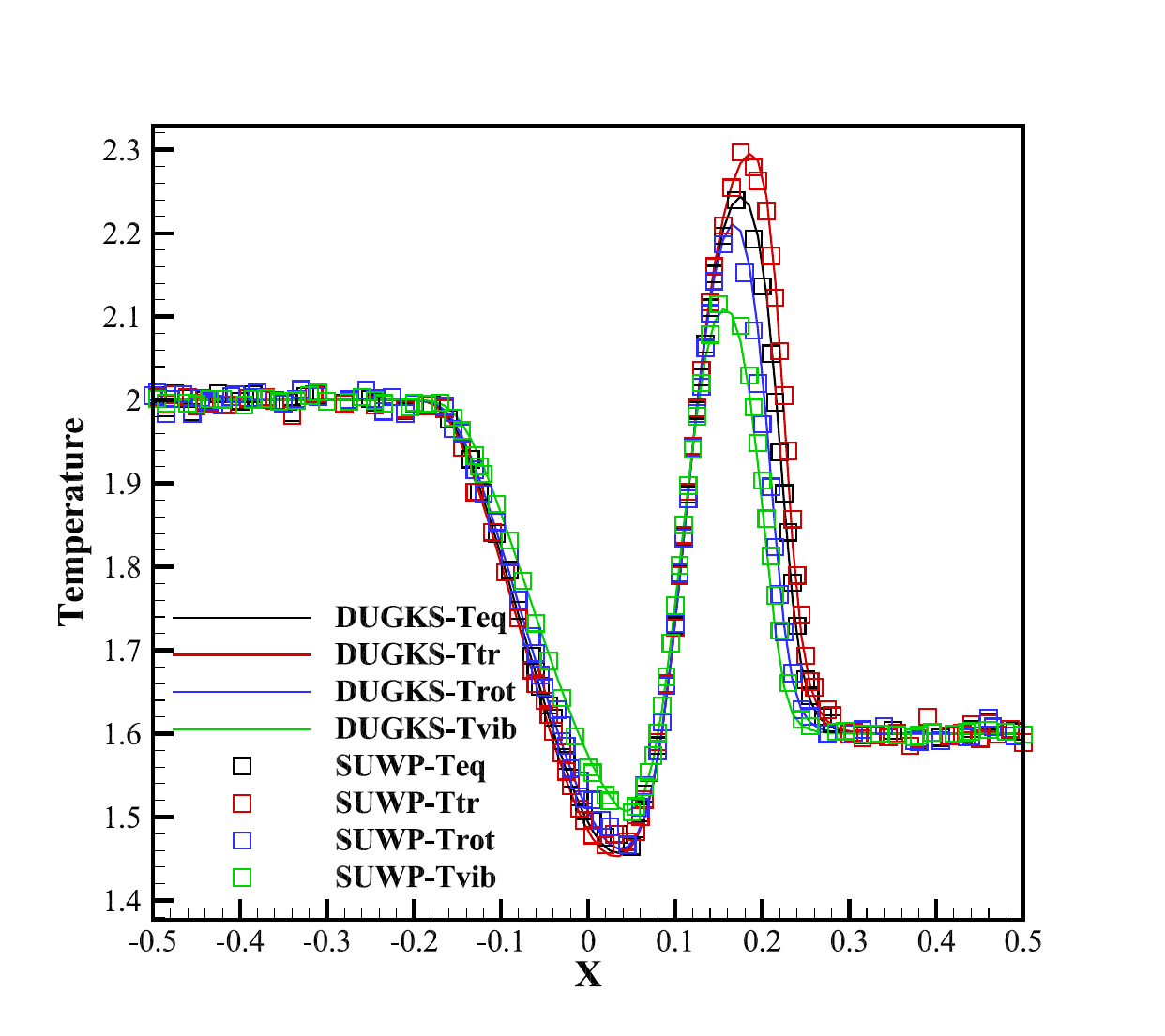}
}\\
\caption{\label{Fig:case1D_sod_kn0.001}The (a) desity, (b) velocity and (c) temperature profiles of the Sod's shock tube at Kn=0.001.}
\end{figure*}

Results at time $t=0.12$ are compared. Identical simulations were performed using the DUGKS solver accounting for rotational and vibrational internal energy non-equilibrium from Reference~\cite{ZHANG2023107079}, serving as benchmark solutions to validate the accuracy of the present method. Fig. \ref{Fig:case1D_sod_kn10}, Fig. \ref{Fig:case1D_sod_kn0.1}和Fig. \ref{Fig:case1D_sod_kn0.001}present the simulated density, velocity, equilibrium temperature, translational temperature, rotational temperature, and vibrational temperature. It can be observed that for both $\rm{Kn=10}$ and $\rm{Kn=0.1}$ cases, the vibrational temperature values are slightly higher than the rotational temperature. This trend is in agreement with the results simulated by the DUGKS method. For all cases, the simulation results of the SUWP-vib method exhibit excellent agreement with those obtained using the DUGKS method under the same physical model. This demonstrates that the SUWP-vib 
method can effectively perform numerical simulations of unsteady flows across various Knudsen numbers.

\subsection{Shock structure}
The capability of the present method for simulating strongly non-equilibrium flows is validated by simulations of one-dimensional normal shock wave structures. For diatomic gases considering rotational and vibrational degrees of freedom, the specific heat ratio is not constant across the shock wave. Therefore, the standard Rankine-Hugoniot relations based on a constant specific heat ratio fail to accurately predict the post-wave flow state. Under the assumption that all post-wave temperatures have relaxed to the equilibrium temperature, the generalized Rankine-Hugoniot relations~\cite{wang2017unified, cai2008one} are invoked to determine the flow states upstream and downstream of the shock. To accurately determine the post-wave flow state, Eq. (\ref{eq:RHvib}) must be solved by an iterative method.
\begin{equation}   \label{eq:RHvib}
\begin{aligned} 
& \frac{p_2}{p_1}=\frac{1+\gamma_1 {\rm{Ma}}_1^2}{1+\gamma_2 {\rm{Ma}}_2^2}, \\ & \frac{T_2}{T_1}=\frac{\left[\gamma_1 /\left(\gamma_1-1\right)\right]+\left(\gamma_1 / 2\right) {\rm{Ma}}_1^2}{\left[\gamma_2 /\left(\gamma_2-1\right)\right]+\left(\gamma_2 / 2\right) {\rm{Ma}}_2^2}, \\ & \frac{u_2}{u_1}=\sqrt{\frac{\gamma_2}{\gamma_1}} \frac{{\rm{Ma}}_2}{{\rm{Ma}}_1} \sqrt{\frac{\left[\gamma_1 /\left(\gamma_1-1\right)\right]+\left(\gamma_1 / 2\right) {\rm{Ma}}_1^2}{\left[\gamma_2 /\left(\gamma_2-1\right)\right]+\left(\gamma_2 / 2\right) {\rm{Ma}}_2^2}}, \\ & \frac{\left(1+\gamma_1 {\rm{Ma}}_1^2\right)^2}{\left\{\left[\gamma_1 /\left(\gamma_1-1\right)\right]+\left(\gamma_1 / 2\right) {\rm{Ma}}_1^2\right\} \gamma_1 {\rm{Ma}}_1^2}=\frac{\left(1+\gamma_2 {\rm{Ma}}_2^2\right)^2}{\left\{\left[\gamma_2 /\left(\gamma_2-1\right)\right]+\left(\gamma_2 / 2\right) {\rm{Ma}}_2^2\right\} \gamma_2 {\rm{Ma}}_2^2},
\end{aligned}
\end{equation} 
where the subscript '1' denotes upstream and '2' denotes downstream of the shock wave, respectively. The expression for the specific heat ratio $\gamma$ is given by:
\begin{equation}  
\gamma=\frac{K_{\rm{tr}}+K_{\rm{rot}}+K_{\rm{vib}}(T)+2}{K_{\rm{tr}}+K_{\rm{rot}}+K_{\rm{vib}}(T)}.
\end{equation} 

Two sets of shock structure simulations of $\rm{Ma}=10$ and $\rm{Ma}=15$ were conducted. The flow parameters for these test cases are presented in Table \ref{table:1Dcase_Shock_Table}.
\begin{table}[ht] 
    \centering
    \caption{The parameters of shock structure cases.} \label{table:1Dcase_Shock_Table}
    \begin{tabular}{ c   c  c  c  c }
        \hline
        $L_{\rm{ref}}$ & $\rho_{1}(\rm{kg/m^{3}})$ & $T_{1}(\rm{K})$ & $Z_{\rm{rot}}$ & $Z_{\rm{vib}}$ \\ 
        \hline
        $\lambda_{1}$ & $1.7413 \times 10^{-2}$ & 226.149 & 4 & 50 \\
        $\lambda_{1}$ & $1.7413 \times 10^{-2}$ & 226.149 & 5 & 25 \\
        \hline
    \end{tabular}
\end{table}

\begin{figure*}[h!t]
\centering
\subfigure[\label{Fig:case1d_shock_ma10_rho}]{
\includegraphics[width=0.3\textwidth]{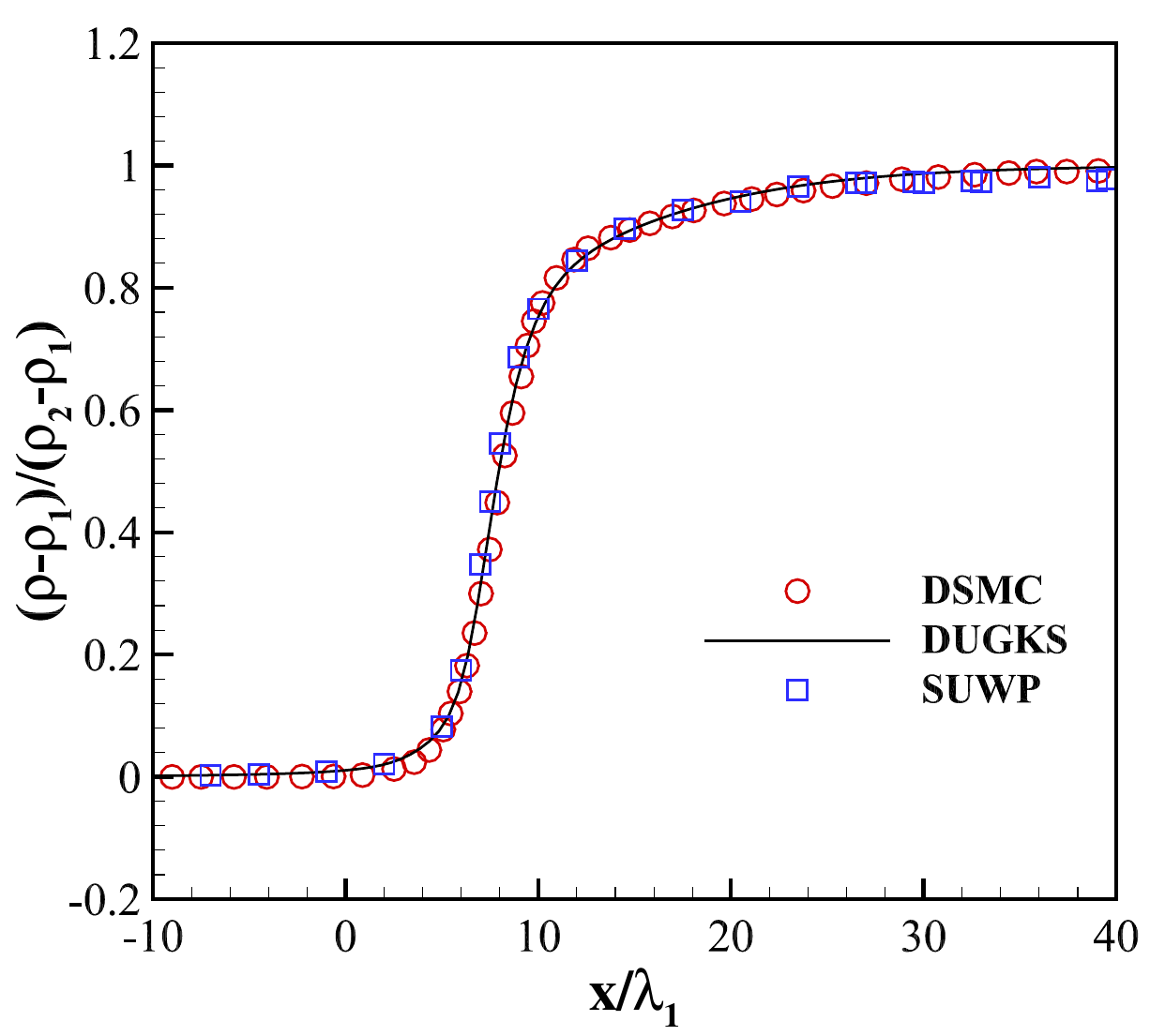}
}\hspace{0.01\textwidth}%
\subfigure[\label{Fig:case1d_shock_ma10_Ttr}]{
\includegraphics[width=0.3\textwidth]{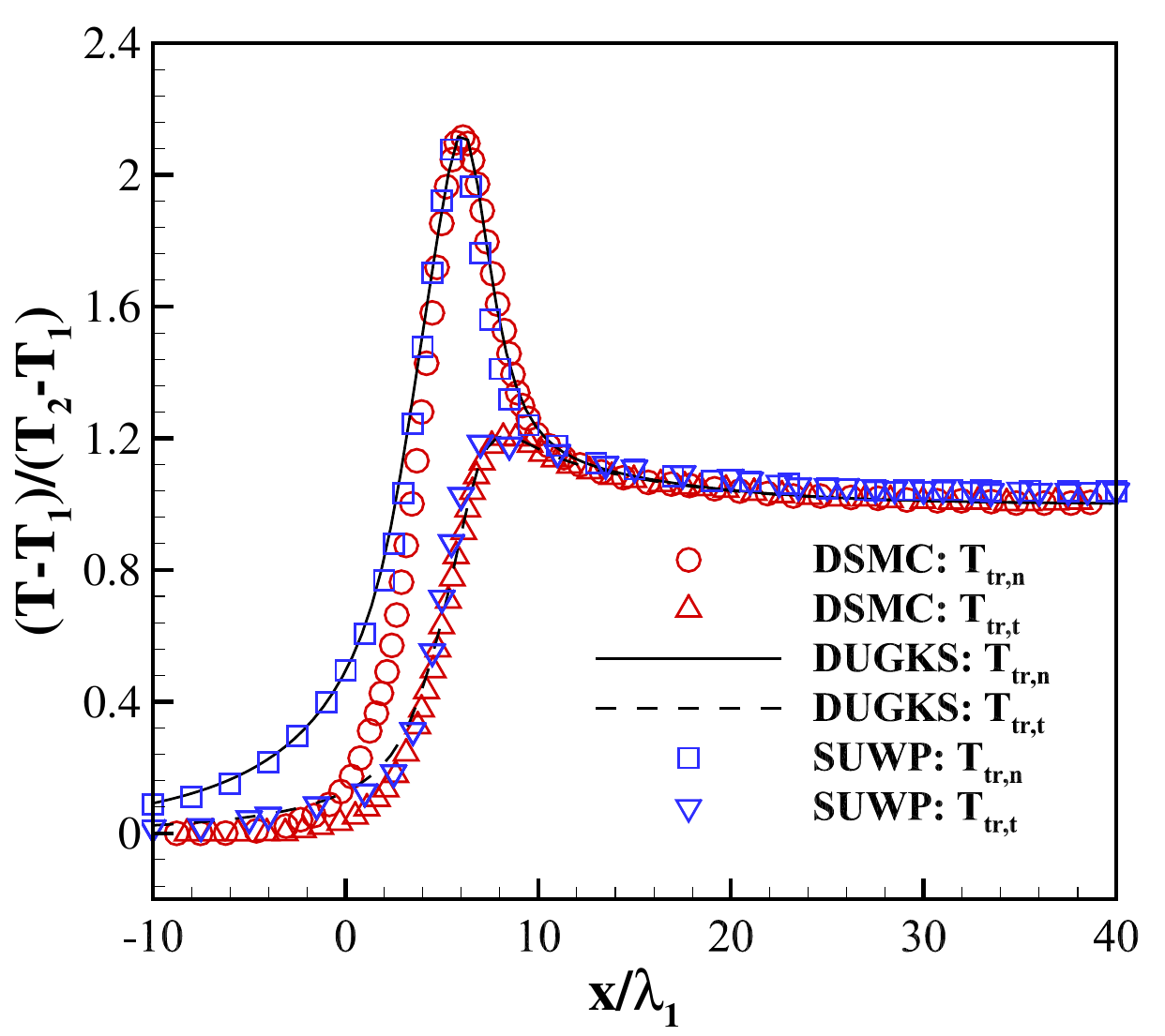}
}\hspace{0.01\textwidth}%
\subfigure[\label{Fig:case1d_shock_ma10_Trot}]{
\includegraphics[width=0.3\textwidth]{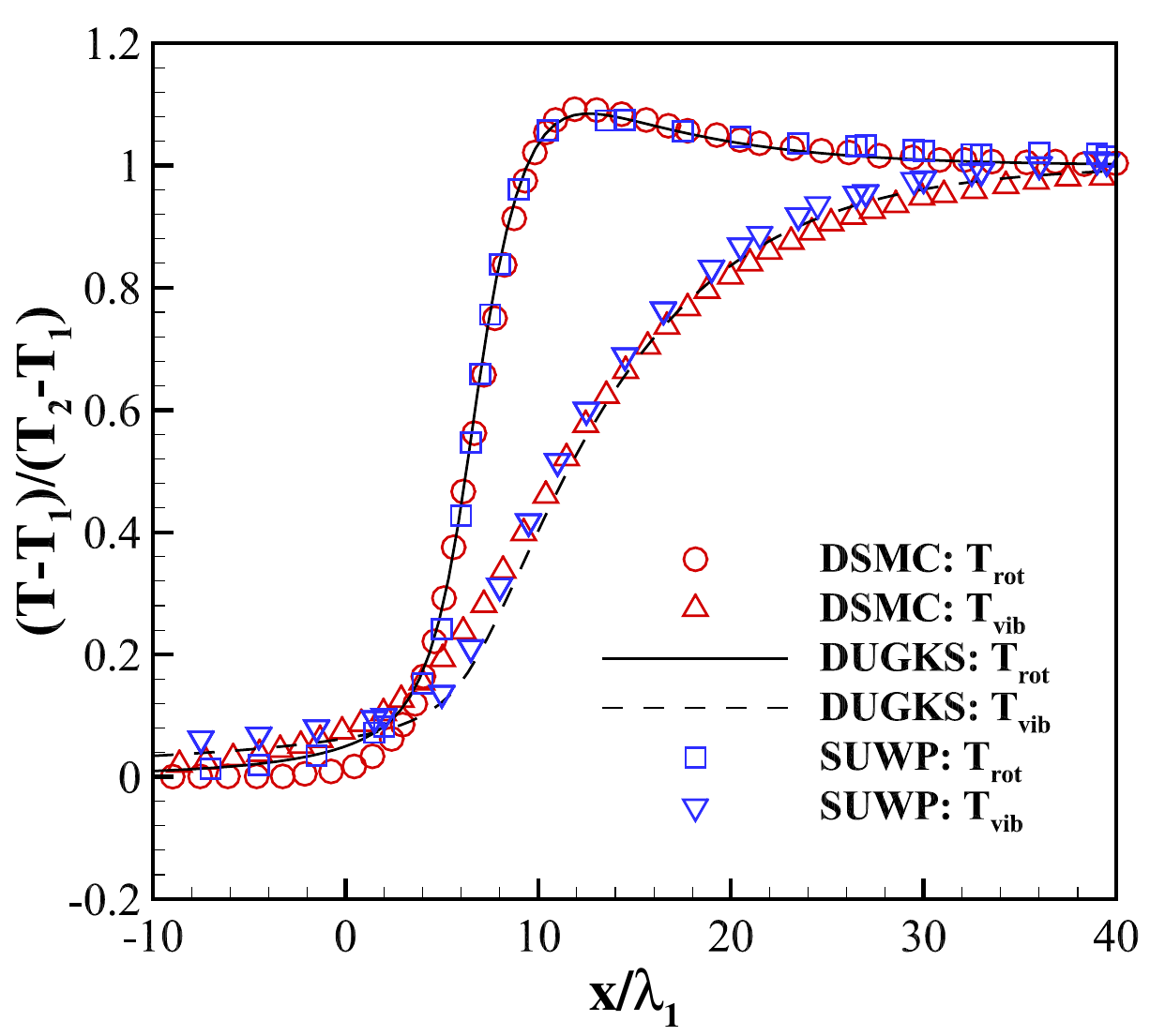}
}\\
\caption{\label{Fig:case1d_shock_ma10}The (a) desity, (b) translational temperature, (c) rotational temperature and vibrational temperature profiles of the Nitrogen gas shock structure at Ma=10.}
\end{figure*}

\begin{figure*}[h!t]
\centering
\subfigure[\label{Fig:case1d_shock_ma15_rho}]{
\includegraphics[width=0.3\textwidth]{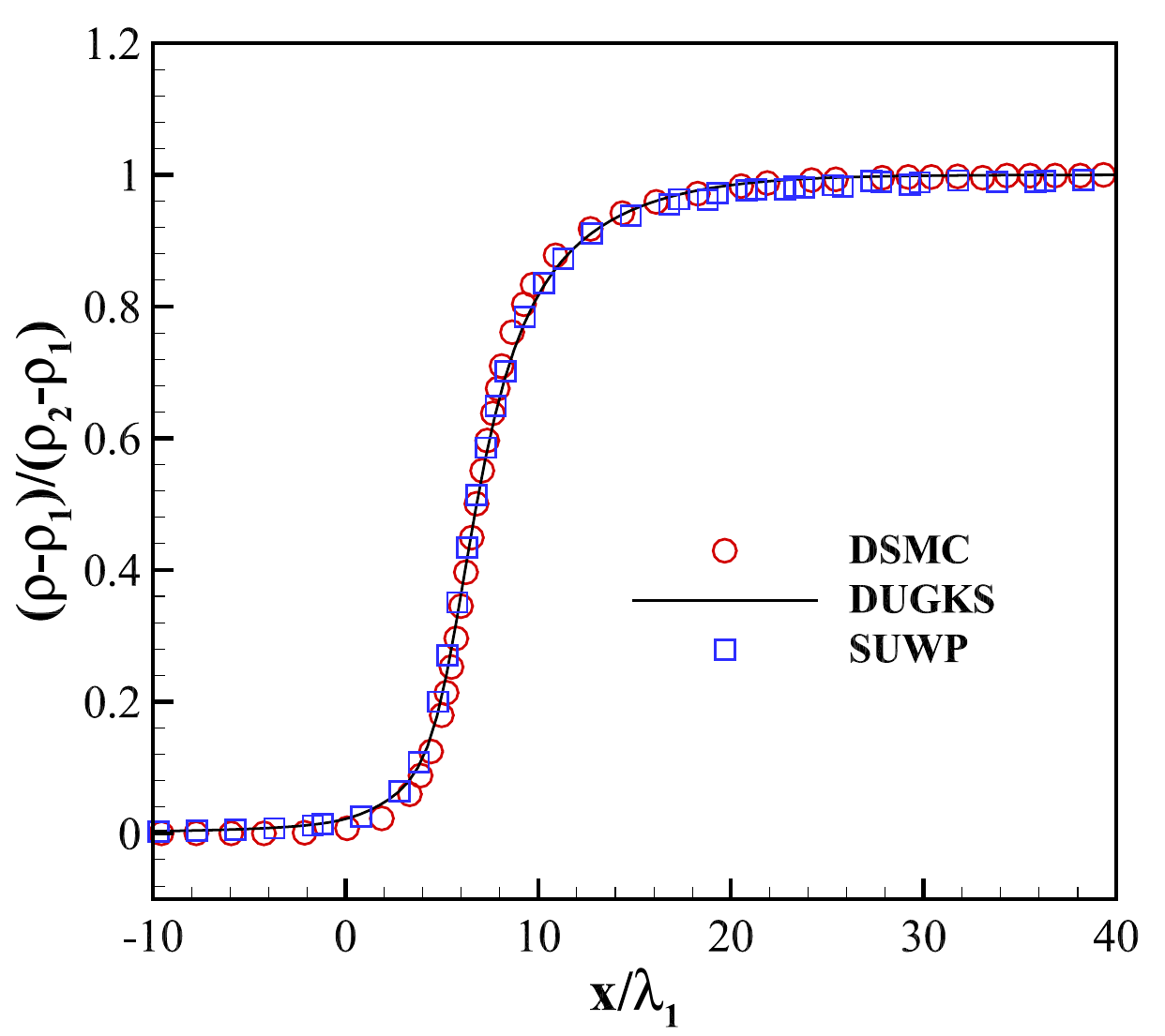}
}\hspace{0.01\textwidth}%
\subfigure[\label{Fig:case1d_shock_ma15_Ttr}]{
\includegraphics[width=0.3\textwidth]{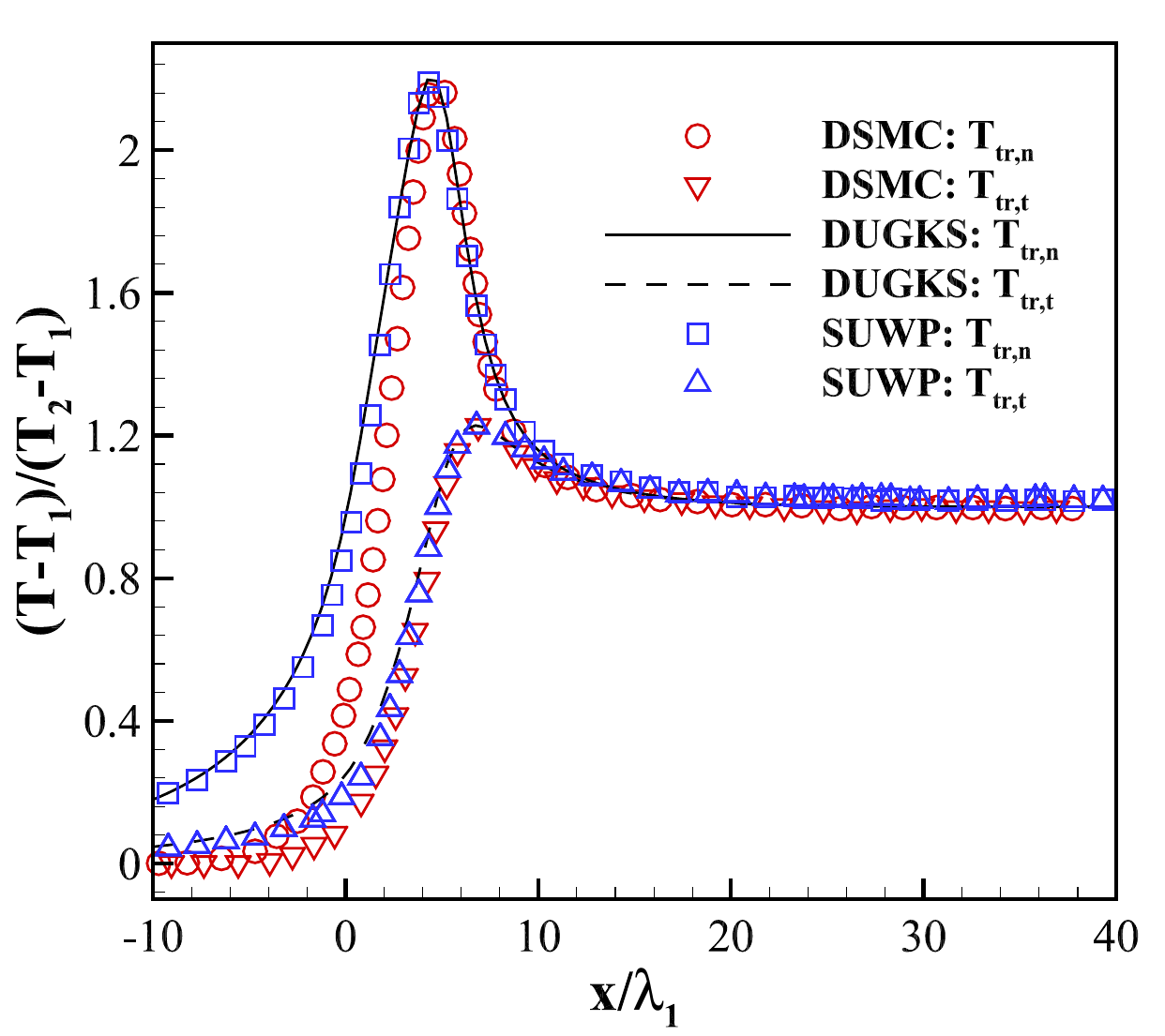}
}\hspace{0.01\textwidth}%
\subfigure[\label{Fig:case1d_shock_ma15_Trot}]{
\includegraphics[width=0.3\textwidth]{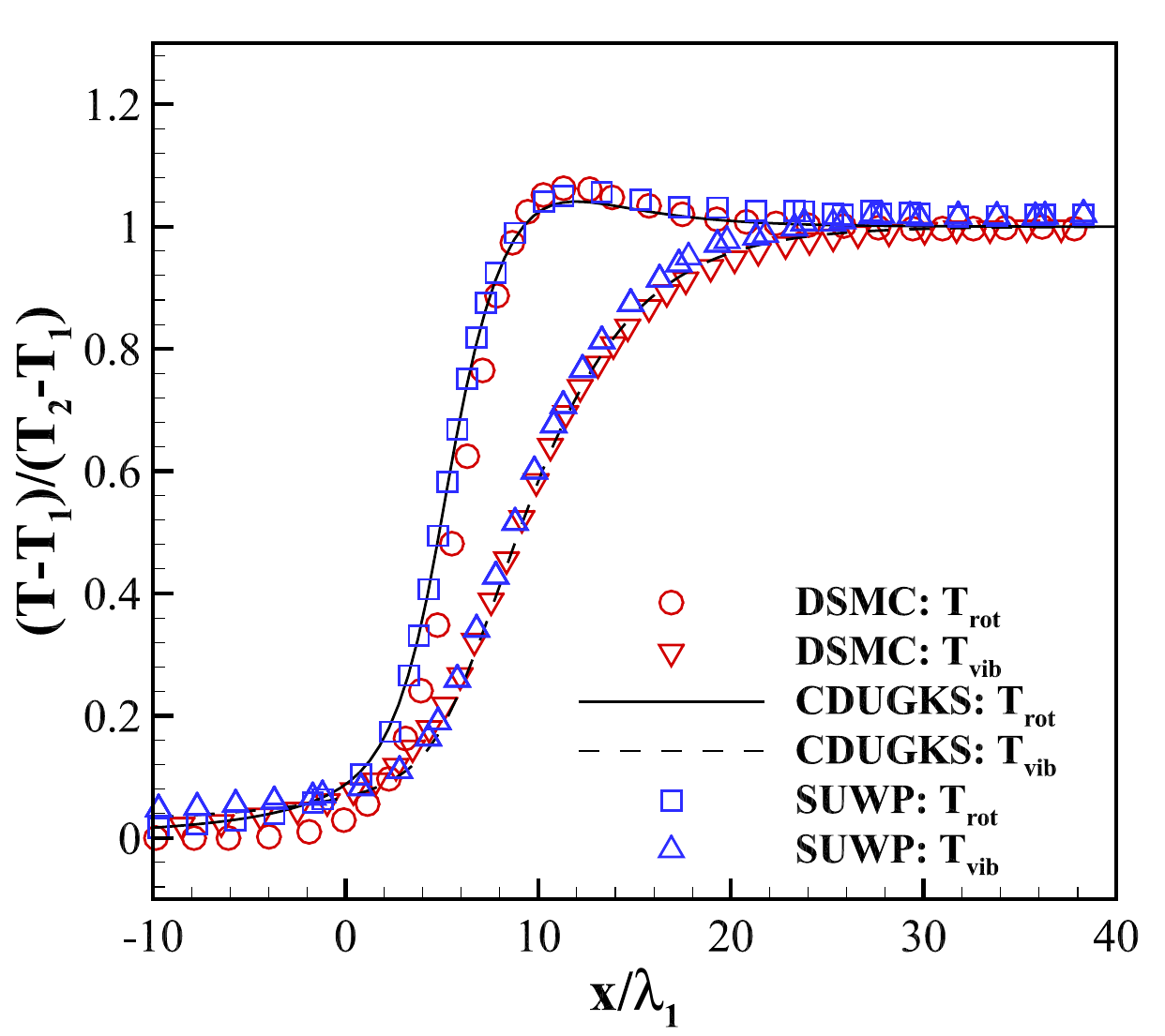}
}\\
\caption{\label{Fig:case1d_shock_ma15}The (a) desity, (b) translational temperature, (c) rotational temperature and vibrational temperature profiles of the Nitrogen gas shock structure at Ma=15.}
\end{figure*}

Non-dimensional variables are employed throughout the computations. The non-dimensionalization of cases is performed by Eq. (\ref{eq:case_noDim02}). The upstream mean free path $\lambda_{1}$ is calculated by Eq. (\ref{eq:shock_lambda1}). The $\mu_{1}$ is determined by substituting the upstream temperature into Eq. (\ref{eq:muEqn}).
The non-dimensional flow conditions for both upstream and downstream states are presented in Table \ref{table:1Dcase_preAndpost_Shock_Table}. The computational domain is set to $[-100\lambda_{1}, 100\lambda_{1}]$ and uniformly discretized into 400 cells. Therefore, each cell has a size of $\Delta x = 0.5\lambda_{1}$. The particle number of cell $N_{\rm{p}} =3 \times 10^{3}$. The results of the present method are compared against DSMC and DUGKS results from Ref.~\cite{ZHANG2023107079} to validate the accuracy of the SUWP-vib method. Fig. \ref{Fig:case1d_shock_ma10} and Fig. \ref{Fig:case1d_shock_ma15} present the computed density, translational temperature, rotational temperature, and vibrational temperature profiles. It is observed that the SUWP-vib method shows good agreement with the DUGKS method and correctly captures the temperature evolution in the shock structure. The translational temperature normal component $T_{\rm{tr,n}}=P_{xx}/(\rho R)$ and tangential component $T_{\rm{tr,t}}=P_{yy}/(\rho R)$ are obtained from the components of the pressure tensor $P_{ij}$. Compared to DSMC results, both the SUWP-vib and DUGKS methods exhibit a noticeable 'early rise' in the normal component of translational temperature in the shock structure. In Ref.~\cite{wei2024unified}, Wei et al. mitigated this phenomenon by modifying the relaxation time, resulting in enhanced agreement between simulations and DSMC benchmarks. To thoroughly assess the inherent characteristics of the SUWP-vib method, no modifications were made to the relaxation time in the current work. However, this technique can be incorporated in future studies to improve the 'early rise' problem.

\begin{table}[ht] 
    \centering
    \caption{The parameters of pre-shock equilibrium state and post-shock equilibrium state for shock structure.} \label{table:1Dcase_preAndpost_Shock_Table}%
    \begin{tabular}{l p{1.3cm} p{1.3cm} p{1.3cm} p{1.3cm} p{1.3cm} p{1.3cm} p{1.3cm} p{1.3cm} p{1.3cm} p{1.3cm} }
        \hline
        & $\rm{Ma}_{1}$ & $\rho_{1}$ & $T_{1}$ & $U_{1}$ & $\gamma_{1}$ & $\rm{Ma}_{2}$ & $\rho_{2}$ & $T_{2}$ & $U_{2}$ & $\gamma_{2}$ \\ 
        \hline
        & 10 & 1.0000 & 1.0000 & 8.3666 & 1.4000 & 0.3532 & 7.0544 & 17.174 & 1.1860 & 1.3127 \\
        & 15 & 1.0000 & 1.0000 & 12.550 & 1.4000 & 0.3426 & 7.5345 & 36.391 & 1.6656 & 1.2987 \\
        \hline
    \end{tabular}
\end{table}

\subsection{Flow past a cylinder}
\begin{figure*}[h!t]
\centering
\includegraphics[width=0.45\textwidth]{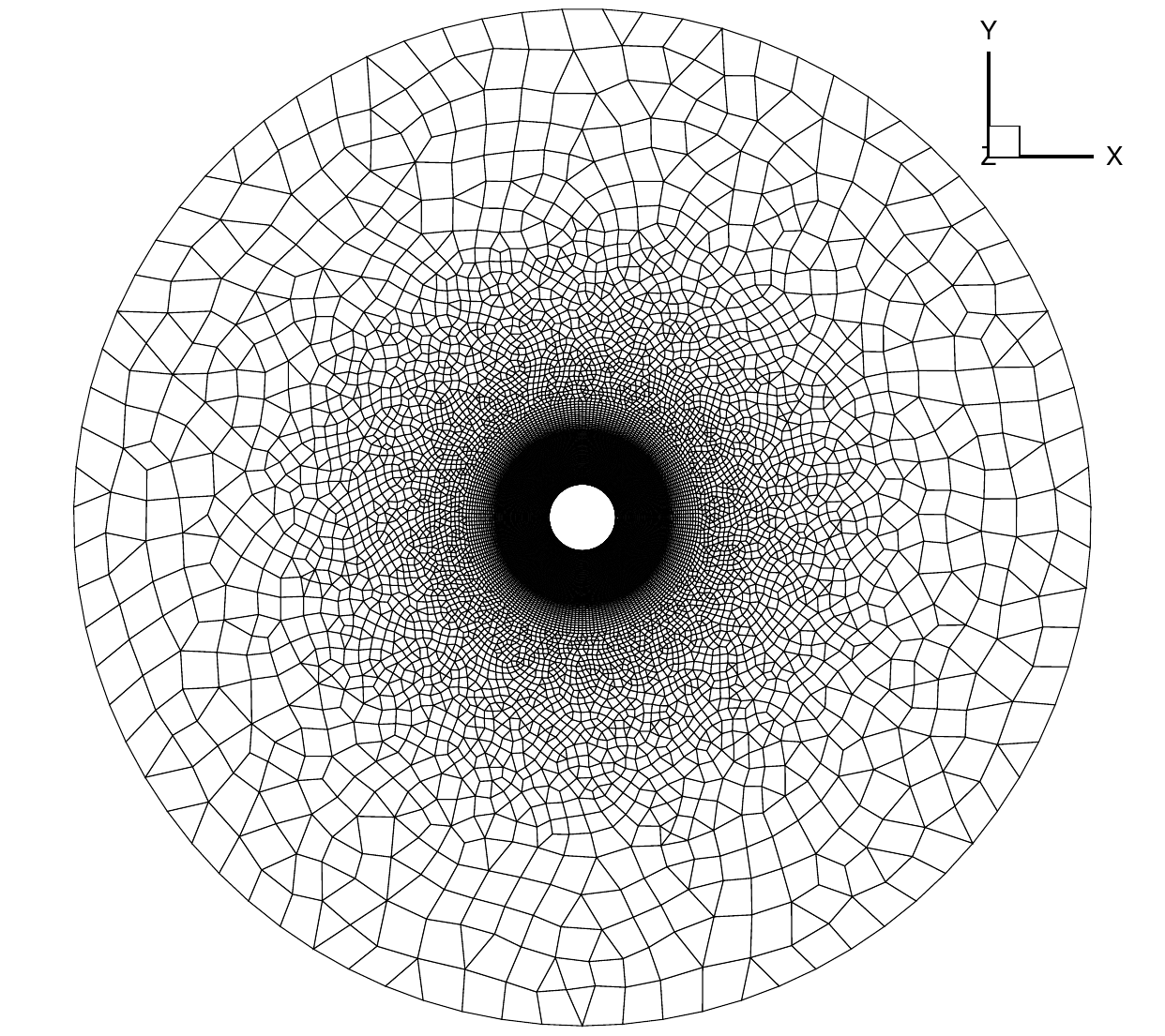}
\caption{\label{Fig:case2d_cyl_mesh}The mesh of the flow past a cylinder.}
\end{figure*}

The hypersonic flow past a cylinder is employed to validate the reliability of the SUWP-vib method for simulating high-speed viscous flows. When high-speed rarefied gas flows over an object, distinct wall slip phenomena arise due to translational non-equilibrium. Additionally, the separation between translational, rotational, and vibrational temperatures within shock layers and boundary layers induces significant thermodynamic non-equilibrium. Following simulations of the cylinder flow using the SUWP-vib method, the results were systematically compared against DSMC and DUGKS solutions computed under identical conditions~\cite{ZHANG2023107079}.
In the simulation, the cylinder diameter $D_{\rm{cyl}}$ is set to 1 m. The computational domain spans approximately 15 m in diameter and is discretized into 29,657 unstructured mesh cells, as shown in Fig. \ref{Fig:case2d_cyl_mesh}. The height of the cell adjacent wall for the cylinder is set to $4 \times 10^{-4}$ m. Unstructured meshing is implemented in off-wall regions to minimize cell number. Two simulation sets were established for freestream Mach numbers $\rm{Ma}_{\infty} = 5$ and $20$. The flow parameters for these test cases are detailed in Table \ref{table:2Dcase_cyl_Table}. The rotational collision number was computed using Parker's formula~\cite{parker1959rotational}. This test case employs dimensional variables for computation and comparison. The particle number in cell $N_{\rm{p}} =1.2 \times 10^{2}$.

\begin{figure*}[h!t]
\centering
\subfigure[\label{Fig:case2d_cyl_ma5_con_Ptr}]{
\includegraphics[trim=30 25 30 60, clip, width=0.45\textwidth]{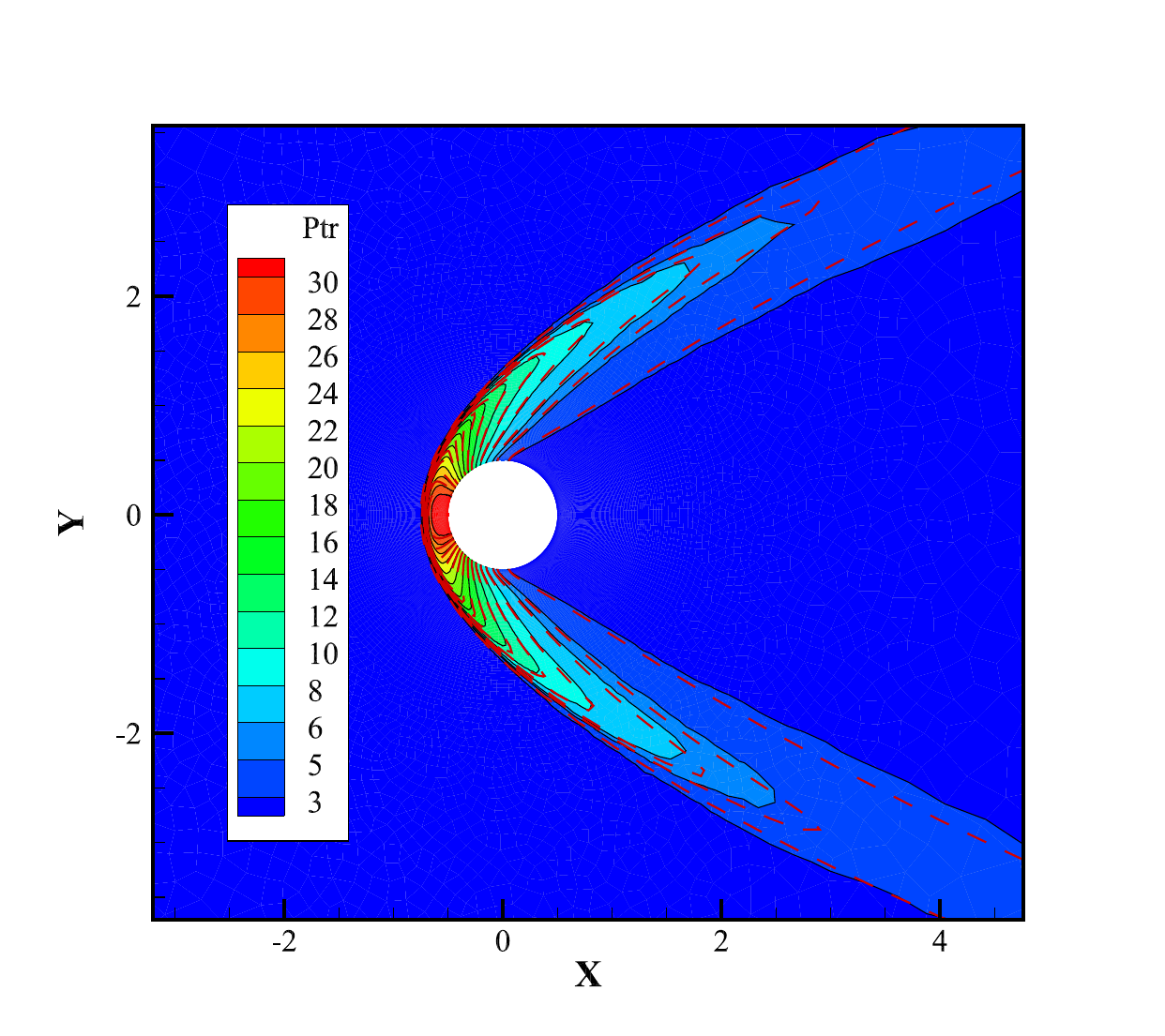}
}\hspace{0.01\textwidth}%
\subfigure[\label{Fig:case2d_cyl_ma5_con_Ma}]{
\includegraphics[trim=30 25 30 60, clip, width=0.45\textwidth]{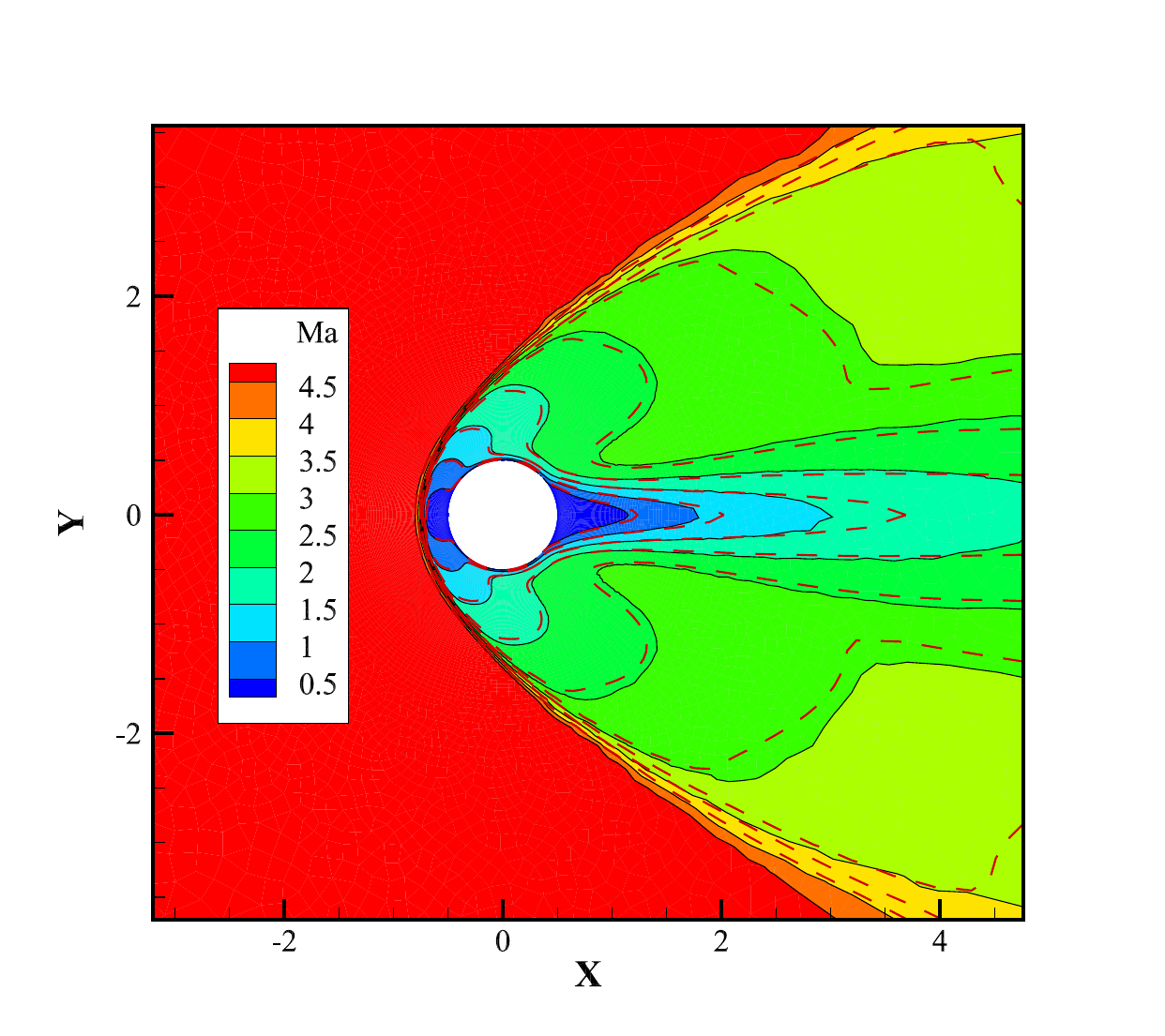}
}\\
\subfigure[\label{Fig:case2d_cyl_ma5_con_Teq}]{
\includegraphics[trim=30 25 30 60, clip, width=0.45\textwidth]{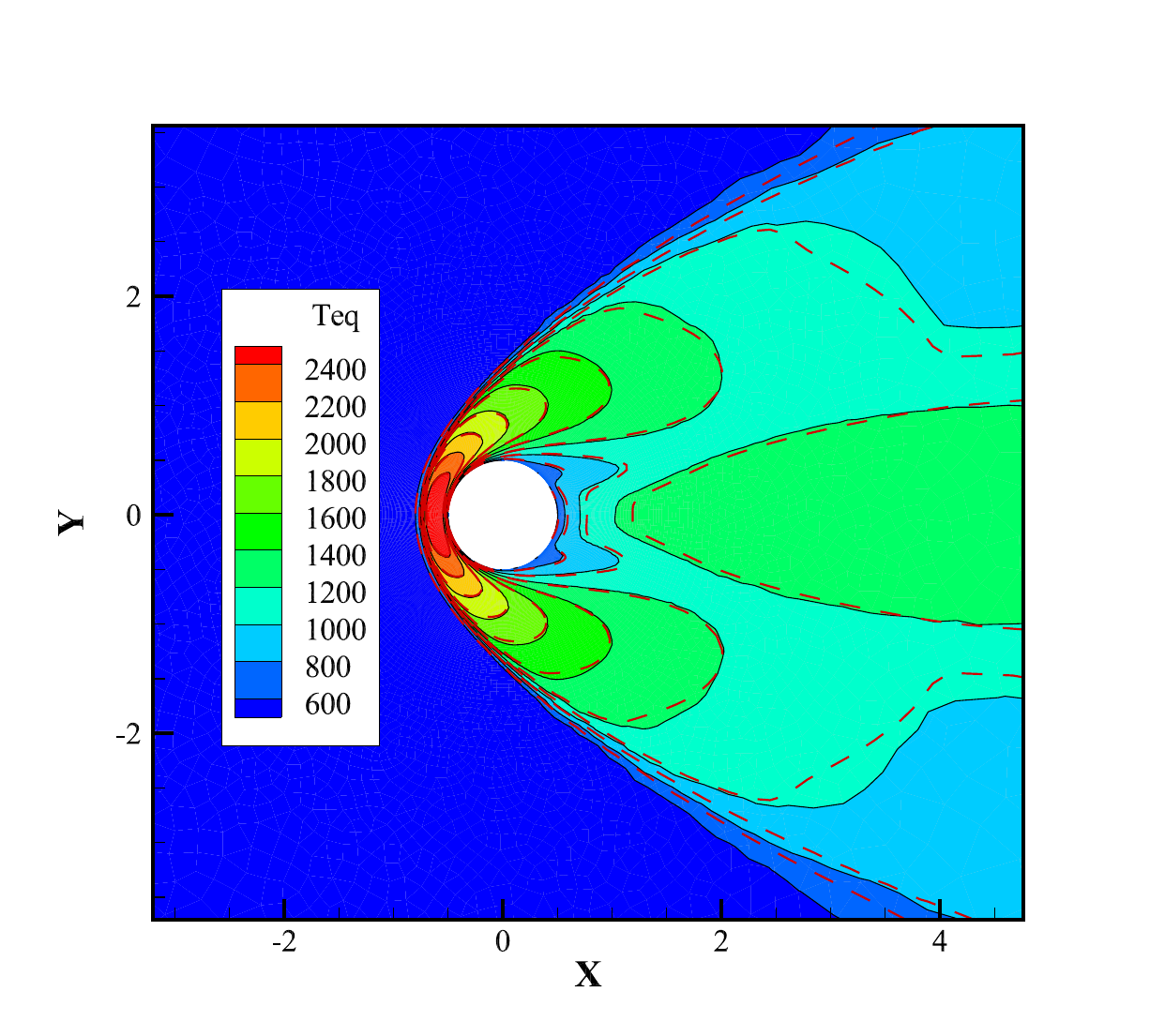}
}\hspace{0.01\textwidth}%
\subfigure[\label{Fig:case2d_cyl_ma5_con_Ttr}]{
\includegraphics[trim=30 25 30 60, clip, width=0.45\textwidth]{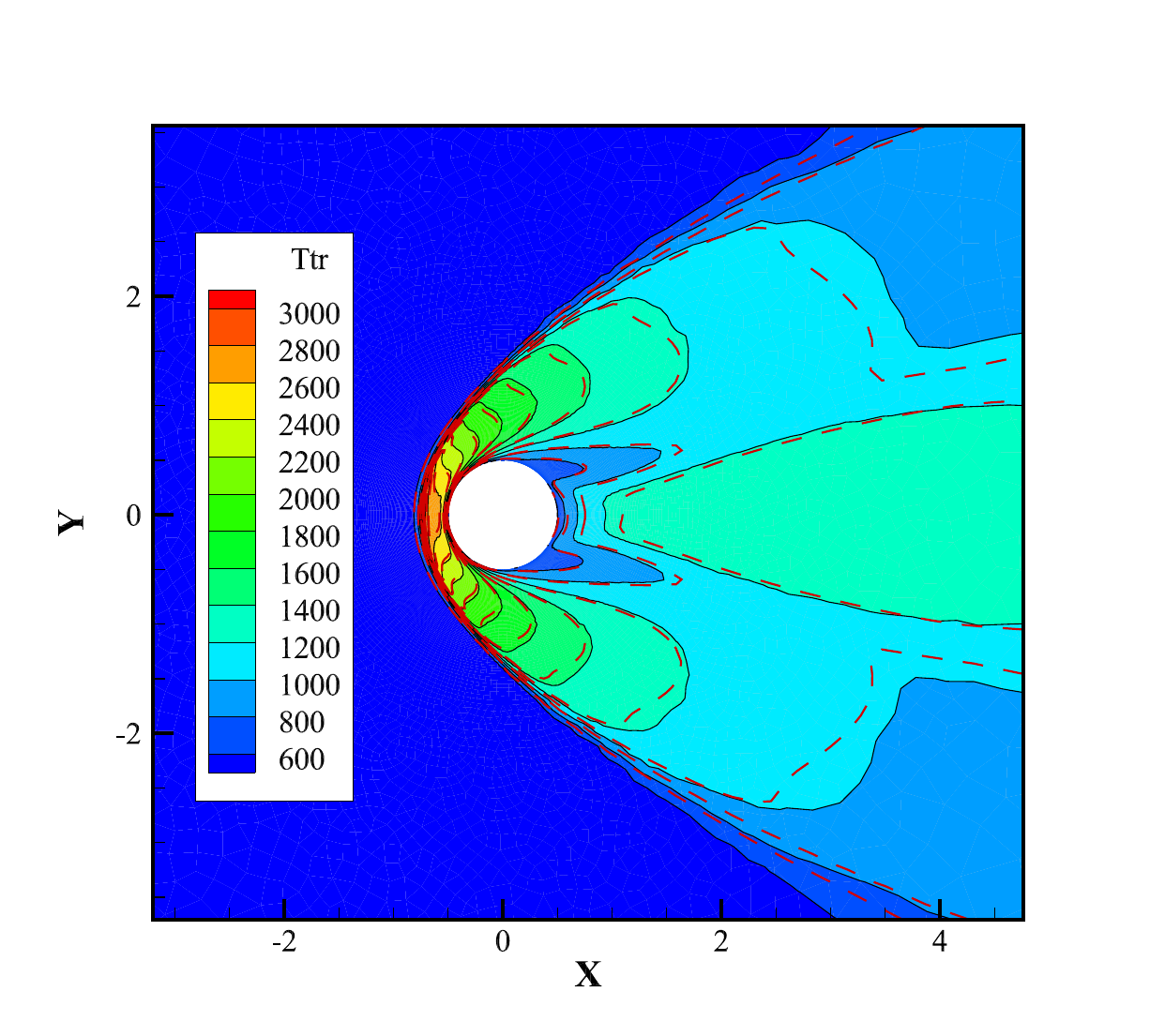}
}\\
\subfigure[\label{Fig:case2d_cyl_ma5_con_Trot}]{
\includegraphics[trim=30 25 30 60, clip, width=0.45\textwidth]{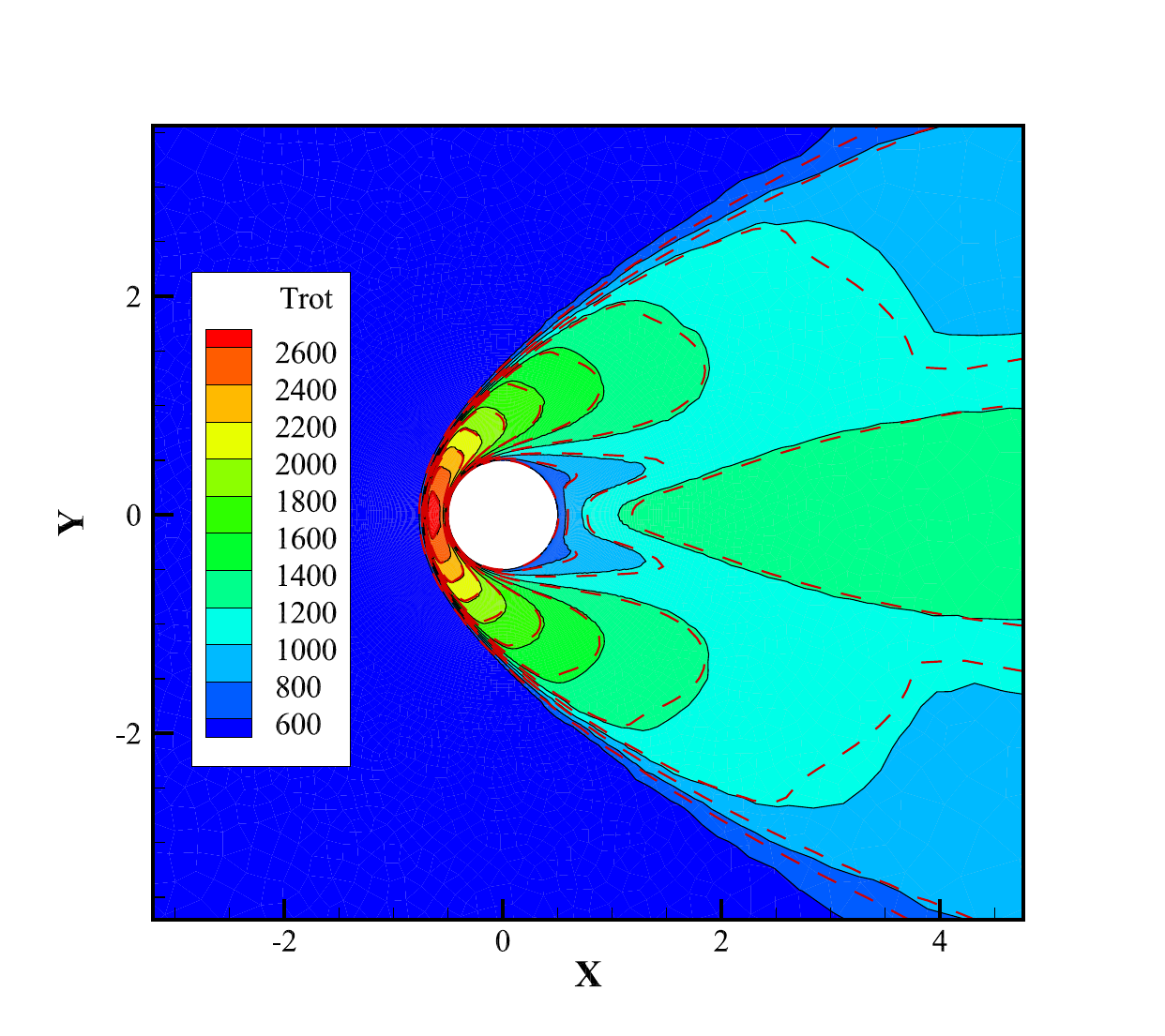}
}\hspace{0.01\textwidth}%
\subfigure[\label{Fig:case2d_cyl_ma5_con_Tvib}]{
\includegraphics[trim=30 25 30 60, clip, width=0.45\textwidth]{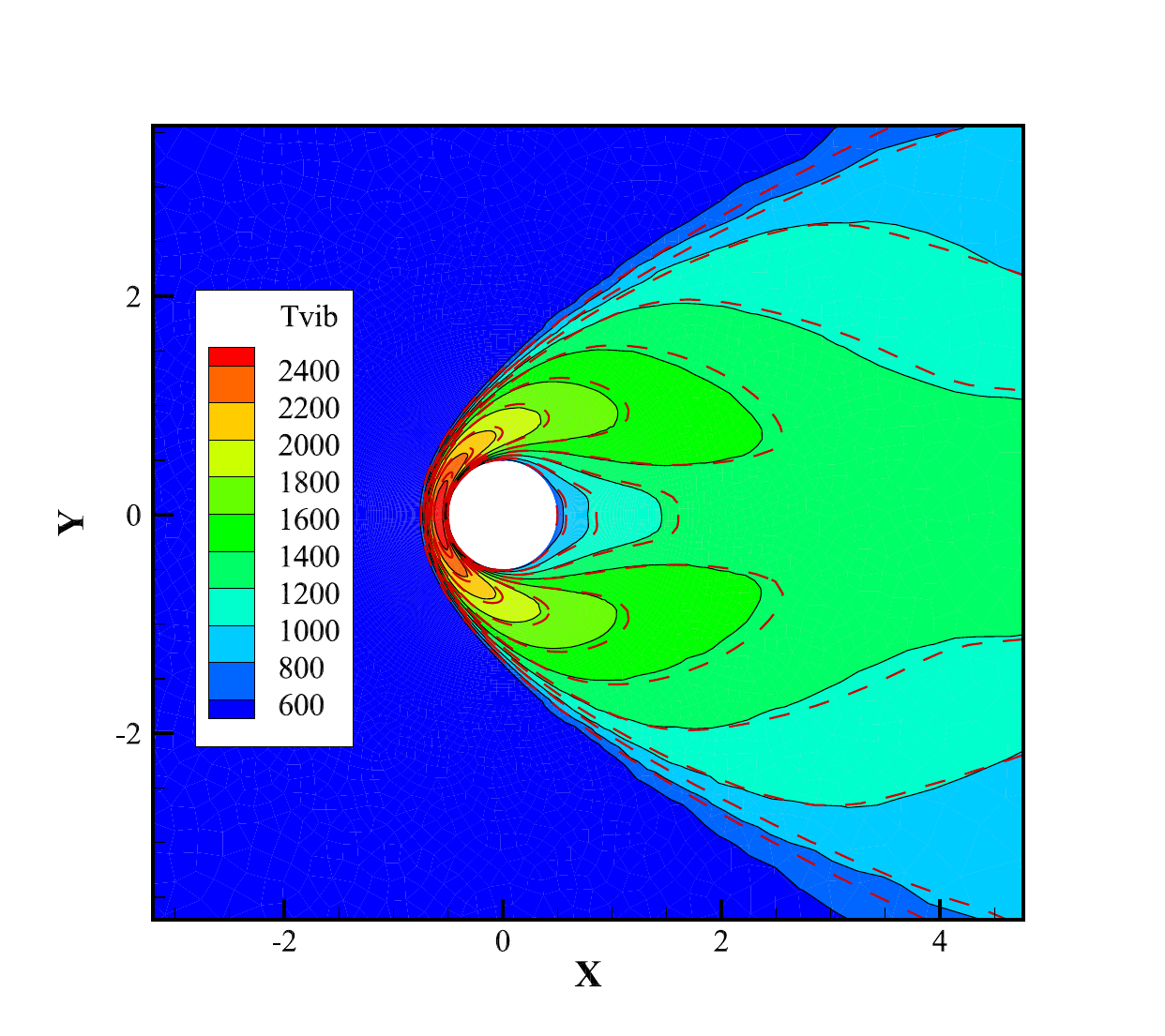}
}\\
\caption{\label{Fig:case2d_cyl_ma5_con}The (a) pressure, (b) Mach number, (c) equilibrium temperature, (d) translational temperature, (e) rotational temperature, and (f) vibrational temperature contours of the flow past a cylinder at Ma=5 (The color band: SUWP, red dash line: DUGKS).}
\end{figure*}

\begin{figure*}[h!t]
\centering
\subfigure[\label{Fig:case2d_cyl_ma5_sl_rho}]{
\includegraphics[width=0.45\textwidth]{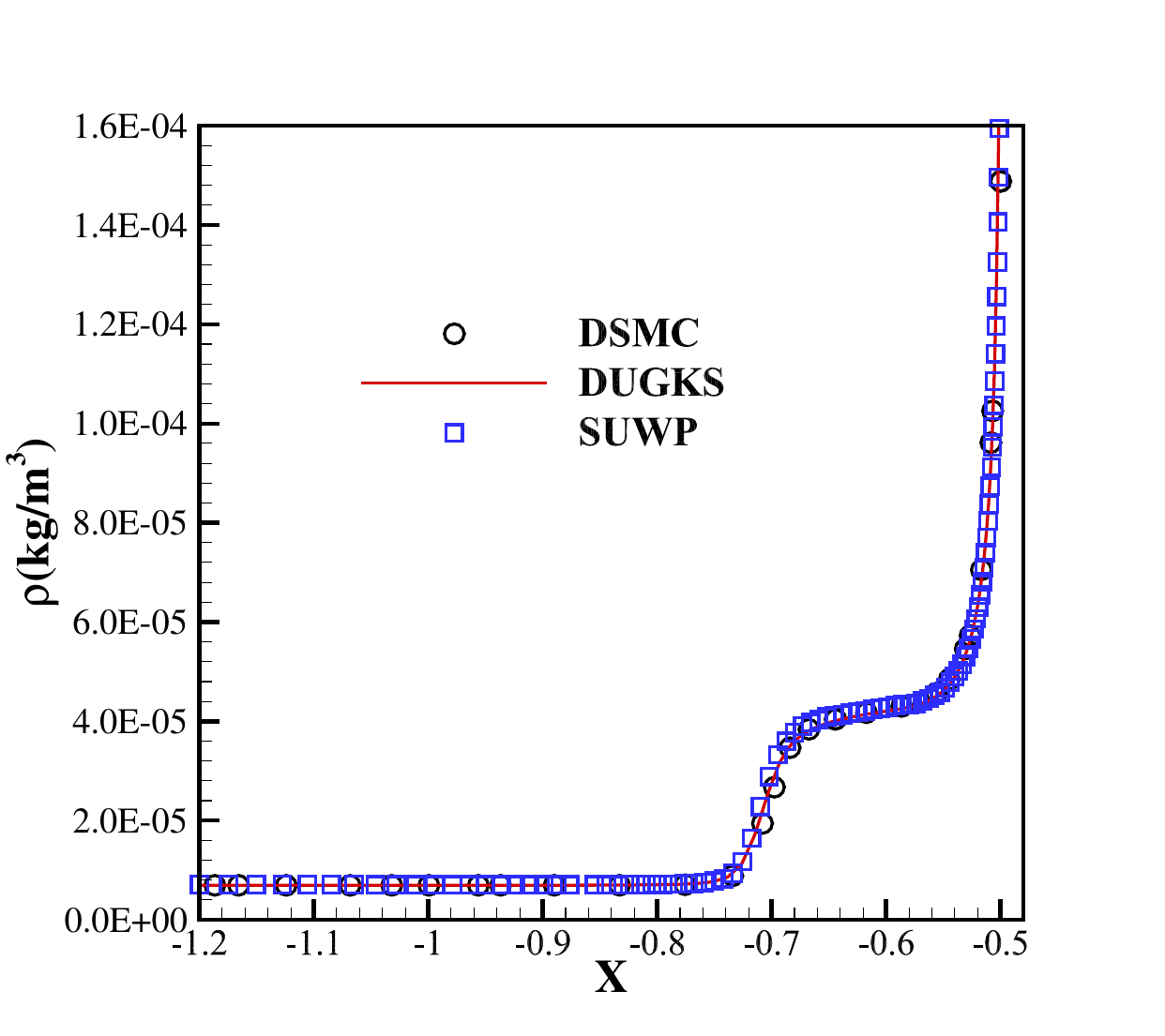}
}\hspace{0.01\textwidth}%
\subfigure[\label{Fig:case2d_cyl_ma5_sl_T}]{
\includegraphics[width=0.45\textwidth]{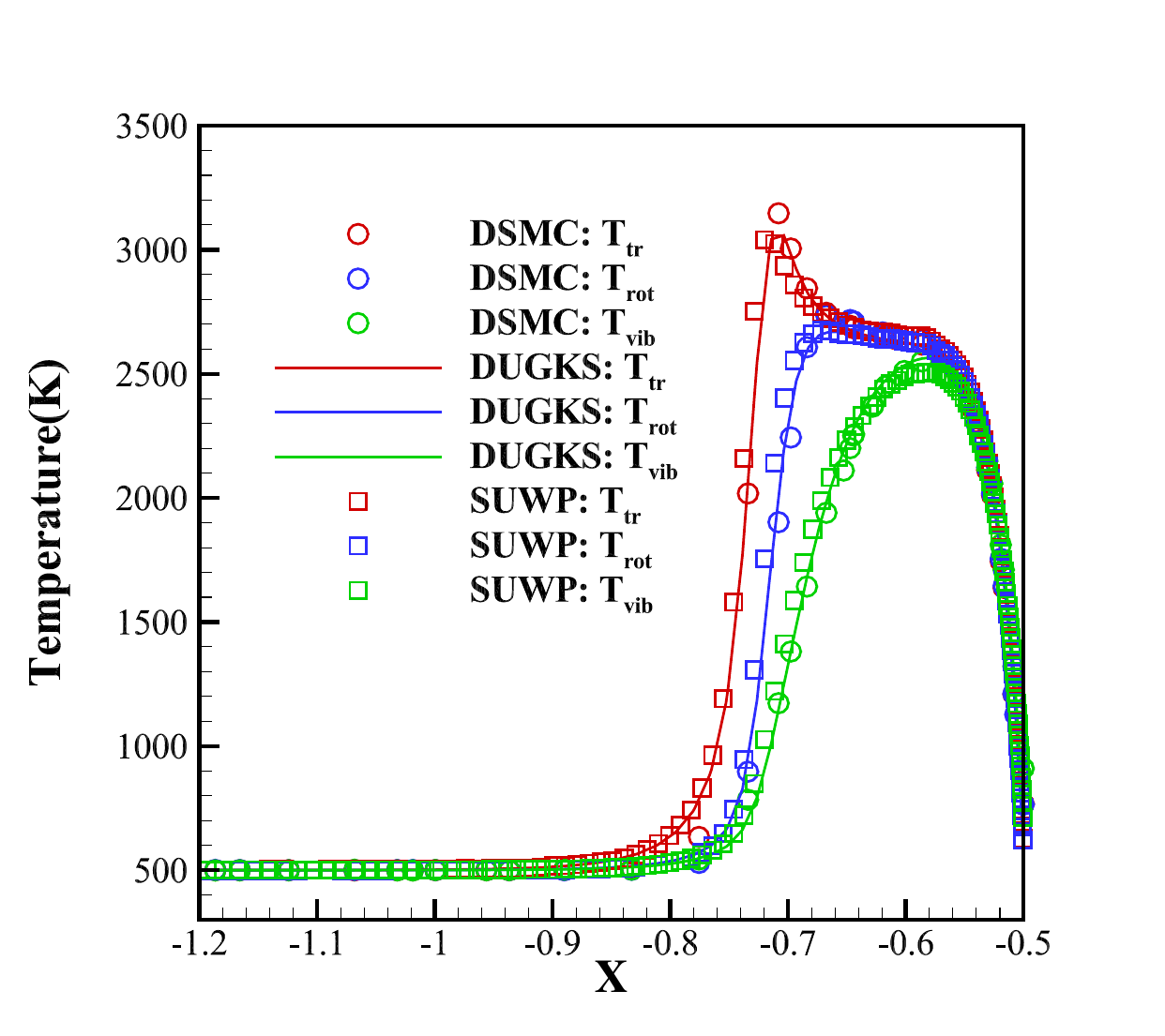}
}\\
\caption{\label{Fig:case2d_cyl_ma5_sl}The (a) density and (b) temperature along the forward stagnation line of the cylinder at Ma=5.}
\end{figure*}

\begin{figure*}[h!t]
\centering
\subfigure[\label{Fig:case2d_cyl_ma5_wall_cp}]{
\includegraphics[width=0.45\textwidth]{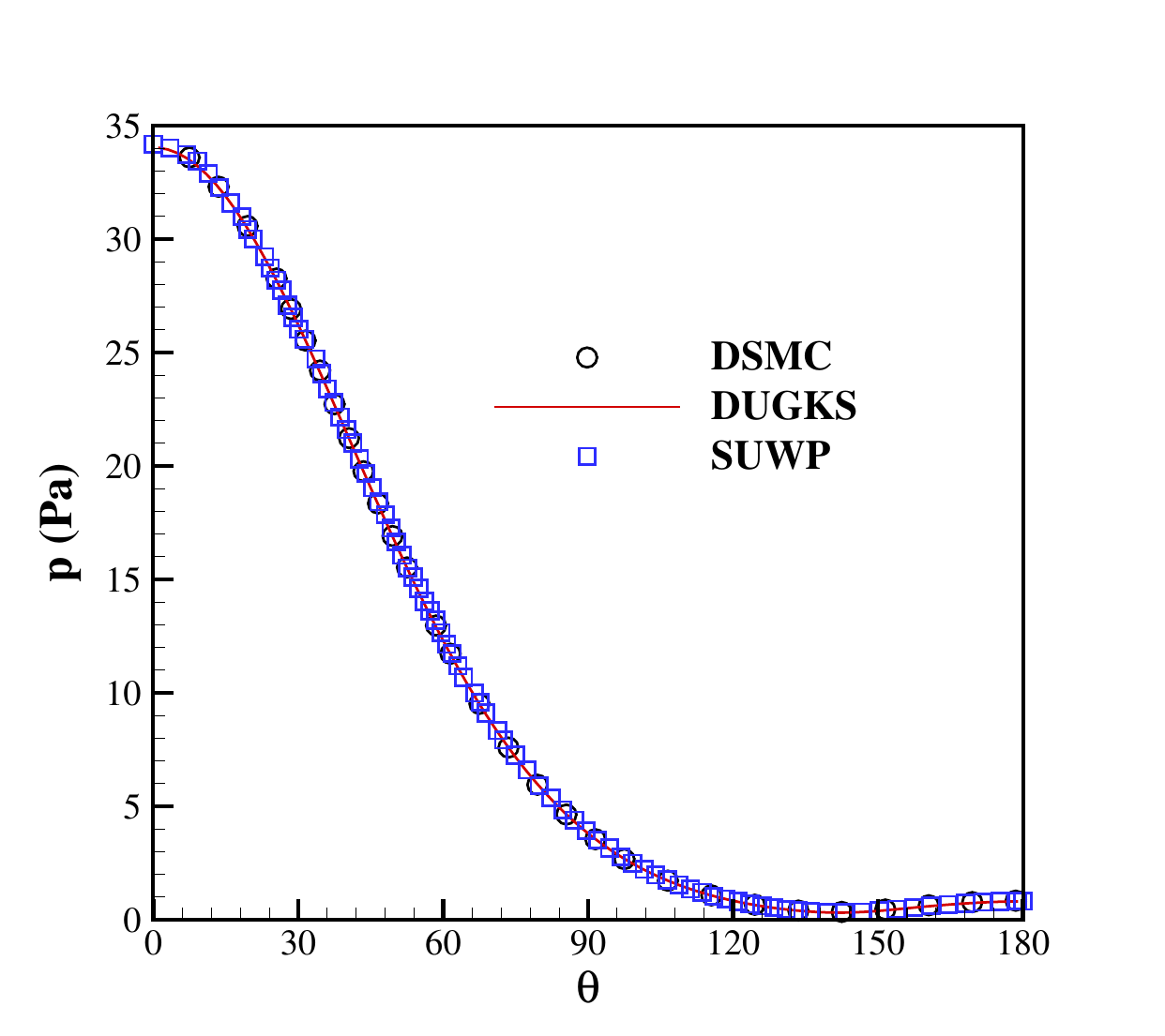}
}\hspace{0.01\textwidth}%
\subfigure[\label{Fig:case2d_cyl_ma5_wall_ch}]{
\includegraphics[width=0.45\textwidth]{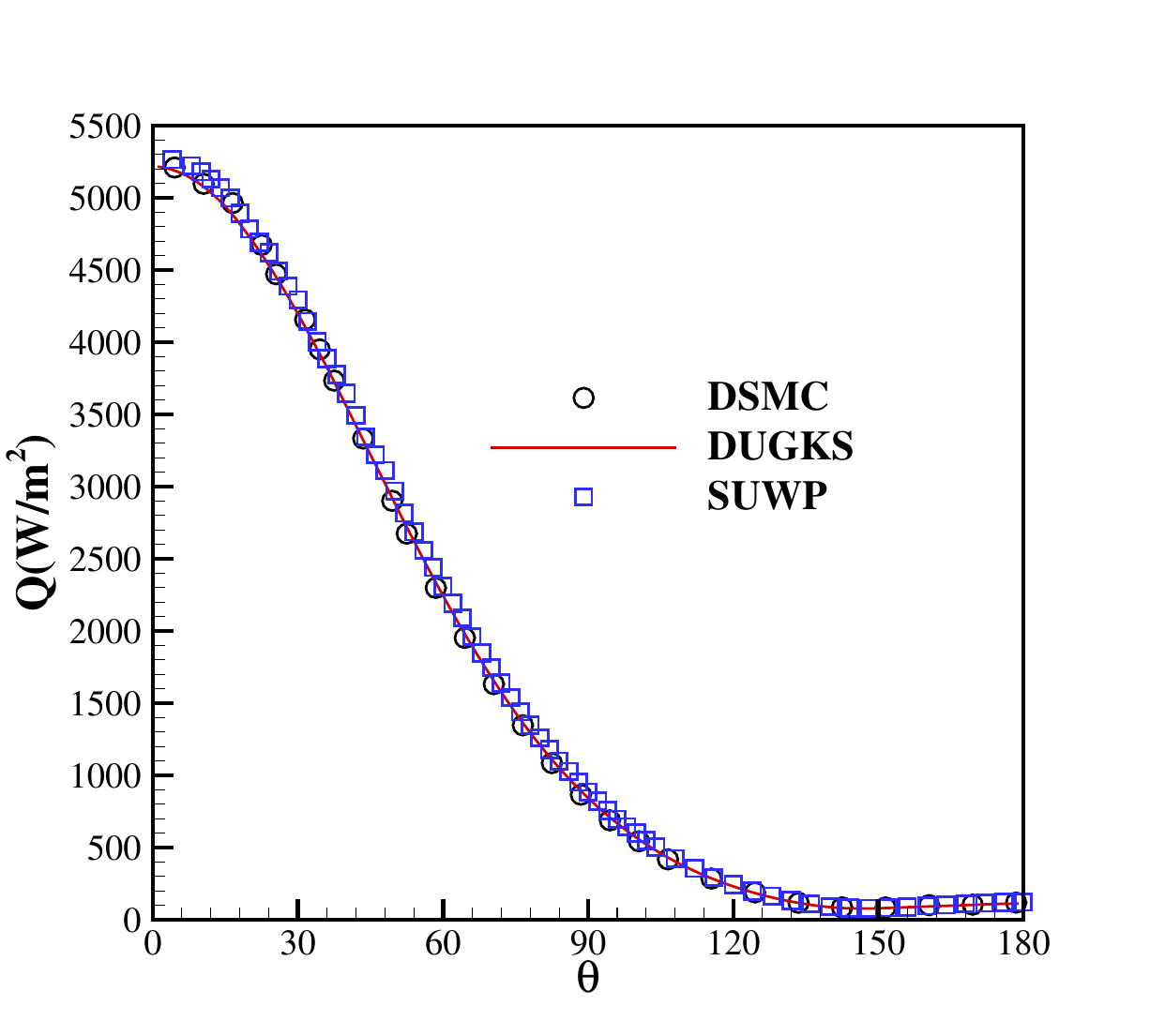}
}\\
\caption{\label{Fig:case2d_cyl_ma5_wall}The (a) pressure and (b) heat flux on the wall surface of the cylinder at Ma=5.}
\end{figure*}

\begin{figure*}[h!t]
\centering
\subfigure[\label{Fig:case2d_cyl_ma20_con_Ptr}]{
\includegraphics[trim=30 25 30 60, clip, width=0.45\textwidth]{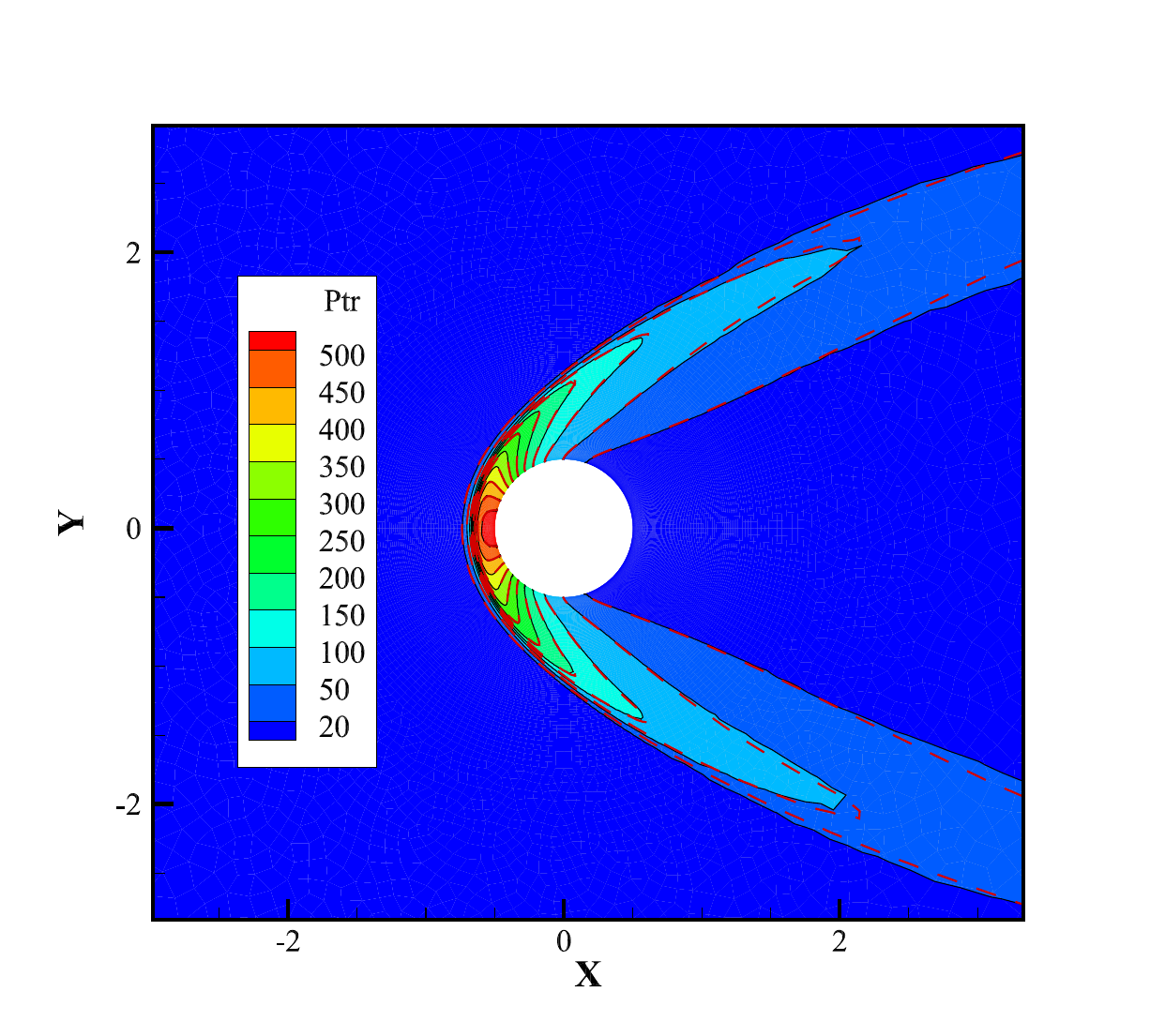}
}\hspace{0.01\textwidth}%
\subfigure[\label{Fig:case2d_cyl_ma20_con_Ma}]{
\includegraphics[trim=30 25 30 60, clip, width=0.45\textwidth]{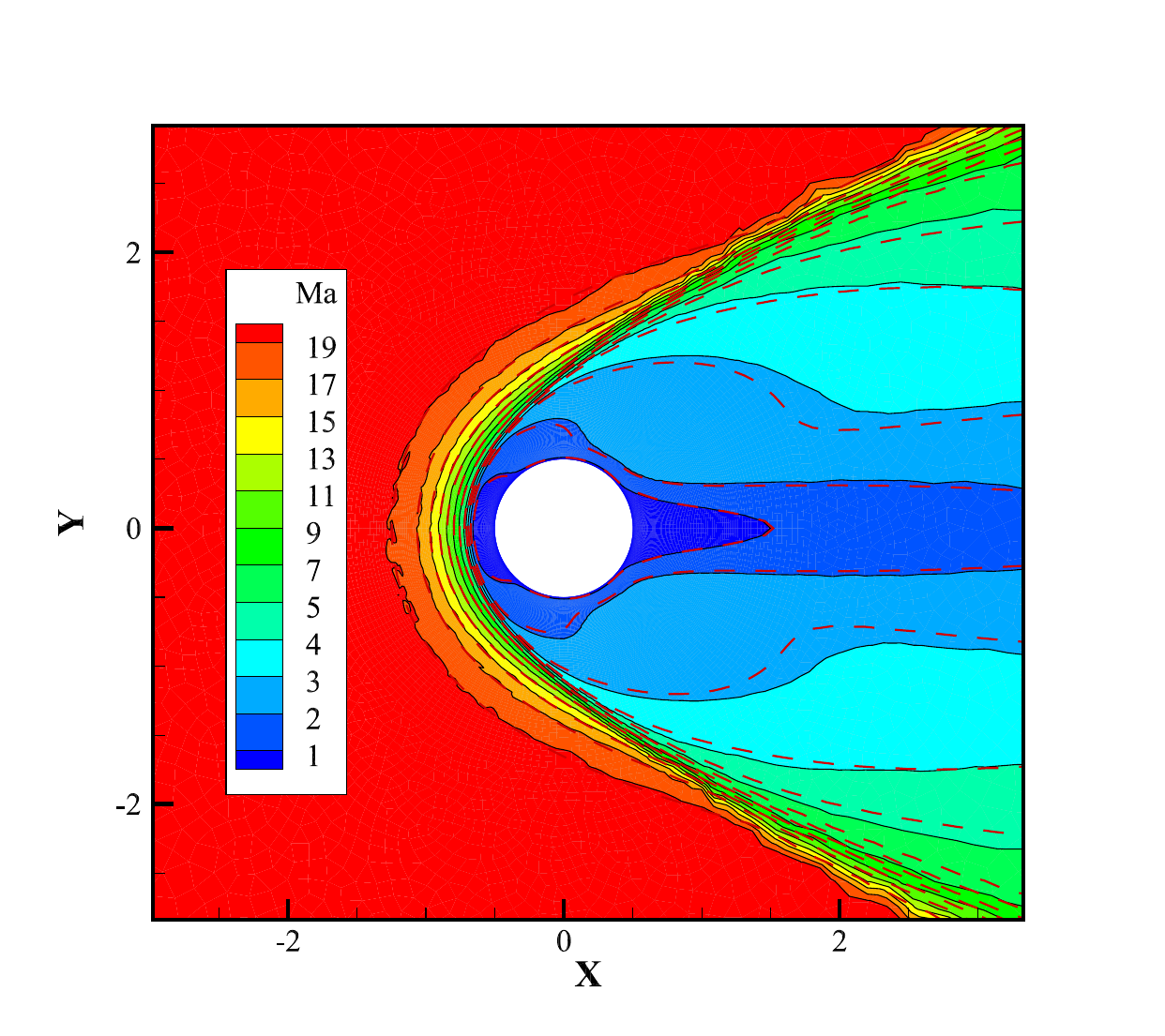}
}\\
\subfigure[\label{Fig:case2d_cyl_ma20_con_Teq}]{
\includegraphics[trim=30 25 30 60, clip, width=0.45\textwidth]{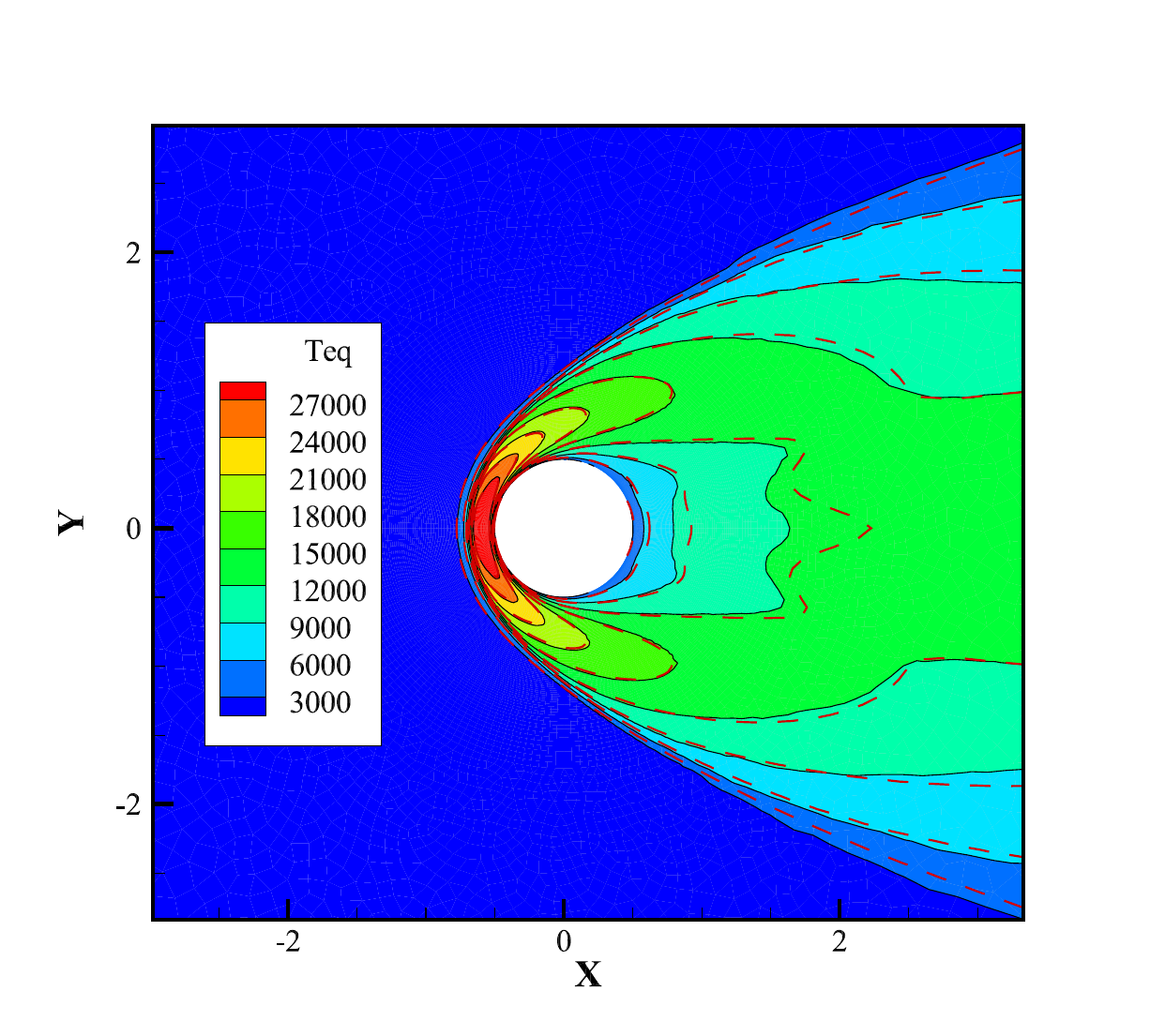}
}\hspace{0.01\textwidth}%
\subfigure[\label{Fig:case2d_cyl_ma20_con_Ttr}]{
\includegraphics[trim=30 25 30 60, clip, width=0.45\textwidth]{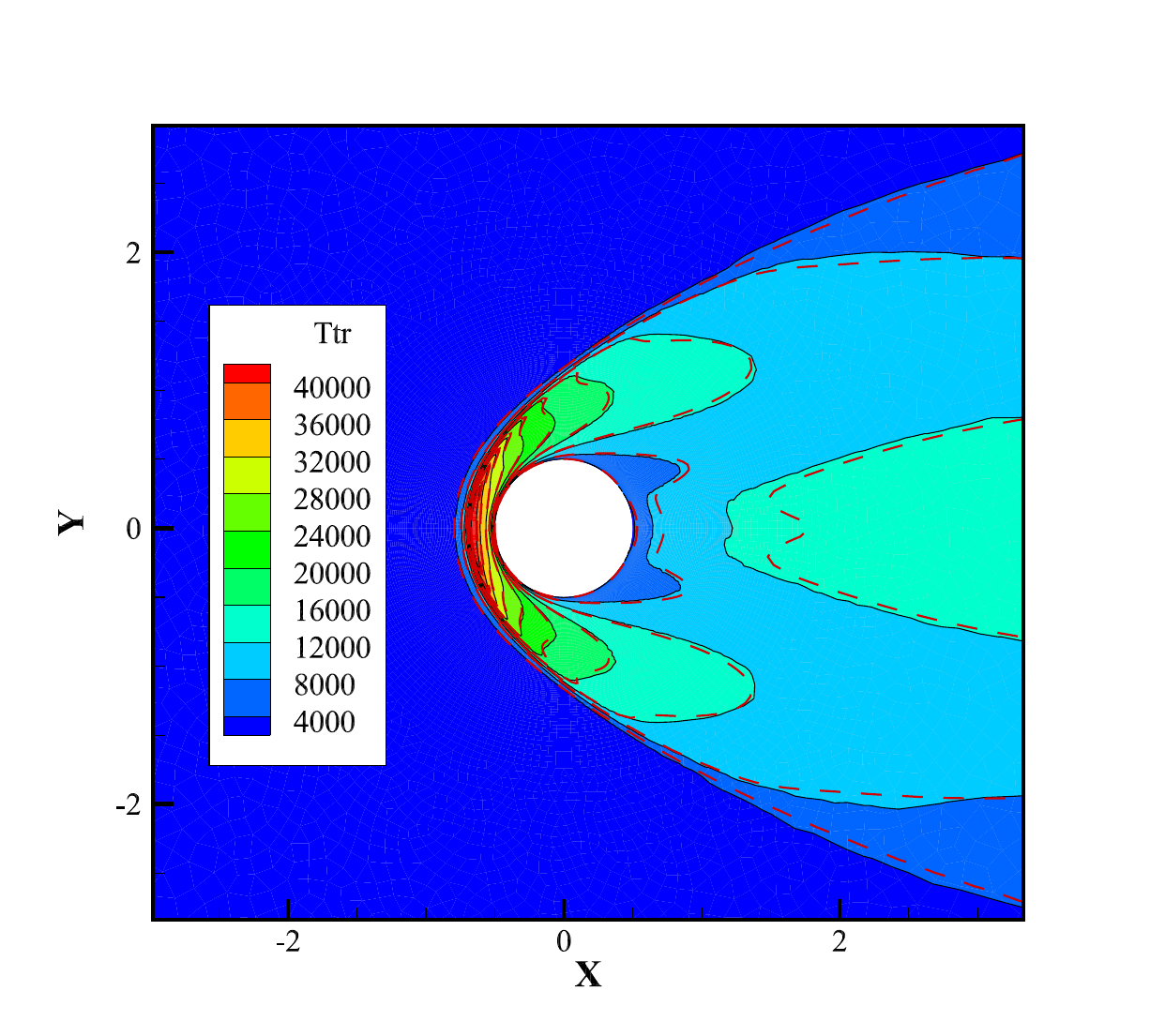}
}\\
\subfigure[\label{Fig:case2d_cyl_ma20_con_Trot}]{
\includegraphics[trim=30 25 30 60, clip, width=0.45\textwidth]{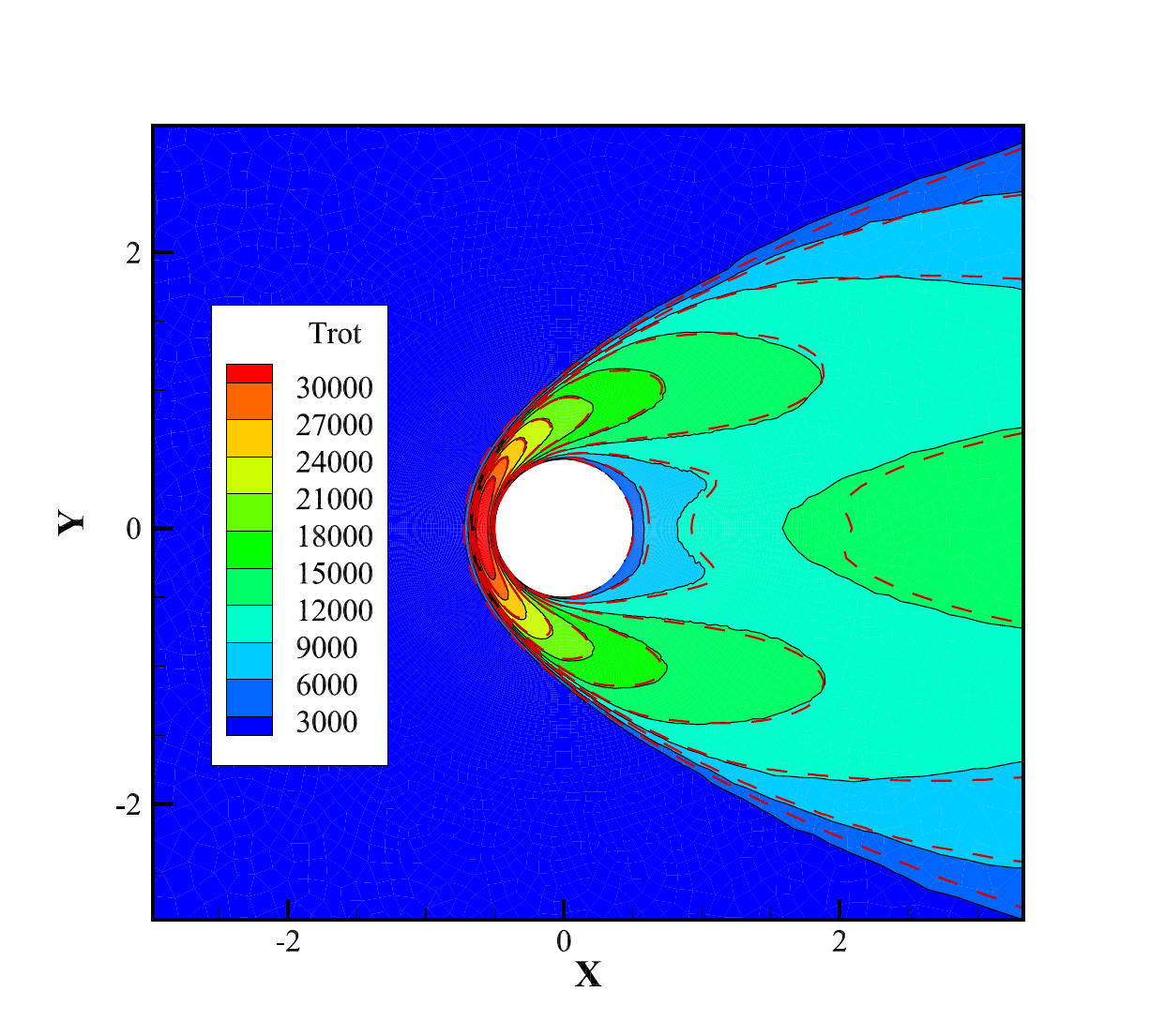}
}\hspace{0.01\textwidth}%
\subfigure[\label{Fig:case2d_cyl_ma20_con_Tvib}]{
\includegraphics[trim=30 25 30 60, clip, width=0.45\textwidth]{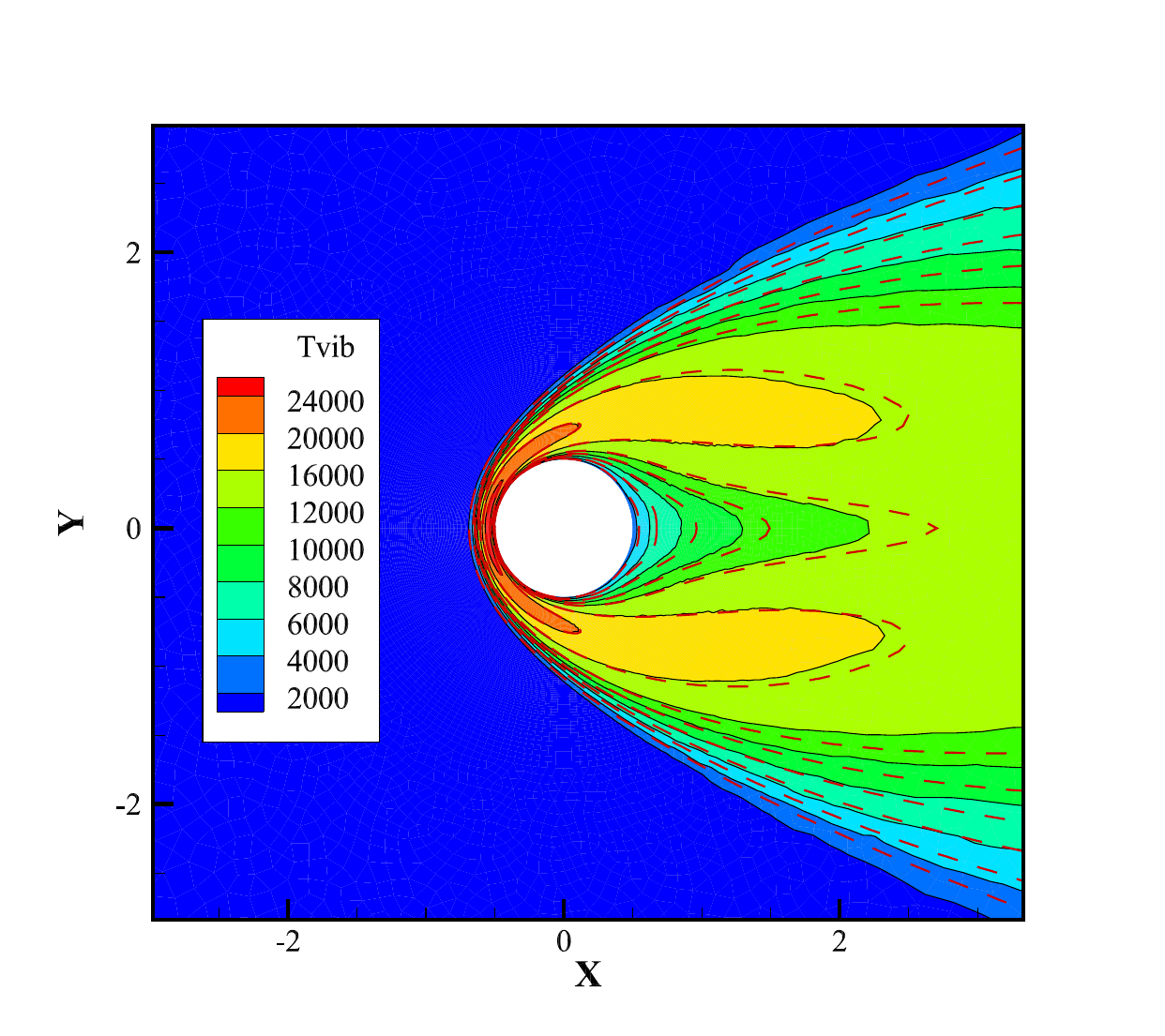}
}\\
\caption{\label{Fig:case2d_cyl_ma20_con}The (a) pressure, (b) Mach number, (c) equilibrium temperature, (d) translational temperature, (e) rotational temperature, and (f) vibrational temperature contours of the flow past a cylinder at Ma=20 (The color band: SUWP, red dash line: DUGKS).}
\end{figure*}

\begin{figure*}[h!t]
\centering
\subfigure[\label{Fig:case2d_cyl_ma20_sl_rho}]{
\includegraphics[width=0.45\textwidth]{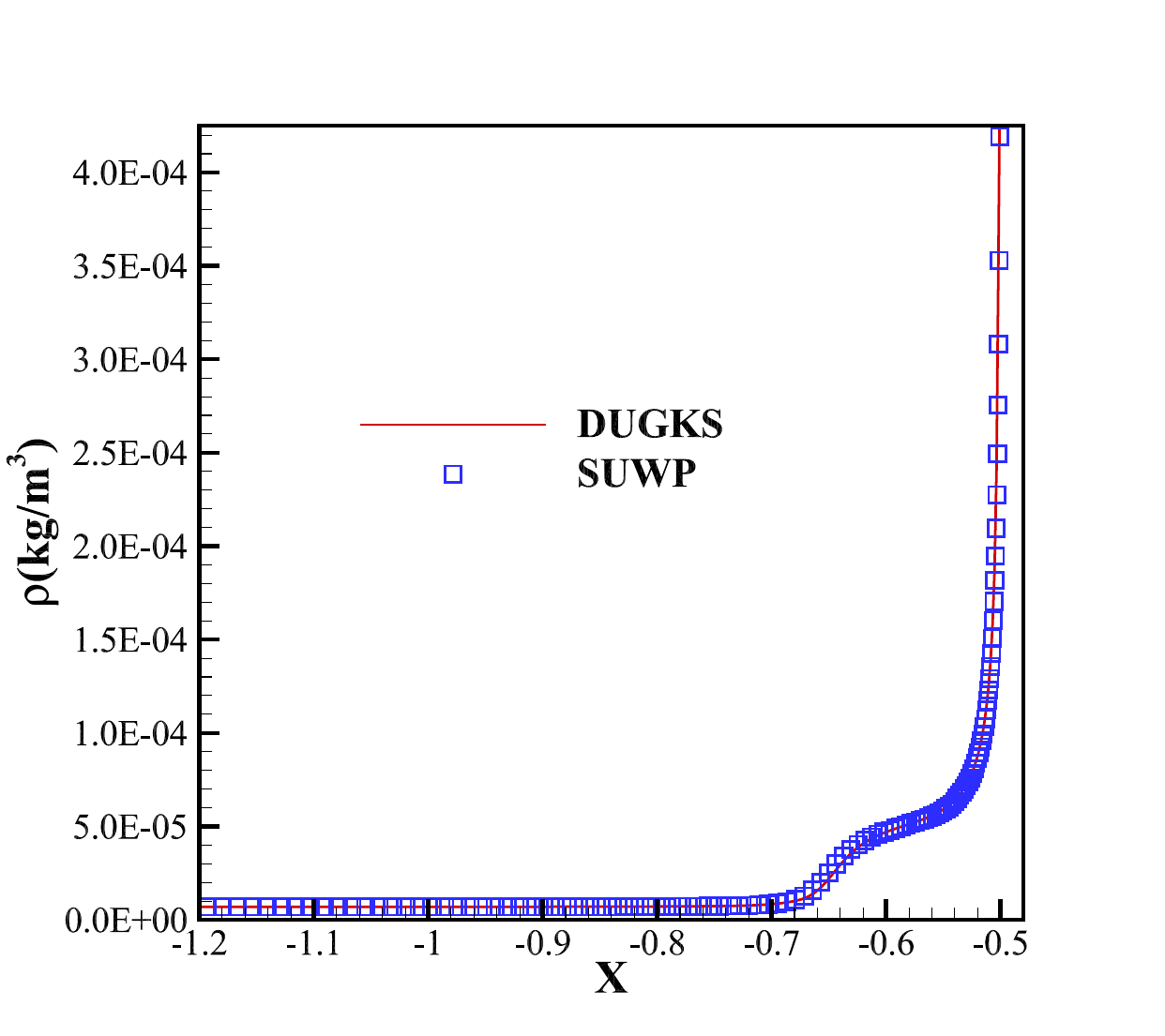}
}\hspace{0.01\textwidth}%
\subfigure[\label{Fig:case2d_cyl_ma20_sl_T}]{
\includegraphics[width=0.45\textwidth]{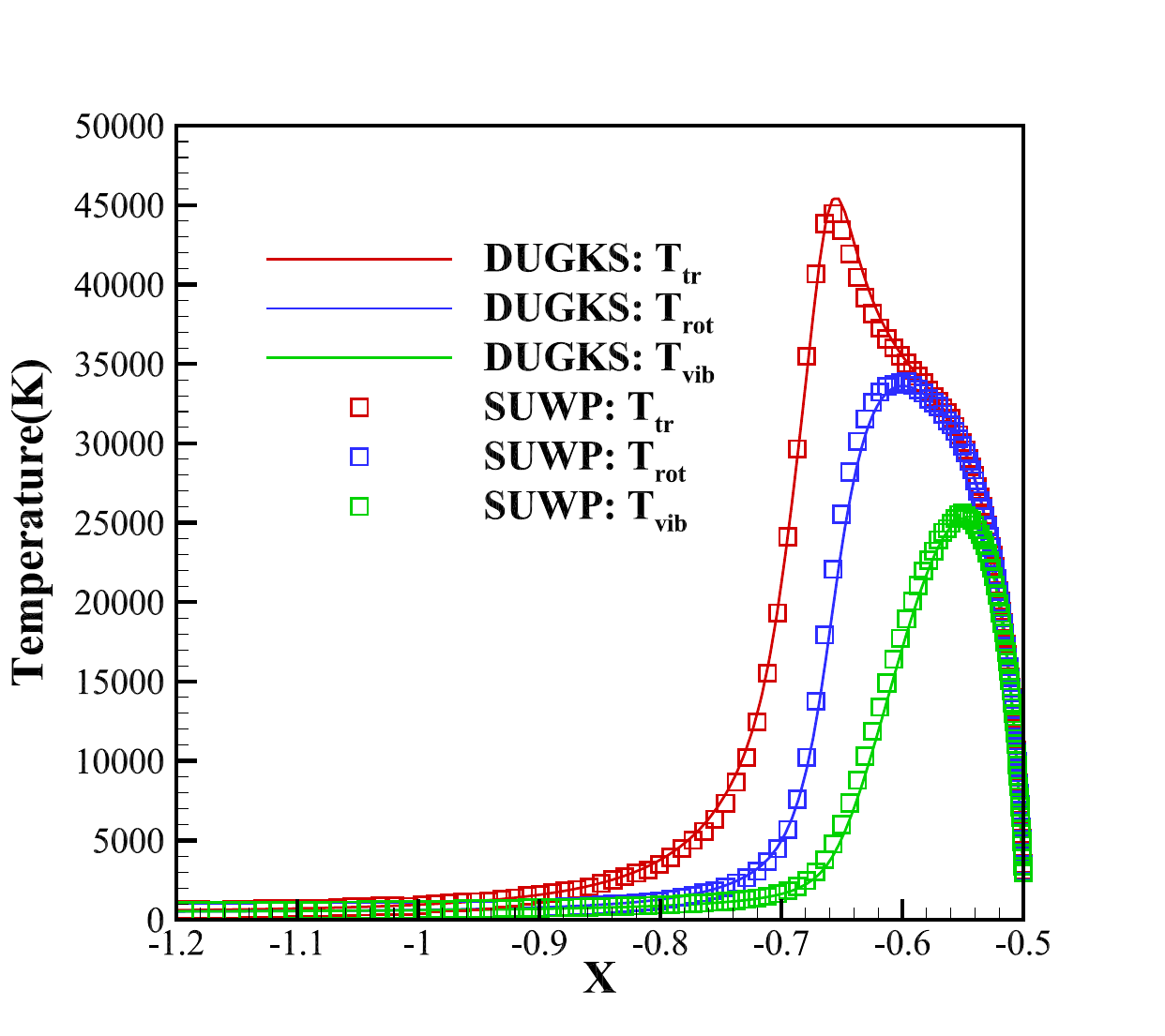}
}\\
\caption{\label{Fig:case2d_cyl_ma20_sl}The (a) density and (b) temperature along the forward stagnation line of the cylinder at Ma=20.}
\end{figure*}

\begin{figure*}[h!t]
\centering
\subfigure[\label{Fig:case2d_cyl_ma20_wall_cp}]{
\includegraphics[width=0.45\textwidth]{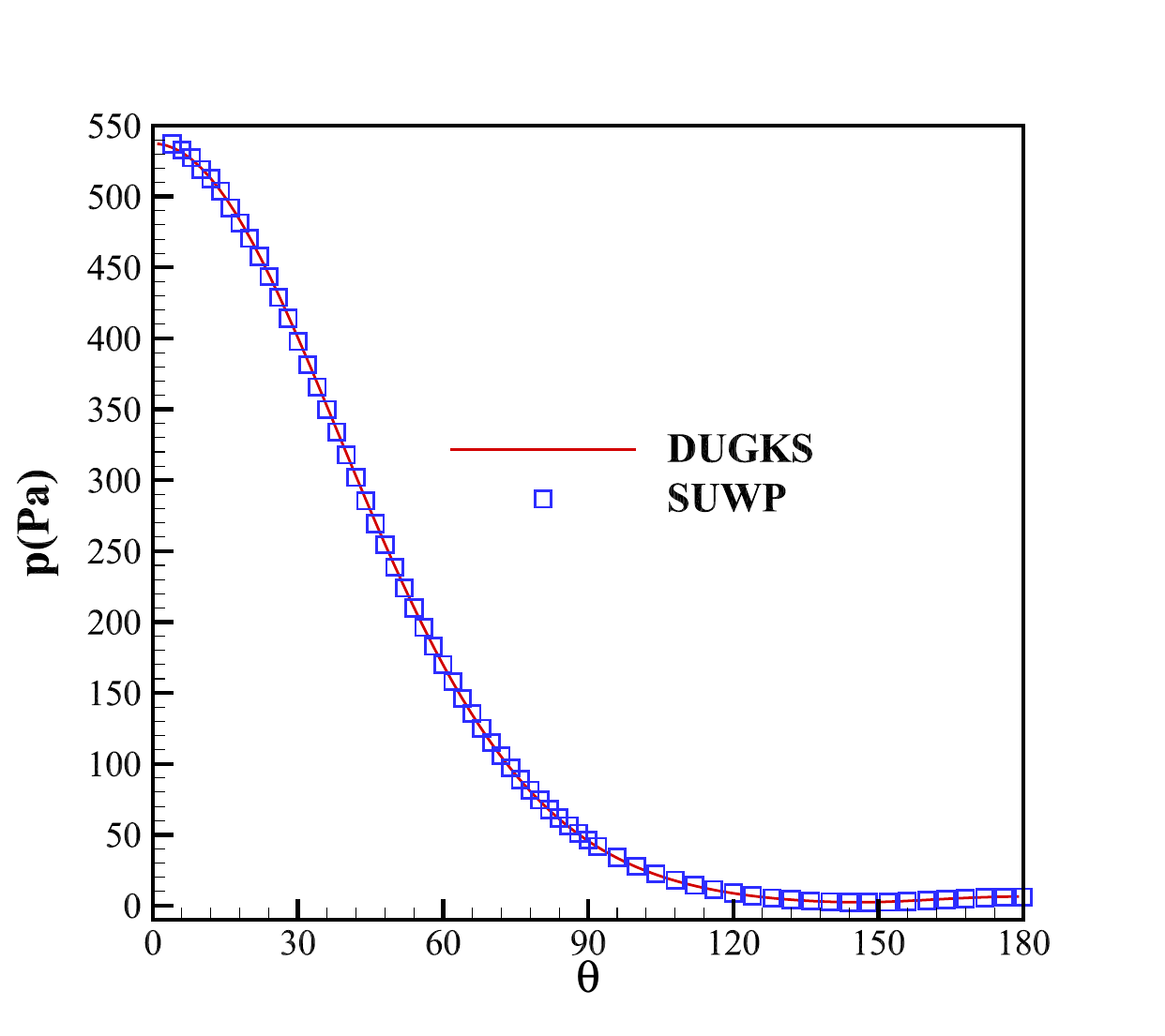}
}\hspace{0.01\textwidth}%
\subfigure[\label{Fig:case2d_cyl_ma20_wall_ch}]{
\includegraphics[width=0.45\textwidth]{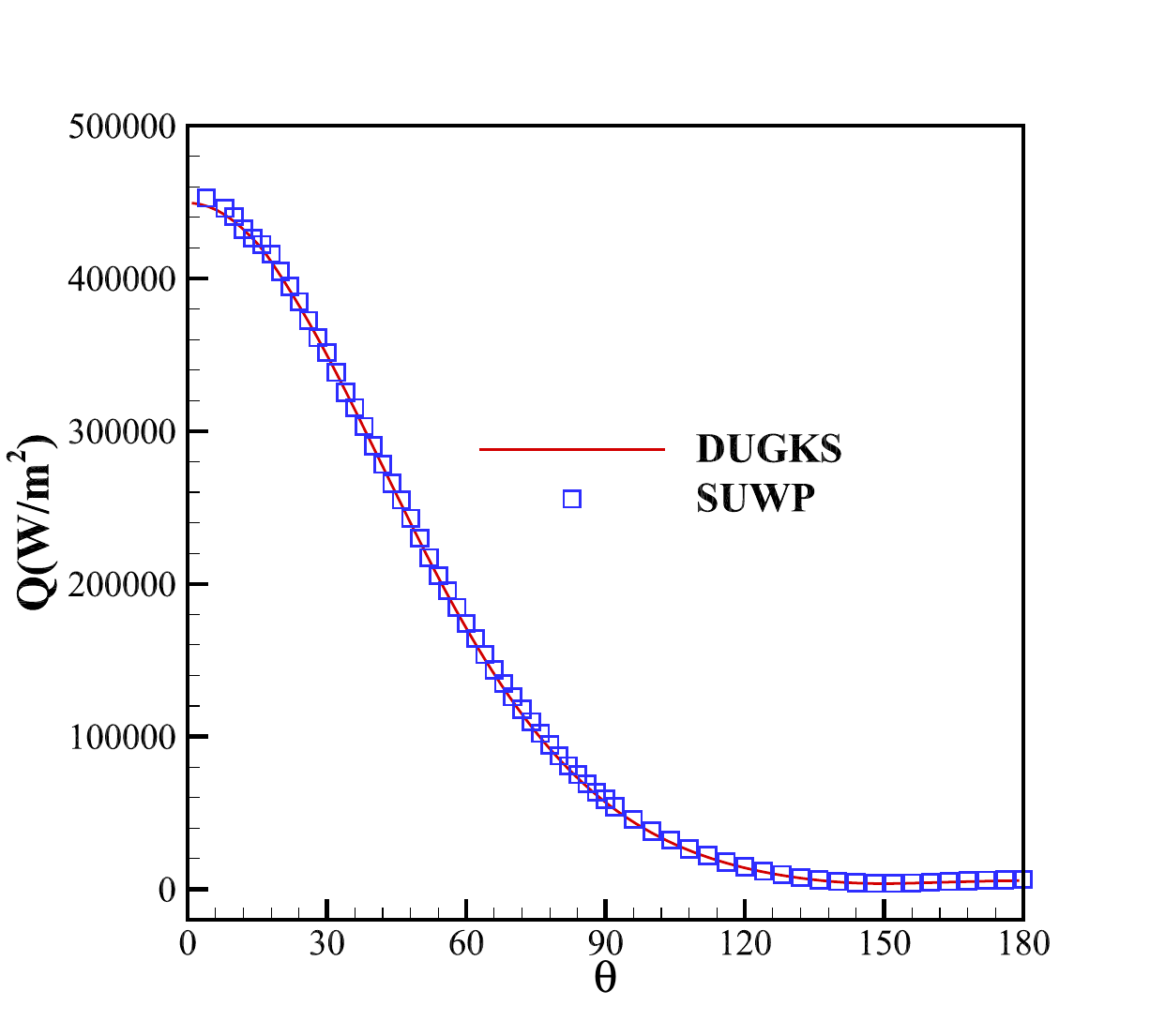}
}\\
\caption{\label{Fig:case2d_cyl_ma20_wall}The (a) pressure and (b) heat flux on the wall surface of the cylinder at Ma=20.}
\end{figure*}

Fig.~\ref{Fig:case2d_cyl_ma5_con} presents contour plots of pressure, Mach number, equilibrium temperature, translational temperature, rotational temperature, and vibrational temperature for the Ma=5 flow case. The SUWP-vib simulations demonstrate essential agreement with the DUGKS results, particularly in the region where $x < 0$. As $x$ increases near the leeward side of the cylinder, the translational, rotational, and vibrational temperatures exhibit marginally faster increase rates compared to the DUGKS. In the lower-right region of the contour, the post-shock temperature decay occurs more gradually in the SUWP-vib than in the DUGKS. Testing confirms this discrepancy does not stem from inadequate grid resolution or lack of convergence. It likely arises from inherent algorithmic differences between SUWP-vib and the DUGKS approach. Fig. \ref{Fig:case2d_cyl_ma5_sl} and Fig. \ref{Fig:case2d_cyl_ma5_wall} The density and temperature distributions along the stagnation line, as well as the pressure and heat flux distributions along the cylinder wall for Ma=5 cylinder flow, are presented respectively. Simulation results demonstrate that the SUWP-vib method correlates well with the DUGKS method using the same model. When compared with DSMC results, general agreement is observed except for the "early-raise" in translational temperature. Fig. \ref{Fig:case2d_cyl_ma20_con} presents flowfield contours for Ma=20 conditions, including pressure, Mach number, equilibrium temperature, translational temperature, rotational temperature, and vibrational temperature. The Ma=20 flow exhibits stronger non-equilibrium effects with reduced shock standoff distance from the cylinder surface. Nevertheless, SUWP-vib simulations maintain excellent agreement with DUGKS results. The temperature contour evolution resembles that observed in the Ma=5 case. Fig. \ref{Fig:case2d_cyl_ma20_sl} and Fig. \ref{Fig:case2d_cyl_ma20_wall} present the stagnation line profiles and wall-surface distributions, respectively, for Ma=20 cylinder flow. It is evident that the SUWP-vib method accurately captures the peak translational temperature.  Wall pressure and heat flux exhibit minimal deviations from DUGKS results.  For Ma=20 conditions, SUWP-vib simulations demonstrate consistent agreement with DUGKS under identical modeling frameworks.  This validates SUWP-vib's capability to simulate hypersonic viscous flow phenomena.

\begin{table}[ht] 
    \centering
    \caption{The parameters of the flow past a cylinder.} \label{table:2Dcase_cyl_Table}
    \begin{tabular}{ c  c  c  c  c  c  c  c }
        \hline
        Ma & $L_{\rm{ref}}$ & Kn & $\rho_{\infty}(\rm{kg/m^{3}})$ & $T_{\infty}(\rm{K})$ & $T_{\rm{wall}}(\rm{K})$ & $Z_{\rm{rot}}$ & $Z_{\rm{vib}}$ \\ 
        \hline
        5  & 1m & 0.01 & $6.9592 \times 10^{-6}$ & 500 & 500  & Parker  & 30 \\
        20 & 1m & 0.01 & $6.9592 \times 10^{-6}$ & 500 & 2000 & Parker  & 35 \\
        \hline
    \end{tabular}
\end{table}

\subsection{Flow past a blunt wedge}
\begin{figure*}[h!t]
\centering
\subfigure[Shape\label{Fig:case2d_blunt_shape}]{
\includegraphics[trim=130 25 240 60, clip, width=0.45\textwidth]{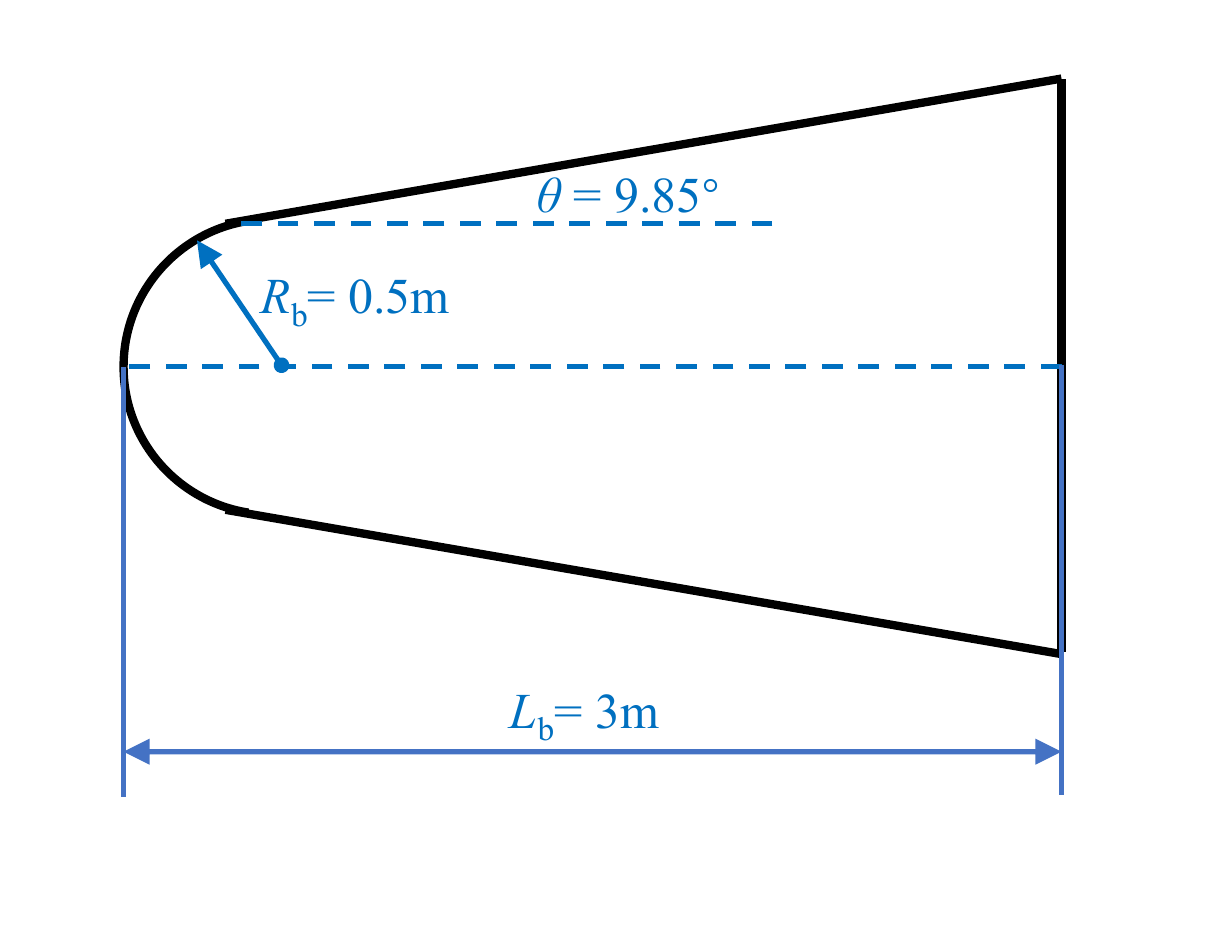}
}\hspace{0.01\textwidth}%
\subfigure[Mesh\label{Fig:case2d_blunt_mesh_self}]{
\includegraphics[width=0.45\textwidth]{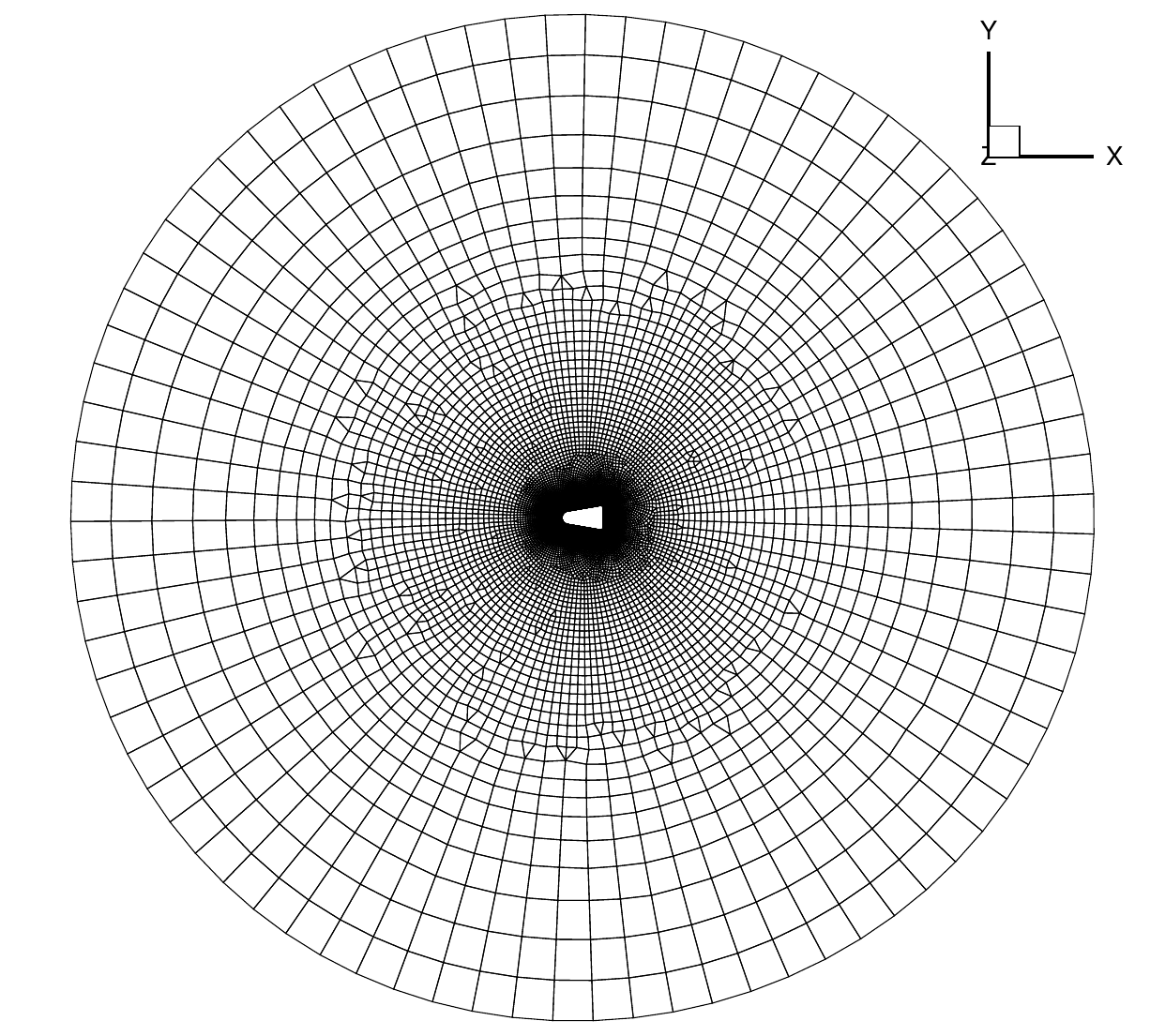}
}
\\
\caption{\label{Fig:case2d_blunt_mesh}The mesh of the blunt wedge at Ma=15.}
\end{figure*}

\begin{figure*}[h!t]
\centering
\subfigure[\label{Fig:case2d_blunt_ma15_kn0.1_con_Ptr}]{
\includegraphics[trim=30 25 30 60, clip, width=0.45\textwidth]{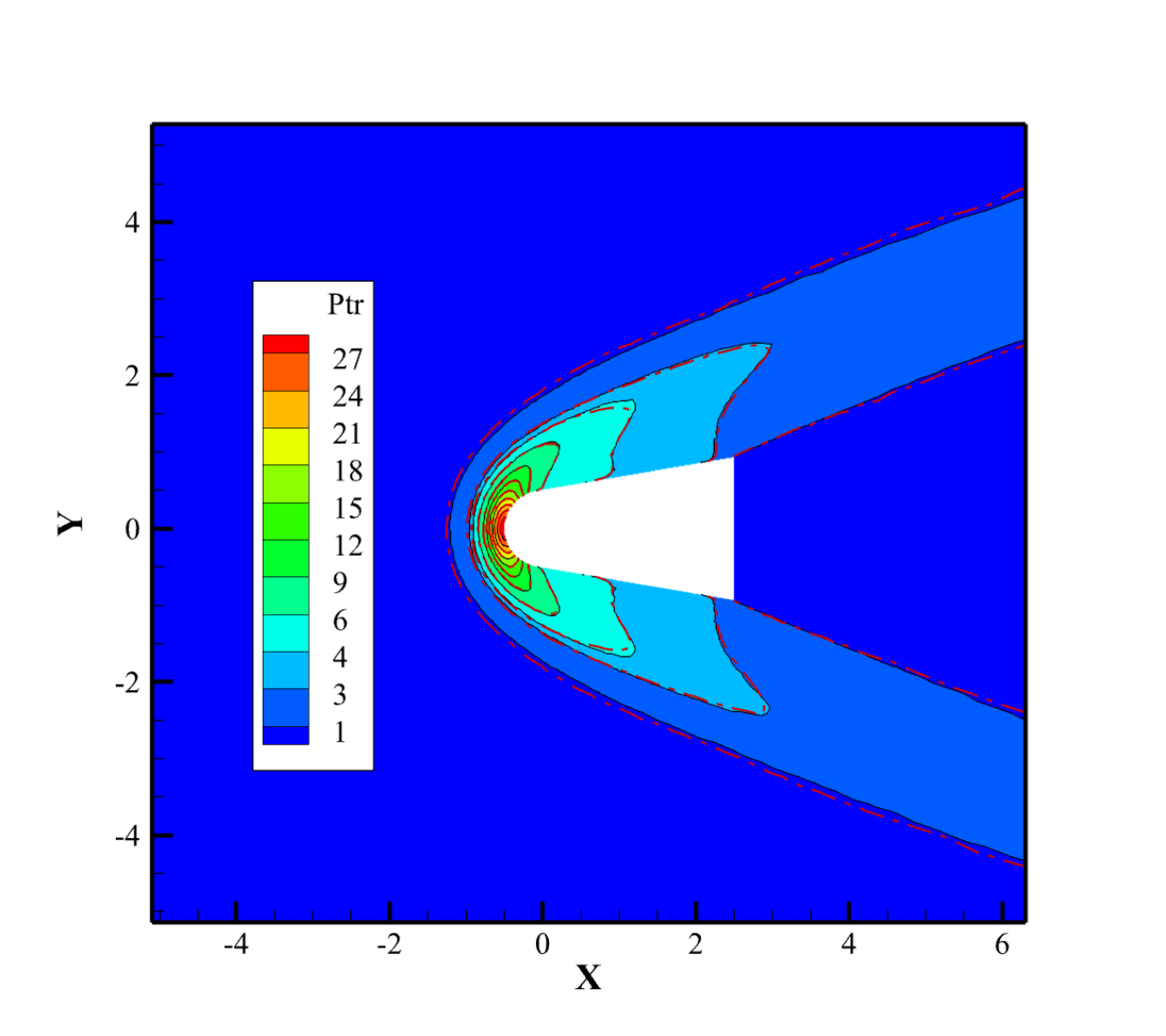}
}\hspace{0.01\textwidth}%
\subfigure[\label{Fig:case2d_blunt_ma15_kn0.1_con_Ttr}]{
\includegraphics[trim=30 25 30 60, clip, width=0.45\textwidth]{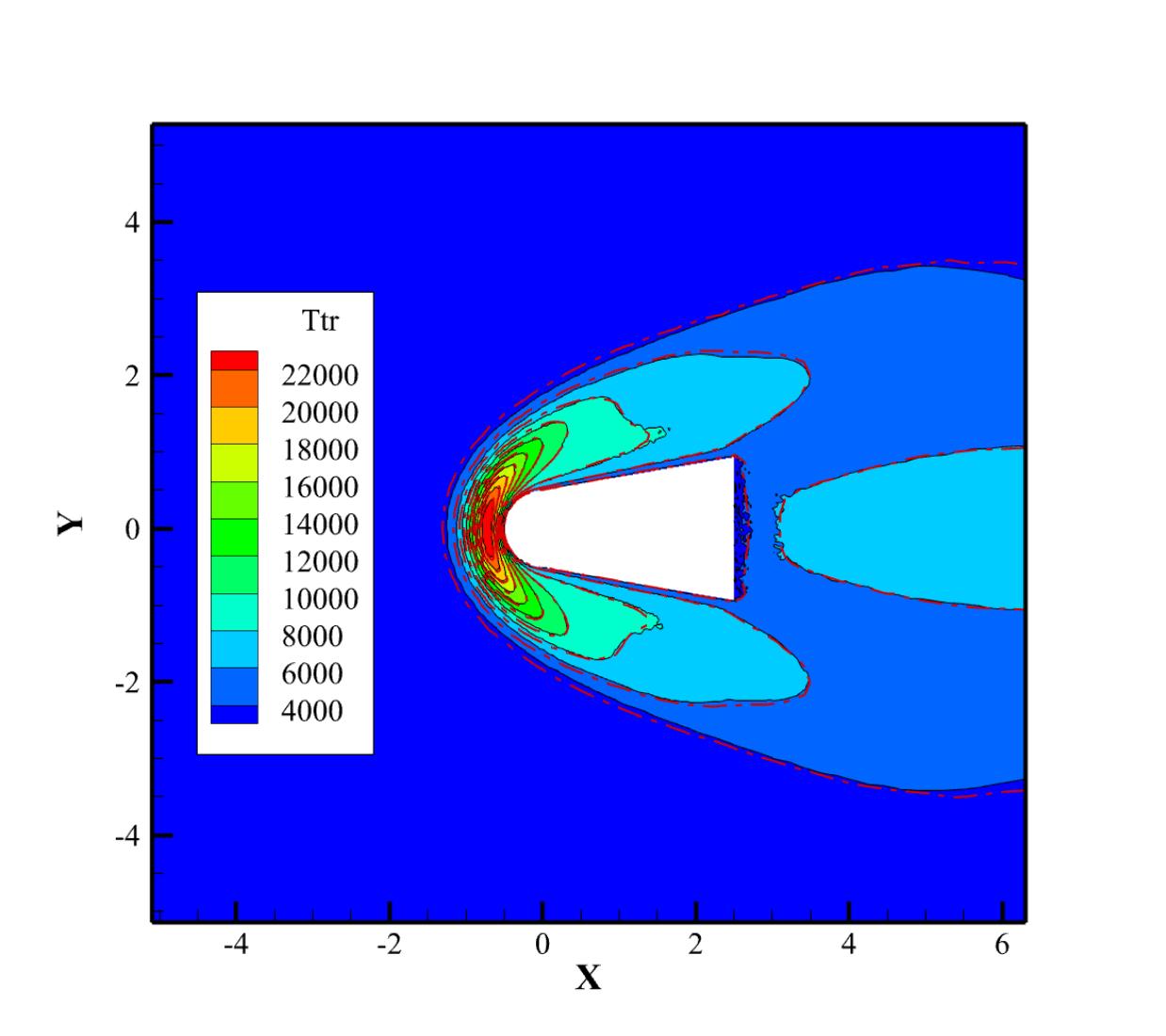}
}\\
\subfigure[\label{Fig:case2d_blunt_ma15_kn0.1_con_Trot}]{
\includegraphics[trim=30 25 30 60, clip, width=0.45\textwidth]{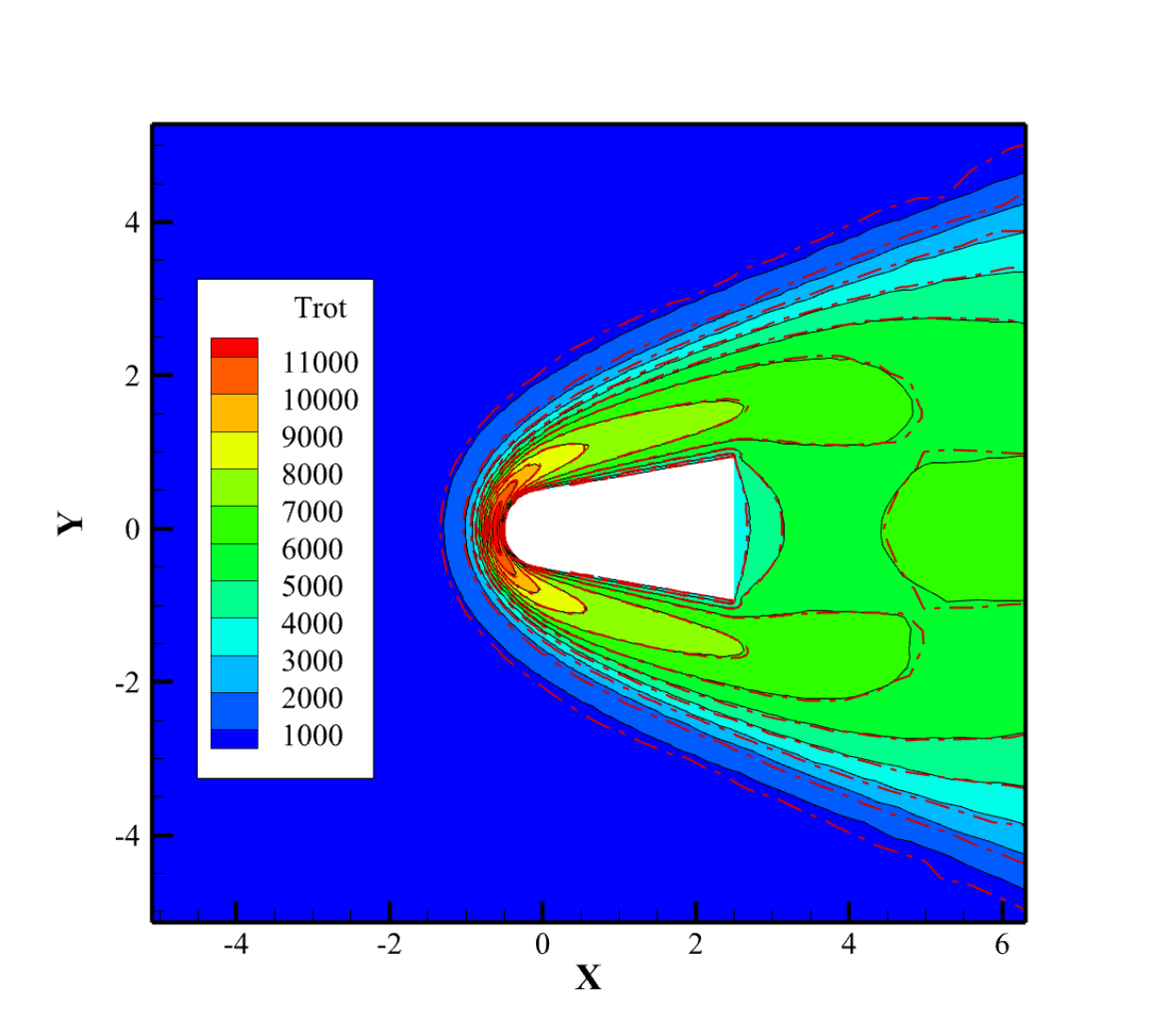}
}\hspace{0.01\textwidth}%
\subfigure[\label{Fig:case2d_blunt_ma15_kn0.1_con_Tvib}]{
\includegraphics[trim=30 25 30 60, clip, width=0.45\textwidth]{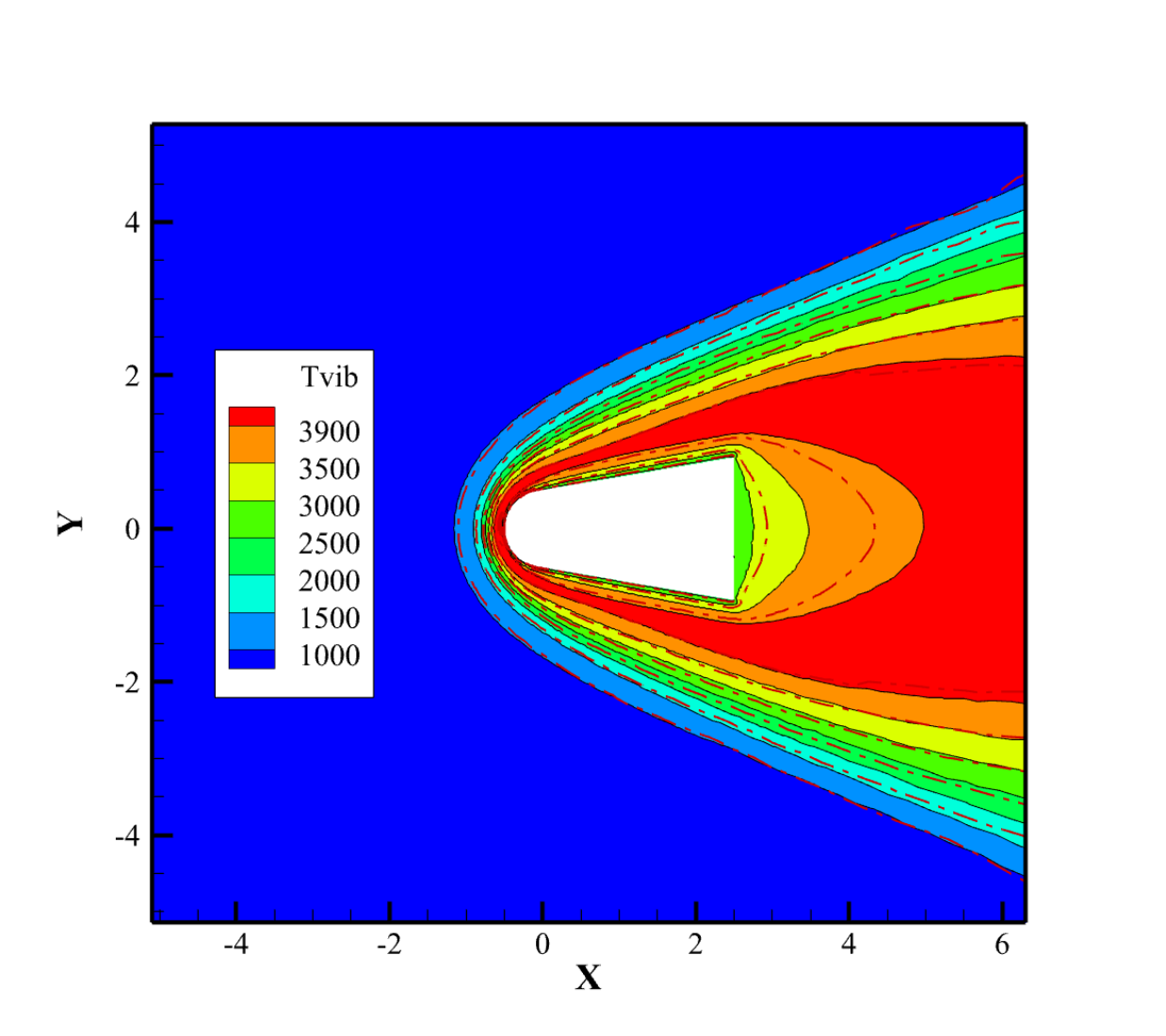}
}\\
\caption{\label{Fig:case2d_blunt_ma15_kn0.1_con}The (a) pressure, (b) translational temperature, (c) rotational temperature, and (d) vibrational temperature contours of the flow past a blunt wedge at Ma=15, Kn=0.1 (The color band: SUWP, red dash line: DUGKS).}
\end{figure*}

\begin{figure*}[h!t]
\centering
\subfigure[\label{Fig:case2d_blunt_ma15_kn0.1_sl_p}]{
\includegraphics[width=0.45\textwidth]{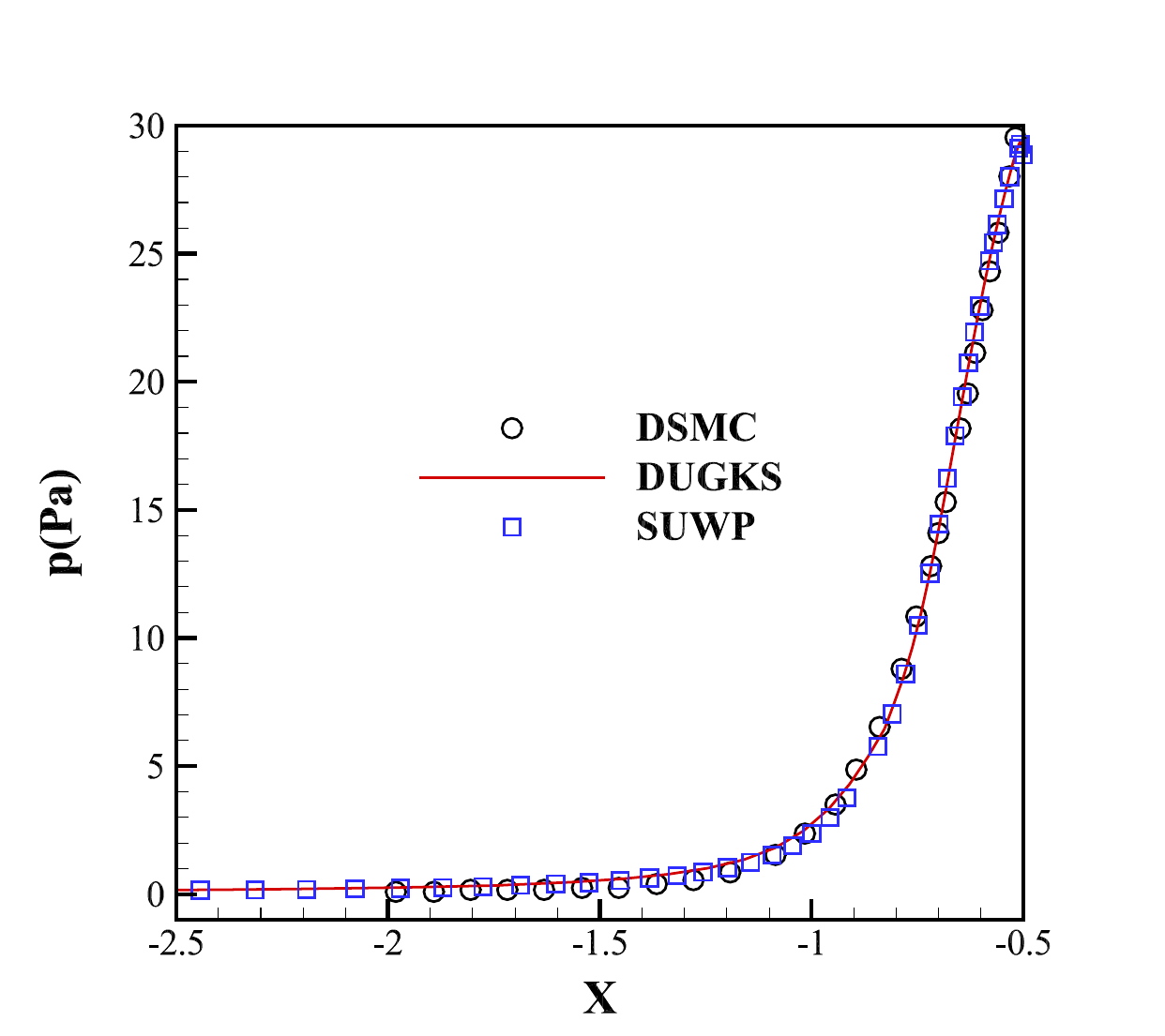}
}\hspace{0.01\textwidth}%
\subfigure[\label{Fig:case2d_blunt_ma15_kn0.1_sl_T}]{
\includegraphics[width=0.45\textwidth]{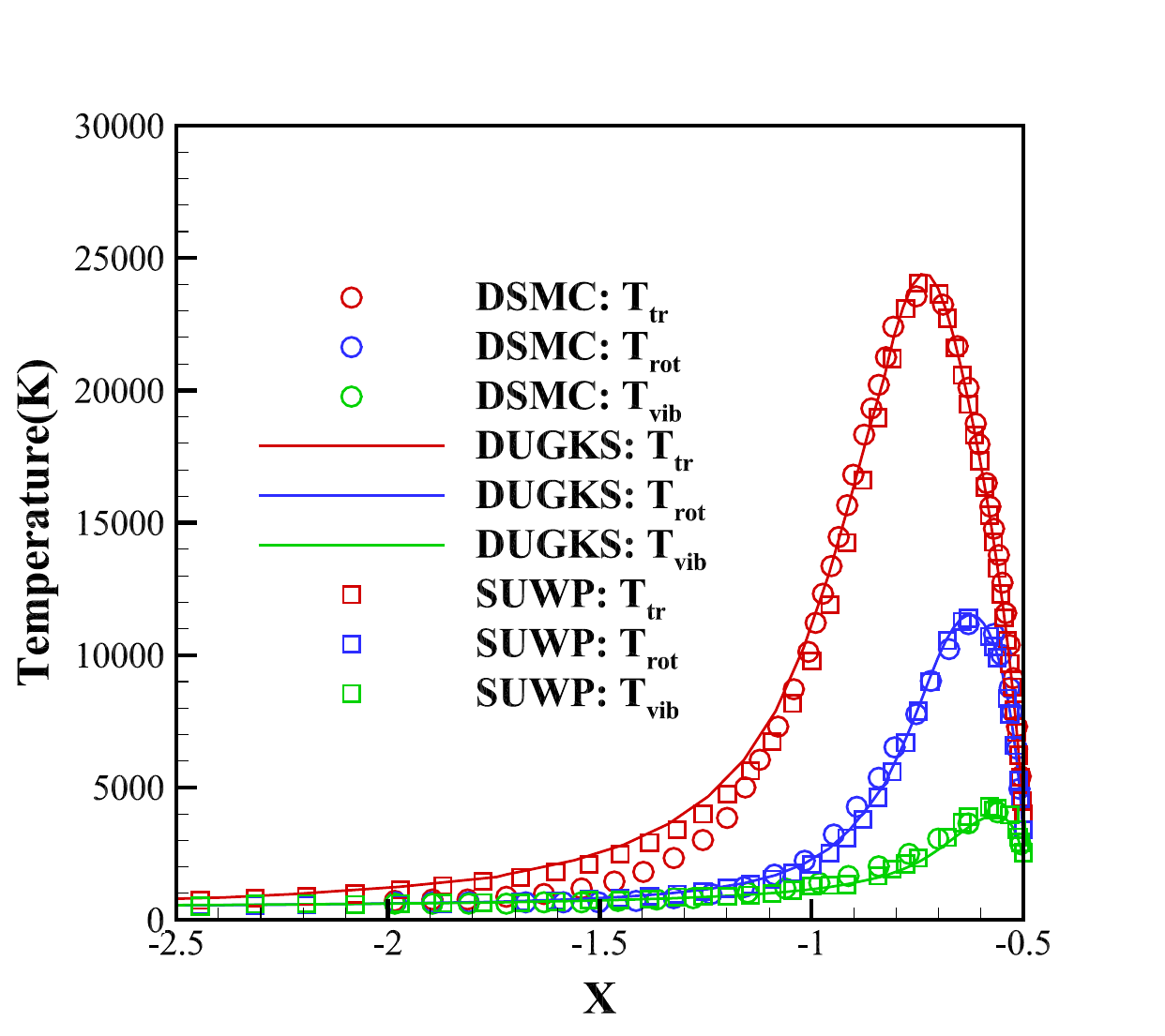}
}\\
\caption{\label{Fig:case2d_blunt_ma15_kn0.1_sl}The (a) pressure and (b) temperature along the forward stagnation line of the blunt wedge at Ma=15, Kn=0.1.}
\end{figure*}

\begin{figure*}[h!t]
\centering
\subfigure[\label{Fig:case2d_blunt_ma15_kn0.1_wall_cp}]{
\includegraphics[width=0.3\textwidth]{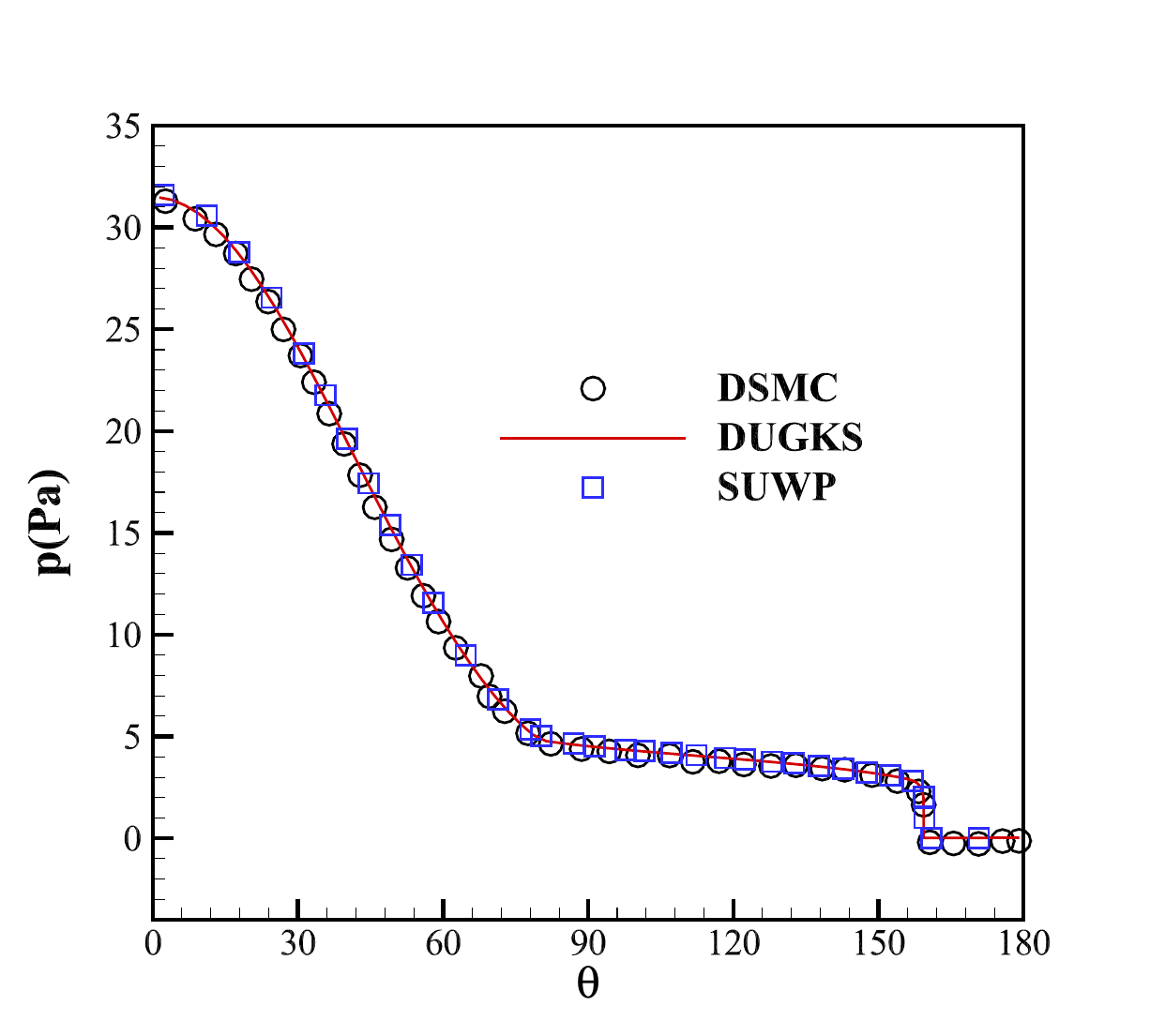}
}\hspace{0.01\textwidth}%
\subfigure[\label{Fig:case2d_blunt_ma15_kn0.1_wall_cf}]{
\includegraphics[width=0.3\textwidth]{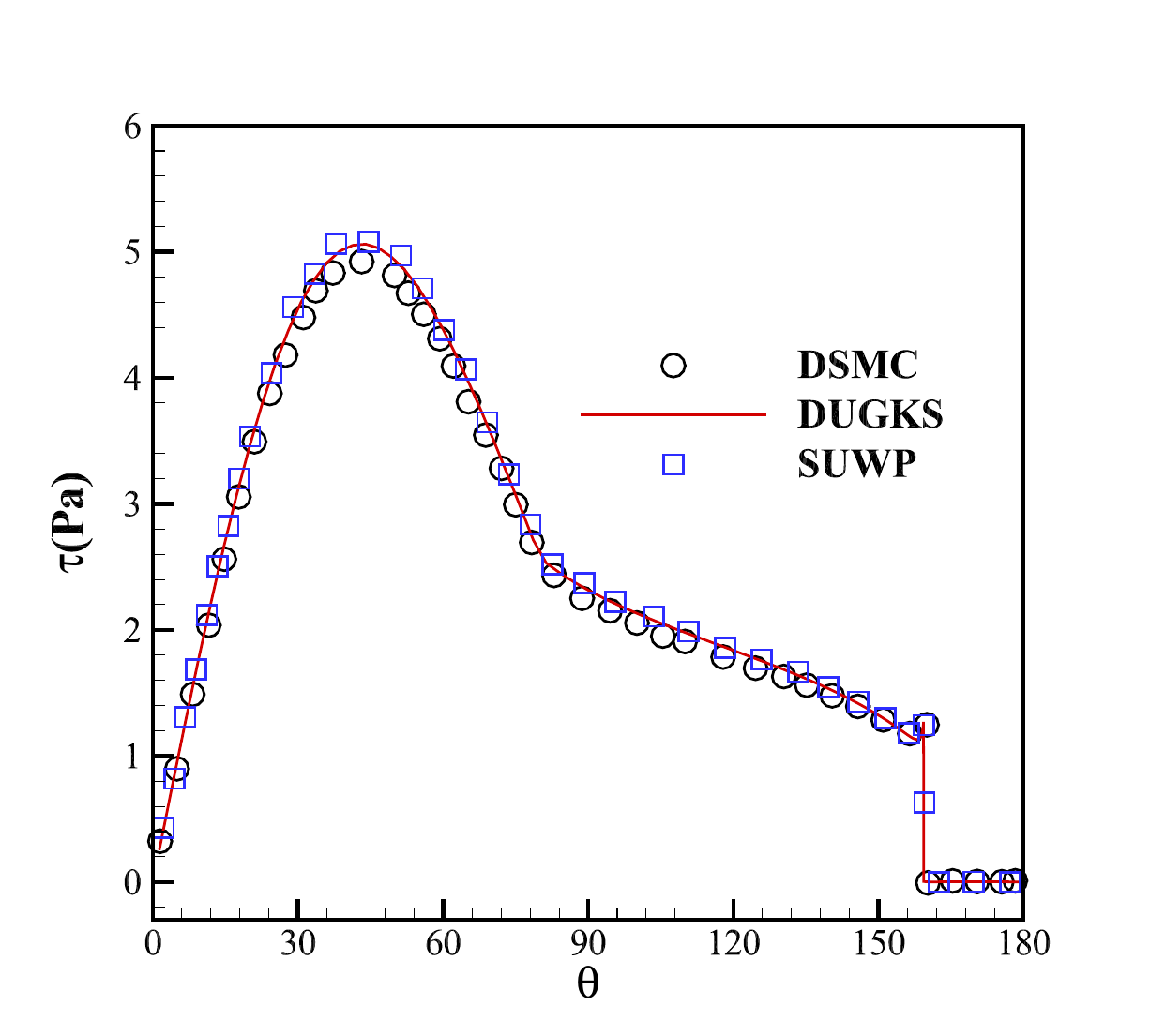}
}\hspace{0.01\textwidth}%
\subfigure[\label{Fig:case2d_blunt_ma15_kn0.1_wall_ch}]{
\includegraphics[width=0.3\textwidth]{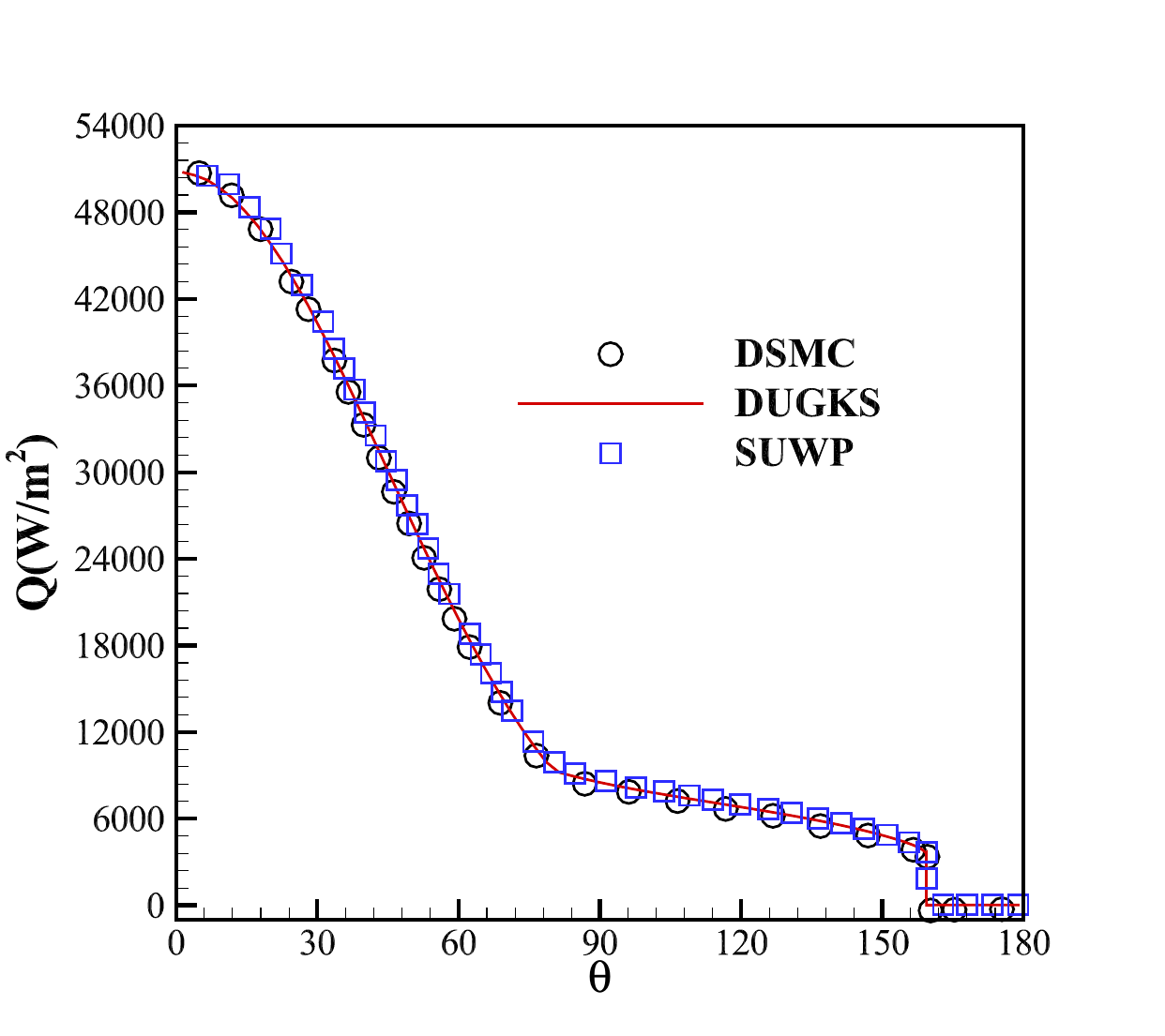}
}\\
\caption{\label{Fig:case2d_blunt_ma15_kn0.1_wall}The (a) pressure, (b) shear stress and (c) heat flux on the wall surface of the blunt wedge at Ma=15, Kn=0.1.}
\end{figure*}

\begin{figure*}[h!t]
\centering
\subfigure[\label{Fig:case2d_blunt_ma15_kn1_con_Ptr}]{
\includegraphics[trim=30 25 30 60, clip, width=0.45\textwidth]{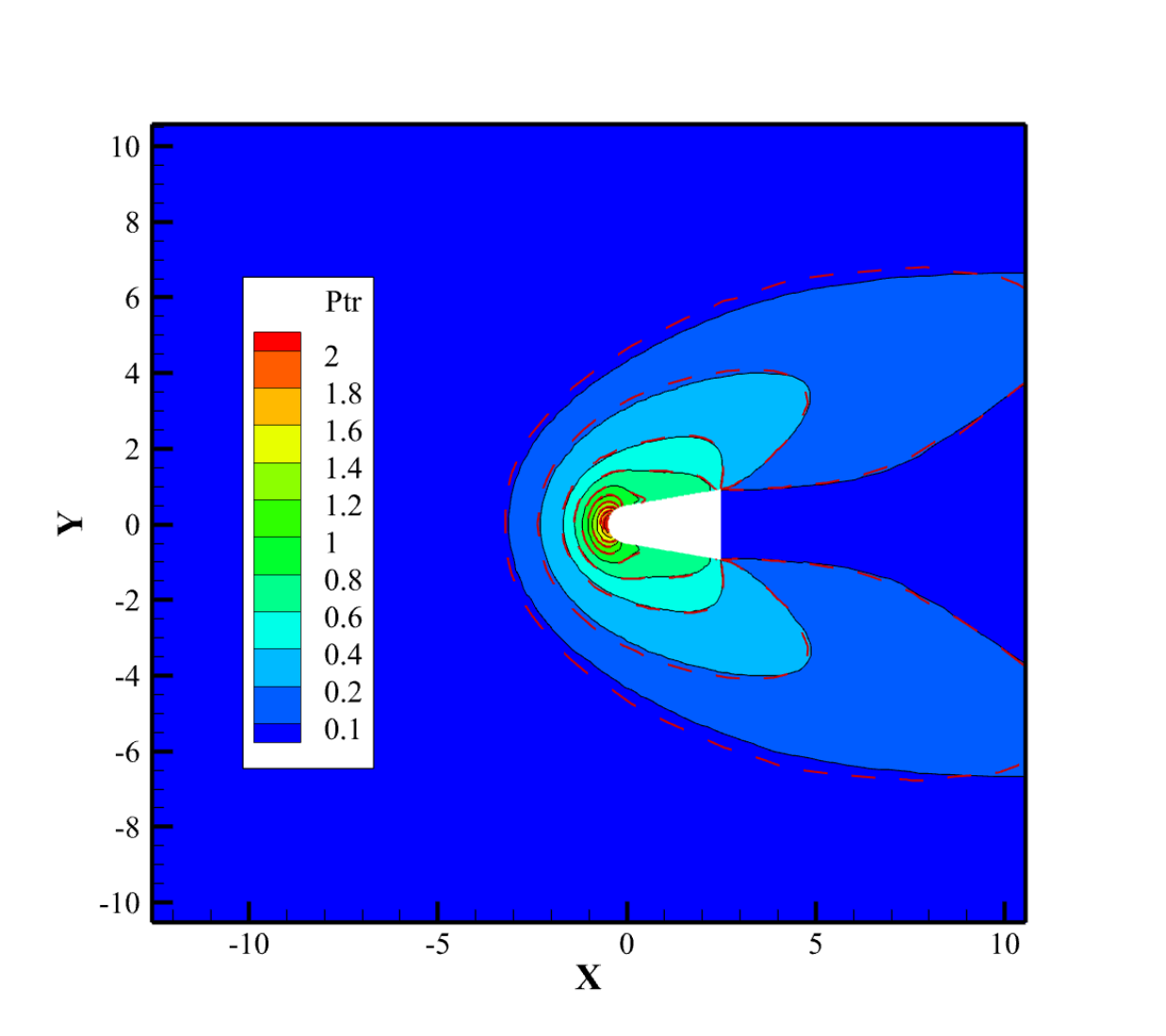}
}\hspace{0.01\textwidth}%
\subfigure[\label{Fig:case2d_blunt_ma15_kn1_con_Ttr}]{
\includegraphics[trim=30 25 30 60, clip, width=0.45\textwidth]{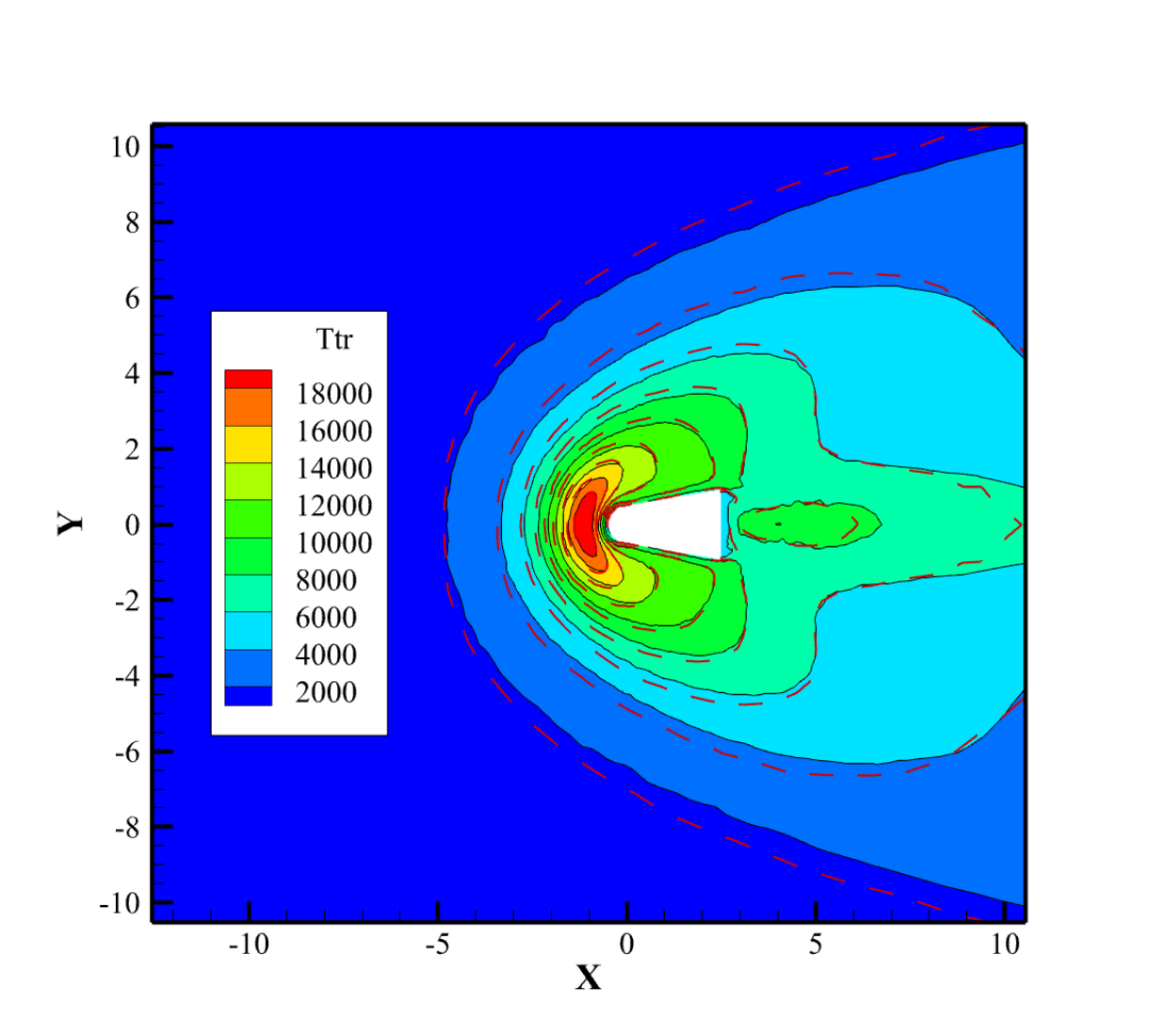}
}\\
\subfigure[\label{Fig:case2d_blunt_ma15_kn1_con_Trot}]{
\includegraphics[trim=30 25 30 60, clip, width=0.45\textwidth]{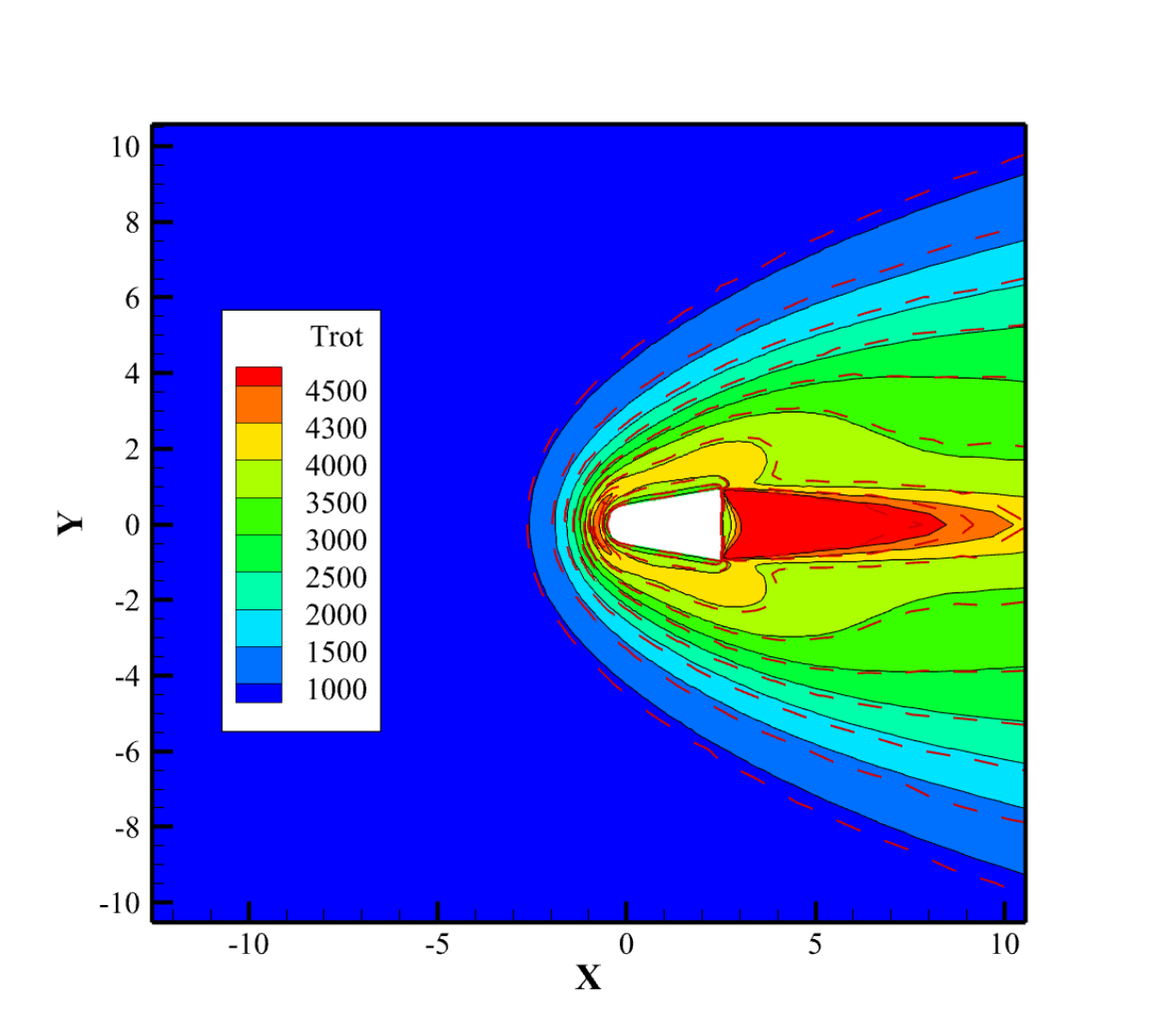}
}\hspace{0.01\textwidth}%
\subfigure[\label{Fig:case2d_blunt_ma15_kn1_con_Tvib}]{
\includegraphics[trim=30 25 30 60, clip, width=0.45\textwidth]{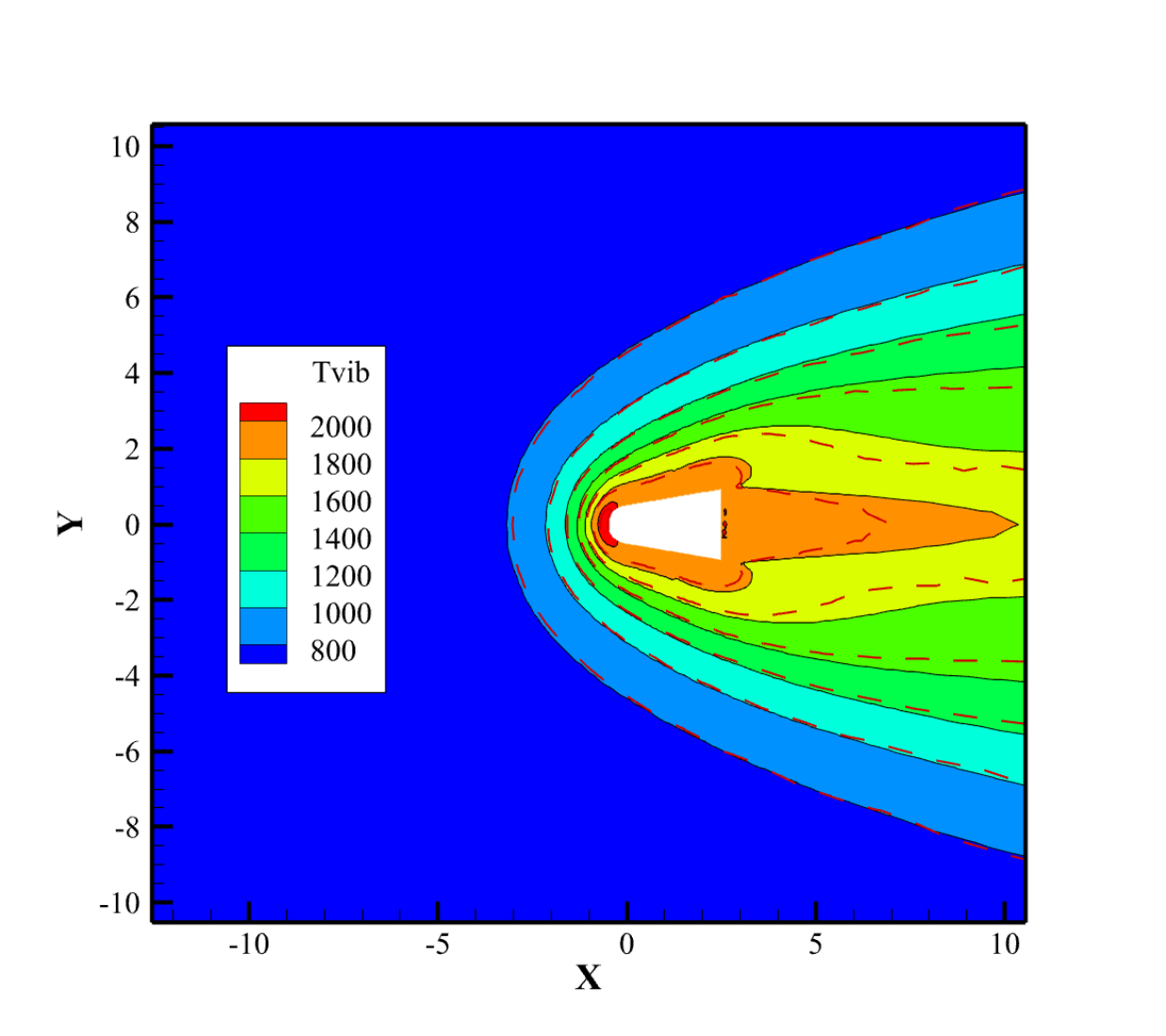}
}\\
\caption{\label{Fig:case2d_blunt_ma15_kn1_con}The (a) pressure, (b) translational temperature, (c) rotational temperature, and (d) vibrational temperature contours of the flow past a blunt wedge at Ma=15, Kn=1 (The color band: SUWP, red dash line: DUGKS).}
\end{figure*}

\begin{figure*}[h!t]
\centering
\subfigure[\label{Fig:case2d_blunt_ma15_kn1_sl_p}]{
\includegraphics[width=0.45\textwidth]{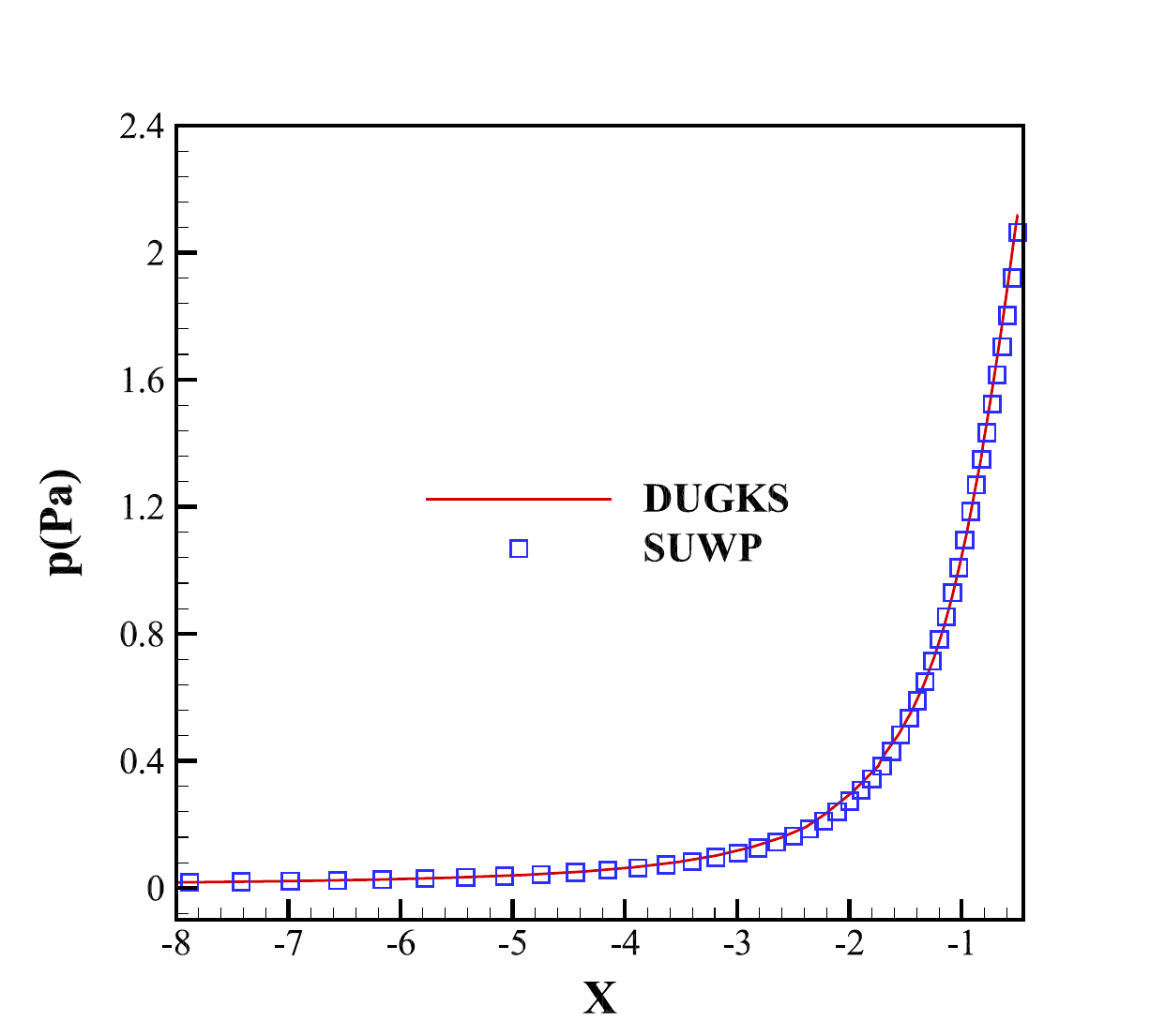}
}\hspace{0.01\textwidth}%
\subfigure[\label{Fig:case2d_blunt_ma15_kn1_sl_T}]{
\includegraphics[width=0.45\textwidth]{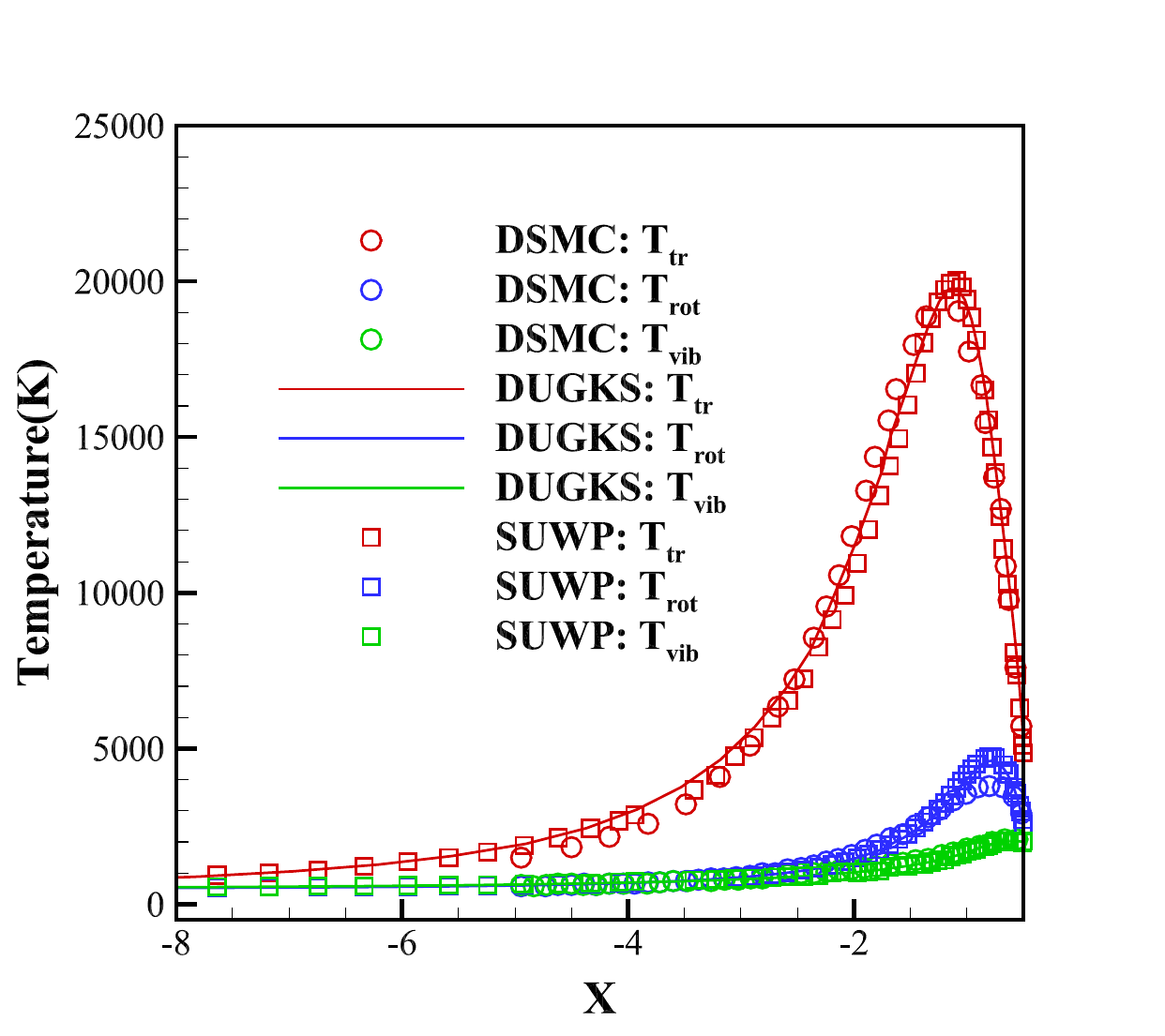}
}\\
\caption{\label{Fig:case2d_blunt_ma15_kn1_sl}The (a) pressure and (b) temperature along the forward stagnation line of the blunt wedge at Ma=15, Kn=1.}
\end{figure*}

\begin{figure*}[h!t]
\centering
\subfigure[\label{Fig:case2d_blunt_ma15_kn1_wall_cp}]{
\includegraphics[width=0.3\textwidth]{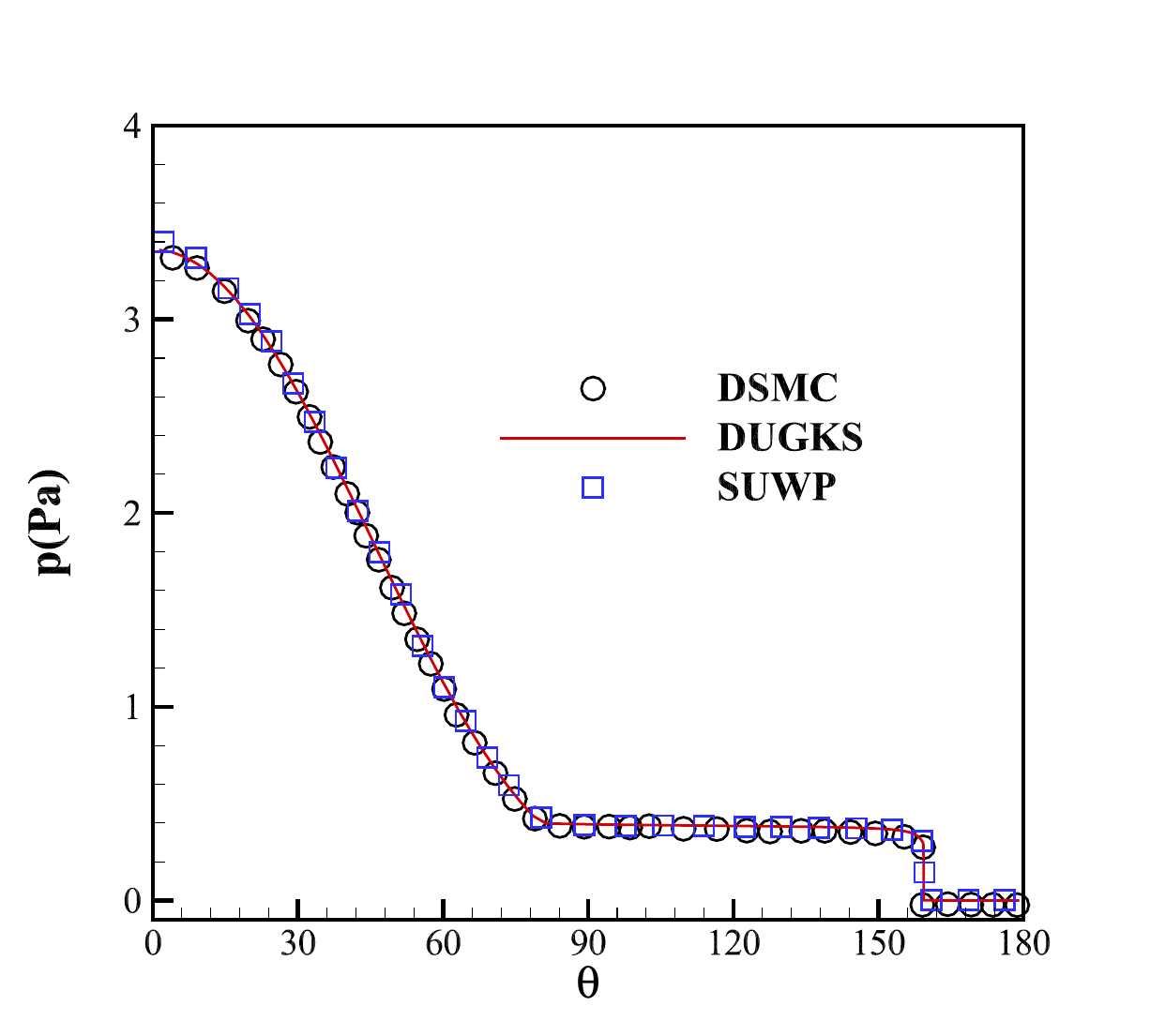}
}\hspace{0.01\textwidth}%
\subfigure[\label{Fig:case2d_blunt_ma15_kn1_wall_cf}]{
\includegraphics[width=0.3\textwidth]{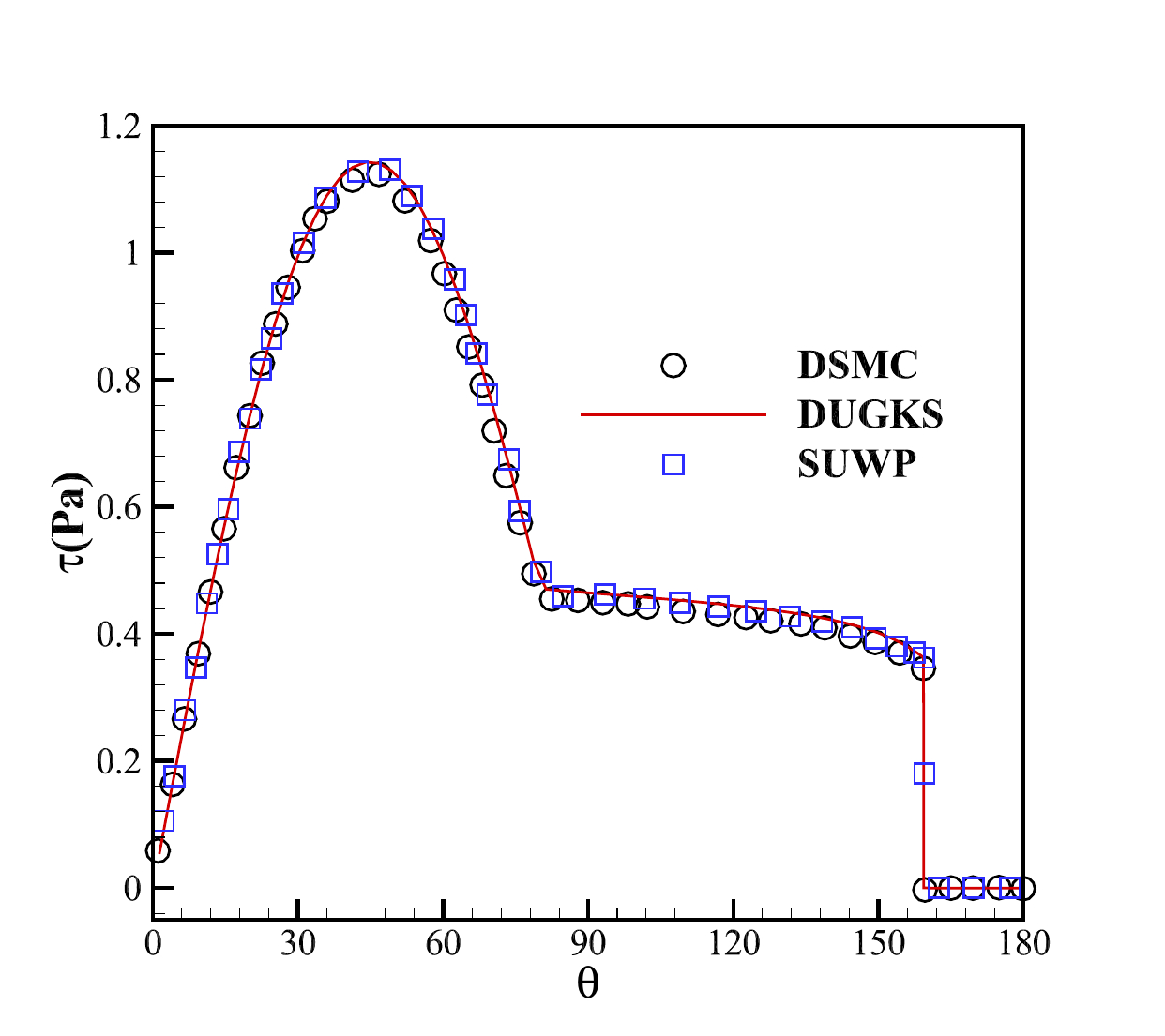}
}\hspace{0.01\textwidth}%
\subfigure[\label{Fig:case2d_blunt_ma15_kn1_wall_ch}]{
\includegraphics[width=0.3\textwidth]{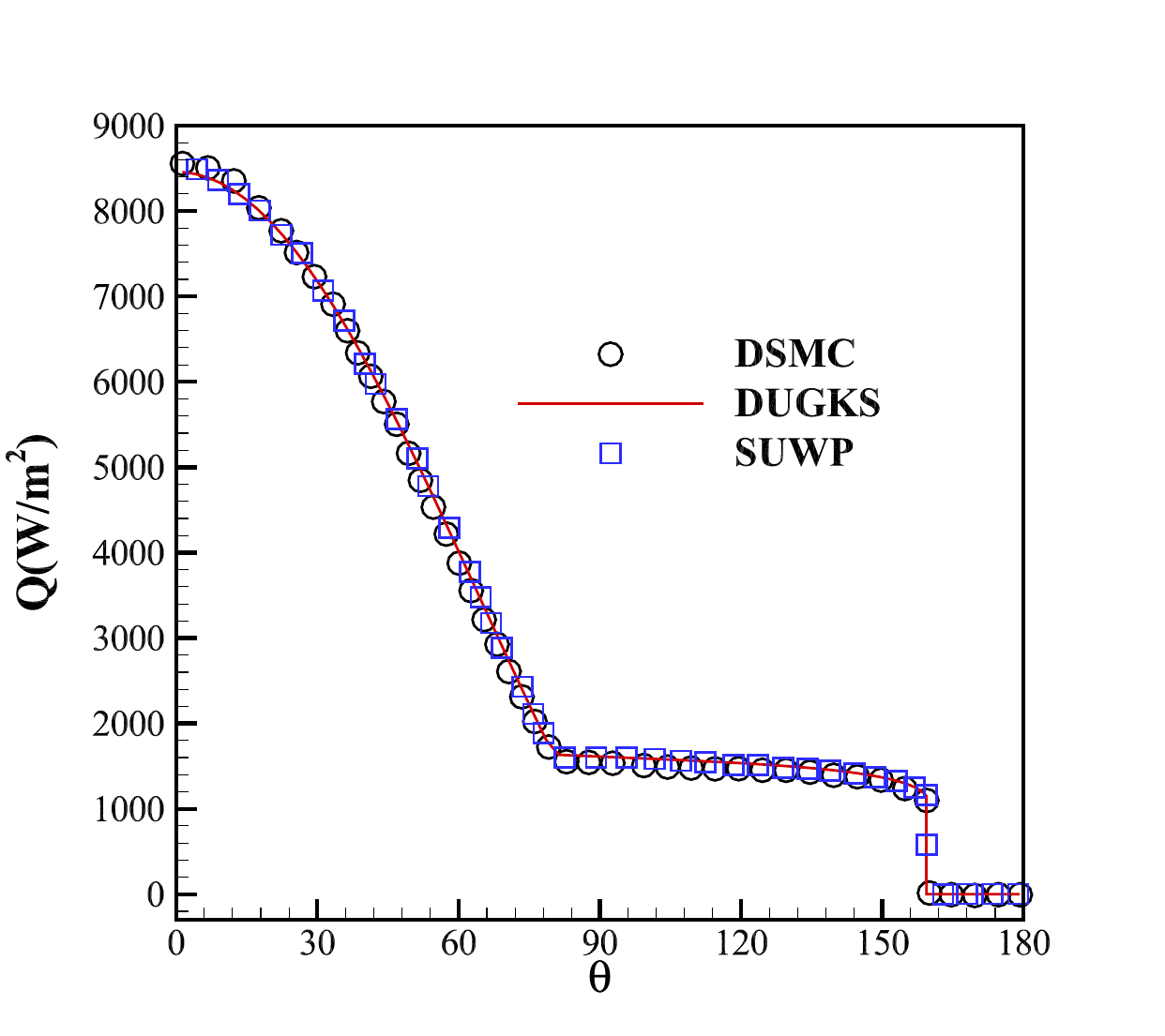}
}\\
\caption{\label{Fig:case2d_blunt_ma15_kn1_wall}The (a) pressure, (b) shear stress and (c) heat flux on the wall surface of the blunt wedge at Ma=15, Kn=1.}
\end{figure*}

To further validate the accuracy of the SUWP-vib method across varying Kn number rarefied flows, we perform a comparative analysis of hypersonic flow past a blunt wedge against DSMC and DUGKS results from Ref.~\cite{zhang2025implicit}. 
The geometry of the blunt wedge and its computational mesh are depicted in Fig. \ref{Fig:case2d_blunt_mesh}. The blunt wedge features a nose radius $R_{\rm{b}} = 0.5$ m and total length $L_{\rm{b}} = 3$ m. The reference length is set to $L_{\rm{ref}} = 2R_{\rm{b}} = 1.0$ m. Similar to the cylinder flow case, the computational domain is discretized into 22988 unstructured mesh cells. The height of the cell adjacent wall for the blunt wedge is set to $3.5 \times 10^{-3}$ m. The unstructured mesh is employed in off-wall regions to reduce cell count. The flow parameters for this test case are specified in Table \ref{table:2Dcase_blunt_Table}. Additionally, for consistent comparison with results in Ref.~\cite{zhang2025implicit}, the parameters in the vibrational kinetic model are specified as follows: $\omega_{1} = 0.75$, $\omega_{2} = 0.25$, $\omega_{3} = 0.75$, and $\omega_{4} = 0.4$. This test case employs dimensional variables for computation and comparison. The particle number in cell $N_{\rm{p}} =2 \times 10^{2}$. Fig.~\ref{Fig:case2d_blunt_ma15_kn0.1_con} and Fig.~\ref{Fig:case2d_blunt_ma15_kn1_con} present contour plots of pressure, translational temperature, rotational temperature, and vibrational temperature for the blunt wedge flow. It is observed that the SUWP-vib simulations exhibit essential agreement with DUGKS results even for test cases featuring sharp corners. Near the afterbody of the blunt wedge, a rapid temperature rise followed by gradual relaxation is observed, exhibiting characteristics consistent with the cylinder flow simulations. Fig.~\ref{Fig:case2d_blunt_ma15_kn0.1_sl} and Fig.~\ref{Fig:case2d_blunt_ma15_kn1_sl} compare pressure and temperature profiles along the stagnation streamline of the blunt cone. The SUWP-vib simulations demonstrate essential agreement with DUGKS results. Compared to DSMC, both DUGKS and SUWP-vib exhibit a slight early rise upstream of the shock front. In the region near the post-shock temperature peak, the temperature increase occurs marginally slower. This trend aligns with observations in Ref.~\cite{zhang2025implicit} and may be attributed to inherent model-level deviations. Fig.~\ref{Fig:case2d_blunt_ma15_kn0.1_wall} and Fig.~\ref{Fig:case2d_blunt_ma15_kn1_wall} present surface distributions of pressure, shear stress, and heat flux along the blunt wedge. It is observed that both pressure and heat flux exhibit good agreement with DSMC and DUGKS results. The DSMC results from Ref.~\cite{zhang2025implicit} exhibit shear stress at the nose of the blunt wedge, likely attributable to insufficient mesh refinement in this region. The shear stress distributions along the forebody show good agreement between SUWP-vib, DUGKS, and DSMC methods. Near the neck region of the blunt wedge, the SUWP and DUGKS simulations show consistent results but exhibit minor discrepancies compared to DSMC data. These deviations may stem from either insufficient statistical averaging or the different of model.

\begin{table}[h!t] 
    \centering
    \caption{The parameters of the flow past a blunt wedge.} \label{table:2Dcase_blunt_Table}
    \begin{tabular}{ c  c  c  c  c  c  c  c }
        \hline
        Ma & $L_{\rm{ref}}$ & Kn & $\rho_{\infty}(\rm{kg/m^{3}})$ & $T_{\infty}(\rm{K})$ & $T_{\rm{wall}}(\rm{K})$ & $Z_{\rm{rot}}$ & $Z_{\rm{vib}}$ \\ 
        \hline
        15  & 1m & 0.1 & $6.9588 \times 10^{-7}$ & 500 & 2000  & 3  & 35 \\
        15  & 1m & 1   & $6.9588 \times 10^{-8}$ & 500 & 2000  & 3  & 35 \\
        \hline
    \end{tabular}
\end{table}

\subsection{Flow past a sphere}
\begin{figure*}[h!t]
\centering
\subfigure[Global\label{Fig:case3d_sphere_mesh_self}]{
\includegraphics[width=0.45\textwidth]{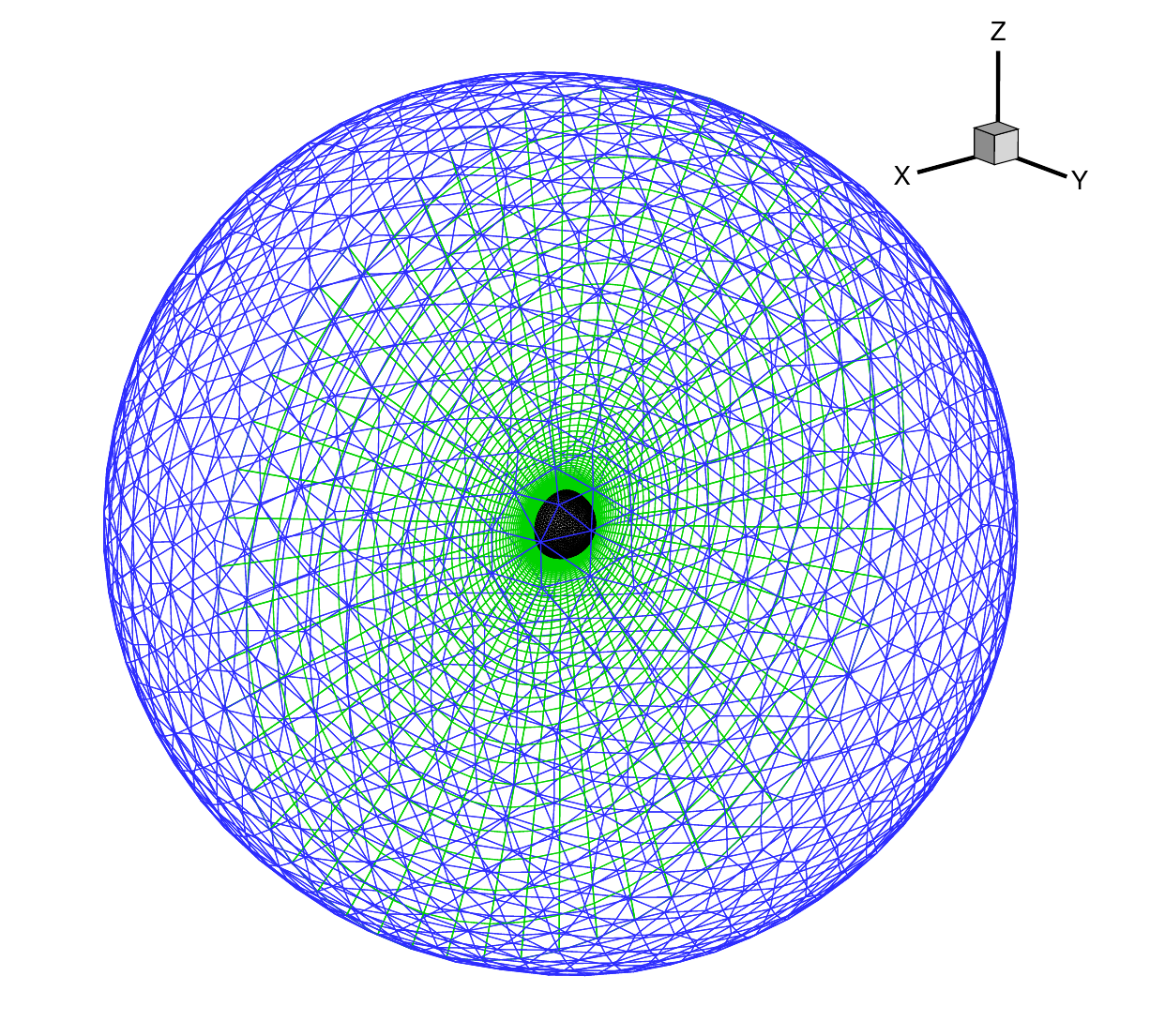}
}\hspace{0.01\textwidth}%
\subfigure[Wall\label{Fig:case3d_sphere_mesh_part}]{
\includegraphics[width=0.45\textwidth]{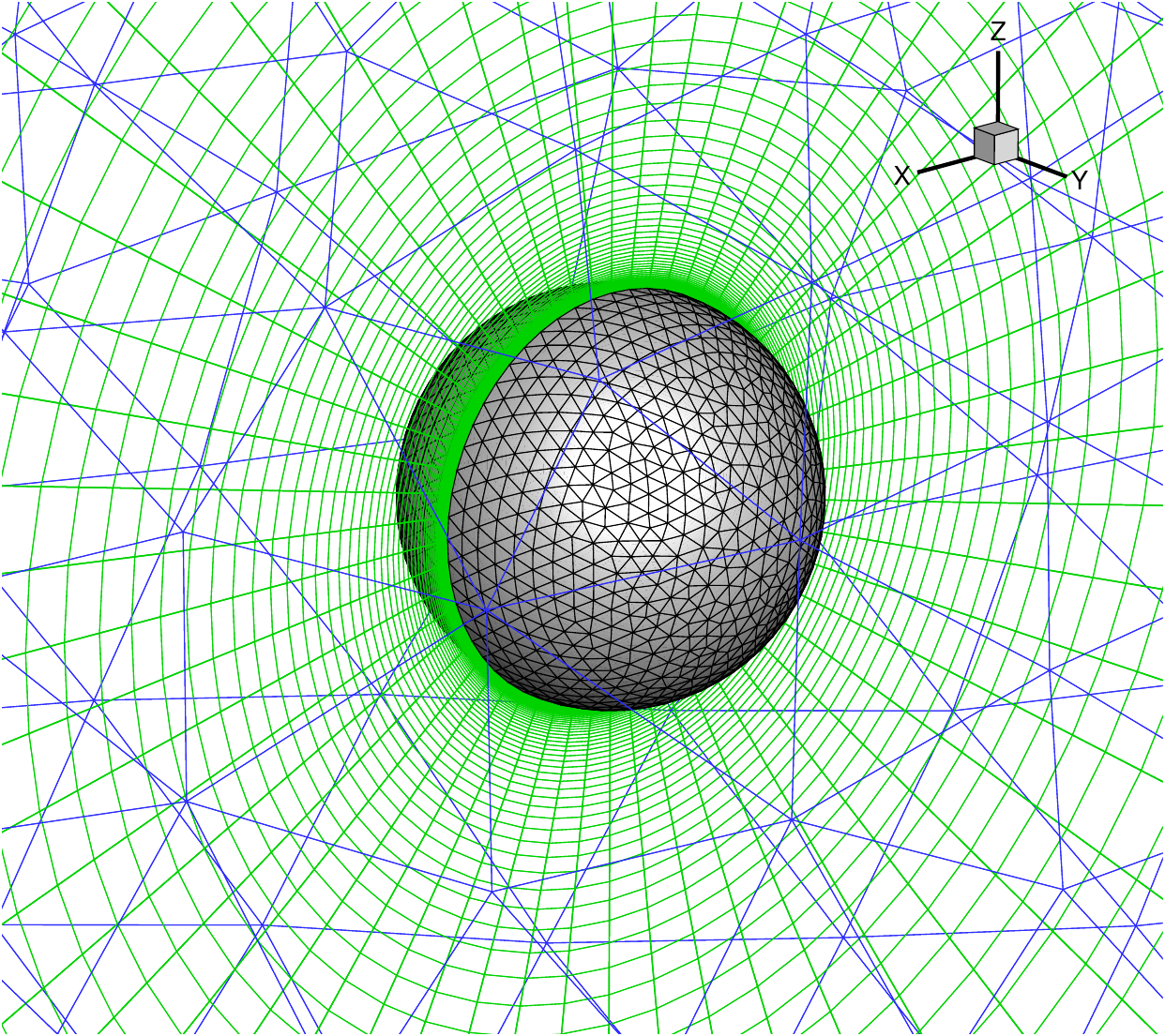}
}\\
\caption{\label{Fig:case3d_sphere_mesh}The mesh of the sphere at Ma=10.}
\end{figure*}

\begin{figure*}[h!t]
\centering
\subfigure[\label{Fig:case3d_sphere_ma10_con_Ptr}]{
\includegraphics[width=0.45\textwidth]{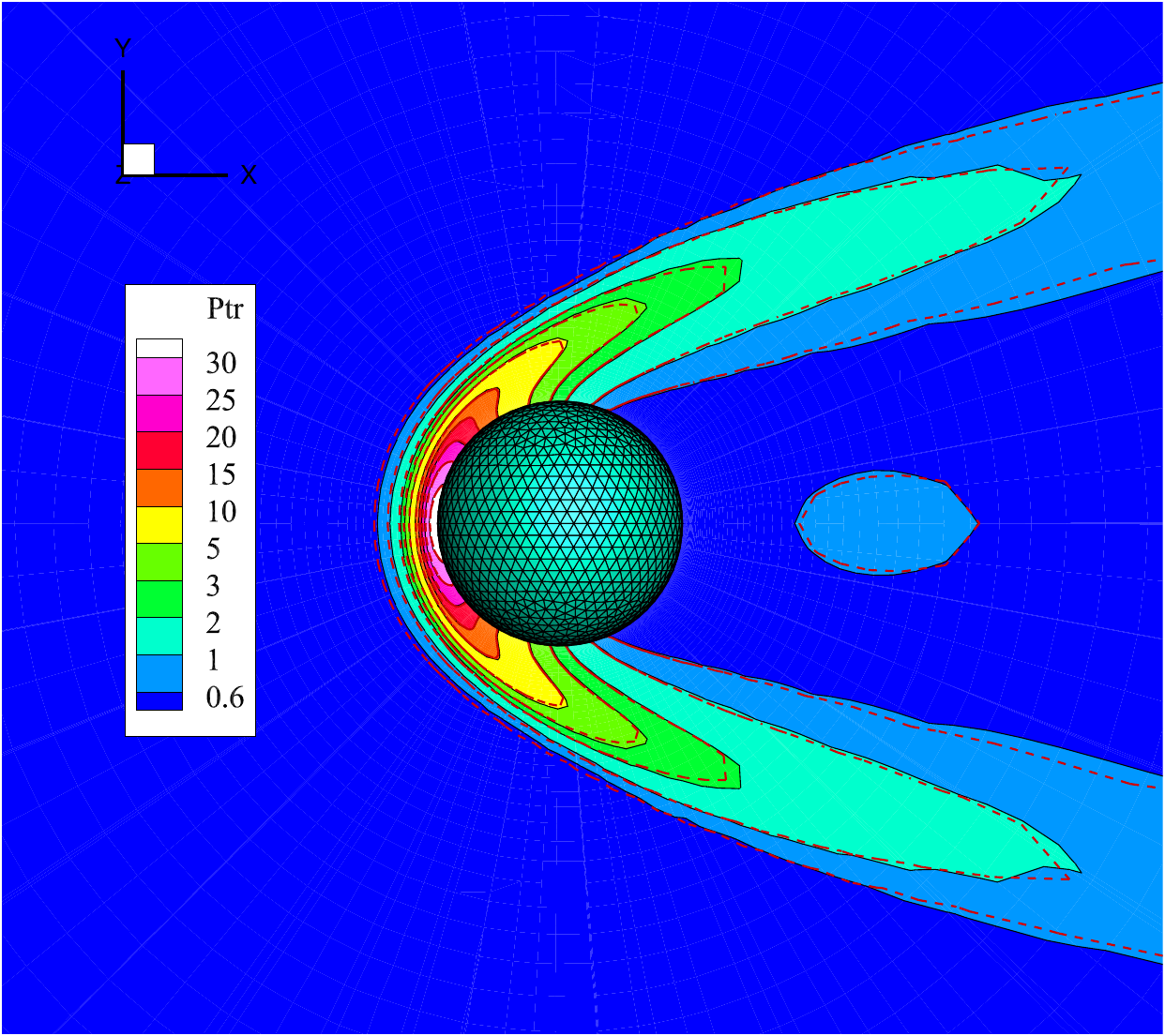}
}\hspace{0.01\textwidth}%
\subfigure[\label{Fig:case3d_sphere_ma10_con_Ttr}]{
\includegraphics[width=0.45\textwidth]{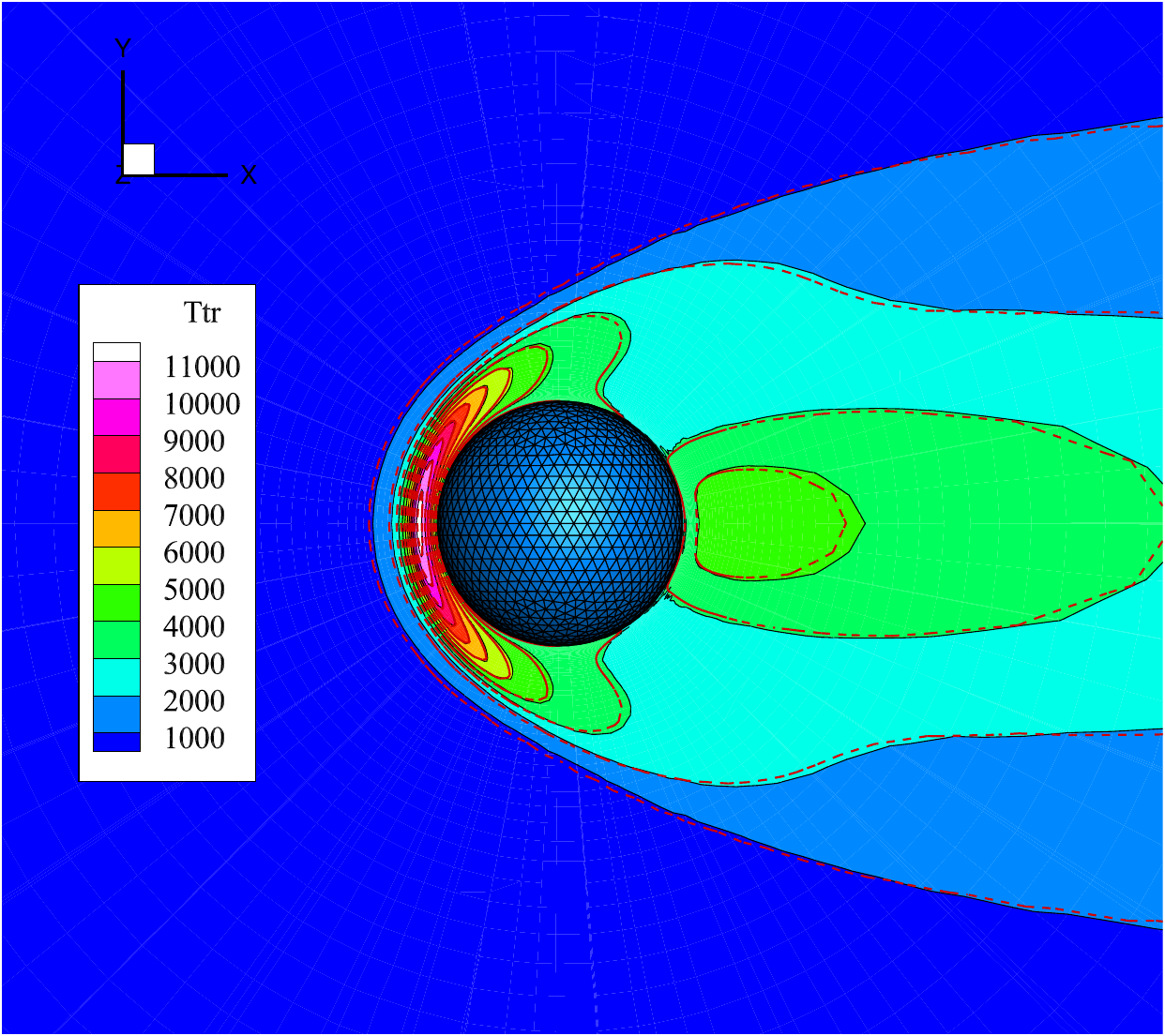}
}\\
\subfigure[\label{Fig:case3d_sphere_ma10_con_Trot}]{
\includegraphics[width=0.45\textwidth]{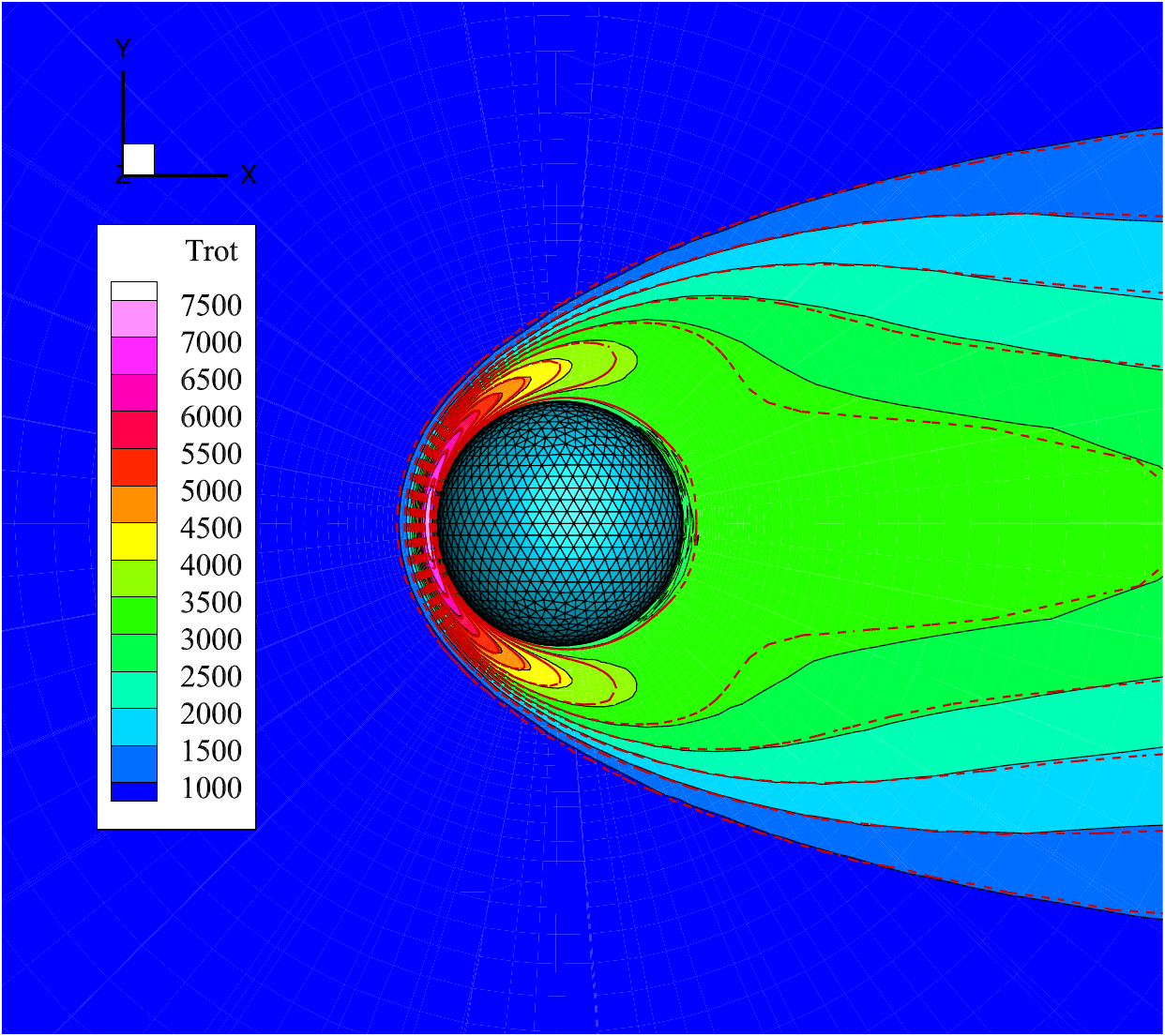}
}\hspace{0.01\textwidth}%
\subfigure[\label{Fig:case3d_sphere_ma10_con_Tvib}]{
\includegraphics[width=0.45\textwidth]{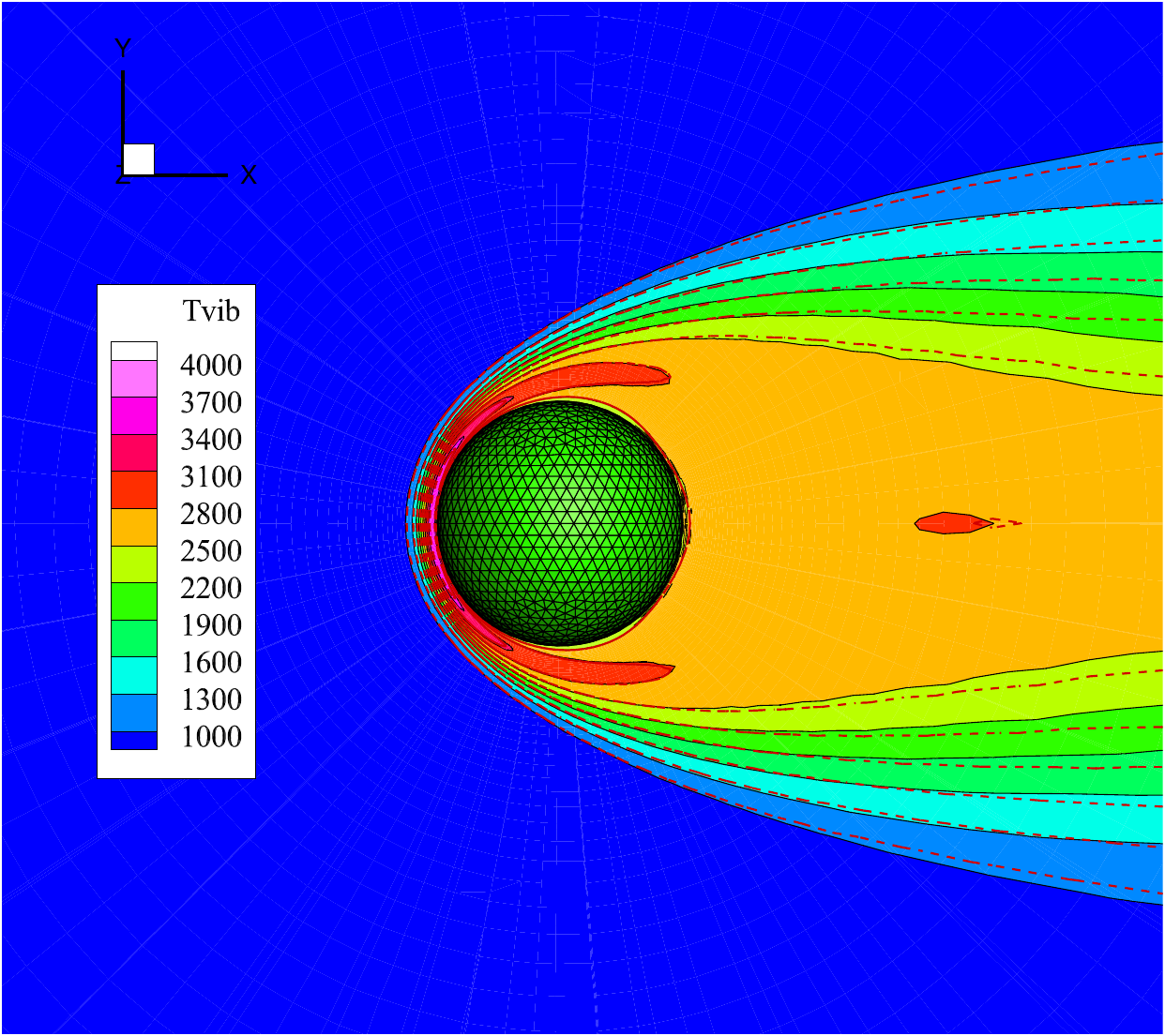}
}\\
\caption{\label{Fig:case3d_sphere_ma10_con}The (a) pressure, (b) translational temperature, (c) rotational temperature, and (d) vibrational temperature contours of the flow past a sphere at Ma=10 (The color band: SUWP, red dash line: DUGKS).}
\end{figure*}

\begin{figure*}[h!t]
\centering
\subfigure[\label{Fig:case3d_sphere_ma10_sl_rho}]{
\includegraphics[width=0.45\textwidth]{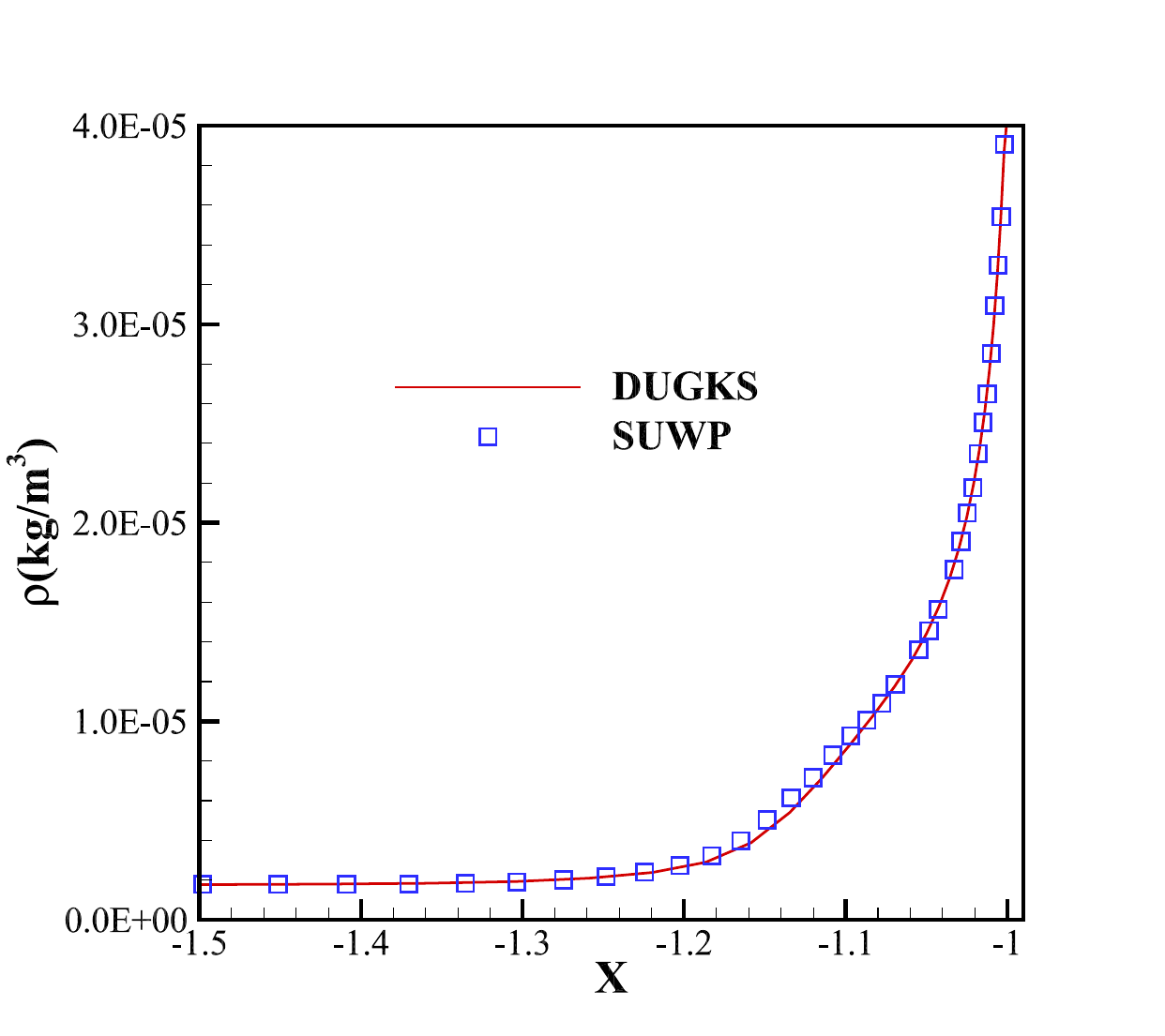}
}\hspace{0.01\textwidth}%
\subfigure[\label{Fig:case3d_sphere_ma10_sl_T}]{
\includegraphics[width=0.45\textwidth]{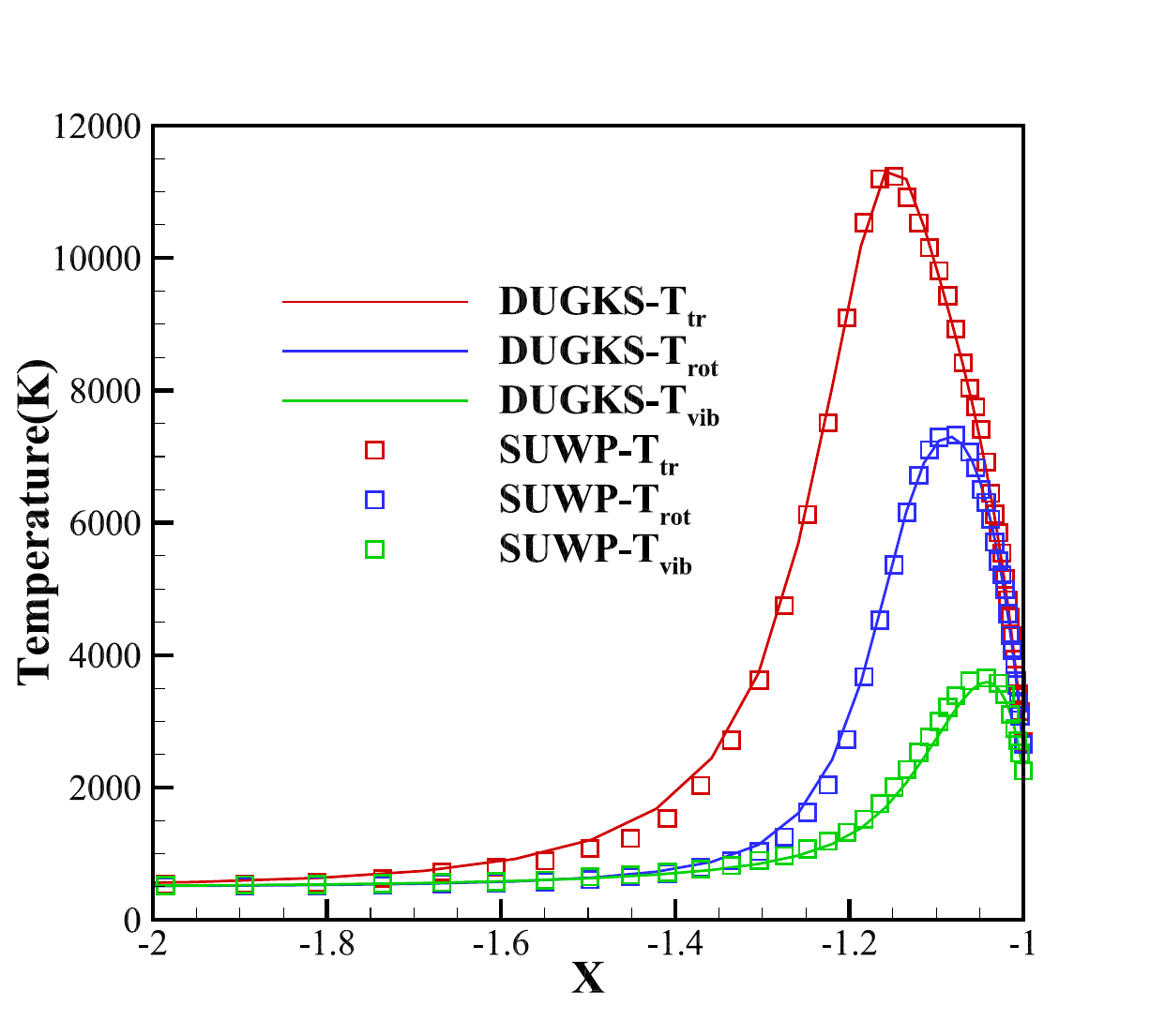}
}\\
\caption{\label{Fig:case3d_sphere_ma10_sl}The (a) density and (b) temperature along the forward stagnation line of the sphere at Ma=10.}
\end{figure*}

\begin{figure*}[h!t]
\centering
\subfigure[\label{Fig:case3d_sphere_ma10_wall_cp}]{
\includegraphics[width=0.45\textwidth]{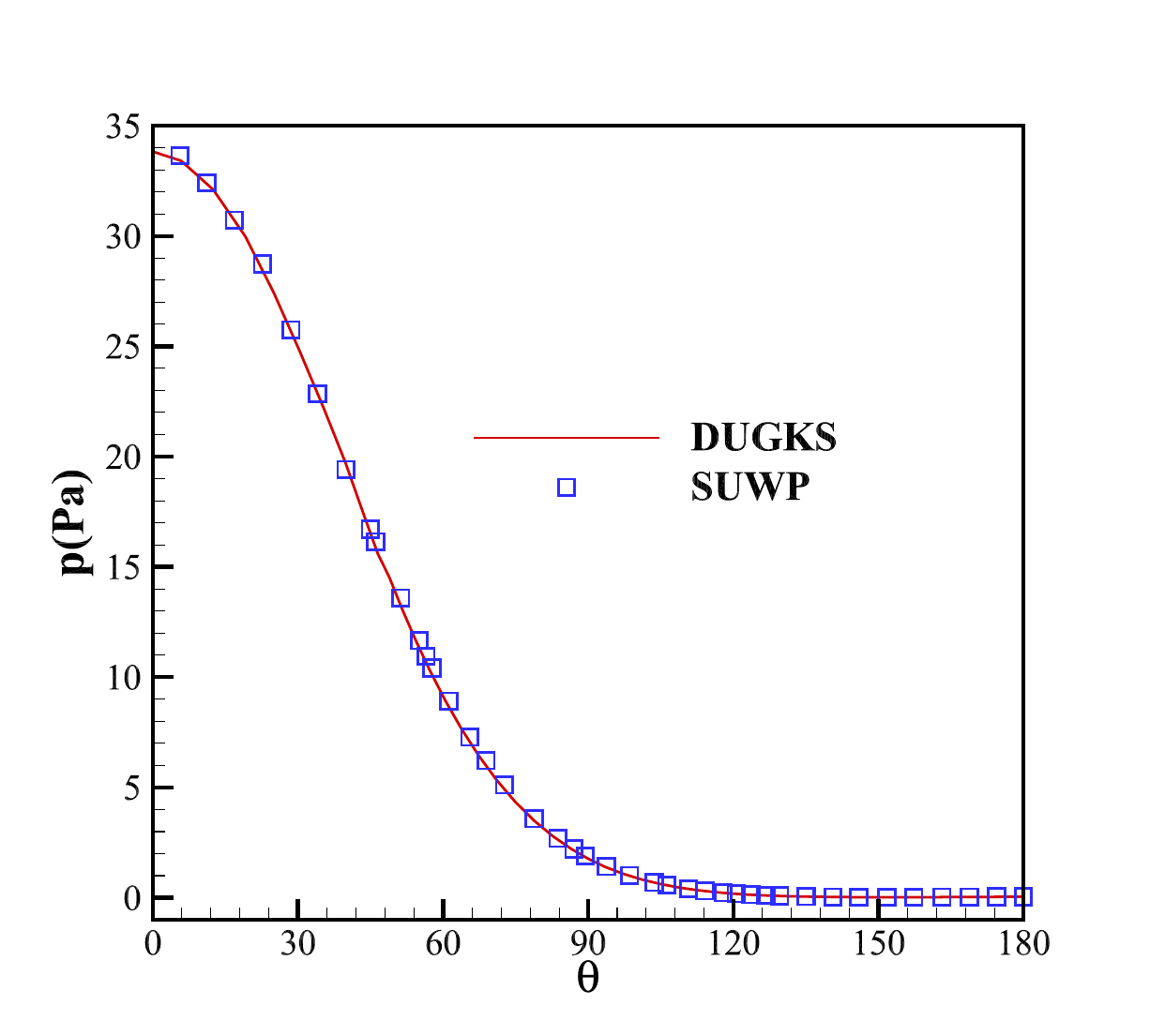}
}\hspace{0.01\textwidth}%
\subfigure[\label{Fig:case3d_sphere_ma10_wall_ch}]{
\includegraphics[width=0.45\textwidth]{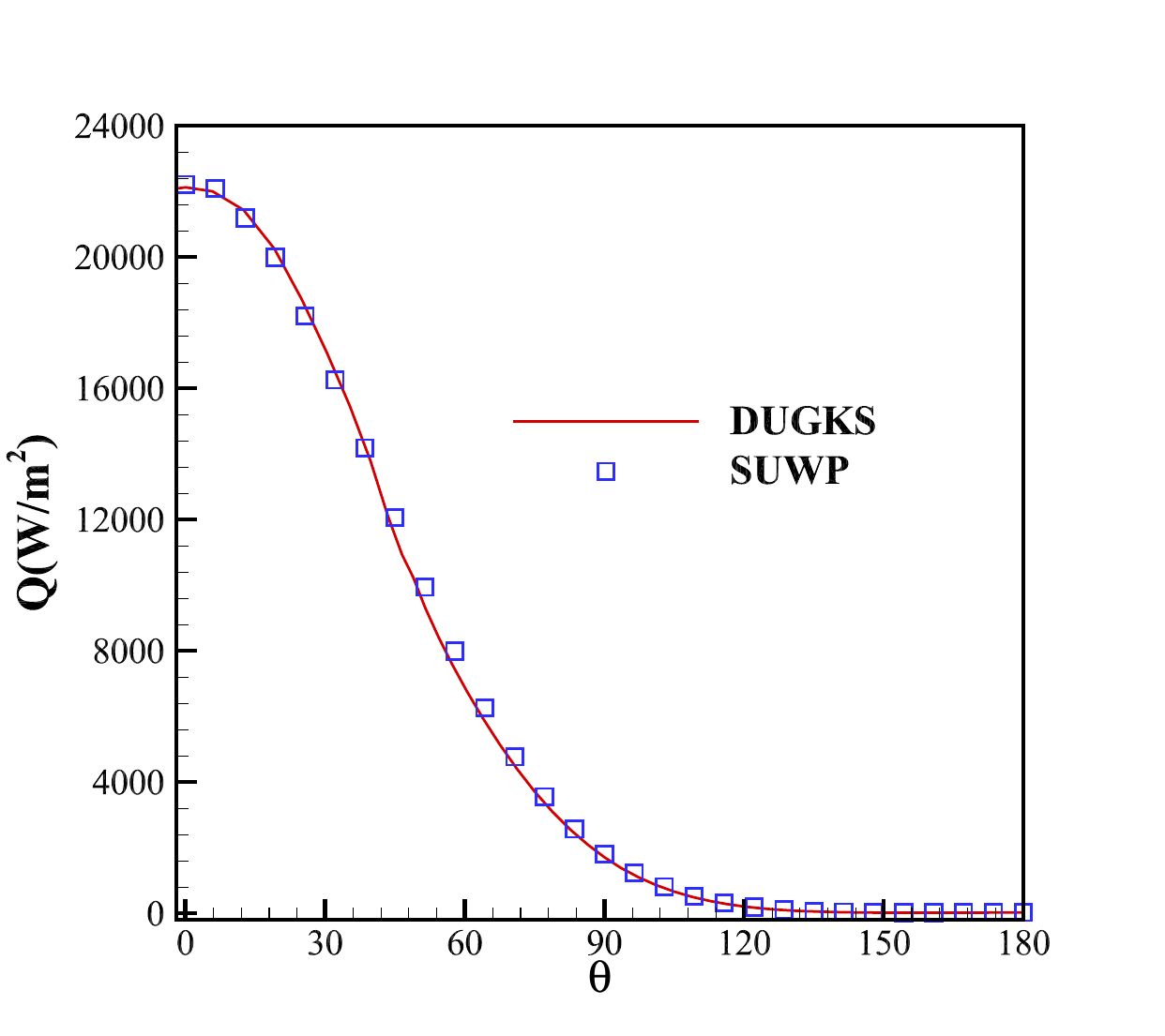}
}\\
\caption{\label{Fig:case3d_sphere_ma10_wall}The (a) pressure and (b) heat flux on the wall surface of the sphere at Ma=10.}
\end{figure*}

The supersonic flow past a sphere serves as a fundamental three-dimensional test case to validate the SUWP-vib method's capability in simulating basic 3D flows. In the simulation, the sphere radius $R_{\rm{sphere}}$ is set to 1 m. The sphere surface is discretized into 2640 triangular surface elements, with the height of the cell adjacent wall set to $2 \times 10^{-3}$ m. The mesh cell height expands from the wall with a growth ratio of 1.1, resulting in a total of 184800 cells. The computational mesh is depicted in Fig. \ref{Fig:case3d_sphere_mesh}. The flow parameters for this test case are specified in Table \ref{table:3Dcase_sphere_Table}. This test case employs dimensional
variables for computation and comparison. The particle number in cell $N_{\rm{p}} =1.2 \times 10^{2}$. 
Fig. \ref{Fig:case3d_sphere_ma10_con} presents contour plots of pressure, translational temperature, rotational temperature, and vibrational temperature for the sphere flow. Fig. \ref{Fig:case3d_sphere_ma10_sl} and Fig. \ref{Fig:case3d_sphere_ma10_wall} respectively show the stagnation line distributions of density and temperature, and the surface distributions of pressure and heat flux. It is observed that the computational results of the SUWP-vib method demonstrate good agreement with simulations of the DUGKS solver from Ref.~\cite{ZHANG2023107079}. It demonstrates that the SUWP-vib method possesses the capability to simulate basic three-dimensional flows.

\begin{table}[ht] 
    \centering
    \caption{The parameters of the flow past a sphere.} \label{table:3Dcase_sphere_Table}
    \begin{tabular}{ c  c  c  c  c  c  c }
        \hline
        Ma & $L_{\rm{ref}}$ & $\rho_{\infty}(\rm{kg/m^{3}})$ & $T_{\infty}(\rm{K})$ & $T_{\rm{wall}}(\rm{K})$ & $Z_{\rm{rot}}$ & $Z_{\rm{vib}}$ \\ 
        \hline
        10  & 1m & $1.73969 \times 10^{-6}$ & 500 & 2000  & 3.5  & 50 \\
        \hline
    \end{tabular}
\end{table}

\subsection{Apollo 6 command module}
\begin{figure*}[h!t]
\centering
\subfigure[Global\label{Fig:case3d_apollo_vib_100km_mesh_self}]{
\includegraphics[width=0.45\textwidth]{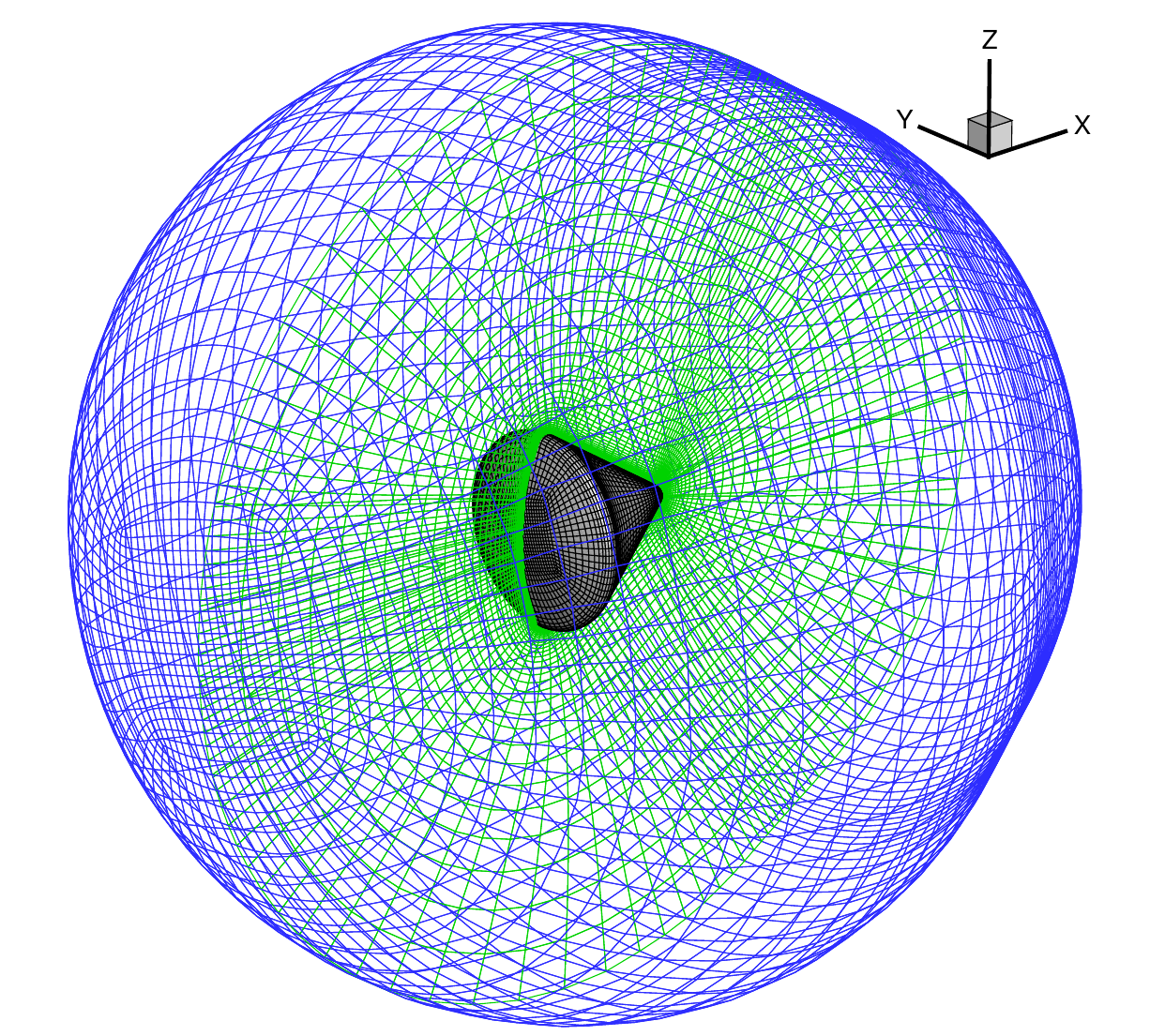}
}\hspace{0.01\textwidth}%
\subfigure[Wall\label{Fig:case3d_apollo_vib_100km_mesh_part}]{
\includegraphics[width=0.45\textwidth]{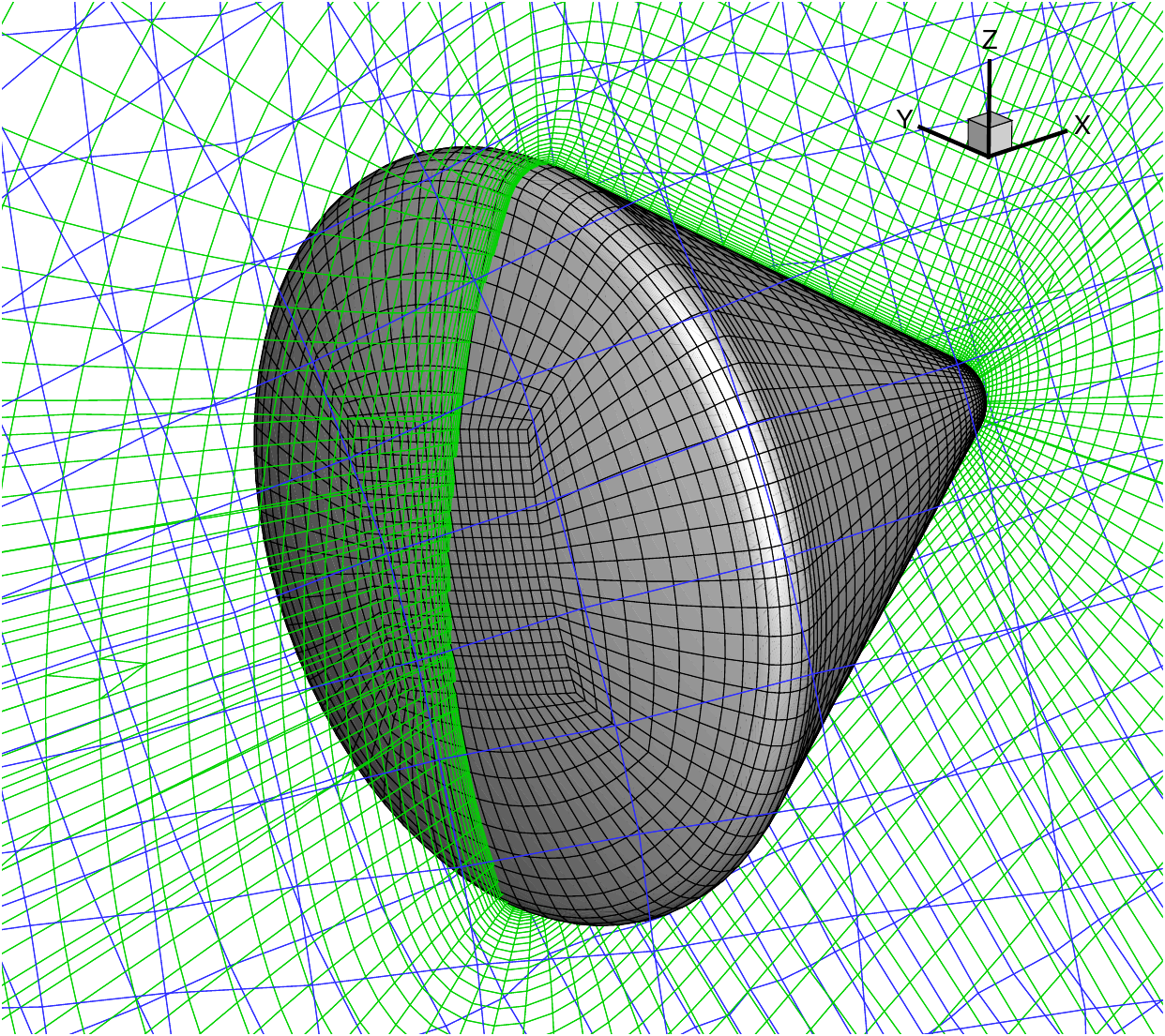}
}\\
\caption{\label{Fig:case3d_apollo_vib_100km_mesh}The mesh of the Apollo 6 command module.}
\end{figure*}

\begin{figure*}[h!t]
\centering
\subfigure[\label{Fig:case3d_apollo_vib_100km_con_Ptr}]{
\includegraphics[width=0.45\textwidth]{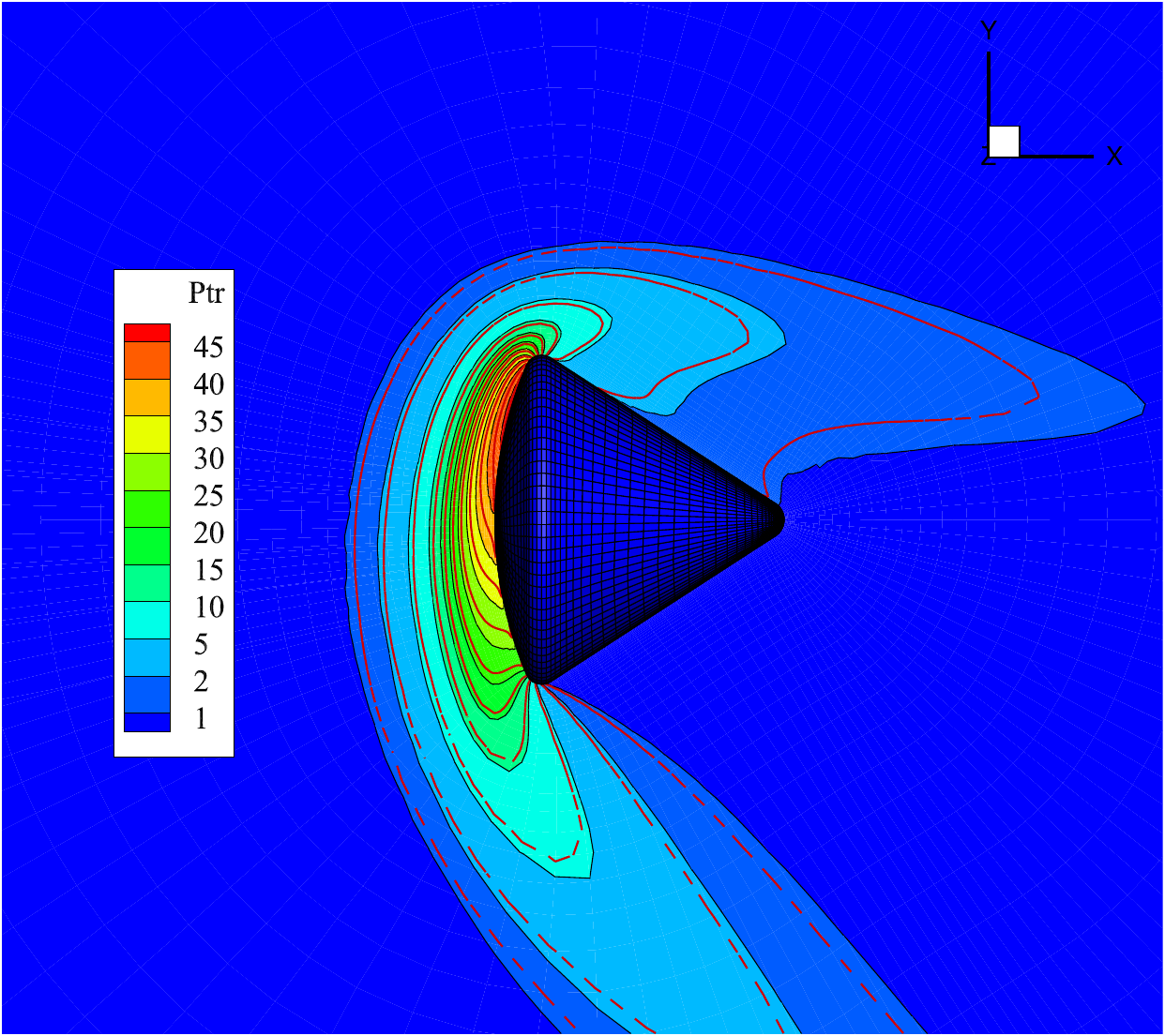}
}\hspace{0.01\textwidth}%
\subfigure[\label{Fig:case3d_apollo_vib_100km_con_Ttr}]{
\includegraphics[width=0.45\textwidth]{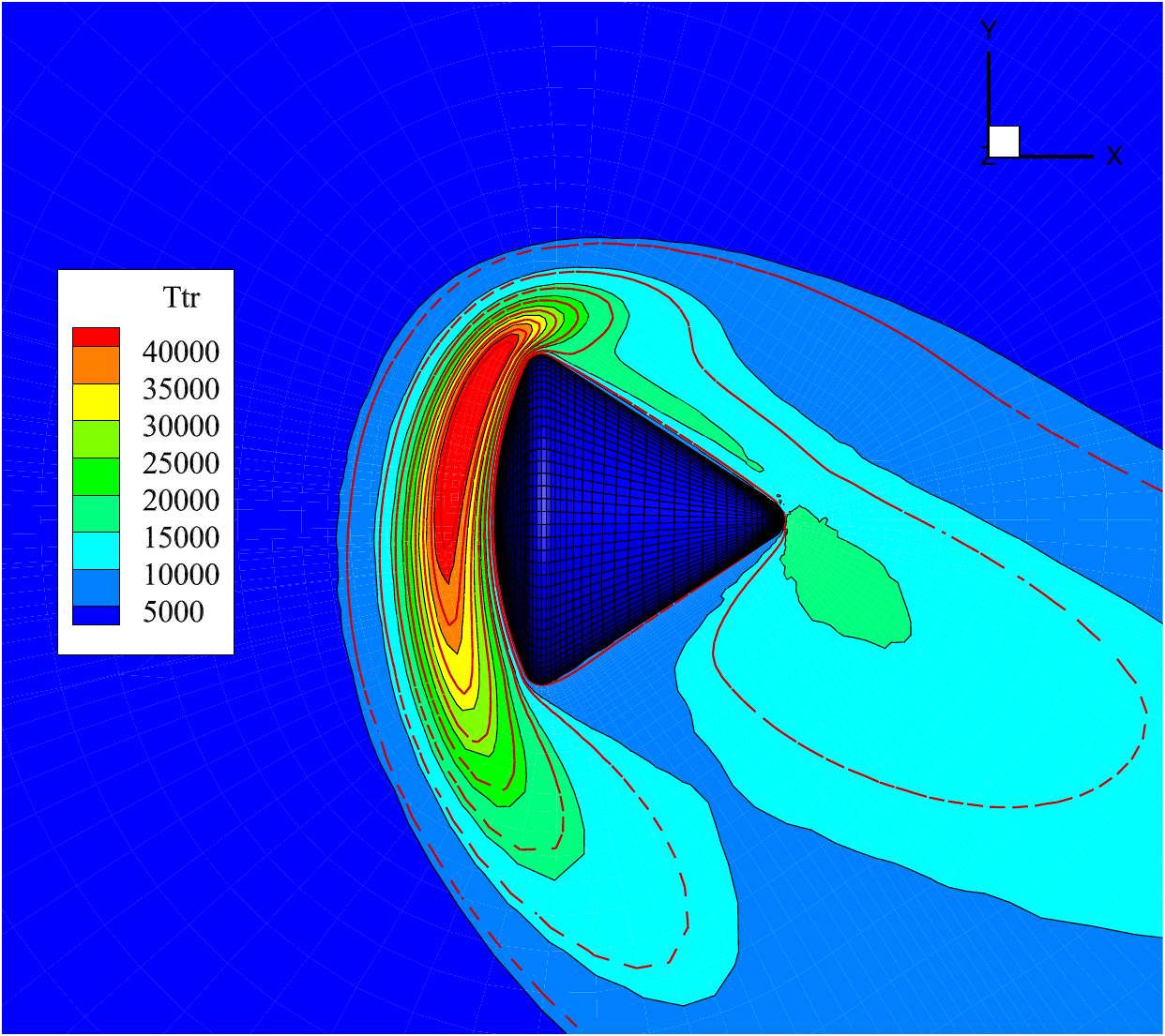}
}\\
\subfigure[\label{Fig:case3d_apollo_vib_100km_con_Trot}]{
\includegraphics[width=0.45\textwidth]{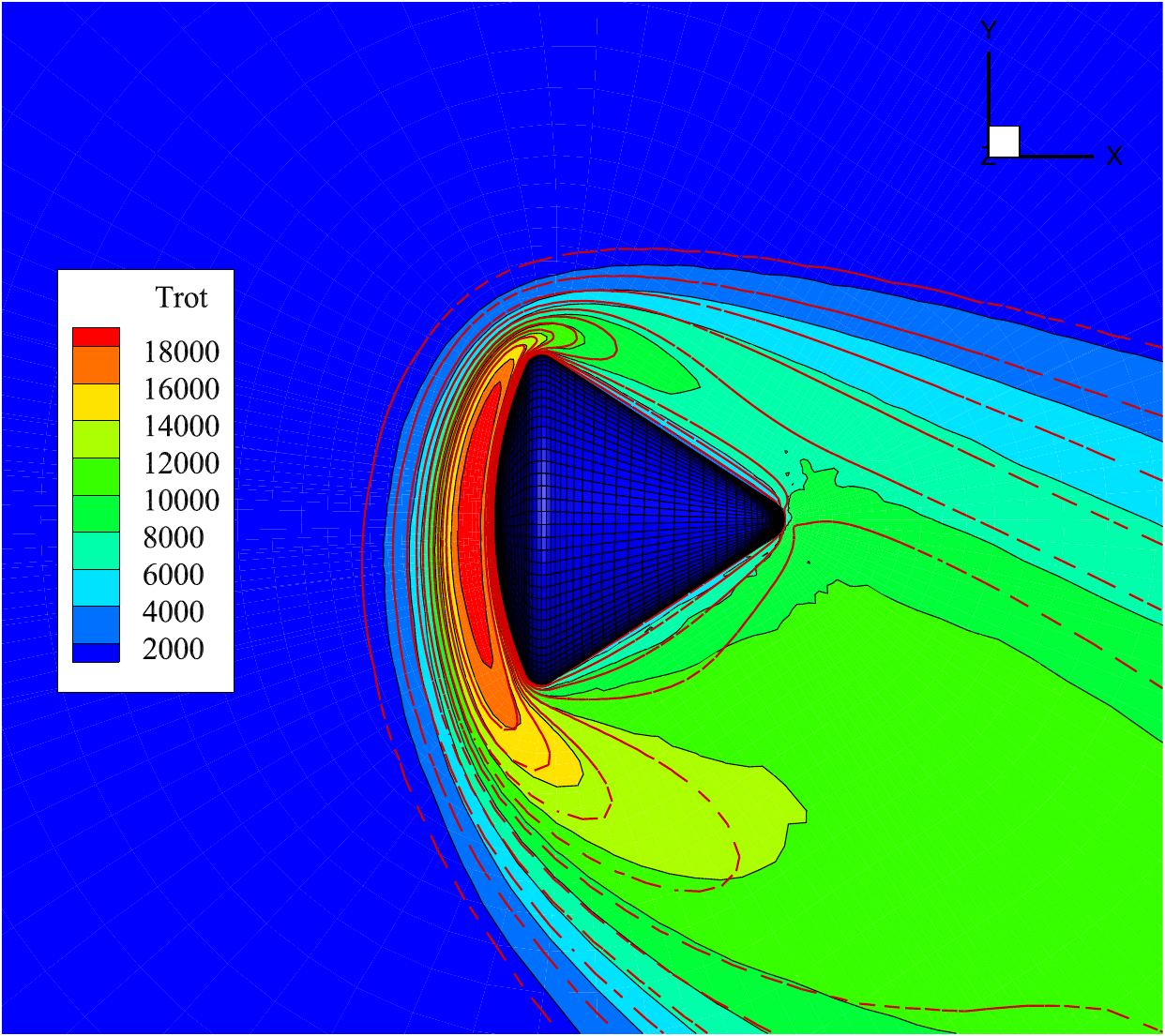}
}\hspace{0.01\textwidth}%
\subfigure[\label{Fig:case3d_apollo_vib_100km_con_Tvib}]{
\includegraphics[width=0.45\textwidth]{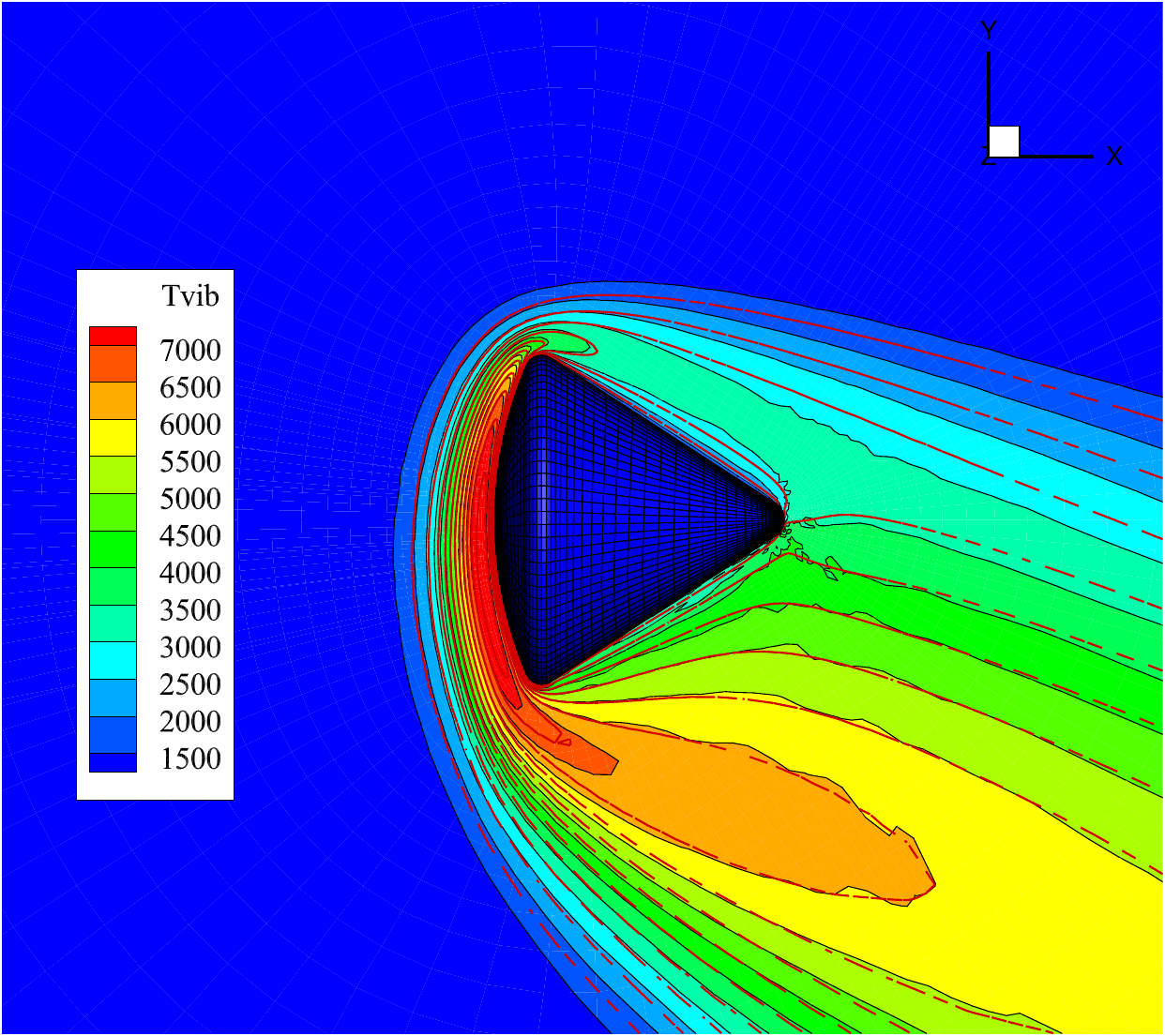}
}\\
\caption{\label{Fig:case3d_apollo_vib_100km_con}The (a) pressure, (b) translational temperature, (c) rotational temperature, and (d) vibrational temperature contours of the Apollo 6 command module (The color band: SUWP, red dash line: DUGKS).}
\end{figure*}

\begin{figure*}[h!t]
\centering
\subfigure[\label{Fig:case3d_apollo_vib_100km_sl_cp}]{
\includegraphics[width=0.45\textwidth]{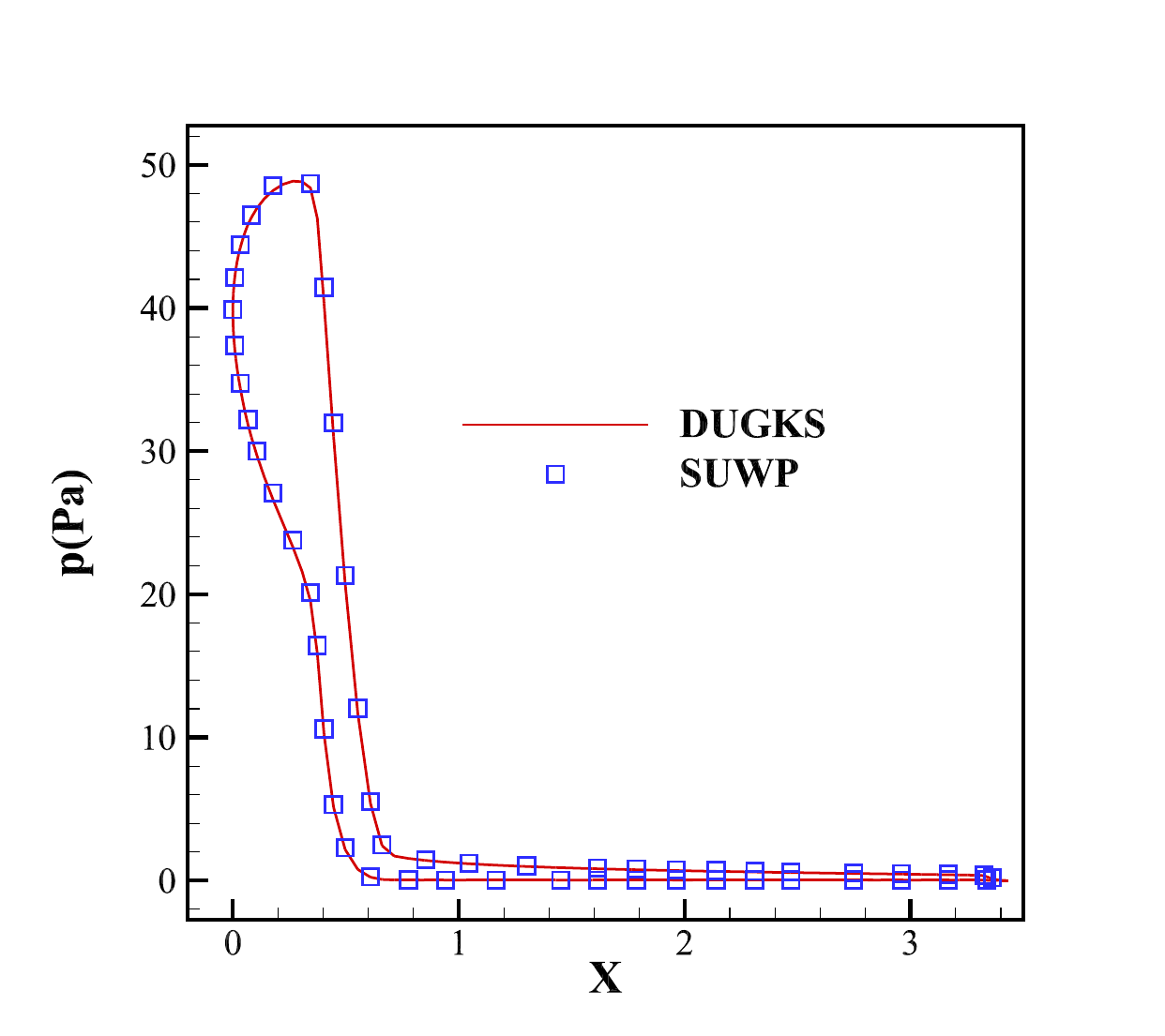}
}\hspace{0.01\textwidth}%
\subfigure[\label{Fig:case3d_apollo_vib_100km_sl_ch}]{
\includegraphics[width=0.45\textwidth]{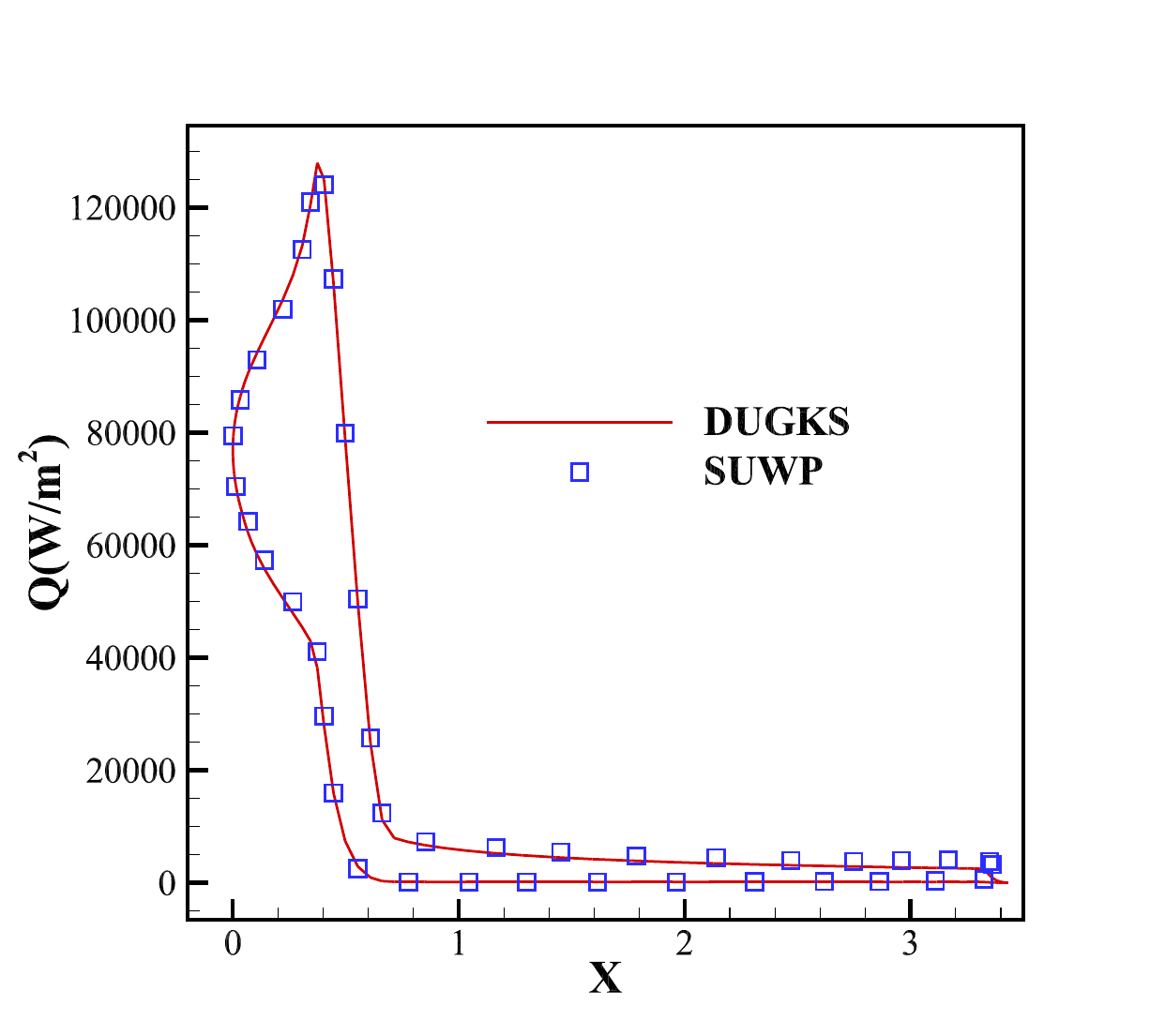}
}\\
\caption{\label{Fig:case3d_apollo_vib_100km_wall}The (a) pressure and (b) heat flux on the wall surface of the Apollo 6 command module.}
\end{figure*}

This section simulates the reentry process of the Apollo 6 command module at 100 km altitude, to further assess the SUWP-vib method's capability for three-dimensional flow simulations and compare its computational efficiency with alternative simulation approaches. We conduct simulations of the Apollo 6 command module using both the DUGKS solver~\cite{ZHANG2023107079} and the SUWP-vib method, followed by comparative analysis of their computational results. In Moss et al.'s study, simulations of the Apollo 6 command module were performed using DSMC, obtaining its lift and drag coefficients~\cite{moss2006dsmc}. The present flow conditions align with those of Moss et al., as specified in Table \ref{table:3Dcase_apollo_Table}. This test case employs dimensional 
variables for computation and comparison. The computational mesh for this test case is identical to that used in Ref.~\cite{rui_zhang_conservative_2024}. The Apollo 6 command module surface is discretized into 4420 elements, and the computational domain contains a total of 154700 elements, as shown in Fig. \ref{Fig:case3d_apollo_vib_100km_mesh}. The particle number of cell $N_{\rm{p}} =1.2 \times 10^{2}$. Fig. \ref{Fig:case3d_apollo_vib_100km_con} presents contour plots of pressure, translational temperature, rotational temperature, and vibrational temperature for the Apollo 6 command module flow. Fig. \ref{Fig:case3d_apollo_vib_100km_wall} shows the surface distributions of pressure and heat flux on the Apollo 6 command module. Good agreement is observed between the SUWP-vib computational results and the DUGKS simulations. Table \ref{table:case3D_apollo_force} compares the drag and lift coefficients obtained by SUWP-vib and other numerical methods with Moss et al.'s benchmark data, along with their relative errors. The SUWP-vib method exhibits errors of 0.42\% for drag coefficient and 3.06\% for lift coefficient, demonstrating good accuracy. This validates SUWP-vib's capability to simulate three-dimensional vehicle flows.

\begin{table}[h!t] 
    \centering
    \caption{The parameters of the Apollo 6 command module.} \label{table:3Dcase_apollo_Table}
    \begin{tabular}{ c  c  c  c  c  c  c  c }
        \hline
        Angle of attack($^{\circ}$) & $L_{\rm{ref}}$ & $\rho_{\infty}(\rm{kg/m^{3}})$ & $U_{\infty}(\rm{m/s})$ & $T_{\infty}(\rm{K})$ & $T_{\rm{wall}}(\rm{K})$ & $Z_{\rm{rot}}$ & $Z_{\rm{vib}}$ \\ 
        \hline
        -25  & 3.9116m & $5.5824 \times 10^{-7}$ & 9600 & 194 & 1146  & 5  & 50 \\
        \hline
    \end{tabular}
\end{table}

\begin{table}[h!t]   
\begin{center}   
\caption{Comparison of the drag coefficients for the Apollo 6 command module.}  
\label{table:case3D_apollo_force} 
{
\begin{tabular}{c p{3cm}<{\centering} p{2cm}<{\centering} p{2cm}<{\centering}}
\hline
  & DSMC: Moss  & DUGKS & SUWP  \\ 
\hline
Drag & 1.431  & 1.399 & 1.425 \\
Relative error  & - & 2.24\% & 0.42\% \\
\hline
Lift & 0.359  & 0.350 & 0.348 \\
Relative error  & - & 2.51\% &  3.06\% \\
\hline 
\end{tabular}
}
\end{center}   
\end{table}

Next, we analyze the efficiency of the SUWP-vib method. For the case, the time-averaging starts from 17000 steps and continues for 8000 steps with an initial field computed by 60000 steps from KIF~\cite{KIF2}. The total computation takes 25000 time steps, and runs on a workstation with AMD EPYC 7763 at 2.45GHz with 108 cores. Compared with the implicit DUGKS method~\cite{ZHANG2023107079} with the same mesh and conditions (runs on a workstation with Intel(R) Xeon(R) Gold 6258 @2.70 GHz with 640 cores), the results are shown in Table \ref{table:case3D_apollo_efficient}.
The SUWP-vib method achieves a speedup ratio of 2.36 compared to the implicit DUGKS method while maintaining comparable accuracy. Furthermore, the DUGKS method requires over 20000 velocity space cells~\cite{rui_zhang_conservative_2024} for the case, while the wave-particle method needs only fewer than 200 particles per cell. The memory consumption of the wave-particle method are one to two orders of magnitude smaller than those of the DUGKS method. This demonstrates that SUWP-vib offers significant cost advantages when simulating high-speed flow problems with high mesh cell amounts.

\begin{table}[h!t]   
\begin{center}   
\caption{Comparison of the computational cost in the Apollo 6 command module case.}  
\label{table:case3D_apollo_efficient} 
\begin{tabular}{ccc | c}
\hline
& Time (hour) & CPU hours (GHz $\cdot$ hour) & SUWP speedup ratio \\ 
\hline
Implicit DUGKS & 3.27 & 2092.8 & 2.36\\
SUWP & 8.22 & 887.86 & -\\
\hline
\end{tabular}
\end{center}   
\end{table}

\subsection{Space station Mir}
\begin{figure*}[h!t]
\centering
{
\includegraphics[trim=10 60 10 75, clip, width=0.7\textwidth]{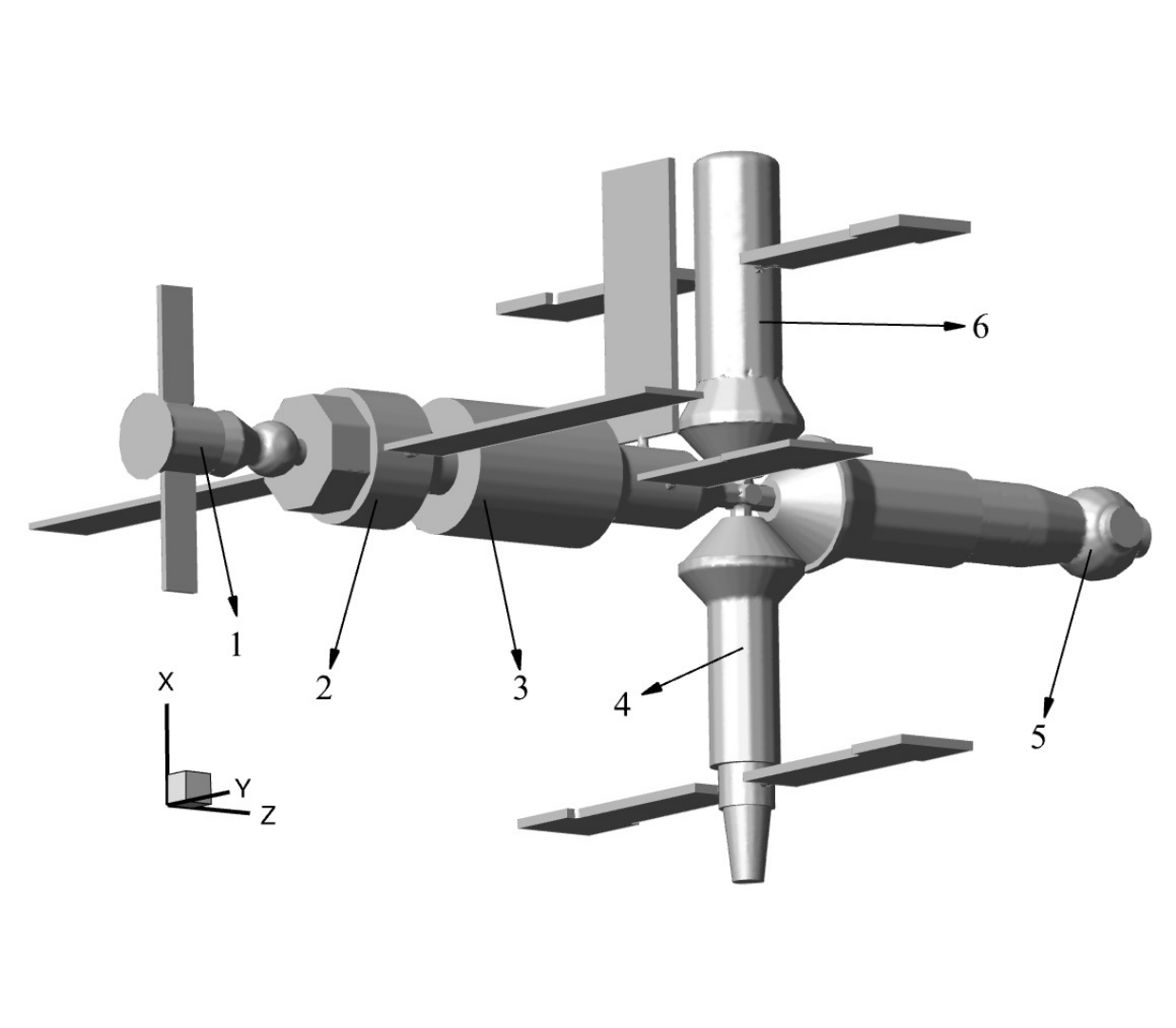}
}
\caption{\label{Fig:case3d_mir_shape}The module composition of the SS Mir: (1) the progress spacecraft; (2) Kvant; (3) the base block; (4)Kvant-2; (5) Kristall; and (6) Spektr.}
\end{figure*}

\begin{figure*}[h!t]
\centering
\subfigure[Global\label{Fig:case3d_mir_mesh_self}]{
\includegraphics[width=0.45\textwidth]{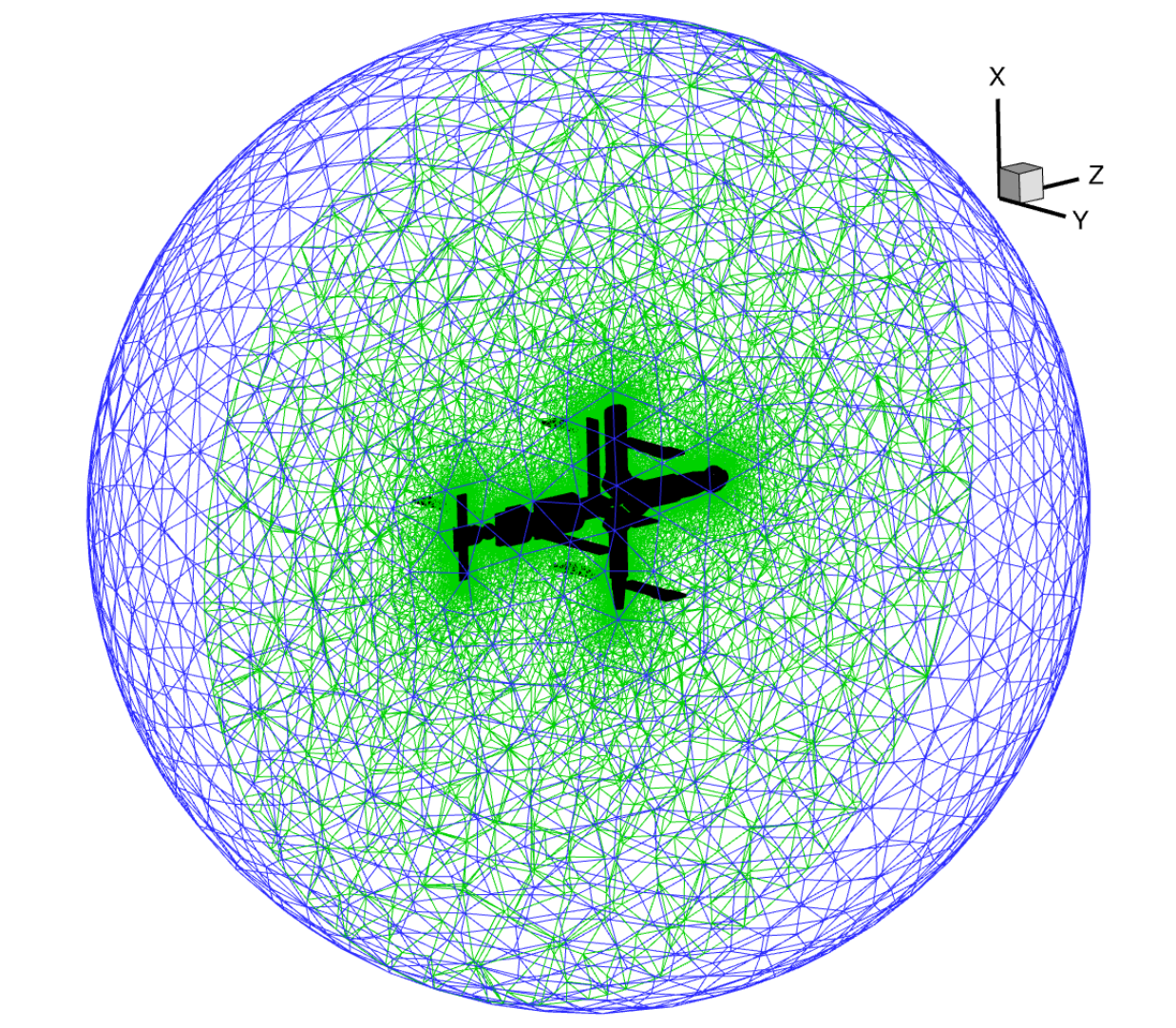}
}\hspace{0.01\textwidth}%
\subfigure[Wall\label{Fig:case3d_mir_mesh_part}]{
\includegraphics[width=0.45\textwidth]{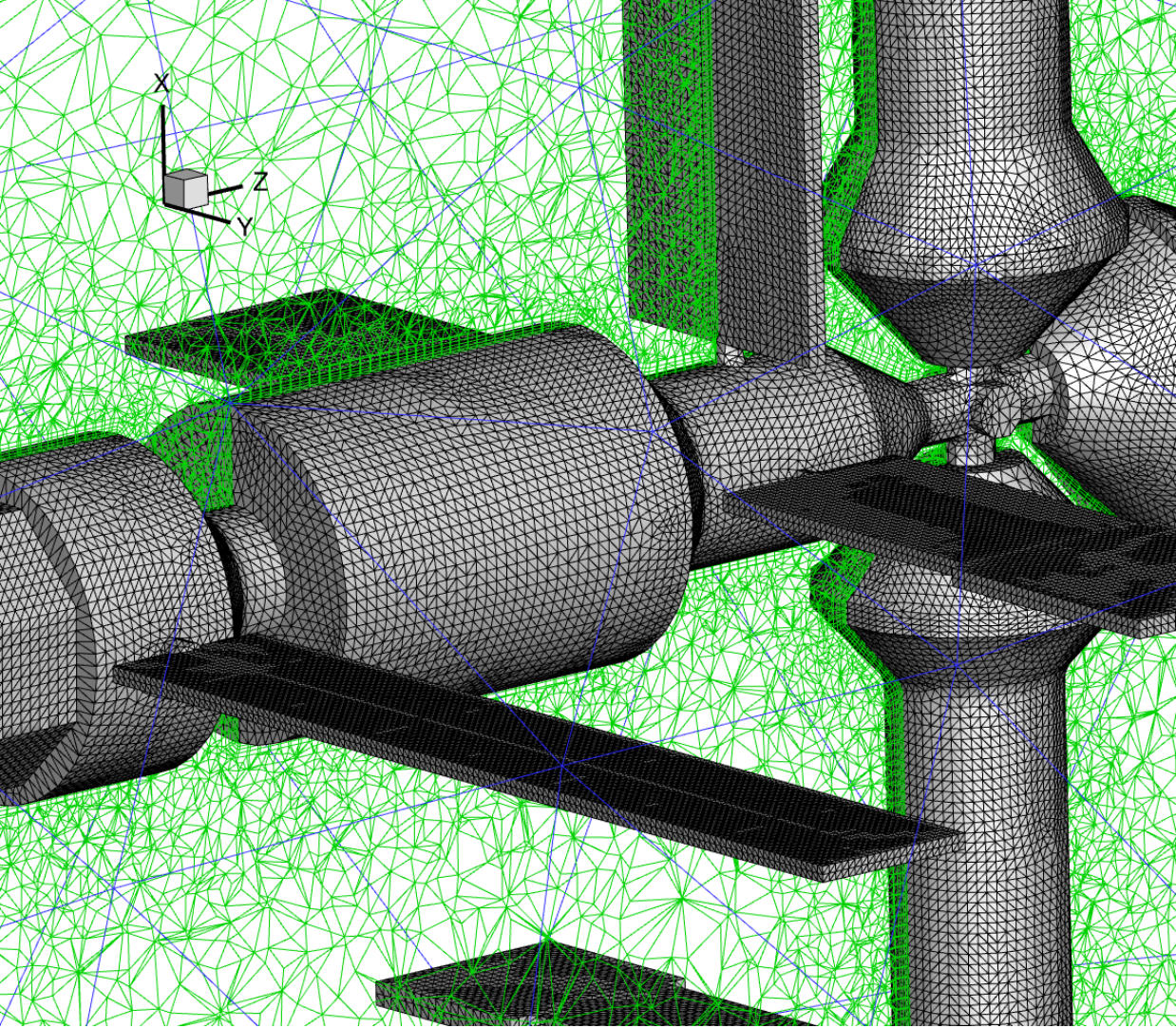}
}\\
\caption{\label{Fig:case3d_mir_mesh}The mesh of the SS Mir.}
\end{figure*}

\begin{figure*}[h!t]
\centering
\subfigure[\label{Fig:case3d_mir_rho_noISO}]{
\includegraphics[width=0.45\textwidth]{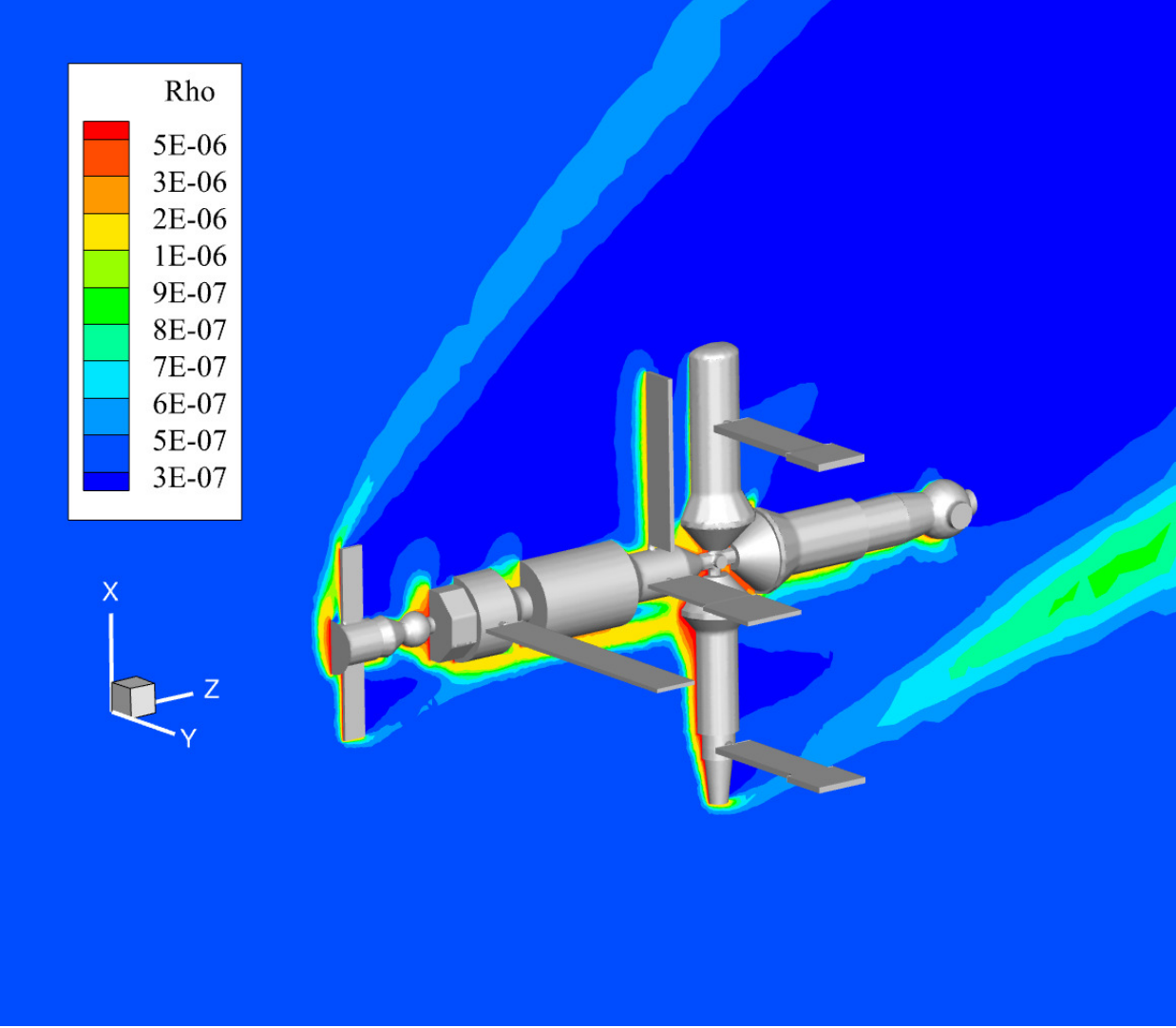}
}\hspace{0.01\textwidth}%
\subfigure[\label{Fig:case3d_mir_rho_isISO}]{
\includegraphics[width=0.45\textwidth]{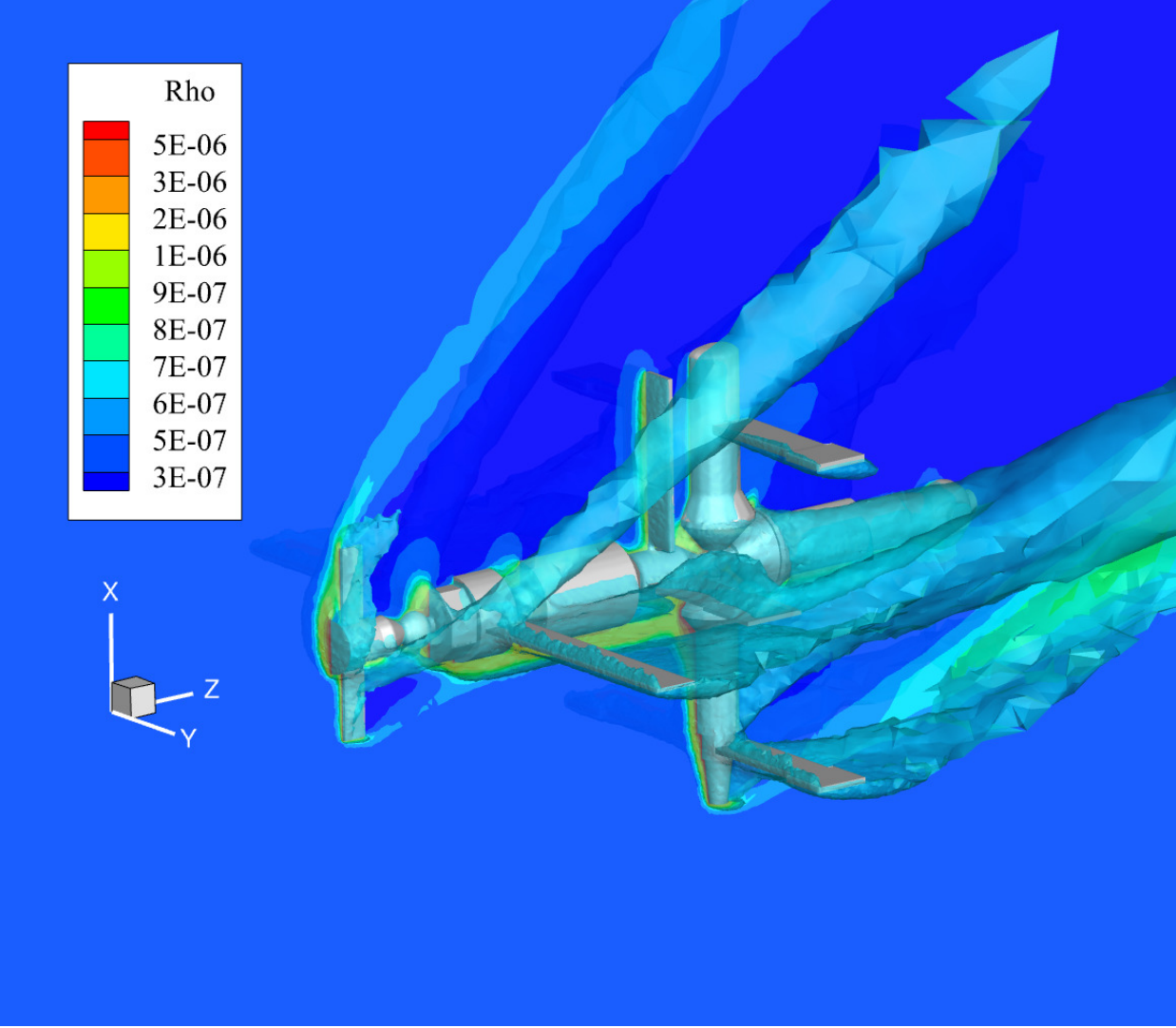}
}\\
\caption{\label{Fig:case3d_mir_ca}The (a) density contours and (b) 3D iso-surface $\rho=7 \times 10^{-7} \rm{kg/m^{3}}$ of the SS Mir.}
\end{figure*}

\begin{figure*}[h!t]
\centering
\subfigure[\label{Fig:case3d_mir_con_Ptr}]{
\includegraphics[trim=1 1 1 1, clip, width=0.45\textwidth]{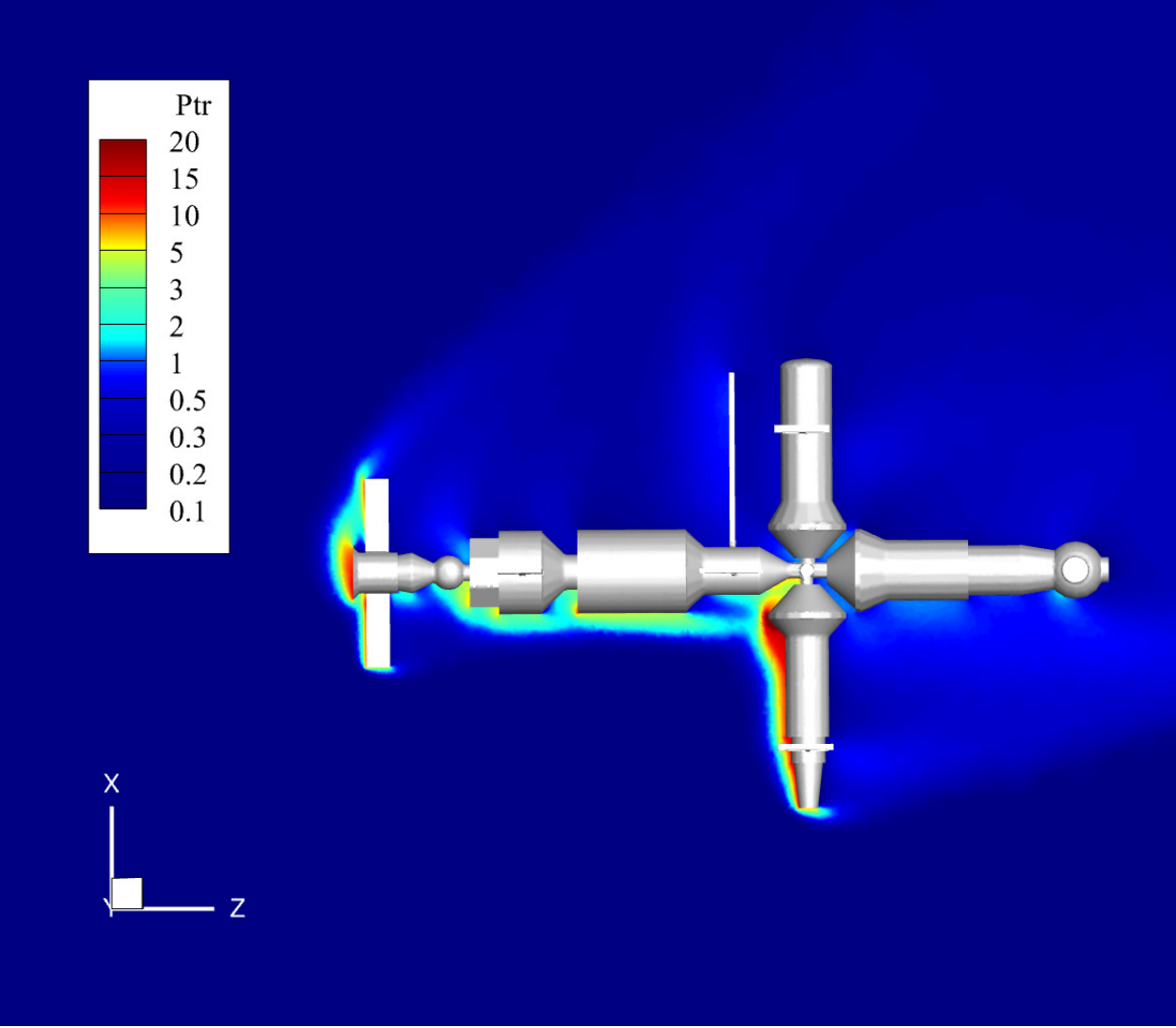}
}\hspace{0.01\textwidth}%
\subfigure[\label{Fig:case3d_mir_con_Ma}]{
\includegraphics[trim=1 1 1 1, clip, width=0.45\textwidth]{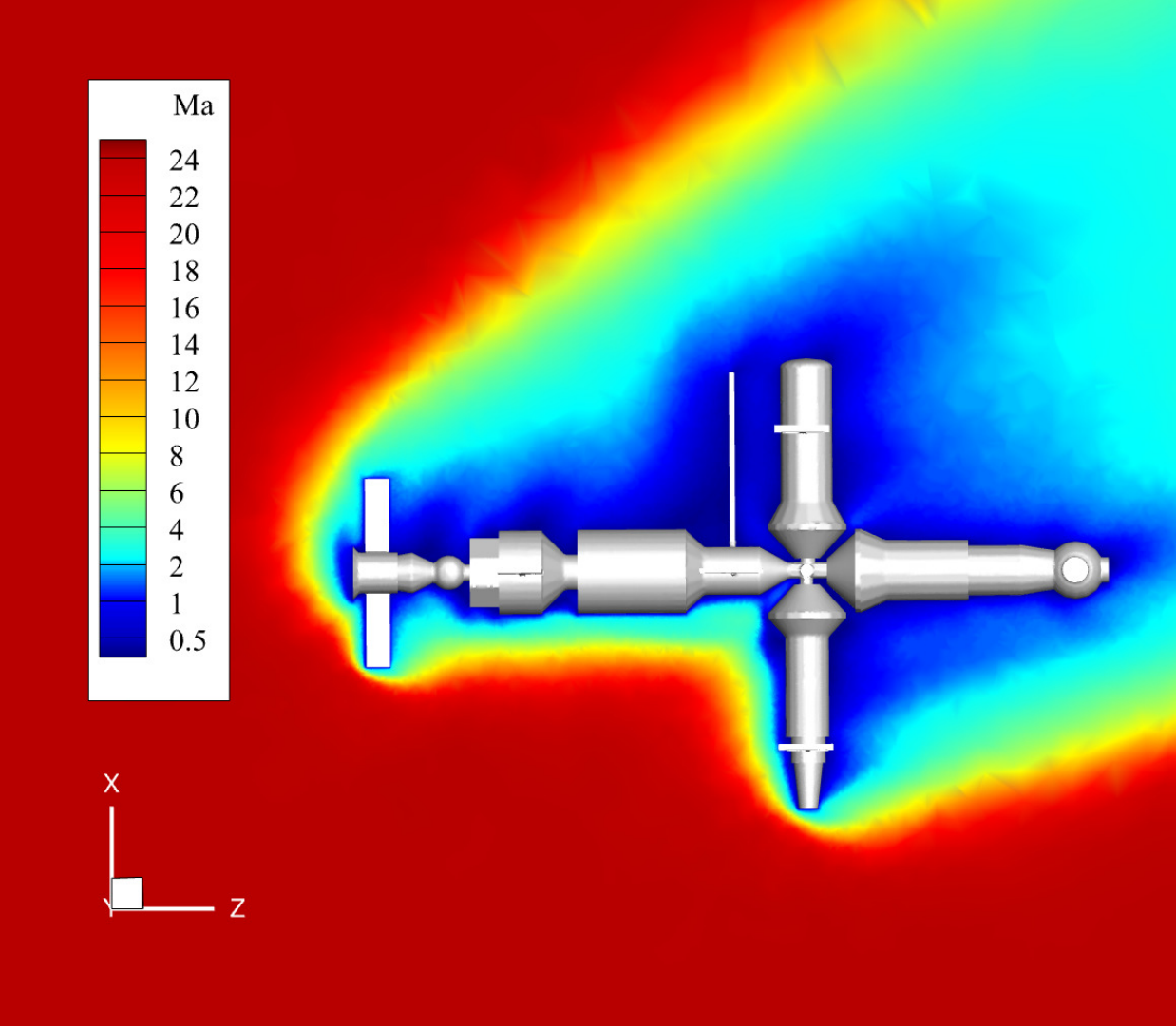}
}\\
\subfigure[\label{Fig:case3d_mir_con_Teq}]{
\includegraphics[trim=1 1 1 1, clip, width=0.45\textwidth]{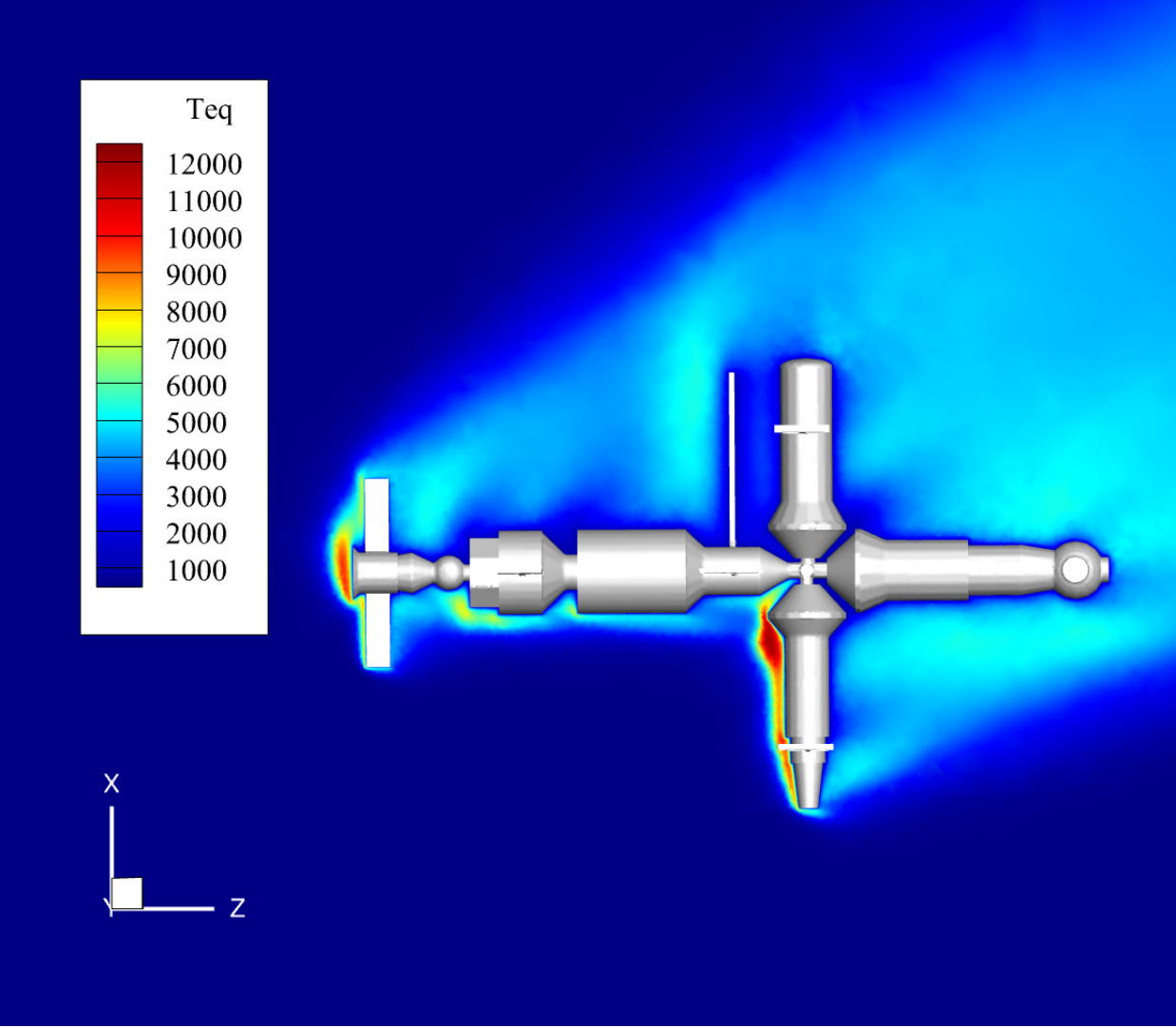}
}\hspace{0.01\textwidth}%
\subfigure[\label{Fig:case3d_mir_con_Ttr}]{
\includegraphics[trim=1 1 1 1, clip, width=0.45\textwidth]{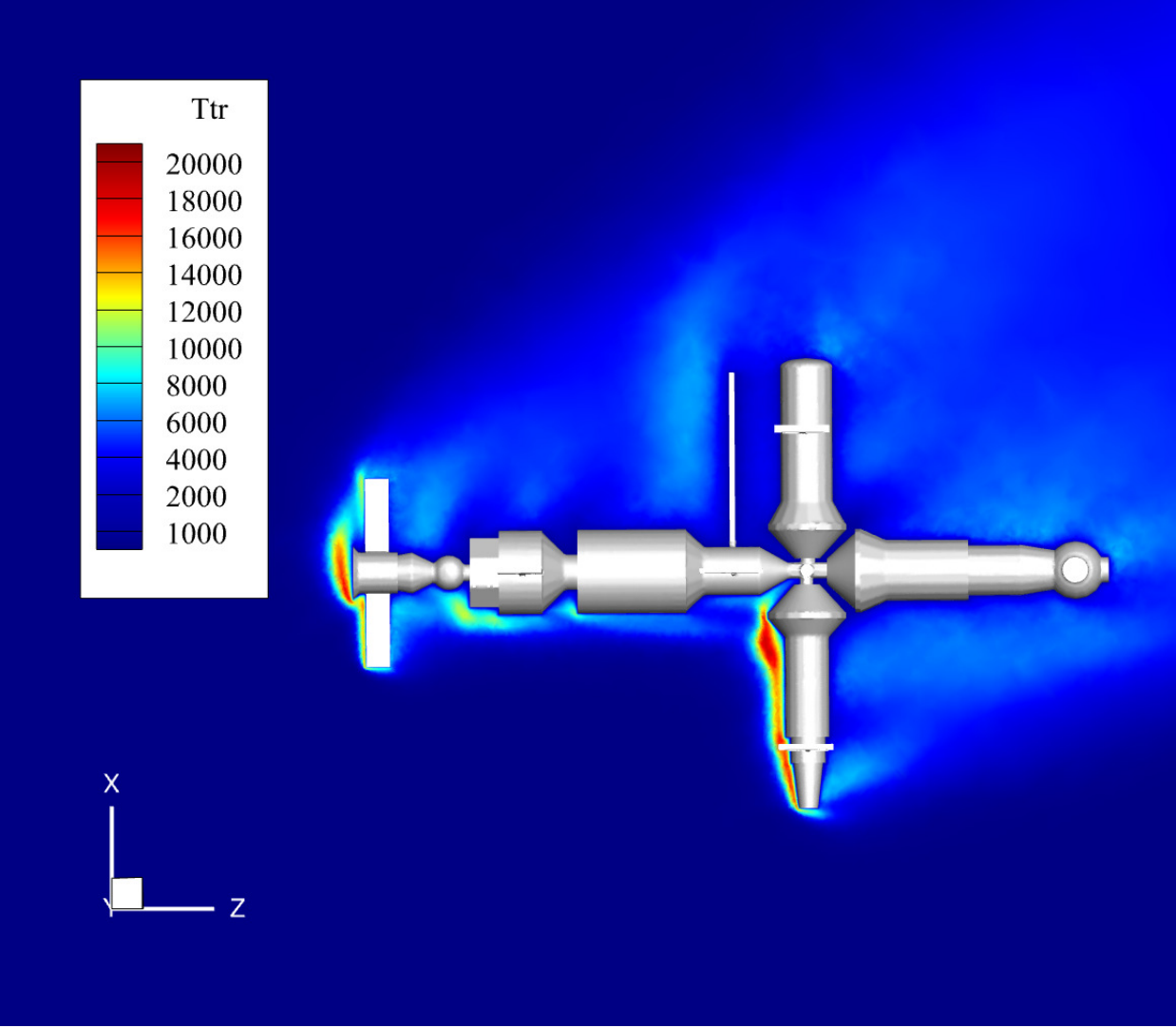}
}\\
\subfigure[\label{Fig:case3d_mir_con_Trot}]{
\includegraphics[trim=1 1 1 1, clip, width=0.45\textwidth]{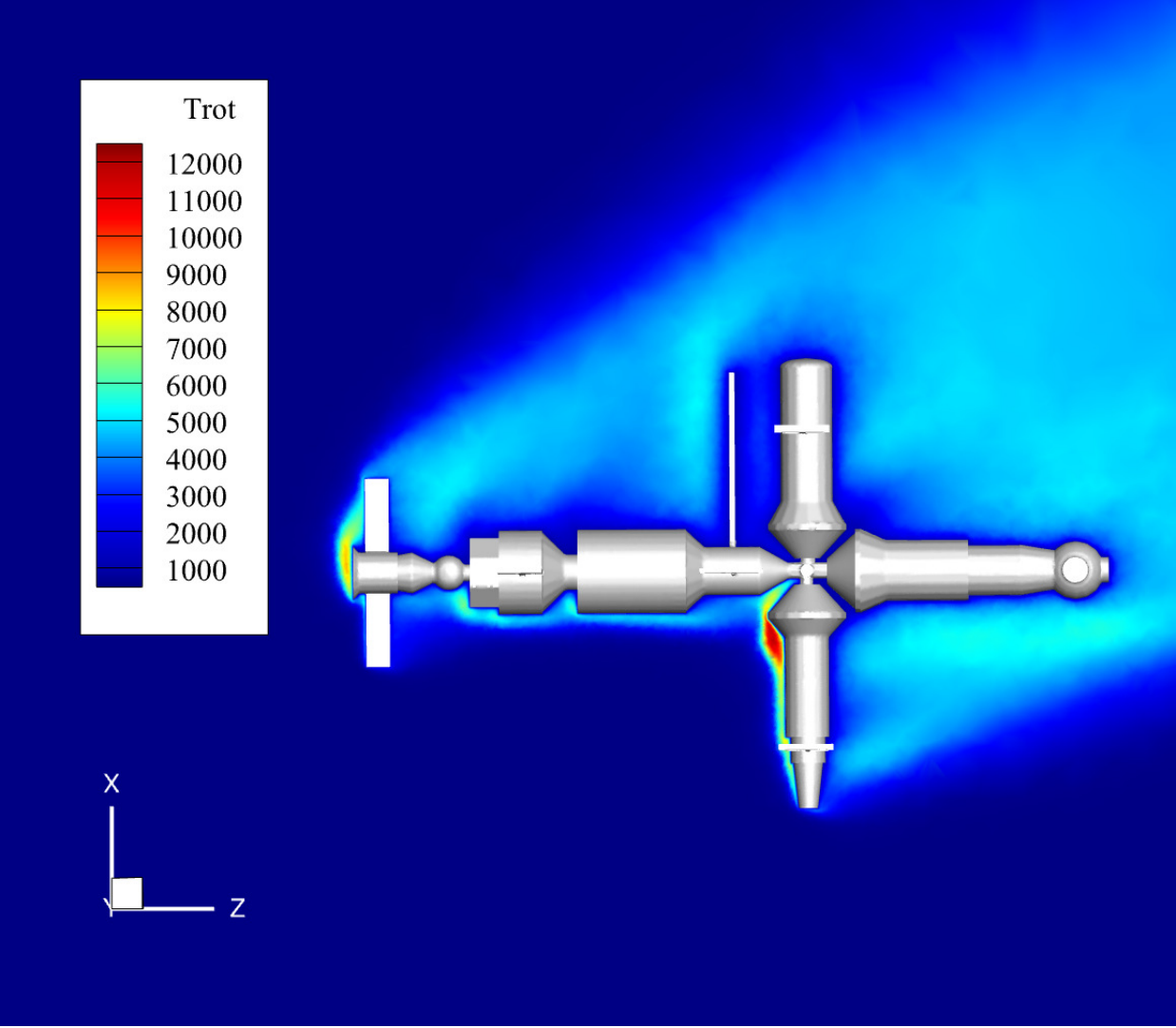}
}\hspace{0.01\textwidth}%
\subfigure[\label{Fig:case3d_mir_con_Tvib}]{
\includegraphics[trim=1 1 1 1, clip, width=0.45\textwidth]{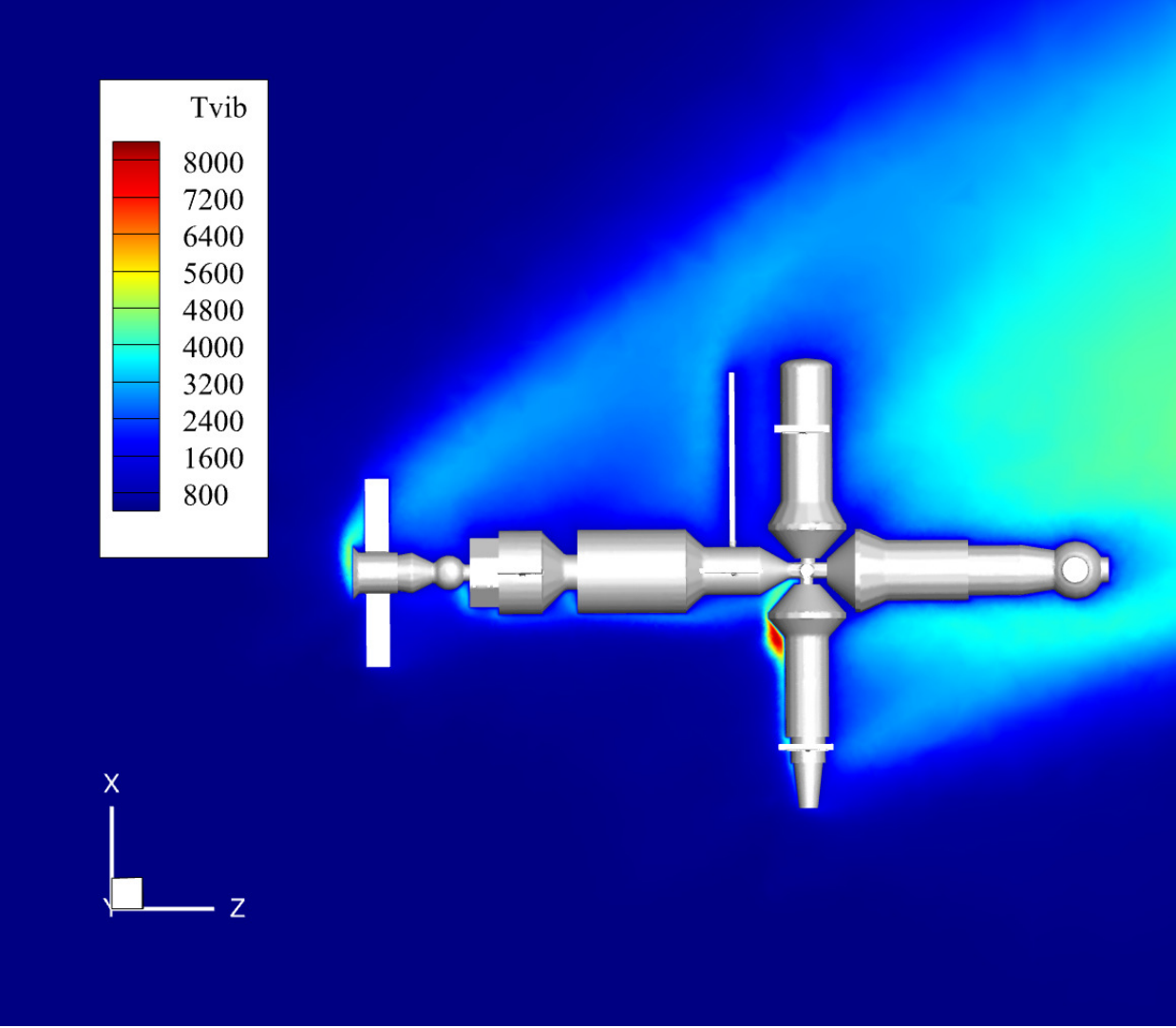}
}\\
\caption{\label{Fig:case3d_mir_con}The (a) pressure, (b) Mach number, (c) equilibrium temperature, (d) translational temperature, (e) rotational temperature, and (f) vibrational temperature contours of the SS Mir.}
\end{figure*}

The Space Station Mir (SS Mir) was assembled in orbit over several years from multiple modules. It was the first long-term habitable space research center developed by humanity, and also the first third-generation space station~\cite{markelov2001space}. In 2001, due to aging components and a lack of maintenance funding, the SS Mir re-entered the Earth's atmosphere. The atmospheric re-entry of the SS Mir is simulated to demonstrate the capability of the SUWP-vib method in handling complex-geometry vehicles. The geometry of the space station Mir is obtained from the NASA 3D Resources website~\cite{NASA3D}, and mainly consists of six modules (excluding the Priroda module and the docking compartment), as shown in Fig.~\ref{Fig:case3d_mir_shape}. As shown in Fig.~\ref{Fig:case3d_mir_mesh}, the computational mesh was generated based on a simplified geometry of the SS Mir. The surface was discretized into 141694 cells. The height of the cell adjacent wall is 0.03 m and the total cell count is 2593915. The total length of the SS Mir is approximately 31 meters. The width of the solar panels is adopted as the reference length $L_{\rm{ref}}=2$m. The simulation conditions for this case are based on previous aerodynamic studies of space stations~\cite{markelov2001space,long2024implicit} as listed in Table~\ref{table:3Dcase_mir_Table}. The particle number of cell $N_{\rm{p}} =1 \times 10^{2}$. Fig.~\ref{Fig:case3d_mir_ca} shows the density distribution and the 3D iso-surface around the SS Mir. A bow shock is observed to form in the vicinity of the station's surface. Fig.~\ref{Fig:case3d_mir_con} presents contour plots of pressure, translational temperature, rotational temperature, and vibrational temperature for the SS Mir flows. Despite the predominantly cylindrical modules and the thin solar panels, noticeable temperature increases are observed near the base block and in the after body of the station. In the vicinity of the Kvant-2 module's walls, the flow temperature exceeds 10,000 K. Under such extreme conditions, vibrational energy modes become fully excited. Compared to the case where only rotational internal energy is considered, the post-shock peak temperature is significantly reduced~\cite{bird2013dsmc}. In practical situations, the peak temperature is lower than the simulation results that consider only rotational and vibrational internal energies, due to the dissociation and ionization of gas molecules in the flow.

\begin{table}[h!t] 
    \centering
    \caption{The parameters of the SS Mir.} \label{table:3Dcase_mir_Table}
    \begin{tabular}{ c  c  c  c  c  c  c  c  c }
        \hline
        Angle of attack($^{\circ}$) & $L_{\rm{ref}}$ & Ma & $\rho_{\infty}(\rm{kg/m^{3}})$ & $U_{\infty}(\rm{m/s})$ & $T_{\infty}(\rm{K})$ & $T_{\rm{wall}}(\rm{K})$ & $Z_{\rm{rot}}$ & $Z_{\rm{vib}}$ \\ 
        \hline
        30  & 2.0m & 25 & $3.7952 \times 10^{-7}$ & 6078.01 & 142.2 & 500  & 5  & 50 \\
        \hline
    \end{tabular}
\end{table}

For the case, the time-averaging starts from 9000 steps and continues for 5000 steps with an initial field computed by 11000 steps from KIF~\cite{KIF2}. The total computation takes 14000 time steps, and runs on a workstation with AMD EPYC 7763 at 2.45GHz with 108 cores and 77.83GB memory. The total wall-clock time for the simulation is 110.89 hours, resulting in a total computational cost of 11,976.12 core-hours. By comparison, deterministic approaches usually demand over 1 TB of memory for simulating such cases~\cite{long2024implicit}. Therefore, the SUWP-vib method not only demonstrates the capability to simulate complex-shaped aerospace vehicles, but also achieves 
high computational efficiency and reduced memory requirements. Furthermore, by introducing adaptive technique and local time-stepping, the computational efficiency of the SUWP-vib method is expected to be further enhanced.

\section{\label{sec:conclusion}Conclusion}
In this paper, a simplified unified wave-particle (SUWP) method for diatomic gas with rotational and vibrational mode is constructed based on the vibrational kinetic model and a three-temperature KIF scheme. A Chapman-Enskog expansion was performed on the vibrational kinetic model. Based on this, the quantified model-competition mechanism considering rotational and vibrational non-equilibrium was derived, along with the three-temperature N-S equations including the relaxation source terms and update schemes for rotational and vibrational energies. The SUWP-vib method couples the three-temperature N-S solver and the particle solver by the QMC mechanism considering rotational and vibrational non-equilibrium. Consequently, the method can directly utilize the existing research results of N-S and DSMC solvers. This study extend the SUWP method to diatomic gases with vibrational mode, enabling it to simulate high-speed non-equilibrium problems effectively.

In the numerical tests, the Sod's shock tube problem in different Knudsen number and the normal shock structures with high Mach number are computed, and the present results agree well with the validated DUGKS and DSMC solutions. In the simulation of hypersonic flow around a circular cylinder and a blunt wedge, the thermodynamic non-equilibrium phenomena in the shock wave zone are accurately computed. Besides, the pressure, shear stress and heat flux on the solid wall are directly and accurately captured. In the Apollo 6 case, Compare with the implicit DUGKS, the SUWP method shows great advantages in its high computational efficiency due to the wave and particle representation. The CPU-hours requiring and memory consumption are reduced by 57\% and one orders of magnitude, respectively. In the flow simulation of the space station Mir, the SUWP-vib method exhibited the ability to compute complex-shaped vehicles with large mesh counts. In summary, the numerical results show that the SUWP with involving vibrational excitation has a significant improvement in numerical prediction of the high-speed non-equilibrium flows. Combined with its high computational efficiency and low memory requirements, this makes the method highly promising for engineering applications.

\appendix
\section{\label{sec:CEex}The Chapman-Enskog expansion of the vibrational kinetic model}
For simplicity, the following denotation for integrals in velocity space and substantial derivative are used
\begin{equation}
    {\mathcal{D}}=\frac{\partial}{\partial t} + {U}_{i}\frac{\partial}{\partial x_{i}}.
\end{equation}
And the translational degrees of freedom $K_{\rm{tr}}$ and rotational degrees of freedom $K_{\rm{rot}}$ are defaulted to 3 and 2 respectively.

The governing equations of the vibrational kinetic model is Eq.(\ref{eq:dfuncVib}). The forms of the distribution functions are Eq.(\ref{eq:thefuncTr}), Eq.(\ref{eq:thefuncRot}) and Eq.(\ref{eq:thefuncVib}). Introduce the formal small parameter $\varepsilon$, which indicates the order of the term before which it stands. Eq.(\ref{eq:dfuncVib}) can be rewritten as:
\begin{equation}   \label{eq:epsFuncVib}
\varepsilon \left( \mathcal{D}{{f}_{{}}}+{\boldsymbol{c}}\frac{\partial {{f}_{{}}}}{\partial {\boldsymbol{x}}} \right)=\frac{{{f}^{\text{tr}}}-{{f}_{{}}}}{\tau }+\frac{{{f}^{\text{rot}}}-{{f}^{\rm{tr}}}}{{{Z}_{\rm{rot}}}\tau }+\frac{{{f}^{\text{vib}}}-{{f}^{\rm{tr}}}}{{{Z}_{\rm{vib}}}\tau }.
\end{equation}

By taking moments of Eq.~(\ref{eq:epsFuncVib}), the conservation equations are obtained:
\begin{equation}   \label{eq:conFuncVib}
\begin{aligned}
  & \mathcal{D}\rho +\rho \frac{\partial {{U}_{i}}}{\partial {{x}_{i}}}=0, \\ 
 & \rho \mathcal{D}{{U}_{i}}+\frac{\partial {{p}_{ij}}}{\partial {{x}_{j}}}=0, \\ 
 & \rho \mathcal{D}\left( \frac{3}{2}R{{T}_{\rm{tr}}}+R{{T}_{\mathrm{rot}}}+\frac{{{K}_{\mathrm{vib}}}\left( {{T}_{\mathrm{vib}}} \right)}{2}R{{T}_{\mathrm{vib}}} \right)+\frac{\partial {{q}_{i}}}{\partial {{x}_{i}}}+{{p}_{{\rm{tr}},ij}}\frac{\partial {{U}_{i}}}{\partial {{x}_{j}}}=0, \\ 
\end{aligned}
\end{equation}
where, the energy equation can be separated into the following components:
\begin{equation}   \label{eq:energyFuncDivide}
\begin{aligned}
  & \frac{3}{2}\rho R\mathcal{D}{{T}_{\mathrm{tr}}}+\frac{\partial {{q}_{\mathrm{tr,}i}}}{\partial {{x}_{i}}}+{{p}_{{\rm{tr}},ij}}\frac{\partial {{U}_{i}}}{\partial {{x}_{j}}}={{s}_{\mathrm{tr}}}, \\ 
 & \rho R\mathcal{D}{{T}_{\mathrm{rot}}}+\frac{\partial {{q}_{\mathrm{rot,}i}}}{\partial {{x}_{i}}}={{s}_{\mathrm{rot}}}, \\ 
 & \frac{{{K}_{\mathrm{vib}}}\left( {{T}_{\mathrm{vib}}} \right)}{2}\rho R\mathcal{D}{{T}_{\mathrm{vib}}}+\frac{\partial {{q}_{\mathrm{vib,}i}}}{\partial {{x}_{i}}}={{s}_{\mathrm{vib}}}. \\ 
\end{aligned}
\end{equation}

Expanded the distribution function and the time derivative as follows
\begin{equation}   \label{eq:expansionDandF}
\begin{aligned}
  &\mathcal{D}={{\mathcal{D}}^{\left( 0 \right)}}+\varepsilon {{\mathcal{D}}^{\left( 1 \right)}}+{{\varepsilon }^{2}}{{\mathcal{D}}^{\left( 2 \right)}}+\ldots, \\ 
 & f={{f}^{\left( 0 \right)}}+\varepsilon {{f}^{\left( 1 \right)}}+{{\varepsilon }^{2}}{{f}^{\left( 2 \right)}}+\ldots .\\ 
\end{aligned}
\end{equation}

By neglecting all terms of order higher than the zeroth order in the expansion of Eq.(\ref{eq:expansionDandF}), Eq.(\ref{eq:epsFuncVib}) can be expressed as
\begin{equation}   \label{eq:epsFuncVib0}
\begin{aligned}
\varepsilon \left( \mathcal{D}f_{{}}^{\left( 0 \right)}+{{\boldsymbol{c}}_{{}}}\frac{\partial f_{{}}^{\left( 0 \right)}}{\partial \mathbf{x}} \right)=\frac{{{f}^{\rm{tr}}}-f^{\left( 0 \right)}}{\tau }+\frac{{{f}^{\rm{rot}}}-{{f}^{\rm{tr}}}}{{{Z}^{\rm{rot}}}\tau }+\frac{{{f}^{\rm{vib}}}-{{f}^{\rm{tr}}}}{{{Z}^{\rm{vib}}}\tau },\\ 
\end{aligned}
\end{equation}
where
\begin{equation}   \label{eq:funcF0}
{{f}^{\left( 0 \right)}}=n{{\left( \frac{1}{2\pi R{{T}_{\rm{tr}}}} \right)}^{\frac{3}{2}}}\exp \left( -\frac{{{\boldsymbol{c}}^{2}}}{2R{{T}_{\rm{tr}}}} \right)\frac{1}{mR{{T}_{\rm{rot}}}}{ \exp \left( {-\frac{{{\eta }_{\rm{rot}}}}{mR{{T}_{\rm{rot}}}}} \right) }\Re \left( {{T}_{\rm{vib}}} \right).
\end{equation}

By taking the moments of the distribution function $f=f^{\left(0 \right)}$, one obtains ${{p}_{{\rm{tr}},ij}}-\delta_{ij} p_{\rm{tr}}/3=0$, and ${{q}_{{\rm{tr}},i}}={{q}_{{\rm{rot}},i}}={{q}_{{\rm{vib}},i}}=0$ (The $\delta_{ij}$ is Kronecker symbol). Therefore, Eq.~(\ref{eq:conFuncVib}) can be simplified as:
\begin{equation}   \label{eq:conFuncVibSim}
\begin{aligned}
  & \mathcal{D}\rho +\rho \frac{\partial {{U}_{i}}}{\partial {{x}_{i}}}=0, \\ 
 & \rho \mathcal{D}{{U}_{i}}+\frac{\partial {{p}_{\rm{tr}}}}{\partial {{x}_{i}}}=0, \\ 
 & \rho R\mathcal{D}\left( \frac{3}{2}{{T}_{\rm{tr}}}+{{T}_{\rm{rot}}}+\frac{{{K}_{\rm{vib}}}\left( {{T}_{\rm{vib}}} \right)}{2}{{T}_{\rm{vib}}} \right)+{{p}_{\rm{tr}}}\frac{\partial {{U}_{i}}}{\partial {{x}_{i}}}=0, \\ 
\end{aligned}
\end{equation}
where, the energy equation can be separated into the following components:
\begin{equation}   \label{eq:energyFuncDivideSim}
\begin{aligned}
  & \frac{3}{2}\rho R\mathcal{D}{{T}_{\rm{tr}}}+{{p}_{\rm{tr}}}\frac{\partial {{U}_{i}}}{\partial {{x}_{i}}}=\frac{1}{\tau }\frac{3}{2}\rho R\left( \frac{{{T}_{\rm{eq}}}-{{T}_{\rm{tr}}}}{{{Z}_{\rm{vib}}}}+\frac{{{T}_{\rm{tr,rot}}}-{{T}_{\rm{tr}}}}{{{Z}_{\rm{rot}}}} \right), \\ 
 & \rho R\mathcal{D}{{T}_{\rm{rot}}}=\frac{1}{\tau }\rho R  \left( \frac{{{T}_{\rm{eq}}}-{{T}_{\rm{rot}}}}{{{Z}_{\rm{vib}}}}+\frac{{{T}_{\rm{tr,rot}}}-{{T}_{\rm{rot}}}}{{{Z}_{\rm{rot}}}} \right), \\ 
 & \frac{{{K}_{\rm{vib}}}\left( {{T}_{\rm{vib}}} \right)}{2}\rho R\mathcal{D}{{T}_{\rm{vib}}}=\frac{1}{\tau }\frac{{{K}_{\rm{vib}}}\left( {{T}_{\rm{vib}}} \right)}{2}\rho R\frac{{{K}_{\rm{vib}}}\left( {{T}_{\rm{eq}}} \right){{T}_{\rm{eq}}}-{{K}_{\rm{vib}}}\left( {{T}_{\rm{vib}}} \right){{T}_{\rm{vib}}}}{{{K}_{\rm{vib}}}\left( {{T}_{\rm{vib}}} \right){{Z}_{\rm{vib}}}}. \\ 
\end{aligned}
\end{equation}
Thus, the vibrational kinetic model recovers the Euler equations.

By neglecting all terms of order higher than the first order in the expansion of Eq.(\ref{eq:expansionDandF}), Eq.(\ref{eq:epsFuncVib}) can be expressed as
\begin{equation}   \label{eq:conFuncVibSim1}
\varepsilon \left( {{\mathcal{D}}^{\left( 0 \right)}}f_{{}}^{\left( 0 \right)}+{\boldsymbol{c}}\frac{\partial f_{{}}^{\left( 0 \right)}}{\partial \boldsymbol{x}} \right)=\frac{{{f}^{tr}}-\left( f_{{}}^{\left( 0 \right)}+f_{{}}^{\left( 1 \right)} \right)}{\tau }+\frac{{{f}^{\rm{rot}}}-{{f}^{\rm{tr}}}}{{{Z}_{\rm{rot}}}\tau }+\frac{{{f}^{\rm{vib}}}-{{f}^{\rm{tr}}}}{{{Z}_{\rm{vib}}}\tau }.
\end{equation}

The partial derivative of the distribution function $f^{\left(0 \right)}$ with respect to the macroscopic variables is given by:
\begin{equation}   \label{eq:macroPartialDerivative}
\begin{aligned}
  & \frac{\partial {{f}^{\left( 0 \right)}}}{\partial \rho }=\frac{{{f}^{\left( 0 \right)}}}{\rho }, \\ 
 & \frac{\partial {{f}^{\left( 0 \right)}}}{\partial {{U}_{i}}}=\frac{1}{R{{T}_{\rm{tr}}}}{{c}_{i}}{{f}^{\left( 0 \right)}}, \\ 
 & \frac{\partial {{f}^{\left( 0 \right)}}}{\partial {{T}_{\rm{tr}}}}=\frac{{{f}^{\left( 0 \right)}}}{{{T}_{\rm{tr}}}}\left( -\frac{3}{2}+\frac{1}{2R{{T}_{\rm{tr}}}}{{c}^{2}} \right), \\ 
 & \frac{\partial {{f}^{\left( 0 \right)}}}{\partial {{T}_{\rm{rot}}}}=\frac{{{f}^{\left( 0 \right)}}}{{{T}_{\rm{rot}}}}\left( -1+\frac{1}{mR{{T}_{\rm{rot}}}}{{\eta }_{\rm{rot}}} \right), \\ 
 & \frac{\partial {{f}^{\left( 0 \right)}}}{\partial {{T}_{\rm{vib}}}}=\frac{{{f}^{\left( 0 \right)}}}{{{T}_{\rm{vib}}}}\left( -\frac{{{K}_{\rm{vib}}}\left( {{T}_{\rm{vib}}} \right)}{2}+\frac{1}{mR{{T}_{\rm{vib}}}}{{\eta }_{\rm{vib}}} \right). \\ 
\end{aligned}
\end{equation}

Based on Eq.(\ref{eq:macroPartialDerivative}), the left-hand side of Eq.(\ref{eq:conFuncVibSim1}) can be rewritten as:
\begin{equation}   \label{eq:conFuncVibSim1LHS}
{{\mathcal{D}}^{\left( 0 \right)}}f_{{}}^{\left( 0 \right)}+{{{c}}_{{i}}}\frac{\partial f_{{}}^{\left( 0 \right)}}{\partial {x}_{i}} = \left( 
\begin{aligned}
  & \left( 
  \begin{aligned}
    &\frac{\partial {{f}^{\left( 0 \right)}}}{\partial \rho }{{\mathcal{D}}^{\left( 0 \right)}}\rho 
    +\frac{\partial {{f}^{\left( 0 \right)}}}{\partial {{U}_{i}}}{{\mathcal{D}}^{\left( 0 \right)}}{{U}_{i}} \\
    &+\frac{\partial {{f}^{\left( 0 \right)}}}{\partial {{T}_{\rm{tr}}}}{{\mathcal{D}}^{\left( 0 \right)}}{{T}_{\rm{tr}}}
    +\frac{\partial {{f}^{\left( 0 \right)}}}{\partial {{T}_{\rm{rot}}}}{{\mathcal{D}}^{\left( 0 \right)}}{{T}_{\rm{rot}}}
    +\frac{\partial {{f}^{\left( 0 \right)}}}{\partial {{T}_{\rm{vib}}}}{{\mathcal{D}}^{\left( 0 \right)}}{{T}_{\rm{vib}}} 
  \end{aligned}
  \right)\\  
 & + {{c}_{i}}\left( 
 \begin{aligned}
  &\frac{\partial {{f}^{\left( 0 \right)}}}{\partial \rho }\frac{\partial \rho }{\partial {{x}_{i}}}
  +\frac{\partial {{f}^{\left( 0 \right)}}}{\partial {{U}_{j}}}\frac{\partial {{U}_{j}}}{\partial {{x}_{i}}} \\
  &+\frac{\partial {{f}^{\left( 0 \right)}}}{\partial {{T}_{\rm{tr}}}}\frac{\partial {{T}_{\rm{tr}}}}{\partial {{x}_{i}}}
  +\frac{\partial {{f}^{\left( 0 \right)}}}{\partial {{T}_{\rm{rot}}}}\frac{\partial {{T}_{\rm{rot}}}}{\partial {{x}_{i}}}
  +\frac{\partial {{f}^{\left( 0 \right)}}}{\partial {{T}_{\rm{vib}}}}\frac{\partial {{T}_{\rm{vib}}}}{\partial {{x}_{i}}} 
 \end{aligned}
 \right) \\ 
\end{aligned} \right) = {\rm{X}}.
\end{equation}

According to Eq.(\ref{eq:conFuncVibSim}) and Eq.(\ref{eq:energyFuncDivideSim}), Eq.~(\ref{eq:conFuncVibSim1LHS}) can be simplified as:
\begin{equation}   \label{eq:conFuncVibSim1LHS_Sim}
{\rm{X}} ={{f}^{\left( 0 \right)}}\left( 
\begin{aligned}
  & \left( \begin{aligned}
  & \frac{1}{{{T}_{\rm{tr}}}}\left( -\frac{3}{2}+\frac{1}{2R{{T}_{\rm{tr}}}}{{c}^{2}} \right)\left( \frac{1}{\tau }\left( \frac{{{T}_{\rm{eq}}}-{{T}_{\rm{tr}}}}{{{Z}_{\rm{vib}}}}+\frac{{{T}_{\rm{tr,rot}}}-{{T}_{\rm{tr}}}}{{{Z}_{\rm{rot}}}} \right) \right) \\ 
 & +\frac{1}{{{T}_{\rm{rot}}}}\left( -1+\frac{{{\eta }_{\rm{rot}}}}{mR{{T}_{\rm{rot}}}} \right)\left( \frac{1}{\tau }\left(\frac{{{T}_{\rm{eq}}}-{{T}_{\rm{rot}}}}{{{Z}_{\rm{vib}}}}+\frac{{{T}_{\rm{tr,rot}}}-{{T}_{\rm{rot}}}}{{{Z}_{\rm{rot}}}} \right) \right) \\ 
 & +\frac{1}{{{T}_{\rm{vib}}}}\left( -\frac{{{K}_{\rm{vib}}}\left( {{T}_{\rm{vib}}} \right)}{2}+\frac{{{\eta }_{\rm{vib}}}}{mR{{T}_{\rm{vib}}}} \right)\left( \frac{1}{\tau }\frac{{{K}_{\rm{vib}}}\left( {{T}_{\rm{eq}}} \right){{T}_{\rm{eq}}}-{{K}_{\rm{vib}}}\left( {{T}_{\rm{vib}}} \right){{T}_{\rm{vib}}}}{{{K}_{\rm{vib}}}\left( {{T}_{\rm{vib}}} \right){{Z}_{\rm{vib}}}} \right) \\ 
\end{aligned} \right) \\ 
 & +{{c}_{i}}\left( 
 \begin{aligned}
  & \left( \frac{1}{R{{T}_{\rm{tr}}}}{{c}_{j}} \right)\left( \frac{\partial {{U}_{j}}}{\partial {{x}_{i}}}-\frac{1}{3}\frac{\partial {{U}_{k}}}{\partial {{x}_{k}}}{{\delta }_{ij}} \right) \\ 
 & +\left( \frac{1}{{{T}_{\rm{tr}}}}\left( -\frac{5}{2}+\frac{1}{2R{{T}_{\rm{tr}}}}{{c}^{2}} \right) \right)\frac{\partial {{T}_{\rm{tr}}}}{\partial {{x}_{i}}} \\ 
 & +\left( \frac{1}{{{T}_{\rm{rot}}}}\left( -1+\frac{{{\eta }_{\rm{rot}}}}{mR{{T}_{\rm{rot}}}} \right) \right)\frac{\partial {{T}_{\rm{rot}}}}{\partial {{x}_{i}}} \\ 
 & +\left( \frac{1}{{{T}_{\rm{vib}}}}\left( -\frac{{{K}_{\rm{vib}}}\left( {{T}_{\rm{vib}}} \right)}{2}+\frac{{{\eta }_{\rm{vib}}}}{mR{{T}_{\rm{vib}}}} \right) \right)\frac{\partial {{T}_{\rm{vib}}}}{\partial {{x}_{i}}} \\ 
\end{aligned} 
\right) \\ 
\end{aligned} \right).
\end{equation}

In the final results we take $\varepsilon$ = 1. Therefore, from Eq.~(\ref{eq:conFuncVibSim1}), the following can be obtained
\begin{equation}   \label{eq:finalFunc}
f_{{}}^{\left( 1 \right)}=\frac{{{f}^{\rm{tr}}}-f_{{}}^{\left( 0 \right)}}{\tau }+\frac{{{f}^{\rm{rot}}}-{{f}^{\rm{tr}}}}{{{Z}_{\rm{rot}}}\tau }+\frac{{{f}^{\rm{vib}}}-{{f}^{\rm{tr}}}}{{{Z}_{\rm{vib}}}\tau }-{\rm{X}}.
\end{equation}

By taking moments of the distribution function $f={{f}^{\left( 0 \right)}}+{{f}^{\left( 1 \right)}}$, the following results are derived:
\begin{equation}   \label{eq:finalMoment}
\begin{aligned}
 & {{p}_{{\rm{tr}},ij}}=2\mu \frac{\partial {{U}_{i}}}{\partial {{x}_{j}}}=2\tau {{p}_{\rm{tr}}}\frac{\partial {{U}_{i}}}{\partial {{x}_{j}}}, \\ 
 & {{q}_{i}}={{q}_{{\rm{tr}},i}}+{{q}_{{\rm{rot}},i}}+{{q}_{{\rm{vib}},i}}, \\ 
 & {{q}_{{\rm{tr}},i}}={{\lambda }_{\rm{tr}}}\frac{\partial {{T}_{{\rm{tr}}}}}{\partial {{x}_{i}}}, \\ 
 & {{q}_{{\rm{rot}},i}}={{\lambda }_{\rm{rot}}}\frac{\partial {{T}_{\rm{rot}}}}{\partial {{x}_{i}}},\\
 & {{q}_{{\rm{vib}},i}}={{\lambda }_{\rm{vib}}}\frac{\partial {{T}_{\rm{vib}}}}{\partial {{x}_{i}}}, \\ 
\end{aligned}
\end{equation}
where,
\begin{equation}   \label{eq:finalMomentPara}
\begin{aligned}
 & {{\lambda }_{\rm{tr}}}=\mu R\frac{3}{2}{{B}_{\rm{tr}}}, \, {{\lambda }_{\rm{rot}}}=\mu R {{B}_{\rm{rot}}}, \, {{\lambda }_{\rm{vib}}}=\mu R\frac{{{K}_{\rm{vib}}}\left( {{T}_{vib}} \right)}{2}{{B}_{\rm{vib}}}, \\ 
 & {{B}_{\rm{tr}}}=\frac{5}{3}\frac{1}{{{A}_{\rm{tt}}}}, \, {{B}_{\rm{rot}}}=\frac{1}{{{A}_{\rm{rr}}}}, \, {{B}_{\rm{vib}}}=\frac{1}{{{A}_{\rm{vv}}}}, \\ 
 & {{A}_{\rm{tt}}}=\frac{2}{3}+\frac{1}{3{{Z}_{\rm{rot}}}}\left( 1-{{\omega }_{0}} \right)+\frac{1}{3{{Z}_{\rm{vib}}}}\left( 1-{{\omega }_{2}} \right), \\ 
 & {{A}_{\rm{rr}}}=\delta +\frac{1}{{{Z}_{\rm{rot}}}}\left( 1-{{\omega }_{1}} \right)\left( 1-\delta  \right)+\frac{1}{{{Z}_{\rm{vib}}}}\left( 1-{{\omega }_{3}} \right)\left( 1-\delta  \right), \\ 
 & {{A}_{\rm{vv}}}=1. \\ 
\end{aligned}
\end{equation}

Eq.(\ref{eq:finalMomentPara}) indicates that the higher-order variables including stress and heat flux are nonzero. By substitution of the derived constitutive relations into Eq. (\ref{eq:conFuncVib}), the vibrational kinetic model recovers the N-S equations.

\section*{Acknowledgments}
The authors thank Prof. Kun XU at Hong Kong University of Science and Technology for discussions of the wave-particle method. Sirui YANG thanks Dr. Rui ZHANG at Northwestern Polytechnical University for discussions of the vibrational kinetic model, and Hao JIN for providing DSMC data. This work is supported by the National Natural Science Foundation of China (No. 12172301) and the the Program of Introducing Talents of Discipline to Universities (No. B17037).

\section*{Data Availability}
The data that support the findings of this study are available from the corresponding author upon reasonable request.
\end{CJK*}
\clearpage




\bibliographystyle{elsarticle-num} 
\bibliography{REF_SUWP_Vib}
\clearpage

\end{document}